\documentclass[backref, breaklinks, onecolumn, colorlinks]{aastex701}   
\hypersetup{linkcolor=blue,citecolor=blue,filecolor=cyan,urlcolor=magenta}

\usepackage{enumitem}
\usepackage{mathtools}
\usepackage{graphicx}
\usepackage{amssymb}
\usepackage{amsmath}
\usepackage{ulem}
\usepackage{epstopdf}
\usepackage{color}
\usepackage{xcolor}
\usepackage{needspace}
\usepackage{array}

\def\lambar{\lambda\llap {--}}

\def\dover#1#2{\hbox{${{\displaystyle#1 \vphantom{(} }\over{
    \displaystyle #2 \vphantom{(} }}$}}
{\catcode`\@=11
\gdef\SchlangeUnter#1#2{\lower2pt\vbox{\baselineskip 0pt\lineskip0pt
\ialign{$\m@th#1\hfil##\hfil$\crcr#2\crcr\sim\crcr}}}}
\def\gtrsim{\mathrel{\mathpalette\SchlangeUnter>}}
\def\lesssim{\mathrel{\mathpalette\SchlangeUnter<}}

\newsavebox{\TopHeatmapBox}
\newlength{\TopHeatmapHeight}

\newcommand{\TopCbarPanel}[2]{%
  \begin{minipage}[t]{0.075\linewidth}
    \vspace{0pt}%
    \centering
    \makebox[\linewidth][r]{\small\strut}\par
    \vspace{0.25ex}%
    \makebox[\linewidth][r]{%
      \includegraphics[height=#2,keepaspectratio]{#1}%
    }%
  \end{minipage}%
}

\newcommand{\TopPanel}[2]{%
  \begin{minipage}[t]{0.175\linewidth}
    \vspace{0pt}%
    \centering
    \makebox[\linewidth][c]{\small\strut #1}\par
    \vspace{0.25ex}%
    \includegraphics[width=\linewidth]{#2}%
  \end{minipage}%
}

\newcommand{\TopPanelBlank}[1]{%
  \begin{minipage}[t]{0.21\linewidth}
    \vspace{-8mm}%
    \centering
    \makebox[\linewidth][c]{\small\strut}\par
    \vspace{0.25ex}%
    \includegraphics[width=\linewidth]{#1}%
  \end{minipage}%
}

\newcommand{\HalfPanelT}[1]{%
  \begin{minipage}[t]{0.5\textwidth}
    \vspace{0pt}%
    \centering
    \includegraphics[width=\linewidth]{#1}%
  \end{minipage}%
}

\newcommand{\PolBlock}[4]{%
  \begin{minipage}[t]{0.5\textwidth}
    \vspace{0pt}%
    \centering
    \makebox[\linewidth][c]{%
      \begin{minipage}[t]{0.5\linewidth}
        \vspace{0pt}%
        \centering
        \includegraphics[width=\linewidth]{#1}%
      \end{minipage}%
      \begin{minipage}[t]{0.5\linewidth}
        \vspace{0pt}%
        \centering
        \includegraphics[width=\linewidth]{#2}%
      \end{minipage}%
    }\par
    \makebox[\linewidth][c]{%
      \begin{minipage}[t]{0.5\linewidth}
        \vspace{0pt}%
        \centering
        \includegraphics[width=\linewidth]{#3}%
      \end{minipage}%
      \begin{minipage}[t]{0.5\linewidth}
        \vspace{0pt}%
        \centering
        \includegraphics[width=\linewidth]{#4}%
      \end{minipage}%
    }%
  \end{minipage}%
}

\newsavebox{\BottomLeftBox}
\newsavebox{\BottomRightBox}
\newlength{\BottomRowHeight}

\newcommand{\BottomCenteredPair}[2]{%
  \sbox{\BottomLeftBox}{\includegraphics[width=0.49\textwidth]{#1}}%
  \sbox{\BottomRightBox}{\includegraphics[width=0.49\textwidth]{#2}}%
  \setlength{\BottomRowHeight}{\dimexpr\ht\BottomLeftBox+\dp\BottomLeftBox\relax}%
  \ifdim\dimexpr\ht\BottomRightBox+\dp\BottomRightBox\relax>\BottomRowHeight
    \setlength{\BottomRowHeight}{\dimexpr\ht\BottomRightBox+\dp\BottomRightBox\relax}%
  \fi
  \makebox[\linewidth][c]{%
    \begin{minipage}[c][\BottomRowHeight][c]{0.49\textwidth}
      \centering
      \includegraphics[width=\linewidth]{#1}%
    \end{minipage}%
    \hspace{0.02\textwidth}%
    \begin{minipage}[c][\BottomRowHeight][c]{0.49\textwidth}
      \centering
      \includegraphics[width=\linewidth]{#2}%
    \end{minipage}%
  }%
}

\newcommand{\ResultsFigure}[4]{%
\begin{figure*}[t]
\centering

\sbox{\TopHeatmapBox}{%
  \includegraphics[width=0.18\linewidth]{#1/heatmap_4keV_#2.png}%
}%
\setlength{\TopHeatmapHeight}{\dimexpr\ht\TopHeatmapBox+\dp\TopHeatmapBox\relax}%

\makebox[\linewidth][c]{%
  \TopCbarPanel{#1/heatcbar.pdf}{\TopHeatmapHeight}%
  \TopPanel{4 keV}{#1/heatmap_4keV_#2.png}%
  \TopPanel{12 keV}{#1/heatmap_12keV_#2.png}%
  \TopPanel{36 keV}{#1/heatmap_36keV_#2.png}%
  \TopPanel{108 keV}{#1/heatmap_108keV_#2.png}%
  \TopPanelBlank{#1/parameters_table_#2.png}%
}

\par\vspace{0.4em}

\makebox[\linewidth][c]{%
  \HalfPanelT{#1/StokesI_impulse_response_#2.png}%
  \hspace{0.02\textwidth}%
  \PolBlock
    {#1/QoverI_heatmap_VB_off_#2.png}%
    {#1/UoverI_heatmap_VB_off_#2.png}%
    {#1/QoverI_heatmap_VB_on_#2.png}%
    {#1/UoverI_heatmap_VB_on_#2.png}%
}

\par\vspace{0.4em}

\BottomCenteredPair{#1/QvsU_#2.png}{#1/Spectrum_with_PD_#2.png}

\caption{#4}
\label{#3}
\end{figure*}%
}

\begin{document}

\title{Magnetar Fireballs and Short Bursts: Curved Spacetime Lensing, QED Effects, Spectra, Polarization, and Impulse Responses}

\author[0000-0002-9249-0515]{Zorawar Wadiasingh}
\affiliation{Department of Astronomy, University of Maryland, College Park, MD 20742, USA}
\affiliation{Astrophysics Science Division, NASA Goddard Space Flight Center, Greenbelt, MD 20771, USA}
\affiliation{Center for Research and Exploration in Space Science and Technology, NASA/GSFC, Greenbelt, MD 20771, USA}
\email[show]{zorawar@umd.edu}
\email[show]{zorawar.wadiasingh@nasa.gov}

\author[0000-0001-9268-5577]{{Hoa} {Dinh Thi}}
\affil{Department of Physics and Astronomy - MS 108, Rice University,
6100 Main Street, Houston, TX 77251-1892, USA}
\email{}

\author[0000-0003-1080-5286]{Constantinos Kalapotharakos}
\affiliation{Astrophysics Science Division, NASA Goddard Space Flight Center, Greenbelt, MD 20771, USA}
\email{c.kalapotharakos@gmail.com}

\author[0000-0002-9705-7948]{Kun Hu}
\affil{Department of Physics, McDonnell Center for the Space Sciences, Washington University in St.
Louis, 1 Brookings Dr, Saint Louis, MO 63130, USA}
\email{hkun@wustl.edu}

\author[0000-0003-4433-1365]{Matthew G. Baring}
\affil{Department of Physics and Astronomy - MS 108, Rice University,
6100 Main Street, Houston, TX 77251-1892, USA}
\email{baring@rice.edu}

\author[0000-0001-6119-859X]{Alice K. Harding}
\affiliation{Theoretical Division, Los Alamos National Laboratory, Los Alamos, NM 87545, USA}
\email{ahardingx@yahoo.com}

\author[0000-0002-7991-028X]{George Younes}
\affiliation{Astrophysics Science Division, NASA Goddard Space Flight Center, Greenbelt, MD 20771, USA}
\affiliation{Center for Space Sciences and Technology, University of Maryland, Baltimore County, Baltimore, MD 21250, USA}
\email{gyounes@umbc.edu}

\author[0000-0002-6449-106X]{Sebastien~Guillot}
\affil{University of Toulouse, CNES, CNRS, IRAP, Toulouse, France}
\email{sebastien.guillot@irap.omp.eu}

\author[0000-0002-0118-2649]{Andrea Sanna}
\affiliation{Universit\`a degli Studi di Cagliari, SP Monserrato-Sestu, KM 0.7, Monserrato, 09042, Italy}
\email{andrea.sanna@dsf.unica.it}

\author[0000-0002-6548-5622]{Michela Negro}
\affiliation{Department of Physics \& Astronomy, Louisiana State University, Baton Rouge, LA 70803, USA}
\email{michelanegro@lsu.edu}

\author[0000-0002-2942-8399]{Jeremy D. Schnittman}
\affiliation{Astrophysics Science Division, NASA Goddard Space Flight Center, Greenbelt, MD 20771, USA}
\email{jeremy.d.schnittman@nasa.gov}

\author[0000-0002-7150-9061]{Oliver J. Roberts}
\affiliation{Physics, School of Natural Sciences, University Road, University of Galway, Galway, H91 TK33, Ireland}
\email{oliver.roberts@universityofgalway.ie}

\author[0000-0002-2942-3379]{Eric Burns}
\affiliation{Department of Physics \& Astronomy, Louisiana State University, Baton Rouge, LA 70803, USA}
\email{ericburns@lsu.edu}

\author[0000-0001-8551-2002]{Chin-Ping Hu}
\affiliation{Department of Physics, National Changhua University of Education, Changhua 500207, Taiwan}
\email{cphu0821@gm.ncue.edu.tw}

\author[0000-0002-5274-6790]{Ersin G\"{o}\u{g}\"{u}\c{s}}
\affiliation{Sabanc{\i} University, Faculty of Engineering and Natural Sciences, 34956, Istanbul, T\"{u}rkiye}
\email{ersing@sabanciuniv.edu}

\begin{abstract}
Magnetar short bursts (SBs) are hard X-ray transients of durations $0.01-1$~s peaking at $\sim 10-100$~keV, and are prime targets for new high-energy missions and polarimeters. The recent association of SBs with bright radio bursts in SGR~1935+2154 has broadened interest in SB physics. We present new advanced fireball models combining general relativistic light bending, polarized transport in magnetized photospheres, magnetic photon splitting attenuation, and magnetospheric vacuum birefringence. These models also have relevance to trapped fireballs in magnetar giant flare pulsating tails. We adopt confined flux tube geometries consistent with adiabatic fireballs, and anisotropic/polarized emergent intensities to produce spectra and polarizations, and energy-time Stokes impulse responses. We predict that most fireballs are highly linearly polarized, especially when vacuum birefringence is important. There is rich potential for diagnostics: coexisting direct and lensed delayed images, gaps by occultation of the neutron star surface, and Shapiro+R{\o}mer delay with temporal caustics. These effects can imprint spin phase dependence of the spectral and polarization character of bursts. Predicted signatures depend strongly on viewing geometry, fireball configuration, and photon splitting assumptions, yielding large variance in model high-energy spectral shapes and cutoffs, and energy-dependent polarization. The models can reproduce established double-blackbody SB spectral phenomenology, and we find that the unusual April 2020 radio-associated SB from SGR~1935+2154 is broadly consistent with a footpoint close to the magnetic pole, and possibly near pole-on viewing geometry. Our models motivate reverberation-style analyses for SBs and suggest that high-quality data might constrain source geometry, burst crustal footpoints, and, potentially, neutron star masses and radii. 
\end{abstract}


\section{Introduction}

\label{sec:intro}

Magnetars, the most highly magnetized neutron stars (NSs), emit across the radio, X-ray and soft gamma-ray bands. They exhibit luminous, variable quasi-thermal surface emission, persistent magnetically-powered non-thermal hard X-rays, and impulsive bursts spanning decades in fluence and peak luminosity \citep{2008A&ARv..15..225M,2024FrASS..1188953N}.  Among their defining high-energy manifestations are (i) rare, extreme, spectrally hard giant flares reaching $\gtrsim 10^{47}\,$erg~s$^{-1}$ peaking at $\gtrsim$~MeV energies \citep[e.g.,][]{2005Natur.434.1107P} that comprise a prompt spike and pulsating emission of hundreds of seconds, and (ii) the far more numerous quasi-thermal short bursts \citep[SBs,][]{1979SvAL....5..343M,1987ApJ...320L.105A,1986Natur.322..152L,1987ApJ...320L.111L,1993Natur.362..728K,1994Natur.368..125K}  peaking below $100$ keV with observed durations as short as microseconds \citep[][]{2025arXiv251212291C} to seconds\footnote{The microsecond bursts reported in \citet{2025arXiv251212291C} are shorter than a flux tube light crossing time, precluding plasma thermalization. Thus they are non-thermal, likely related to pair cascades or reconnection in a very localized region.}, typically $ \sim100$~ms. Three giant flares have been observed in the local Galactic neighborhood \citep{1979Natur.282..587M,1999Natur.397...41H,2005Natur.434.1107P} and are now recognized as an emerging channel for extragalactic short gamma-ray transients \citep{2021Natur.589..207R,2021ApJ...907L..28B,2024Natur.629...58M,2024A&A...687A.173T,2025A&A...694A.323T,2025ApJ...979L..25R,2025ApJ...980..211B}.  In contrast, SBs often occur in ``storms" with up to thousands transpiring in a few hours \citep[e.g.,][]{2015ApJS..218...11C,2020ApJ...902L..43L,2022ApJS..260...24C,2026ApJ...997..272P,2025arXiv251212291C} and have repeatedly enabled discovery and association of new Galactic magnetars \citep[e.g.,][]{2013ApJ...770L..24K}.  SBs span power-law distributions in isotropic-equivalent energy and luminosity, from the weakest ones at $10^{36}$~erg to rarer events\footnote{Intermediate flares have also been observed from a few magnetars \citep[e.g.,][]{1999ApJ...519L.151M,2001ApJ...558..237I,2004ApJ...616.1148O,2004A&A...416..297G,2009ApJ...696L..74M,2011ApJ...740...55G,2016MNRAS.460.2008K,2025ApJ...989...63H}. The pulsed tails attendant with giant flares and intermediate bursts, at least in some aspects could also be described as a confined and evaporating fireball \citep{2001ApJ...549.1021F} possibly with radiatively-driven outflows \citep[e.g.,][]{2016MNRAS.461..877V}. These fireball-like pulsating tails in the nearby universe may be observable by next-generation X-ray facilities \citep[e.g.,][]{2025OJAp....8E.159N}.}  attaining $10^{42}$~erg \citep[e.g.,][]{1996Natur.382..518C,1999ApJ...526L..93G,2000ApJ...532L.121G,2004ApJ...607..959G}. The prevalence, brightness, and diversity of SBs render them palpable probes of magnetar environments, radiative transfer in ultrastrong fields, and the coupling between the crust and the magnetosphere \citep{2023ApJ...947L..16L}.

\subsection{Short Bursts as Trapped Fireballs}

Evidence argues that magnetar SBs are fireballs confined in magnetic flux tubes (see below), most within the magnetar's ``closed-zone" magnetosphere. The observed burst durations are typically $10^3-10^5$ NS light crossing times, with plasma densities high enough to quickly thermalize via Coulomb and magnetic Compton processes. The bursts are Compton thick (with large photon diffusion timescales) but at low enough luminosities and energetics that the photon gas and plasma remain magnetically trapped\footnote{Considering $a T^4 \lesssim {\cal B}^2/(8\pi)$, this implies $k_{\rm B} T \lesssim 54 {\cal B}_{14}^{1/2} \varrho_1^{-3/2}$~keV where $\varrho = r/r_\star$, ${\cal B}_{14}$ the surface field in units of $10^{14}$~G, $a$ the radiation constant, and $\varrho_1 = \varrho/10$, see \S\ref{sec:tempspec}.}, rather than combing out field lines. The Compton y-parameter is much greater than unity. A high electron scattering optical depth $\tau_{\rm es} \gg 10^3$ is required by the typical burst duration of $T_{\rm 90} \sim 0.1-1 {\rm\,\, s} \sim t_{\rm evap} \sim \tau_{\rm es} r_\star/c$ for $r_\star/c \sim 10^{-4}$~s, where $r_\star$ is the NS radius, requiring plasma densities $n_e \gtrsim 10^{21}$~cm$^{-3}$ in the magnetic Thomson regime. Here, $T_{\rm 90}$ is the usual observed $90\%$-containment burst duration \citep{1993ApJ...413L.101K}, $t_{\rm evap}$ denotes the characteristic fireball evaporation
(or photon diffusion) timescale, and $c$ the speed of light. The low-altitude, confined picture now has strong observational support and agrees with earlier theoretical arguments \citep{1982AIPC...77..249L,1982ApJ...260..371K,1984NYASA.422..237L,1984Natur.310..121L,1994ApJ...432..742F,1995ApJ...448L..29M,1995MNRAS.275..255T,2001ApJ...561..980T}. The flexible flux tube geometry adopted here is therefore a physically motivated choice.

 \enlargethispage{-4\baselineskip}
 
Observationally, magnetar SB spectra are typically quasi-thermal and are regularly described by double blackbody (BB) models \citep[e.g.,][]{2004ApJ...612..408F,2008ApJ...685.1114I,2010ApJ...716...97K,2012ApJ...756...54L,2011ApJ...739...87L,2012ApJ...749..122V,2014ApJ...785...52Y,2016MNRAS.460.2008K,2017ApJS..232...17K,2023ApJ...950..121R,2024ApJ...969...38R,2024ApJ...965..130K,2025ApJ...988..282D,2025ApJS..276...60R,2026A&A...707A.289T} in time integrated or resolved spectral analyses.  Empirically, these fits often yield characteristic temperatures $k_{\rm B} T_{\rm cool}\sim 3$--$6\,$keV and $k_{\rm B} T_{\rm hot}\sim 10$--$50\,$keV, with apparent emission areas of order $[(0.3$--$5)r_\star]^2$ for the cooler component and $[(0.03$--$0.3)r_\star]^2$ for the hotter component (for $r_\star\approx 12\,$km), and with comparable energy flux in both.  The hotter component is naturally associated with compact, high-temperature regions plausibly anchored near magnetic footpoints, while the cooler component suggests emission from a larger, lower-temperature photospheric surface.  The observed near-equipartition of flux between the two BB components is difficult to reconcile with unrelated emission zones and instead argues for physical coupling of a multicolor BB, whose simple phenomenological description is two BBs.
 
SB durations are much shorter than typical magnetar spin periods, and SBs appear to be randomly distributed with rotational phase, with occasional exceptions \citep[for a summary of searches circa-2018, see][]{2018MNRAS.476.1271E}. In SGR~1935+2154, during the April 2020 storm, time-resolved spectroscopy with phase information suggests that a compact hot component can be suppressed at specific rotational phases, consistent with self-shadowing or obscuration of low-altitude emission by higher-altitude, cool optically thick plasma \citep{2021ApJ...916L...7K}. These observations are thus at odds with magnetic reconnection or solar flare-like twisting events, which can trigger heating and dissipation at high altitudes.  The plasma in the SB emission zone is dense and optically thick, and the radiation is dynamically important for determining plasma motions. Large amplitude low-frequency Alfv\'en-like disturbances from the crust therefore likely cannot propagate far, because the background plasma  (sourced by single-photon pair cascades) prior to the SB may not support such waves triggering explosive pair creation \citep{Wadiasingh2020}, nor within SB plasma clouds due to strong damping.

\citet{2022ApJ...924..136Y} reported similarly strong evidence of phase dependence of SBs in NICER observations of SGR~1830--0645. The bursts are phase-aligned with surface thermal emission, implying very low altitude and possible association with surface hot spots. These SBs are quasi-thermal with apparent area of $\sim 25$~km$^{2}$, i.e., $\sim5-10\%$ of the apparent NS surface area. Likewise, for SGR~1935+2154, \citet{2025ApJ...989...63H} find that phase-folded burst hardness follows persistent pulsed hardness. The correlation is absent below 10~keV, again placing the hottest burst regions close to the surface. Recent time-resolved studies also highlight that SBs exhibit structured spectral evolution on sub-second timescales to hour-scale variations \citep[][]{2025ApJ...989...63H}, including correlations between flux and inferred BB areas, and in some cases, indications of short-timescale spectral oscillations \citep{2023ApJ...956L..27R}.

The appropriate description, therefore, couples magnetospheric geometry to radiative transport in strongly magnetized, scattering-dominated atmospheres, with gravitational redshift and light bending shaping the observer signal.  In this regime, quantum electrodynamical (QED) processes such as photon splitting can selectively attenuate high-energy photons depending on propagation angle relative to the magnetic field vector ${\boldsymbol{B}}$ and polarization mode, introducing observer-dependent spectral softening that is inherently geometric and phase dependent \citep[e.g.,][]{2019MNRAS.486.3327H}. 

Because general relativistic (GR) light bending makes emission from otherwise occulted regions visible, it is generally difficult to suppress bursts that occur during ``off" rotational phases \citep{2018MNRAS.476.1271E}. As we show in this work, even emission from a flux tube with a height below one stellar radius can remain observable. Such phase dependence argues that (at least in some episodes) the emitting volumes are compact and located at very low altitudes, potentially near the surface, and that emission anisotropy with respect to the local field from magnetized photon transport is involved.  These observations motivate models in which: (i) emitting structures occupy fixed or slowly drifting surface/magnetospheric locales over burst storms, and (ii) occultation and lensing are treated self-consistently.

\subsection{Astrophysical Relevance and Physics of Short Bursts}

The astrophysical importance of SBs is multifaceted.  First, as SBs are quasi-thermal in nature, they are roughly calorimetric and can therefore provide an order-of-magnitude proxy to the energy injection physics.\footnote{This energy injection is a complex multi-step process ultimately resulting in the fireball: impulsive motion of the crust injecting energy into the magnetosphere, followed by generation of electric fields and single-photon pair cascades, scattering and Comptonization, pair annihilation and radiative cooling of the fireball \citep{1984AIPC..115..615H,1986A&A...170..179W,1986ApJ...300..167H,1988MNRAS.235...51B,1988MNRAS.235...79B}. Pair annihilation (single or two photon) may sustain the fireball somewhat.  Yet annihilation line spectral features are expected to be weak to non-existent \citep[e.g.,][]{1992NASCP3137..245B}, since the pair density and scattering optical depth are extremely large and destructive for line formation.}  This helps constrain the energetics of magnetar burst triggers, for instance from impulsive magneto-elastic crust yielding or quakes \citep[e.g.,][]{2025ApJ...985...45K}. Since crustal shear velocities occupy a narrow range of values ($\sim 10^8$~cm/s), the energy injection and therefore observed burst energetics directly couple to the amplitude of crust perturbations and involved surface area \citep[e.g.,][]{Wadiasingh2020,2025ApJ...980..211B}. Burst rate, energetics, and clustering encode stress accumulation/release statistics and can be compared with crustal failure and avalanche models \citep[e.g.,][]{1999ApJ...512L.113P,2017ApJ...841...54T,2023ApJ...947L..16L}. An accurate physical model and deeper understanding may even enable better independent distance constraints to magnetars from spectra and temporal signatures which imprint a size-scale, which must be self-consistent with the luminosity.

Second, some SBs are also associated with fast radio bursts (FRBs) from Galactic magnetars SGR~1935+2154 \citep{2020Natur.587...54C,2020Natur.587...59B} and possibly 1E 1547.0-5408 \citep{2021ApJ...907....7I}. In particular, SBs were associated with radio bursts from SGR~1935+2154 in April 2020 \citep{2020ApJ...898L..29M,2021NatAs...5..372R,2021NatAs...5..378L,2021NatAs...5..401T} and in October/November 2022 \citep{wang2026gecamdiscoverysecondfrbassociated,2026ApJ...998L..44X,2026A&A...707A.289T}. However, from intense radio monitoring campaigns it is clear only a small fraction $\lesssim 10^{-3}$ of SBs are associated with radio bursts in SGR~1935+2154. Statistical comparisons also suggest SBs could underpin some extragalactic FRBs \citep[e.g.,][]{2019ApJ...879....4W}. Establishing which peculiar or rare burst properties correlate with bright radio emission \citep[e.g., quasi-polar burst locales for burst emission,][]{2021NatAs...5..408Y} is therefore a plausible pathway to constraining FRB physical conditions or trigger sites. As extragalactic counterparts to FRBs are out of reach except for within a few Mpc, this physics is best studied in nearby magnetars in the hard X-rays and soft gamma-rays. 

Third, as SB emission originates at low altitudes in the closed magnetosphere, GR light bending, occultation, and strong-field radiative transfer effects are unavoidable offering tests of our understanding of magnetars. Yet,  rotational phase dependence of burst detectability and spectral components implies compact emission locales that can be partially obscured, despite gravitational lensing that ordinarily broadens visibility. 

Fourth, since SB emission is confined to compact regions near the surface, observables can depend on NS mass $M$ and radius $r_{\star}$  through lensing magnification, secondary images, redshift, and time delays, as well as on magnetic obliquity and observer geometry through phase-dependent self-occultation and anisotropic beaming. This situates SBs as potential probes of magnetar fundamental parameters and viewing geometry, complementary to persistent emission pulse profile modeling. Importantly, many key parameters are fixed for a given source and therefore might be jointly constrained by ensembles of bursts.

Fifth, for $B\gtrsim B_{\rm cr} \equiv 4.4\times 10^{13}\,$G, photon scattering is strongly modified in the burst photosphere in polarization and angle-dependent ways.  This is a regime where exotic aspects~\citep{2006RPPh...69.2631H} such as photon splitting may cut emission down to $30-50$ keV \citep{1991A&A...249..581B,1995ApJ...440L..69B,2019MNRAS.486.3327H}.  Vacuum birefringence (VB) and polarization-dependent scattering reshape emergent spectra and angular distributions.  SBs routinely access photon energies where these effects become non-negligible, motivating physically grounded radiative transfer models rather than purely phenomenological fits. Additionally, if baryon loading is significant, plasma effects competing with VB (the vacuum resonance) and mode conversion might imprint spectral structure below $\sim 40\,$keV \citep[e.g.,][]{1997MNRAS.288..596B,2003MNRAS.338..233H,2025PhRvD.112l3027W}. However, the requisite plasma densities for generating magnetar SBs are three to four orders of magnitude below values realized in surface atmosphere models where the vacuum resonance is considered. In addition, photon energies for bursts are $10-100$ times higher than realized for atmospheres, and thus plasma dispersion is $\sim 5-8$ orders of magnitude lower than for the atmosphere situation.

Sixth, the observational landscape has shifted: the sample size has grown to the point that population-level inferences might be statistically meaningful, while key diagnostics (time-resolved spectroscopy, rotational phase dependence, and polarization sensitivity) are becoming technically accessible. The excellent time resolution and broad energy coverage of current instruments, along with recent catalogs of magnetar SBs \citep[e.g.,][]{2025arXiv251212291C}, call for accurate physical models. Bright bursts can exceed $10^5$ counts s$^{-1}$ and potentially contain significant as-yet untapped information on the geometry, physics and source properties. At present, most SB analyses focus on phenomenological spectral fits whose physical interpretation and mapping to source parameters are ambiguous. This mismatch between data quality and model fidelity increasingly limits the physical inferences that can be extracted from existing burst archives. 

We detail the first GR ray-traced models of magnetar burst fireballs with magnetized atmospheres that treat the energy and polarization dependent emission anisotropy in full 3D, extending beyond geometric, unmagnetized, 1D and/or semi-analytic models \citep[e.g.,][]{1994ApJ...437L.111U,1995MNRAS.275..255T,2002MNRAS.332..199L,2015ApJ...815...45Y,2017MNRAS.469.3610T,2021ApJ...921...92B,2026arXiv260424750X}. This enables new qualitatively observable quantities. The ray tracing formalism we adopt has advantages, particularly for properly treating anisotropic emission, self-shadowing and lensing signatures. The procedure stores impact positions and surface zenith angles of photons, and angles with respect to the local magnetic field. It also tracks GR redshift, time delays and computes a photon splitting opacity (below 1 MeV) for each trajectory. The zenith and magnetic field angles are particularly important for anisotropies and limb darkening in surface layers of magnetized fireball photosphere. These are implemented from magnetized scattering models \citep[][]{2021MNRAS.500.5369B,2022ApJ...928...82H,2023IAUS..363..297H,2025ApJ...992..188D} which account for these quantities and their photon energy dependence (relative to the local electron cyclotron energy). Originally developed for NS surface atmospheres and emission, they are valid for any scattering layer that is dominated by magnetic Thomson electron scattering.  The activated flux tubes are presumed to be in local thermodynamic equilibrium, possessing temperature gradients that approximately satisfy adiabatic conditions. Local emission anisotropy, and emergent spectra at the photosphere, as well as polarization dependence of the fireball photosphere and photon splitting, are treated on the ray-traced image plane (IP).

The paper is organized as follows. Section \ref{sec:methods} presents the model framework, from the photospheric temperature and geometry prescriptions in \S\ref{sec:tempspec} and \ref{sec:fluxtubes} through the ray tracing, anisotropic and polarization-dependent magnetized atmospheres, photon splitting opacities, polarization transport, and IP summation in \S\ref{sec:raytracing}--\ref{sec:imageplane_sum}. Table~\ref{tab:model_params} provides a compact reference for the principal model parameters. Readers interested primarily in the physical setup and assumptions may focus on \S\ref{sec:tempspec} and \ref{sec:fluxtubes}, whereas those interested in the physical methodology may follow \S\ref{sec:raytracing}--\ref{sec:imageplane_sum} in sequence, since each stage supplies inputs to the next. Readers interested mainly in the observable predictions may review \S\ref{sec:tempspec}, \ref{sec:fluxtubes}, and \ref{sec:imageplane_sum} before proceeding to the results in \S\ref{sec:results}. \S\ref{sec:model_data} presents illustrative comparisons between the models and observed burst spectra, and \S\ref{sec:discussion} discusses the resulting source constraints, limitations of the framework, and directions for future work.

Throughout, we use the dimensionless photon energy
$\omega = \hbar \omega^\prime/(m_e c^2)$ and dimensionless magnetic field
in units of $B = B^\prime/B_{\rm cr}$. We set $G=c=1$ only in the spacetime, field geometry
and geodesic equations and associated calculations. Elsewhere, $c$ is retained
explicitly. Note $\hbar$ is not explicitly set to unity. The quantum critical field $B_{\rm cr} = m_e^2 c^3/(e \hbar)$ is defined by when the electron cyclotron energy $\hbar \omega_B^\prime$ with $\omega_B^\prime = e B^\prime/(m_e c)$ is equal to the rest mass, $\hbar \omega_B^\prime = m_e c^2$, and $\hbar$ is absorbed into the units of $B$ and $\omega$. We adopt metric signature $(-,+,+,+)$.

\section{Setup and Methods}
\label{sec:methods}

We first detail the adiabatic fireball considerations, temperature scaling and bounds in \S\ref{sec:tempspec} where we adopt a parametric prescription for mapping the flux tube's local temperature to altitude (or equivalently, magnetic field). Our method for specifying the geometry parametrically by assuming surface loci of a magnetic flux tube surface, which we identify with the photosphere, is detailed in \S\ref{sec:fluxtubes}. The ray tracing aspect and computation of photon trajectories in curved spacetime, parallel transport of polarization vectors to an observer situated at a particular angle relative to the magnetic axis, are specified in \S\ref{sec:raytracing}. In \S\ref{sec:splitting} we describe how we compute photon splitting opacities on-the-fly during the ray tracing calculation. In \S\ref{sec:magthomscatt}  we describe the magnetized atmospheres above and below the local cyclotron resonance for emission anisotropy in an energy and polarization dependent manner. Vacuum birefringence and its application are described in \S\ref{subsec:vacuum_birefringence}. Computation of spectra, polarization maps, and impulse responses (IRs) by summing and marginalizing over quantities at the IP is described in \S\ref{sec:imageplane_sum}.   

\subsection{The Adiabatic Fireball: Temperature Specification and Bounds}
\label{sec:tempspec}

Deep inside the fireball, the enormous scattering depth implies saturated Comptonization and rapid kinetic equilibration between the radiation and pairs. Local thermodynamic equilibrium additionally requires photon production and destruction processes to drive the photon chemical potential to a small value. Double Compton scattering might efficiently drive radiation to a Planckian distribution at large optical depth \citep[e.g.,][]{1984MNRAS.209..175S,2002MNRAS.332..199L}. Pair creation/annihilation further couples photons and pairs, and at sufficiently high temperatures, photon splitting may also contribute to photon redistribution (depending on allowed modes).

Hence, an equilibrium description is most justified deep in the fireball and at low altitudes and high densities, while Comptonization-like deviations might be expected nearer the photosphere, at high altitudes (low temperatures) where the plasma might become optically thin, and for the shortest bursts \citep{2025arXiv251212291C}. The characteristic Spitzer/Coulomb timescale for thermal relaxation is of order milliseconds following an energy injection event that produces pairs \citep{2025ApJ...980..211B} at low altitudes.

As noted in \S\ref{sec:intro},  time integrated \citep{2012ApJ...749..122V} and resolved \citep[e.g.,][]{2014ApJ...785...52Y,2025ApJ...988..282D} spectroscopic studies of SBs with two-BB models empirically obey similar luminosities (with considerable scatter) in the two components, ${\cal A}_{\rm hot} T_{\rm hot}^4 \sim {\cal A}_{\rm cool} T_{\rm cool}^4$ with apparent areas $\cal A$. For an extended magnetic flux tube, if their filling factors and projection effects are comparable, then magnetic flux conservation implies that the proper tube cross section scales as ${\cal A} \propto B^{-1}$ so that $B_{1} {\cal A}_{\rm hot} \sim B_{2} {\cal A}_{\rm cool}$ with $B_{1} > B_{2}$ which implies $T_{\rm cool} \sim T_{\rm hot} (B_{2}/B_{1})^{1/4}$ (i.e., opposite to the equipartition in sunspots where cooler regions have higher magnetic fields). On the other hand, \citet{2008ApJ...685.1114I} suggest that ${\cal A}_{\rm hot} T_{\rm hot}^3 \approx {\cal A}_{\rm cool} T_{\rm cool}^3$ is commensurate with spectral data, which implies $T_{\rm cool} \sim T_{\rm hot} (B_{2}/B_{1})^{1/3}$. This proportionality is generated when photon number is conserved along a flux tube, since the Planck distribution number density scales as $T^3$.  Other temperature-field strength couplings can be envisaged, for example if the plasma equation of state controls the adiabatic variations with locale.

Accordingly, to encompass a variety of possibilities, we parameterize the local photon temperature as,
\begin{equation}
    \Theta_i = \Theta_{\rm max} \left(\frac{B_i}{B_p}\right)^\alpha   ,
    \label{eq:BT_relation}
\end{equation}
where we write the temperature in dimensionless form $\Theta\equiv k_{\rm B}T/(m_e c^2)$. Here $i$ labels a local emitting element on the fireball photosphere, and $B_i$ and $B_p$ are respectively the local and polar magnetic field strengths (i.e., $B_i \leq B_p$), both dimensionless in units of $B_{\rm cr}$. Representative values of the scaling index are $\alpha\sim1/4-1/3$. This phenomenological prescription is motivated by radiation-dominated adiabatic expansion and flux conservation, and could be revised in the future. The value of $\alpha$ is bounded by physical considerations, and as we show (\S\ref{sec:results}), $\alpha \sim 1/3$ appears to reproduce the ${\cal A}-k_{\rm B}T$ correlations observed in bursts. The plasma confinement in the fireball is given by the condition \citep{1982AIPC...77..249L,2001ApJ...561..980T}:
\begin{equation}
 p_{e^\pm} + p_{\rm rad} \sim p_{\rm tot}  \sim a  \left( \frac{m_e c^2 \Theta_i}{k_{\rm B}} \right)^4 \lesssim \frac{(B_{\rm cr} B_i)^2 }{8\pi} = p_B   ,
 \label{eq:pressures}
\end{equation}
where $a$ is the radiation constant, and $p$ are the pressures for various species. This ought to be satisfied throughout the flux tube, and particularly at the outer boundary. Assuming $B_i \sim B_p (r_\star/r_{\rm max})^3$ (with $r_{\rm max}$ the maximum flux tube or loop height), and Eq.~(\ref{eq:BT_relation}), implies
\begin{equation}
\Theta_{\rm max} \lesssim \left(\frac{r_{\rm max}}{r_\star}\right)^{-\frac{3}{2}+3\alpha} \left( \frac{15 B_p^2}{8\pi^3\alpha_f}\right)^{1/4} \approx 1.7 \sqrt{B_p} \left(\frac{r_{\rm max}}{r_\star}\right)^{-\frac{3}{2}+3\alpha} \quad ,
\label{eq:tmax_lim}
\end{equation}
and
\begin{equation}
\frac{r_{\rm max}}{r_\star} \lesssim  \left(\frac{8\pi^3 \alpha_f \Theta_{\rm max}^4}{15 B_p^2} \right)^{\frac{1}{12\alpha-6}} \approx \left(\frac{8.2\, B_p^2 }{\Theta_{\rm max}^4} \right)^{\frac{1}{6-12\alpha}} \quad {\rm if} \quad \alpha < \frac{1}{2}.
\label{eq:rmax_lim}
\end{equation}
Here $\alpha_f$ is the fine structure constant. For $\alpha<1/2$, the most restrictive condition occurs at the weakest-field location, i.e., near the loop apex.  For $\alpha>1/2$, the ratio $p_{\rm tot}/p_B \propto B^{4\alpha-2}$ decreases outward, so confinement is guaranteed everywhere along the loop provided the footpoint region satisfies Eqs.~(\ref{eq:pressures})--(\ref{eq:tmax_lim}). Inverted temperature hierarchies, $\alpha < 0$, where the hottest regions are at high altitudes (a la reconnection events), impose more stringent requirements on the maximum altitude and limit the luminosity of bursts for observed temperatures and timescales. As they are disfavored by observations (see \S\ref{sec:intro}), we do not consider that regime. The pair density is of order $n_{e^\pm} \sim 10^{22}-10^{26}$~cm$^{-3}$ if in pressure equilibrium with the photon gas of temperature $m_e c^2 \Theta \sim 10-50$~keV. This implies that if $n_{e^\pm} \propto B$ and the cross section is close to Thomson, the fireballs may become optically thin, or enter the unsaturated Comptonization regime, at tens to hundreds of stellar radii. For fixed $\Theta_{\rm max}$ and size of the flux tube, small values of $\alpha \sim 0 $ can lead to relatively high photon luminosities, while $\alpha \gtrsim 1$ values can dramatically decrease the luminosity and narrow spectral extent by weighting footpoint regions more in the total spectrum.

For slow rotators such as magnetars, the light cylinder is large compared to $r_{\rm max}/r_\star$ ($\Omega/(c r_\star) \gtrsim 10^{3}$, $\Omega$ the angular spin frequency of the magnetar), even for flux tubes extending up to several hundred stellar radii. For such cases, the footpoints must be highly localized as spectral fits suggest apparent emission areas do not significantly exceed $(10r_{\star})^{2}$. 
We do not include the possible presence of resonant cyclotron reprocessing and high-altitude scattering layers, which might broaden the observed burst spectral shape at higher energies  \citep[$\gtrsim 50$ keV, e.g.,][]{2020MNRAS.498..484Y}. The existence of such dense unconfined high altitude charge screens is not guaranteed, given strong (resonant) radiative pressure from past bursts in a storm.

\subsection{Specification of Geometric Locales of Fireball Photospheres}
\label{sec:fluxtubes}

As mentioned above, we consider confined fireball geometries and, as a first step, adopt geometry described by a GR static vacuum dipole that is effectively symmetric about the magnetic equator. For this dipole specialization, because the burst dynamical timescale is much shorter than the observed durations, and because resonant Compton scattering can effectively couple photons to leptons in the presence of strong radiation anisotropy and pressure along $\boldsymbol{B}$ inside the fireball, we expect the fireball to be effectively symmetric about the magnetic equator, except for activations near the pole that either violate the conditions set out in \S\ref{sec:tempspec} or become optically thin at large height (``polar adiabatic funnel''). Higher-order multipolar fields and less symmetric activation geometries may alter the fireball morphology and the resulting spectra and polarization signals, but we defer such generalizations to future work.

An analytic specification of the shape of flux tubes is advantageous computationally, for ascertaining which photon geodesics originate from the flux tubes, and for quantities such as local surface normal $\boldsymbol{n}_{\cal S}$ directions relative to the local magnetic field $\boldsymbol{B}$ and photon 3-momentum $\boldsymbol{k}$ required for the radiative transfer through the burst photosphere. An expedient parameterization is also essential for the physical interpretation of fitted burst parameters.

The loci of surface footpoints on the NS completely define a field flux tube surface (or volume) in our construction. Below, we detail this analytic approach.

\begin{figure}[t]
\centering
\includegraphics[width = 0.38\textwidth]{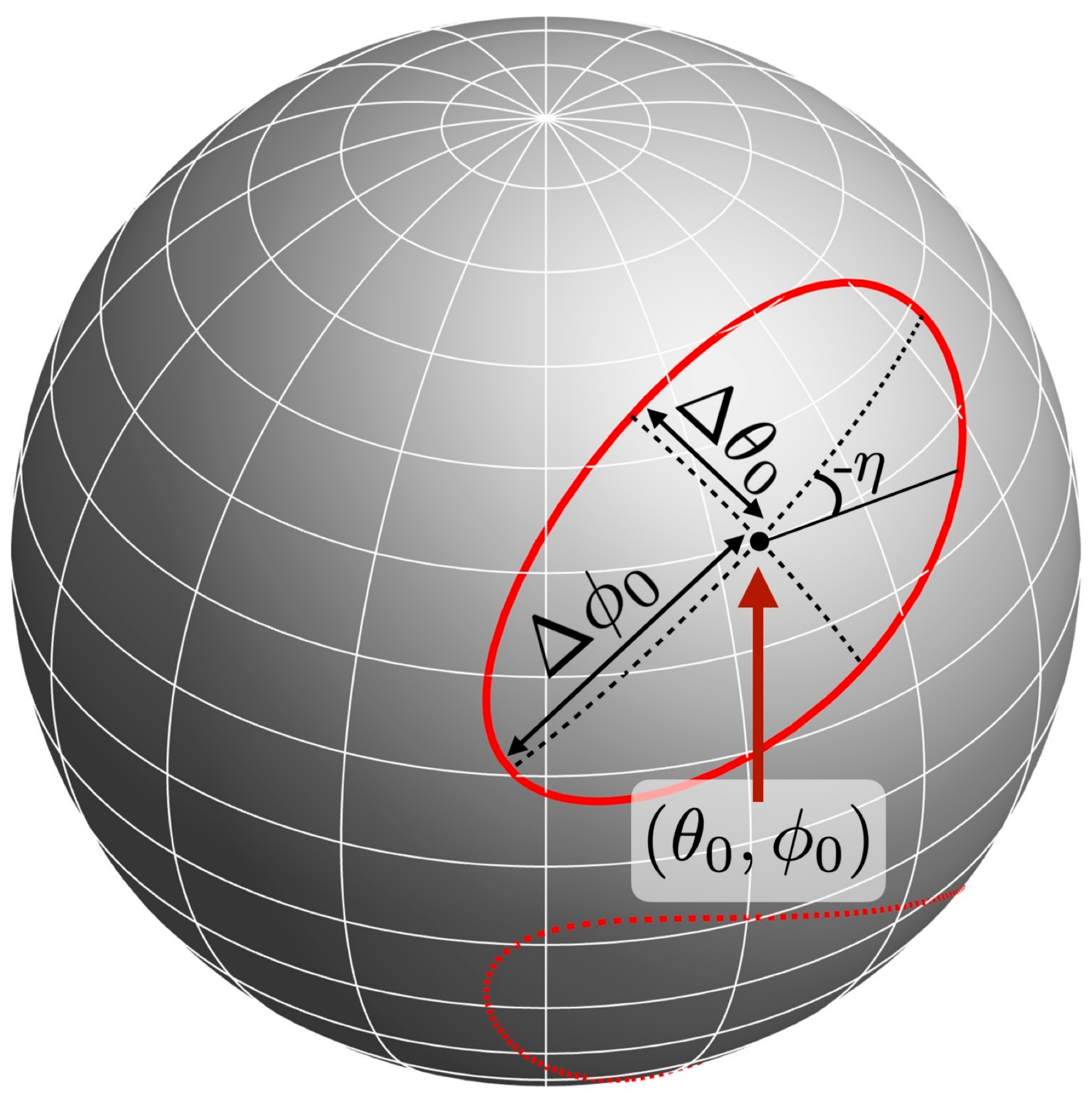} 
\caption{Geometric angular coordinate parameters describing the loci of field line footpoints that define the fireball. The ellipse is centered at magnetic-frame coordinates $(\theta_0,\phi_0)$, has angular semiaxes $(\Delta\theta_0,\Delta\phi_0)$, and is rotated by $\eta$ on the stellar surface. Field lines anchored on this contour define the fireball flux tube boundary discussed in \S\ref{sec:fluxtubes}. Symmetry about the magnetic equator is assumed, given the confined nature of fireballs.}
\label{fig:schematic_ellipse}
\end{figure}

The exterior magnetic field embedded in Schwarzschild geometry is \citep{1974PhRvD..10.3166P,1983ApJ...265.1036W,1986SvA....30..567M,2013MNRAS.433..986P},
\begin{equation}
    \boldsymbol{B} = \bigl\{ c_r (\Psi) B_{r}, \,  {c}_\Omega (\Psi) B_{\theta}, \, {c}_\Omega (\Psi)B_{\phi} \bigr\}  ,
    \label{Bfull}
\end{equation}
where for an untwisted curl-free dipole $B_{\phi} = 0$. Here $\Psi = 2M/r$ and 
\begin{eqnarray}
c_r (\Psi) &=& -\frac{3}{\Psi^3}\left[ \ln (1-\Psi) +\Psi+\frac{\Psi^2}{2} \right] ,\\
c_\Omega (\Psi) &=& \frac{6}{\Psi^3\sqrt{1-\Psi}}\left[ (1-\Psi)\ln(1-\Psi)+\Psi-\frac{\Psi^2}{2} \right].
\end{eqnarray}
For a static dipole, $\boldsymbol{B} = B_p \{ 2 \cos\theta, \sin \theta , 0\}/(2 \varrho^3)$ where $\theta\in [0,\pi]$ is the magnetic colatitude and $\varrho = r/r_\star$.  The field strength on the surface at the pole is enhanced by a factor 
\begin{equation}
    B_p \approx B_{p}^{\rm flat} \left[\frac{3 \left|\Psi(2+\Psi) + 2 \ln(1-\Psi)\right|}{2\Psi^3} \right] \approx  B_{p}^{\rm flat} \left(1+ \frac{3}{4}\Psi + {\cal O}(\Psi^2) \right),
\end{equation}
which can attain increases of order $30-50\%$ for compact NSs for which $\Psi \sim 0.4$. Note that the same scaling applies to dipole moment $\mu \equiv (B_p r_\star^3)/2$.
Poloidal field lines are established by the solution of 
\begin{equation}
    \frac{1}{r}\frac{dr}{d\theta} = \frac{B_r}{B_\theta}.
    \label{fieldloopode}
\end{equation}
Eq.~(\ref{Bfull}) does not admit a closed-form solution in elementary functions required to compactly parameterize flux tube surfaces. At the lowest nontrivial order in compactness, the radial solution is expressible in closed form,
\begin{equation}
    r_{\rm loop} (\theta) \approx \frac{M}{2} - \left(\frac{M}{2}-r_\star \right)\frac{\sin^2 \theta}{\sin^2 \theta_{\rm fp}} = \frac{M}{2} - \left(\frac{M}{2}-r_{\rm max} \right)\sin^2 \theta ,
    \label{rloop}
\end{equation}
where $\theta_{\rm fp}$ is the magnetic colatitude at which the
particular field line intersects the stellar surface (generalized below). Compared to numerical integration of the ODE Eq.~(\ref{fieldloopode}), this approximation is better than $2\%$ for the maximum loop height $r_{\rm max}$ for the extreme case of $M= 2.2 \,M_\odot $ ($3.25$ km geometric mass) and  $r_\star = 12$ km radius NS, and better than $0.5\%$ for $M= 1.4 \,M_\odot$.

Flux tubes may be specified by closed boundaries on the surface of the NS, with the shape's perimeter specifying the exterior surface of the flux tube. The activation regions on the surface of the magnetar, which participate in bursts, are unknown, but likely irregular. To retain some generality, we consider angular coordinate ellipses with arbitrary tilt and aspect ratio. This is a coordinate space boundary rather than
a geodesic ellipse on the spherical surface. Consequently, its physical
azimuthal scale varies as $\sin\theta\,d\phi$, and its apparent shape is distorted close to a magnetic pole. In this work, contours or parameter regimes that pass through or inscribe a pole are excluded as they would not be physically realized as confined fireballs. Instead they would encompass solutions that contain outflows \citep[e.g.,][]{1999ApJ...525L.125H,2016MNRAS.461..877V} which lie outside the scope of this paper. A bound on footpoint colatitude can also be estimated from Eq.~(\ref{eq:rmax_lim}) in Eq.~(\ref{rloop}).
The generic form here is intended to capture a wide range of burst phenomenology, analogous to the success of NICER teams \citep[e.g.,][]{2019ApJ...887L..21R,2019ApJ...887L..24M,2024ApJ...974..295D} in modeling hot spots on the surfaces of millisecond pulsars with shapes. By construction, the flux tubes exhibit reflection symmetry about the magnetar equator and, naturally, do not cross field lines. The introduction of nonvanishing tilt breaks left-right symmetry of the flux tubes in azimuth. 
Consider the angular ellipse defining the perimeter of flux tube footpoints, centered at angular coordinates relative to the magnetic pole $\{\theta_{0},\phi_{0} \}$, with angular axes $\{\Delta \theta_{0}, \Delta\phi_{0}\}$ tilted by angle $\eta$ -- see Figure~\ref{fig:schematic_ellipse}. This ellipse algebraically satisfies
\begin{eqnarray}
\left[\frac{\left({\cal R}{\cal T}\right)_\phi}{\Delta \phi_{0}}\right]^2+ \left[\frac{\left({\cal R}{\cal T}\right)_\theta}{\Delta \theta_{0}}\right]^2 = 1, 
\label{ellipse}
\end{eqnarray}
where 
\begin{equation}
{\cal R} = \begin{Bmatrix}
\cos \eta & \sin \eta  \\
-\sin \eta  & \cos \eta 
\end{Bmatrix} \qquad , \qquad   {\cal T} = \begin{bmatrix}
\phi \\
\theta \\
\end{bmatrix} - \begin{bmatrix}
\phi_{\rm 0} \\
\theta_{\rm 0} \\
\end{bmatrix}.
\end{equation}
An infinitesimal footpoint value for a loop is specified by the identification
\begin{equation}
\theta_{\rm fp} =\arcsin \sqrt\frac{M -2 r_\star}{M -2 r_{\rm max}}.
\end{equation}
Substitution of this expression into Eq.~(\ref{ellipse}), solving for $r_{\rm max}$, and replacement of $r_{\rm max}$ in Eq.~(\ref{rloop}) results in the specification of the full flux tube. These solutions for Eq.~(\ref{rloop}) are,
\begin{equation}
r_{\rm loop} = \frac{1}{4} \left(2 r_\star - M \cos 2 \theta_{\rm fp}^{(s)}(\phi) + (M-2r_\star) \cos 2\theta \right) \csc^2 \theta_{\rm fp}^{(s)}(\phi)  ,
\label{eq:rloop}
\end{equation}
where
\begin{equation}
   \theta_{\rm fp}^{(s)}(\phi) = \theta_0 +  \frac{{\cal T}_{\phi} \kappa\sin (2 \eta )-\sqrt{2} \Delta \phi_{0} \Delta \theta_{0} s
   \sqrt{\kappa\cos (2 \eta )+\sigma-2 {\cal T}_{\phi}^2}}{\kappa\cos (2 \eta
   )+\sigma},
\end{equation}
with ${\cal T}_{\phi} \equiv {\rm wrap}_{[-\pi,\pi]}(\phi-\phi_0)$, and $\kappa = \Delta \phi_{0}^2-\Delta \theta_{0}^2 $, $\sigma =  \Delta \phi_{0}^2 + \Delta \theta_{0}^2 $ and $s = \pm1$ specifying the upper or lower branches (relative to the semi-major axis defined by $\Delta \phi_{0}$)  of the closed footpoint boundary at fixed azimuth.

The surface normal (germane for outgoing photons) is then readily obtained analytically by the gradient evaluated on the surface (in the tetrad defined by a timelike stationary observer, see \S\ref{sec:raytracing}),
\begin{equation}
    \boldsymbol{\hat n}_{\cal S} = s \left. \frac{\boldsymbol{\nabla}(r-r_{\rm loop})}{\left|\boldsymbol{\nabla}(r-r_{\rm loop})\right|} \right|_{\cal S} \qquad \rm (outgoing \, \, photons).
    \label{eq:surfnormal}
\end{equation}
The factor $s=\pm 1$ here also sets the pertinent outward surface normal for the fireball\footnote{For example, an accretion column-like lensing geometry \citep[e.g.,][]{2024MNRAS.530.3051M} may be obtained by selecting only one branch and reversing the sign in Eq.~(\ref{eq:surfnormal}).}. Equation~(\ref{eq:rloop}) defines the event condition for terminating GR ray tracing trajectories' integration in \S\ref{sec:raytracing}.

\begin{deluxetable}{cl}
\tablecaption{Model parameters and notation used throughout this work.}
\label{tab:model_params}
\tablehead{
\colhead{Variable} & \colhead{Description} 
}
\startdata
$M$ & Neutron star gravitational mass\\
$r_{\star}$ & Neutron star radius \\
$B_{p}^{\rm flat}$ & Polar field value in flat spacetime in units of $B_{\rm cr}$ \\
$\theta_{v}$ & Observer viewing angle relative to the magnetic axis \\
$\theta_{0}$ & Polar coordinate (relative to magnetic axis) of surface footpoint ellipse center \\
$\phi_{0}$ & Azimuthal coordinate (relative to magnetic axis) of surface footpoint ellipse center \\
$\Delta\theta_{0}$ & Polar ellipse (semiminor) axis radius  \\
$\Delta\phi_{0}$ & Azimuth ellipse (semimajor) axis radius \\
$\eta$ & Orientation angle of the footpoint ellipse in magnetic coordinates  \\
$\Theta_{\rm max}$  & Maximum dimensionless fireball temperature, $\Theta_{\rm max}=k_{\rm B}T_{\rm max}/(m_{e} c^{2})$ \\
$\alpha$ & Adiabatic fireball temperature-scaling exponent (dimensionless) \\
\hline \hline
\enddata
\end{deluxetable}

\subsection{General Relativistic Ray Tracing}
\label{sec:raytracing}

\begin{figure}[t]
  \centering
\begin{tabular*}{0.85\textwidth}{@{\extracolsep{\fill}}*{7}{c}}
$\Delta t_{\rm obs}$ & $\boldsymbol{\hat{k}\cdot \hat{n}_S}$  & $\boldsymbol{\hat{k}\cdot \hat{B}}$ & $\cos \phi_{kB}$ & $\log \tau_{\gamma\rightarrow\gamma\gamma}$ & $1+z$ & $|B|$ \\
\hline
\end{tabular*}
    \includegraphics[width=0.94\textwidth]{RayTracingSection_Figures/panels_2026-02-23_1408.png}  \\
    \includegraphics[width=0.94\textwidth]{RayTracingSection_Figures/panels_2026-02-23_1719.png}  \\
    \includegraphics[width=0.94\textwidth]{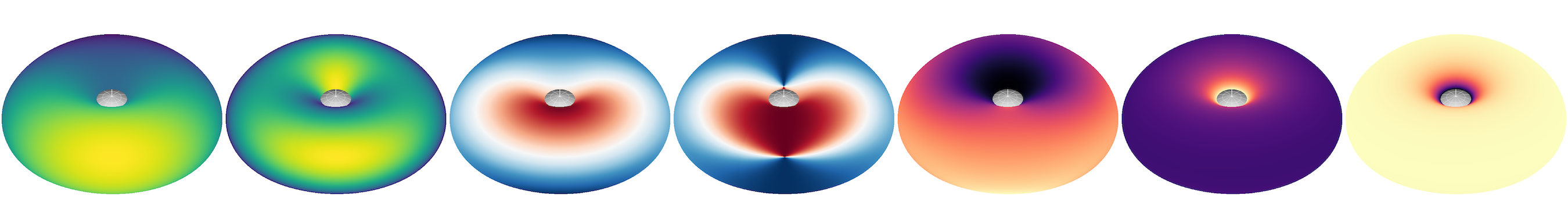}  \\
    \includegraphics[width=0.94\textwidth]{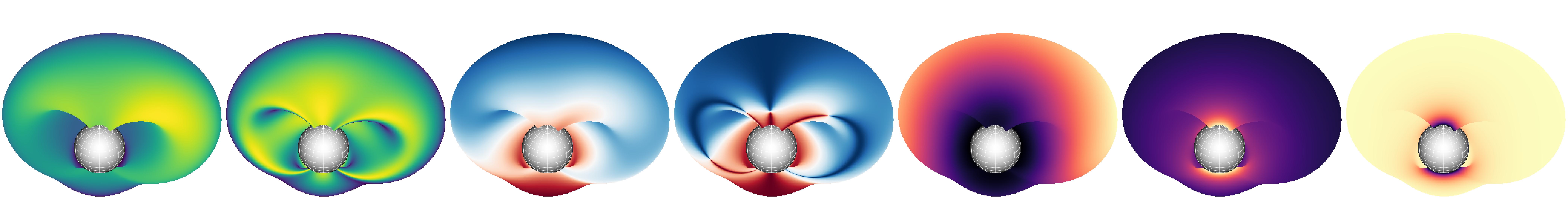}  \\
    \includegraphics[width=0.94\textwidth]{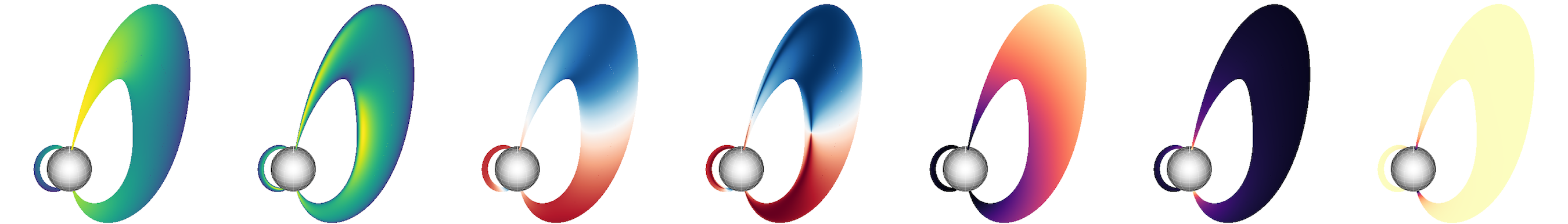}  \\
    \includegraphics[width=0.94\textwidth]{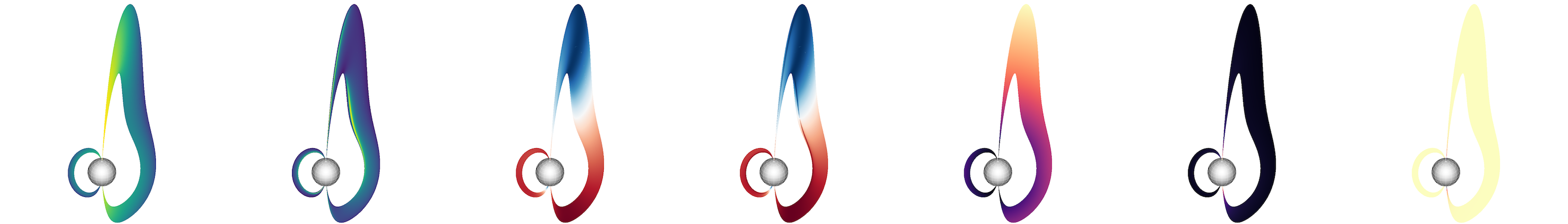}  \\
    \includegraphics[width=0.94\textwidth]{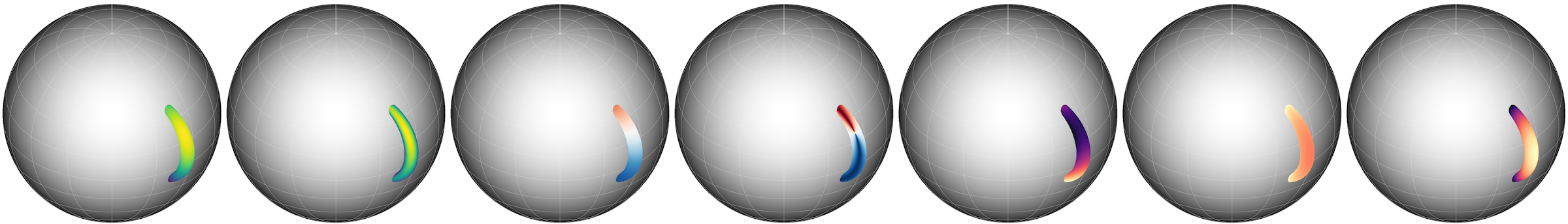}  \\
    \includegraphics[width=0.94\textwidth]{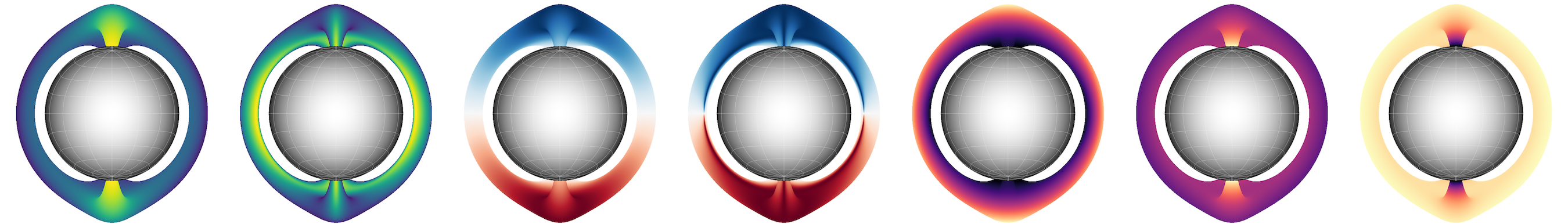}  \\
    \includegraphics[width=0.94\textwidth]{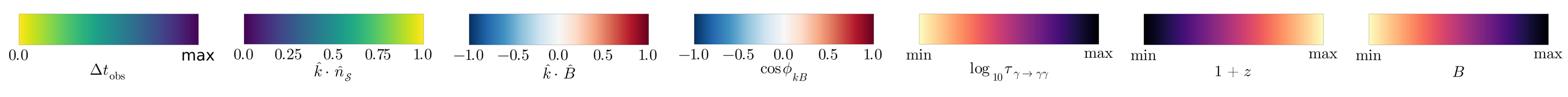}
    \vspace{-3mm}
  \caption{Diverse mosaic of ray-traced flux tubes, illustrating how geometry controls local transfer quantities and observables. In each row (one fireball setup), the columns show normalized color maps of observed delay $\Delta t_{\rm obs}$, surface normal zenith cosine $\boldsymbol{\hat{k}\cdot \hat{n}_{\cal S}}$,  photon magnetic cosine $\boldsymbol{\hat{k}\!\cdot\!\hat{B}}$, photon magnetic azimuth angle $\cos\phi_{kB}$, accumulated polarization-averaged splitting optical depth $\log \tau_{\rm ave}$, gravitational redshift factor $1+z$, and local field strength $|\boldsymbol{B}|$. Across rows, only the surface footpoint ellipse parameters are varied, with $M=1.5\,M_\odot$ and $r_\star=12$ km fixed. The first seven rows are viewing angle $\theta_v=60^\circ$, and the last row is $\theta_v=90^\circ$. The 5th and 6th cases with limited footpoint area near the pole seem to be most compatible with spectral phenomenology of magnetar SBs (see \S\ref{sec:results}).}
  \label{fig:archetypes}
\end{figure}

 We developed an updated parallel C++ version of {\tt GIKS} \citep{2021ApJ...907...63K}, a ray tracing code that computes photon geodesics in curved spacetime. This code is similar to that formulated in other works \citep{2010ApJ...718..446J,2012ApJ...754..133K,2012ApJ...745....1P,2012ApJ...753..175B,2013ApJ...777...11S,2014ApJ...792...87P}. Trajectories are propagated backward in time on a grid from a distant IP toward the compact object, with hits recorded at either the surface or flux tube surfaces as defined in \S\ref{sec:fluxtubes}. The external spacetime included in {\tt GIKS} is Kerr, but given the negligible spin of magnetars, it is effectively Schwarzschild. 
We assume geometric optics and that the flux tube medium is at rest with respect to the NS (although this can be relaxed for future investigations of dynamic bursts). That is, its 4-velocity is only timelike $u^\mu = (1/\sqrt{-g_{tt}},0,0,0)$, given the normalization condition $u^\mu u_\mu = -1$. The spatial path length measured by static observers is $d\ell^2 = (g_{\mu\nu} + u_\mu u_\nu) dx^\mu dx^\nu$ or equivalently $dl = -(u_\mu k^\mu) d\lambda$ \citep[e.g.,][]{2007ApJ...654..458C} with photon 4-vectors $k^\mu = \{k^t, k^r, k^\theta, k^\phi \}$ and coordinates $x^\mu = \{t, r, \theta, \phi \}$.

Plasma propagation/transfer effects and refraction are neglected for the X-rays and gamma-rays considered here, whereas VB and polarization effects may be considered (see \S\ref{subsec:vacuum_birefringence}). Thus, while geometric optics is considered for the photon momenta, polarization vectors may adiabatically rotate before freezing at a polarization limiting surface. We also assume the radiation is incoherent, and for ensembles of photons, cross terms associated with wave phases vanish in the polarization transport and ensemble Stokes parameters for photons. 

We compute photon geodesics in Kerr spacetime using global coordinates $x^\mu=(t,r,\theta,\phi)$ and the metric $g_{\mu\nu}$. Each ray is initialized on a distant IP at radial coordinate distance $r=d_{\rm ip}=256r_\star$ with Cartesian impact parameters $(x_o,y_o)$ and inclination $\theta_v$ relative to the magnetic axis. The initial point and direction on the IP are obtained by mapping $(x_o,y_o,\theta_v,d_{\rm ip})$ to $(r_o,\theta_o,\phi_o)$ and to an initial wavevector $k^\mu_0$ that satisfies the null constraint $g_{\mu\nu}k^\mu k^\nu=0$ \citep[see Appendix A of ][]{2010ApJ...718..446J}. Conserved quantities, related to the energy and angular momentum of each photon, are enforced through the impact
parameter
$b$, detailed in Eqs.~(7)-(8) of \citet{2012ApJ...745....1P}. This is encoded in the $(t,\phi)$ components of the $k$-vector via
\begin{align}
k^t &= \frac{-g_{\phi\phi}-b\,g_{t\phi}}{g_{\phi\phi}g_{tt}-g_{t\phi}^2}, \qquad
k^\phi = \frac{b\,g_{tt}+g_{t\phi}}{g_{\phi\phi}g_{tt}-g_{t\phi}^2},
\label{eq:kvec1}
\end{align}
while $k^r$ and $k^\theta$ are evolved directly\footnote{Eq.~(\ref{eq:kvec1}) corrects a typo in Eq.~(10) of \citet{2012ApJ...745....1P}.}. The geodesic equations are integrated in the affine parameter $\lambda$,
\begin{align}
\frac{dx^\mu}{d\lambda} &= k^\mu, \qquad
\frac{dk^\mu}{d\lambda} = -\Gamma^\mu_{\alpha\beta}k^\alpha k^\beta,
\end{align}
where $\Gamma^\mu_{\alpha\beta}$ are the Christoffel symbols. In practice, we advance $(r,\theta)$ and their affine derivatives, with $k^t$ and $k^\phi$ recomputed from $b$ at each step, and we monitor the numerical null constraint through the diagnostic,
\begin{equation}
\xi \equiv \frac{g_{rr}(k^r)^2+g_{\theta\theta}(k^\theta)^2+g_{\phi\phi}(k^\phi)^2+2g_{t\phi}k^t k^\phi}{g_{tt}(k^t)^2},
\end{equation}
which equals $-1$ for an exactly null trajectory \citep[Eq.~(14) in][]{2012ApJ...745....1P}. We store $\xi$ as a diagnostic output for every ray. Integration is terminated when the ray reaches the stellar surface $r \leq r_{\star}$, when $\lambda$ exceeds $2d_{\rm ip}$, or when an event crossing is detected. Near event surfaces, we implement adaptive stepping to ensure accurate crossing detection.

New to {\tt GIKS}, we propagate polarization information by parallel-transporting two orthonormal IP basis vectors $\varepsilon_{(1)}^\mu$ and $\varepsilon_{(2)}^\mu$ along each photon geodesic,
\begin{equation}
\frac{d\varepsilon^\mu_{(i)}}{d\lambda} = -\Gamma^\mu_{\alpha\beta}k^\alpha \varepsilon^\beta_{(i)}, \qquad i\in\{1,2\},
\end{equation}
with initial conditions constructed at the observer by projecting a fixed ``up'' direction into the local orthonormal tetrad and forming a right-handed triad with the photon propagation direction (see below). The ray tracing code computes local directional cosines using a static-observer orthonormal tetrad. 

Given the 4-vector $k^\mu$, specializing now explicitly to Schwarzschild, the tetrad components are\footnote{A more general tetrad set in Kerr-like metrics for arbitrary motion fluid emitters is detailed in Appendix A of \citet{2008MNRAS.390...21B}.}
\begin{equation}
k^{\hat t}=\sqrt{-g_{tt}}\,k^t,
\quad
k^{\hat r}=\sqrt{g_{rr}}\,k^r,
\quad
k^{\hat\theta}=\sqrt{g_{\theta\theta}}\,k^\theta,
\quad
k^{\hat\phi}=\sqrt{g_{\phi\phi}}\,k^\phi
\, ,
\label{eq:k_tetrad_components}
\end{equation}
followed by the explicit spherical-to-Cartesian map,
\begin{align}
k^{\hat x} &= \sin\theta\cos\phi\,k^{\hat r}+\cos\theta\cos\phi\,k^{\hat\theta}-\sin\phi\,k^{\hat\phi},\nonumber\\
k^{\hat y} &= \sin\theta\sin\phi\,k^{\hat r}+\cos\theta\sin\phi\,k^{\hat\theta}+\cos\phi\,k^{\hat\phi},\nonumber\\
k^{\hat z} &= \cos\theta\,k^{\hat r}-\sin\theta\,k^{\hat\theta}.
\label{eq:k_cart_components}
\end{align}
All dot products presented are evaluated in this local tetrad frame, not by directly contracting coordinate-basis spatial components. In particular, for any unit 3-vector $\hat{\boldsymbol q}$ represented in the local Cartesian tetrad basis of the static flux tube surface, $\hat{\boldsymbol k}\cdot\hat{\boldsymbol q}=k^{\hat x} q_x+k^{\hat y}q_y+k^{\hat z} q_z$. For instance,
$\hat{\boldsymbol B}\equiv\boldsymbol B/|\boldsymbol B|$ is the local unit magnetic field vector measured in the
static observer orthonormal tetrad.

The local unit propagation vector is $\hat{\mathbf{k}}=-\mathbf{k}_{\rm loc}/|\mathbf{k}_{\rm loc}|$ given our backward in time integration, and products involving $\hat{\mathbf{k}}$ are Euclidean dot products using these tetrads in a Cartesian basis. The photon zenith cosine relative to the local photosphere normal direction is
\begin{equation}
\mu_{kn} \equiv \cos\theta_{kn}=\hat{\boldsymbol k}\cdot\hat{\boldsymbol n}_{\cal S},
\label{eq:thetakn}
\end{equation}
where for the flux tube we adopt Eq.~(\ref{eq:surfnormal}). Likewise,
\begin{equation}
\cos\theta_{kB}=\hat{\boldsymbol k}\cdot\hat{\boldsymbol B},
\label{eq:thetakB}
\end{equation}
and the azimuth of the photon direction in the local slab plane, measured from the magnetic field direction, is established from
\begin{equation}
\tan \phi_{kB}= \frac{\hat{\boldsymbol k}\cdot(\hat{\boldsymbol n}_{\cal S}\times\hat{\boldsymbol B})}{\hat{\boldsymbol k}\cdot\hat{\boldsymbol B}}.
\label{eq:phi_kB}
\end{equation}

Examples of stored quantities for various flux tube geometries are shown in Figure~\ref{fig:archetypes}, which also highlights possible geometries enabled by the scheme of \S\ref{sec:fluxtubes}.

For polarization transport, two IP basis 4-vectors $\varepsilon_{(1)}^\mu$ and $\varepsilon_{(2)}^\mu$ are parallel transported along the photon geodesic \citep[see, e.g.,][]{2012ApJ...754..133K}.
At the observer, a reference ``up'' direction is projected onto the IP orthogonal to $\hat{\boldsymbol k}$, orthonormalized, and converted to spherical tetrad components before integration. At the emission point, we construct the local polarization vectors
\begin{equation}
\hat{\boldsymbol \varepsilon}_{\perp}=\frac{\hat{\boldsymbol B}\times\hat{\boldsymbol k}}{|\hat{\boldsymbol B}\times\hat{\boldsymbol k}|},
\qquad
\hat{\boldsymbol \varepsilon}_{\parallel}= \hat{\boldsymbol \varepsilon}_{\perp}\times\hat{\boldsymbol k},
\label{eq:epar_eperp}
\end{equation}
and project them onto the transported IP basis using metric inner products. 

\begin{figure}[t]
\centering
\begin{tabular}{cc|c|c|c}
  & \large $M=0$ & \large $M=1.2\,M_\odot$ & \Large $M=1.7\,M_\odot$ & \Large $M=2.2\,M_\odot$ \\
\includegraphics[width=0.08\textwidth]{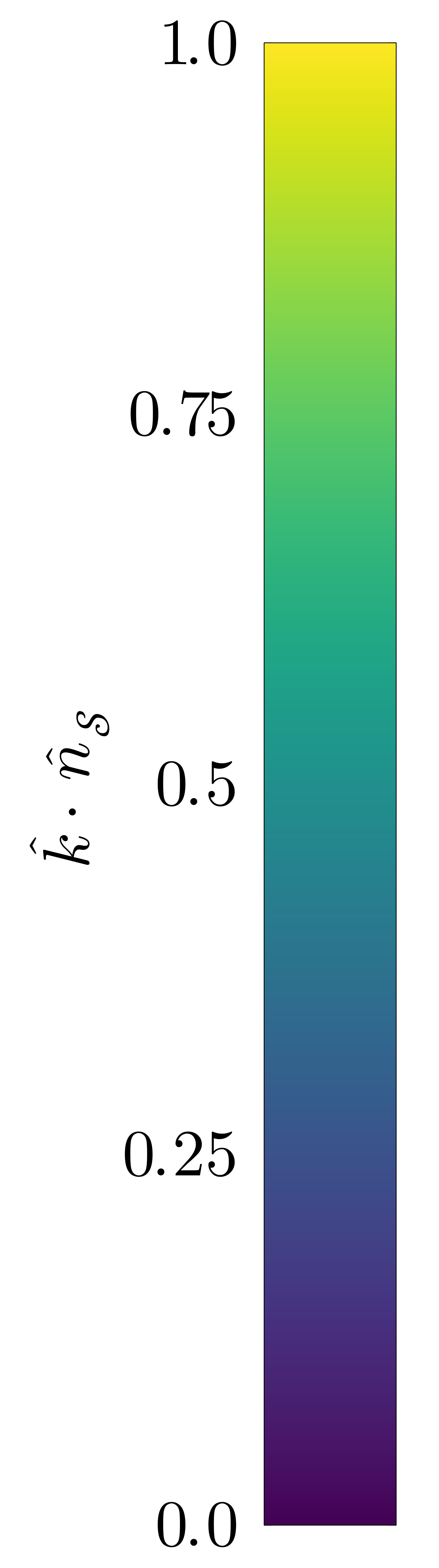} & 
\includegraphics[width=0.21\textwidth]{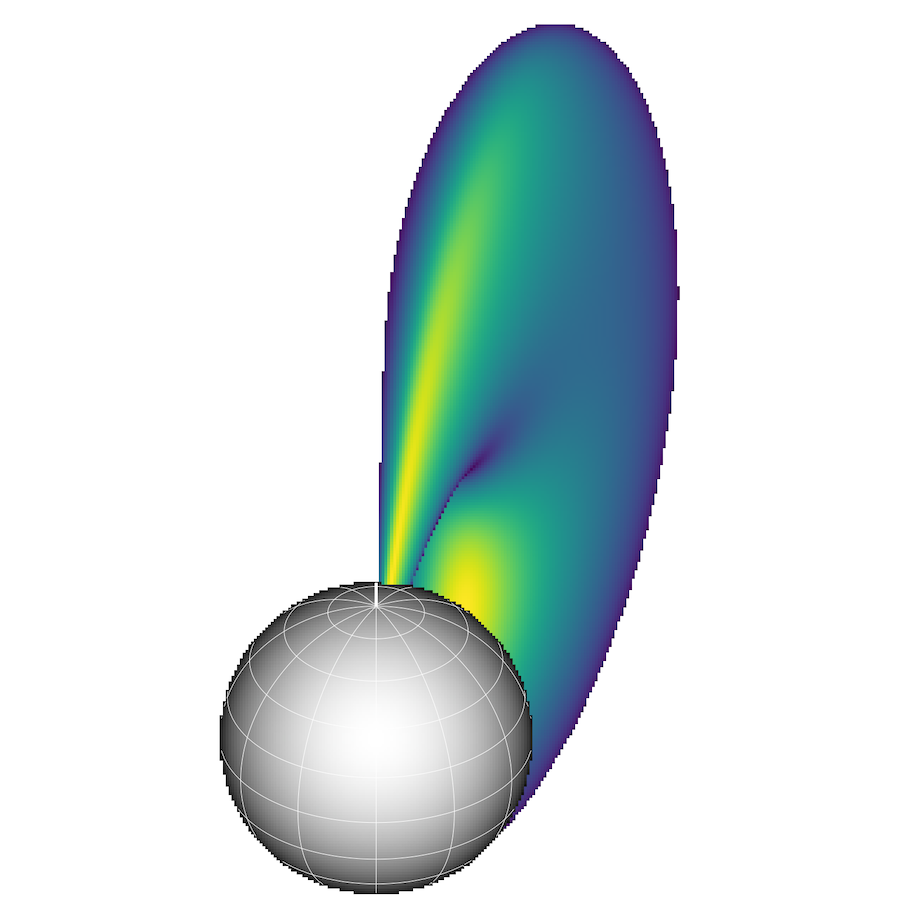} &
\includegraphics[width=0.21\textwidth]{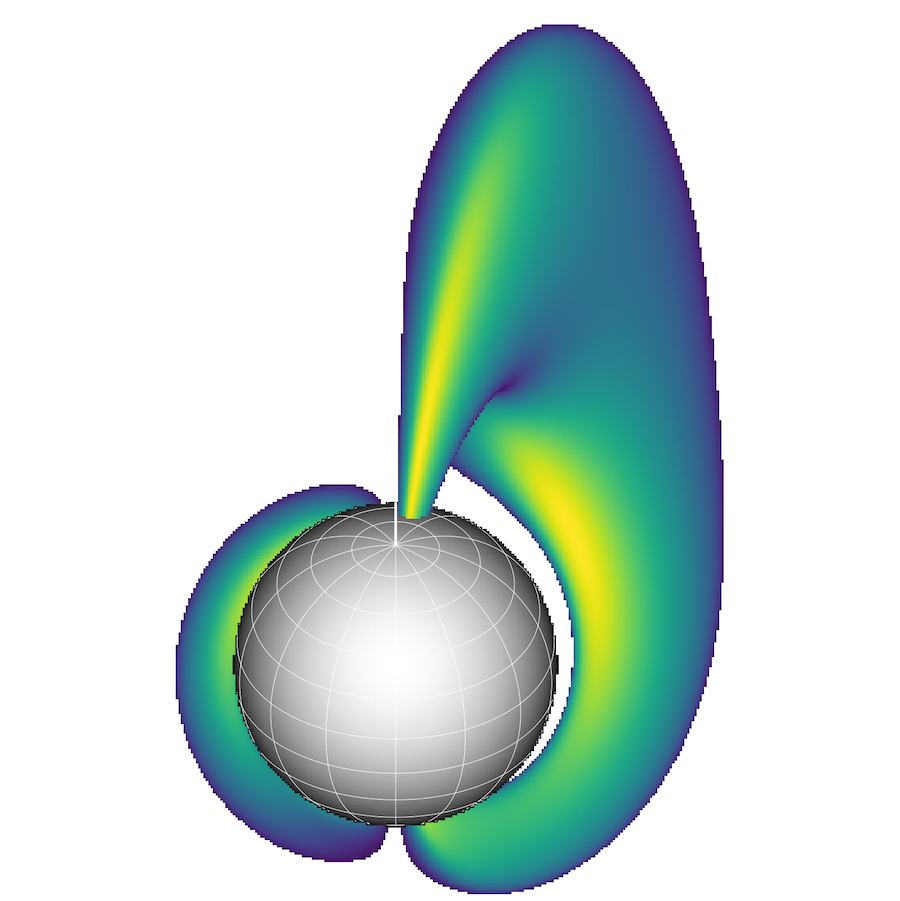} &
\includegraphics[width=0.21\textwidth]{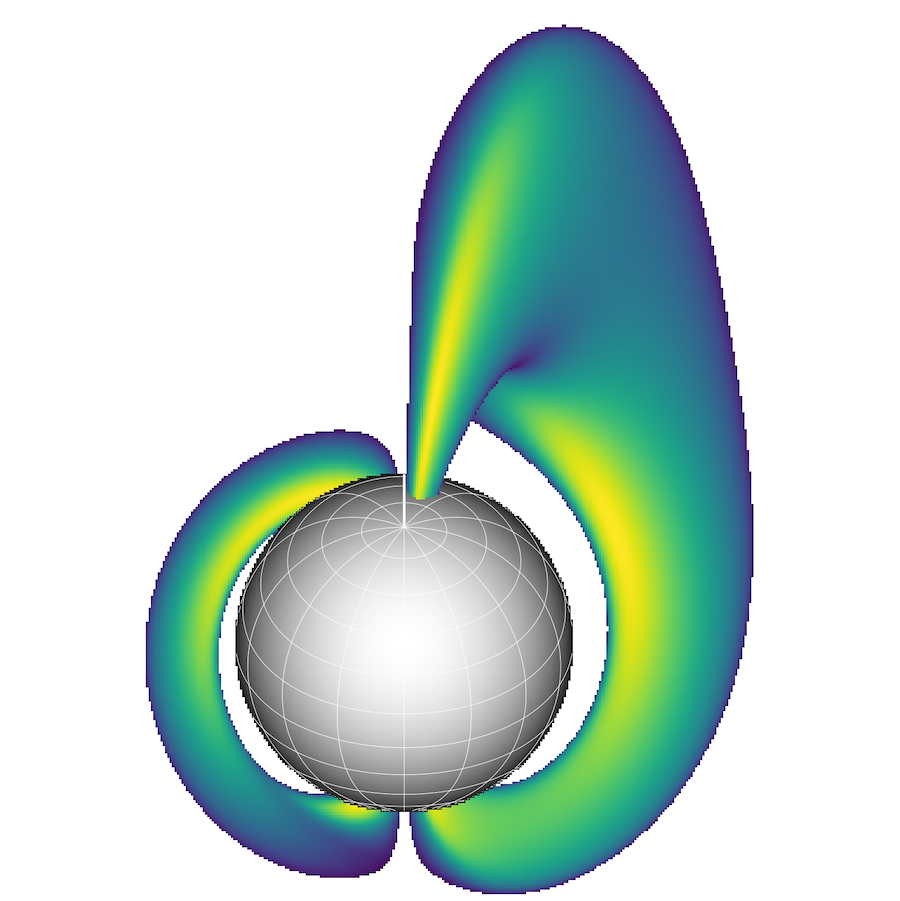} &
\includegraphics[width=0.21\textwidth]{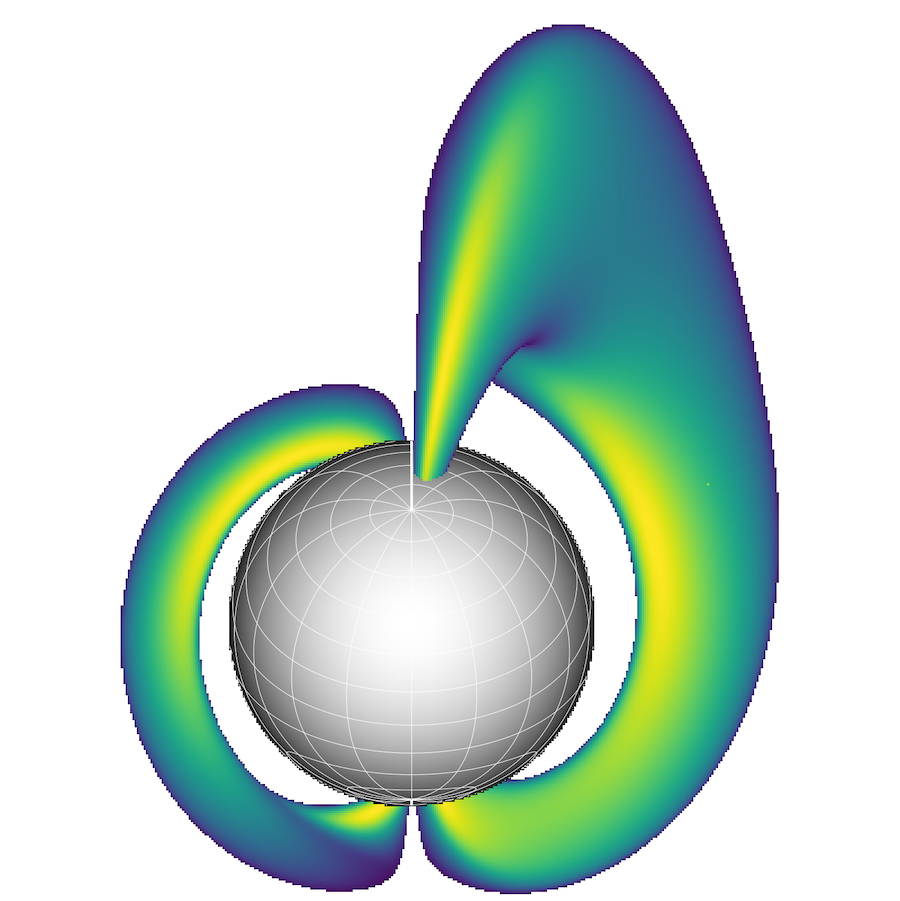} \\
\includegraphics[width=0.08\textwidth]{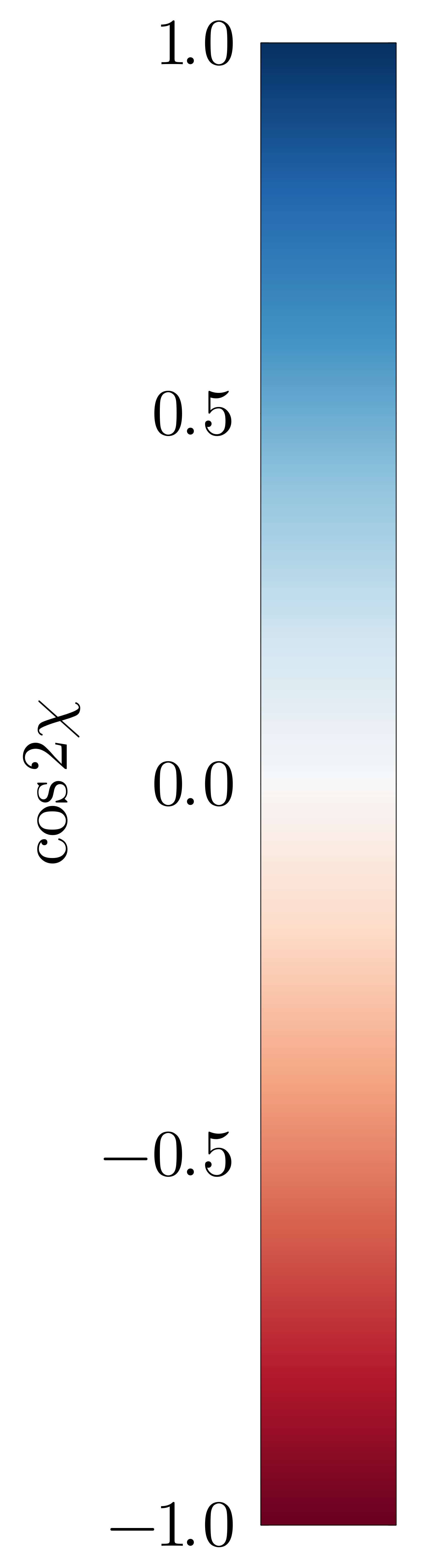} & 
\includegraphics[width=0.21\textwidth]{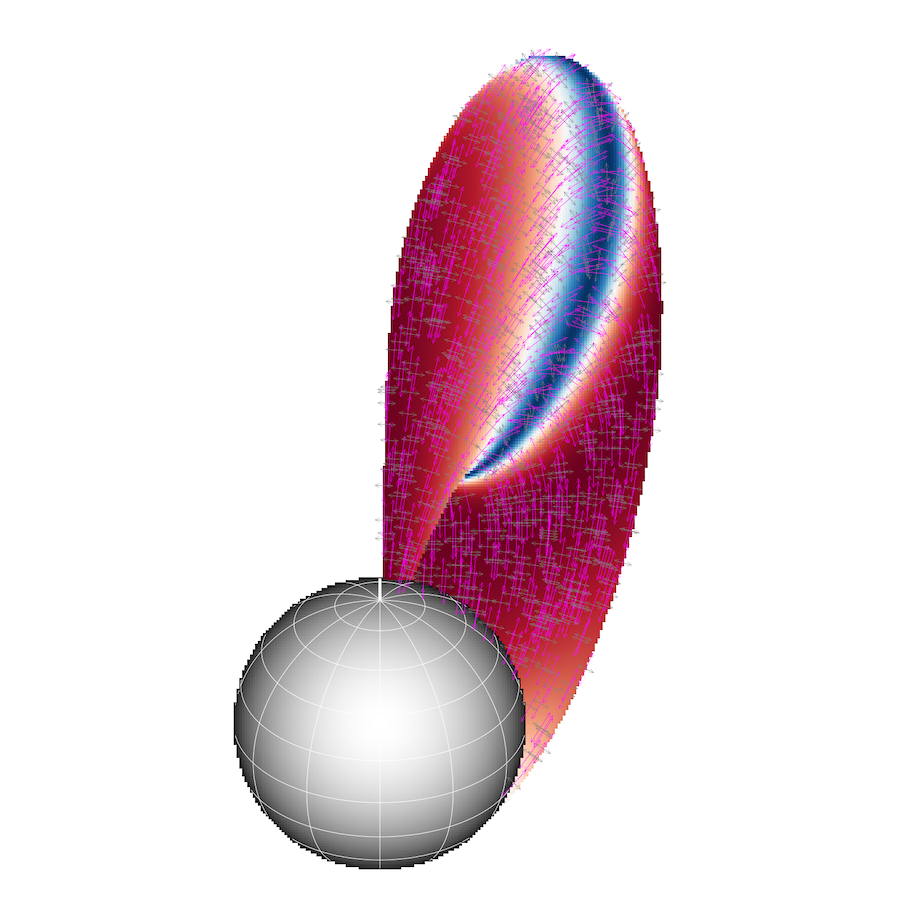} &
\includegraphics[width=0.21\textwidth]{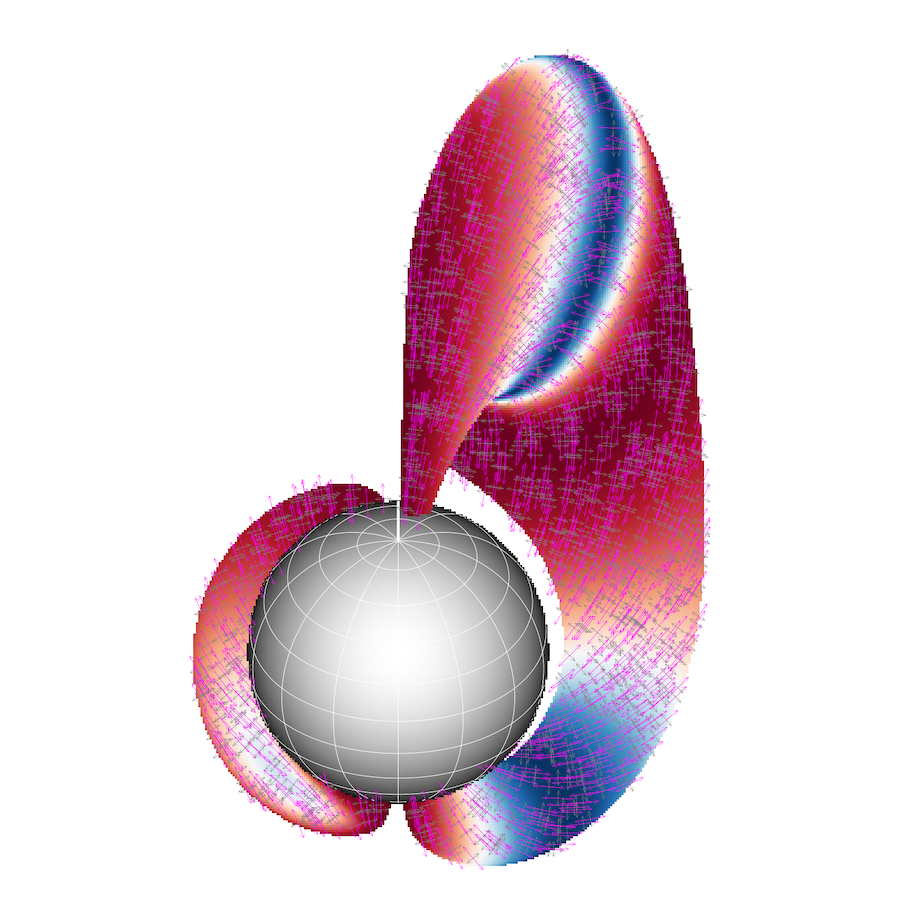} &
\includegraphics[width=0.21\textwidth]{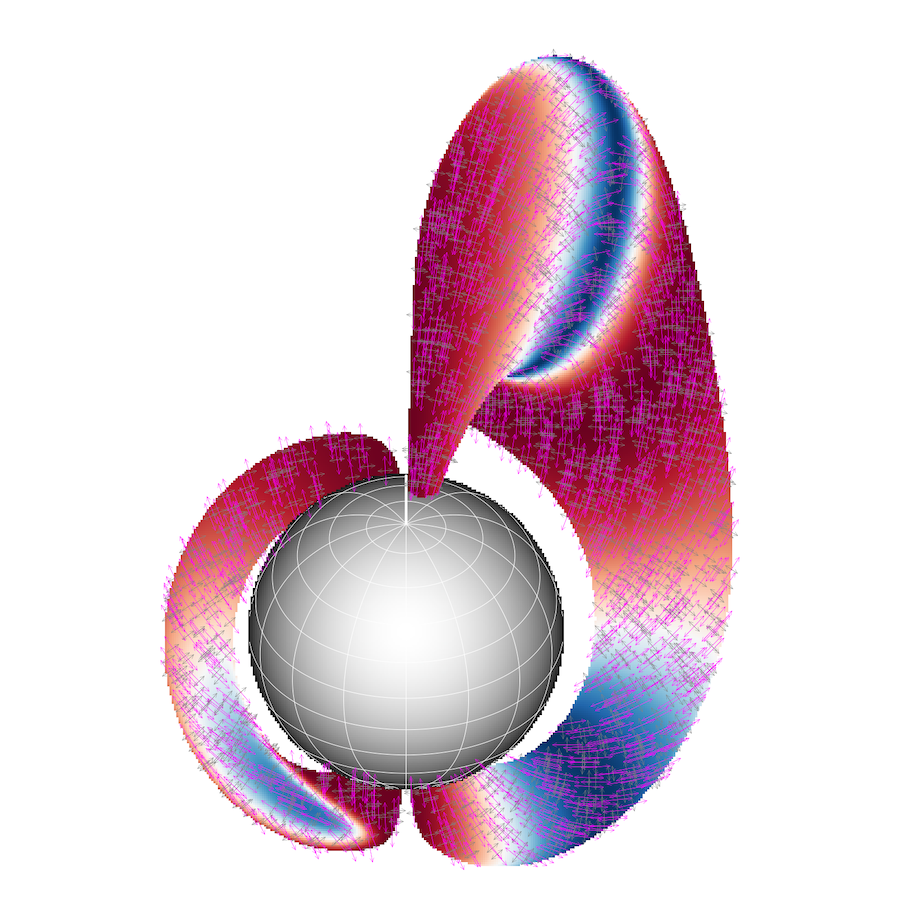} &
\includegraphics[width=0.21\textwidth]{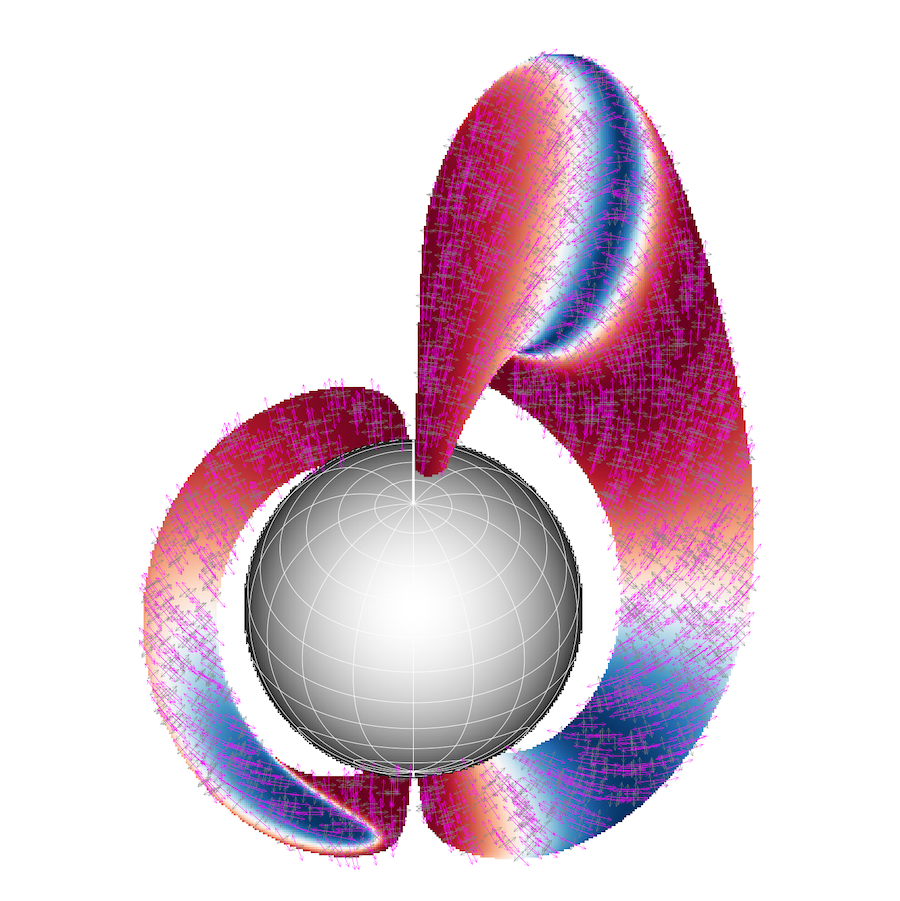} \\
\end{tabular}%
\caption{Mass (compactness) dependence of morphology and polarization angle structure for the same emitting geometry and viewing angle. Columns show $M=0$, $1.2\,M_\odot$, $1.7\,M_\odot$, and $2.2\,M_\odot$ for $r_\star = 12$~km, surveying low to high compactness regimes. Top row: surface-normal projection maps $\boldsymbol{\hat{k}\cdot\hat{n}_{\cal S}}$, depicting how lensing/visibility patterns change with stellar compactness. Bottom row: corresponding VB-off $\cos 2\chi$ maps, showing progressive distortion of polarization position angle structure. The color scale in the lower row indicates $\cos 2\chi$, along with a random sample of overlaid polarization angle vectors.}
\label{fig:Mdep}
\end{figure}

At hits on the flux tube, we compute local physical mode vectors $\hat{\boldsymbol \varepsilon}_{\perp}$ and $\hat{\boldsymbol \varepsilon}_{\parallel}$ and convert them back to contravariant components in the coordinate basis, and project with metric contractions. Here $(a_{1x},a_{1y})$ and $(a_{2x},a_{2y})$ are the IP basis components of $\hat{\boldsymbol \varepsilon}_{\perp}$ and $\hat{\boldsymbol \varepsilon}_{\parallel}$, respectively:
\begin{equation}
a_{1x}=g_{\mu\nu}\varepsilon_{(1)}^\mu \varepsilon_\perp^\nu,
\quad
a_{1y}=g_{\mu\nu}\varepsilon_{(2)}^\mu \varepsilon_\perp^\nu,
\quad
a_{2x}=g_{\mu\nu}\varepsilon_{(1)}^\mu \varepsilon_\parallel^\nu,
\quad
a_{2y}=g_{\mu\nu}\varepsilon_{(2)}^\mu \varepsilon_\parallel^\nu.
\label{eq:a1a2_projections}
\end{equation}
We then compute and record polarization angles on the IP,
\begin{equation}
\cos2\chi = a_{2x}^2-a_{2y}^2,
\qquad
\sin2\chi = 2a_{2x}a_{2y},
\label{eq:cos2chi_from_a2}
\end{equation}
since the $a_{1,2}$ are orthonormal. An example of polarization transport, and its distortion by lensing on the IP as mass increases, is depicted in Figure~\ref{fig:Mdep}.

We also compute the approximate polarization-limiting surface for VB (see \S\ref{subsec:vacuum_birefringence}) for a grid of photon energies $ \{10^{-1},10^{-0.5},\ldots,10^{3}\}\,\mathrm{keV}$ and record the IP basis projections of $\hat{\boldsymbol \varepsilon}_{\perp}$ and $\hat{\boldsymbol \varepsilon}_{\parallel}$ at the first point where the ray crosses that radius\footnote{For VB surface radii larger than $d_{\rm ip}$, these are evaluated at $d_{\rm ip}$. Given the small transverse size of the IP at $d_{\rm ip}$ compared to $d_{\rm ip}$, this flat projection of the self-similar surface at large distances is a reasonable approximation.}. These quantities are subsequently used to construct energy-dependent polarization maps.

At the location of trajectory termination\footnote{Reflection from and irradiation on the NS surface is potentially important, as recently noted by \cite{2023MNRAS.518..810D}, and may impart interesting and potentially informative signals on the IR or polarization at lower energies. This aspect is deferred to future work.}, we store all relevant quantities, including position, local surface zenith angle, angle with respect to magnetic field and photon splitting opacity (see \S\ref{sec:splitting}). At the emission point, we compute the local magnetic-field direction using the GR dipole prescription in \S\ref{sec:fluxtubes} (see Figure~\ref{fig:archetypes}). Each trajectory stands in for a multitude of photons in reality (i.e., the trajectories constitute ``macrophotons"). Gravitational redshift, emission anisotropy and normalization are incorporated by adopting weights for assigned energies of macrophotons on the IP \citep{2014ApJ...792...87P} -- see \S\ref{sec:imageplane_sum}. 

\subsection{Photon Splitting}
\label{sec:splitting}

Photon splitting $\gamma\rightarrow \gamma\gamma$ is a third-order QED process in which, in a strong magnetic field, a photon splits into two nearly collinear photons while conserving energy and momentum. Furry's theorem (charge conjugation symmetry) prevents this process in field-free conditions. Photon splitting in strong fields has no threshold and will operate well below the single-photon pair threshold of 1 MeV. This makes photon splitting particularly relevant for magnetar bursts, both deep inside the fireball, where it redistributes energy and photon number \citep[e.g.][]{1992herm.book.....M,1995MNRAS.275..255T}, and for the polarization structure of emergent radiation through splitting attenuation in the magnetosphere. The latter is what we implement in our model.

In the weakly dispersive limit of VB (\S\ref{subsec:vacuum_birefringence}), charge-parity invariance of QED admits three splitting channels, $\perp \rightarrow \parallel \parallel$, $\parallel\rightarrow\parallel\perp \mathrm{ or } \perp \parallel$, and $\perp\rightarrow \perp \perp$. Kinematic selection rules in the weak dispersion limit further limit splitting to $\perp \rightarrow \parallel \parallel$ below pair threshold \citep{1971AnPhy..67..599A}. In the non-linear regime of QED where VB is strong and weak dispersion no longer holds, i.e., when locally $B \gtrsim 100$ (see \S\ref{subsec:vacuum_birefringence}), it is not known which modes split below pair creation threshold (e.g., all allowed by CP symmetry). For this reason, we allow for a switch between no photon splitting (for controlled comparisons), $\perp \rightarrow \parallel \parallel$ only, and all three (see below).

This polarization selectivity is potentially important for magnetar SBs, because attenuation by photon splitting modifies the spectrum and polarization of emission reaching a distant observer \citep[e.g.,][]{1991A&A...249..581B,1995ApJ...440L..69B}. This comes with the proviso of neglecting plasma effects to dispersion. In particular, resonant corrections at very large $B$, or plasma contributions in dense surface layers and near cyclotron resonance, may modify the propagation eigenmodes and hence the effective splitting selection rules. Such circumstances are relevant near the base of a bursting magnetosphere, where vacuum and plasma dispersion might become comparable (if baryons are present).

We implement photon splitting as an energy and polarization state dependent on-the-spot opacity, i.e., taking it as only an absorption term in polarized radiative transfer of Stokes vectors. For each trajectory or ray $i$ from the IP to the burst, we compute an optical depth through attenuation coefficients $\cal{R}$ evaluated locally along the ray from the IP $l_{\rm ip}$ to the emission point $l_{\rm e}$,
\begin{equation}
   \tau_i  \;=\; \int_{l_{\rm ip}}^{l_{\rm e}} {\cal{R}} \,dl \quad ,   
 \label{eq:opt_depth}
\end{equation}
with attenuation coefficients \citep{1991A&A...249..581B} that apply well below pair threshold (i.e., $\omega\ll 2$) as follows:
\begin{equation}
{\cal R}_{\perp\to\parallel\parallel}
=
\frac{\alpha_{\rm f}^3}{60\pi^2 \lambar_c}\,
\omega^5\, B^6\, {\cal M}_1^2\, \sin^6\theta_{kB}
\label{eq:split_pol_perp}
\end{equation}
for the $\perp \rightarrow \parallel \parallel$ channel, and 
\begin{equation}
{\cal R}_{\rm ave}
=
\frac{\alpha_{\rm f}^3}{120\pi^2\lambar_c}\,
\omega^5\, B^6\,
\left( 3{\cal M}_1^2 + {\cal M}_2^2 \right)\sin^6\theta_{kB}
\, .
\label{eq:split_pol_ave}
\end{equation}
for all three modes allowed by CP symmetry. Here $\lambar_c=\hbar/(m_e c)$ is the reduced Compton wavelength. This form, separable in energy $\omega$ and $B$, is convenient -- we set $\omega = 1$ and rescale the optical depth by the emergent photon energy as measured at the burst photosphere.   We note that in highly supercritical field domains, $B\gg 1$, the attenuation coefficients exhibit a weaker dependence on $B$, a nuance that does not significantly impact the emphases of this paper.

The one-loop amplitudes ${\cal M}_\sigma$ in a strong field below pair threshold are
\begin{equation}
{\cal M}_{\sigma}
=
\frac{1}{B^4}\int_{0}^{\infty} \frac{dy}{y} e^{-y/B}\,\Lambda_{\sigma}(y) \, \qquad \sigma \in \{1,2\},
\label{eq:calM_i_form}
\end{equation}
with
\begin{eqnarray}
\Lambda_1(y) &=&
\left(-\frac{3}{4y}+\frac{y}{6}\right)\frac{\cosh y}{\sinh y}
+\frac{3+2y^2}{12\sinh^2 y}
+\frac{y\cosh y}{2\sinh^3 y}
\, ,
\nonumber\\[-5.5pt]
\label{eq:Lambda_s_def}\\[-5.5pt]
\Lambda_2(y) &=&
\frac{3}{4y}\frac{\cosh y}{\sinh y}
+\frac{3-4y^2}{4\sinh^2 y}
-\frac{3y^2}{2\sinh^4 y}
\, .
\nonumber
\end{eqnarray}
We implement lookup tables for the splitting amplitude functions ${\cal{M}}_1(B)$ and ${\cal{M}}_2(B)$ with asymptotic forms for $B \geq 100$, setting $\omega = 1$. The path-length element in Eq.~(\ref{eq:opt_depth}) is computed with the spatial projector for static observers, $u^\mu=(1/\sqrt{-g_{tt}},0,0,0)$ and $h_{\mu\nu}=g_{\mu\nu}+u_\mu u_\nu$, giving $dl^2=h_{\mu\nu}dx^\mu dx^\nu$. The cumulative splitting depth for each ray is accumulated on-the-fly during the geodesic integration over the whole trajectory. Numerically, this is a trapezoidal sum over successive geodesic integration steps,
\begin{equation}
\tau \approx \sum_n \frac{1}{2}\left[ {\cal R}(x_n)+{\cal R}(x_{n-1})\right]\Delta l_n,
\qquad
\Delta l_n = \left[h_{\mu\nu}\Delta x^\mu_n\Delta x^\nu_n\right]^{1/2}.
\label{eq:tau_trap}
\end{equation}

We store two dimensionless path integrals, one for the effective average mode and one for the pure $\perp\rightarrow\parallel\parallel$ channel, both evaluated at the
dimensionless reference energy $\omega=1$, corresponding to
$\hbar \omega^\prime=m_e c^2 \approx 511$~keV. Because the attenuation coefficients scale as $\omega^5$, the physical optical depth for an emitted photon of energy $\omega_{\rm e}$ is recovered in post-processing via $\tau(\omega_{\rm e}) = \tau(1)\,\omega_{\rm e}^5$. This is the form used when computing emergent spectra and Stokes parameters in \S\ref{sec:imageplane_sum}. Here $\omega_e$ is the photon energy evaluated locally by a static observer at the fireball position -- incorporating decreasing photon redshift along the trajectory is a small effect that slightly increases transparency, but breaks the convenient rescaling procedure.

\begin{figure}[htbp]
  \centering
  \setlength{\tabcolsep}{0pt}
  \renewcommand{\arraystretch}{0.5} 
  \begin{tabular}{ccc}
    \includegraphics[width=0.333\textwidth]{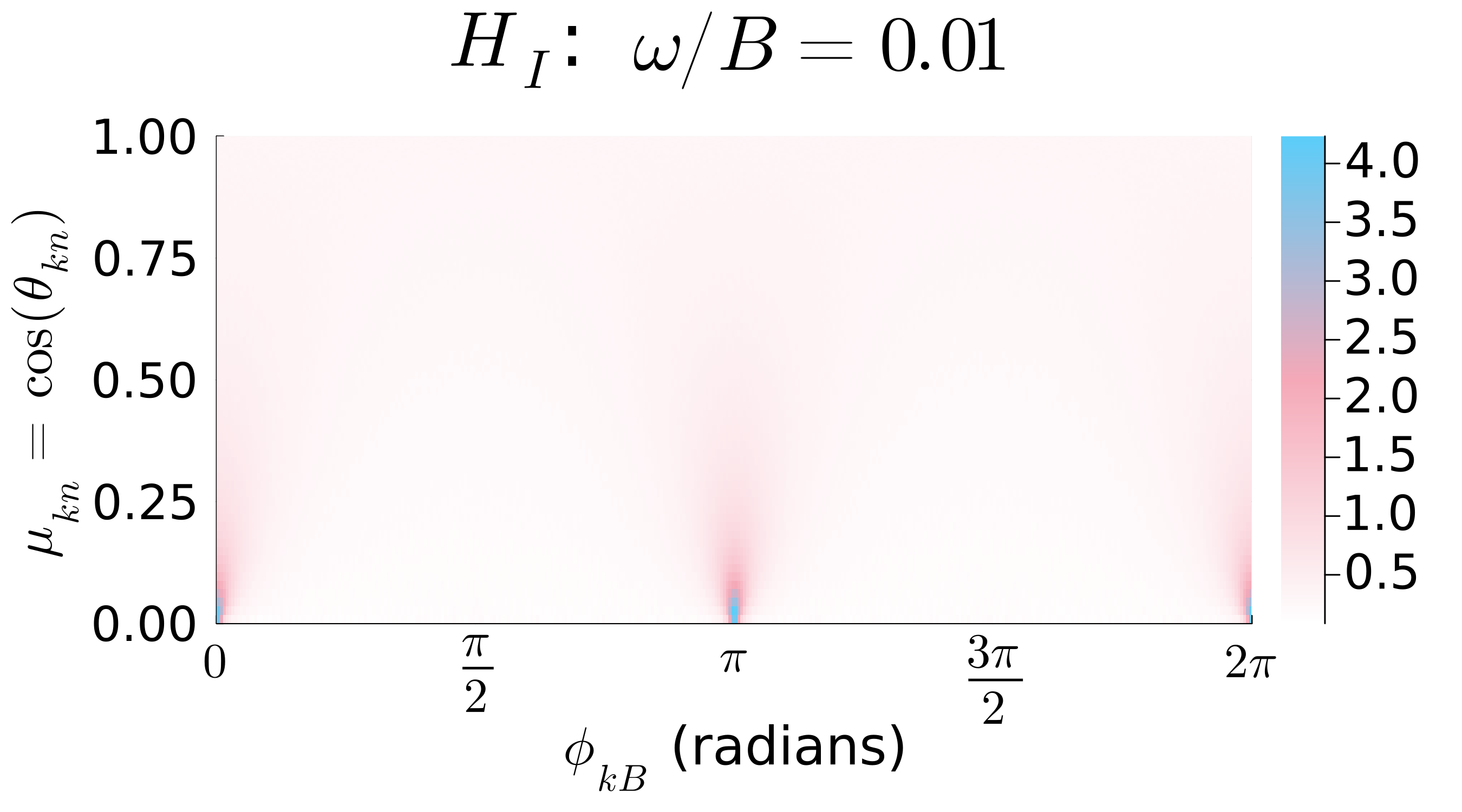} &
    \includegraphics[width=0.333\textwidth]{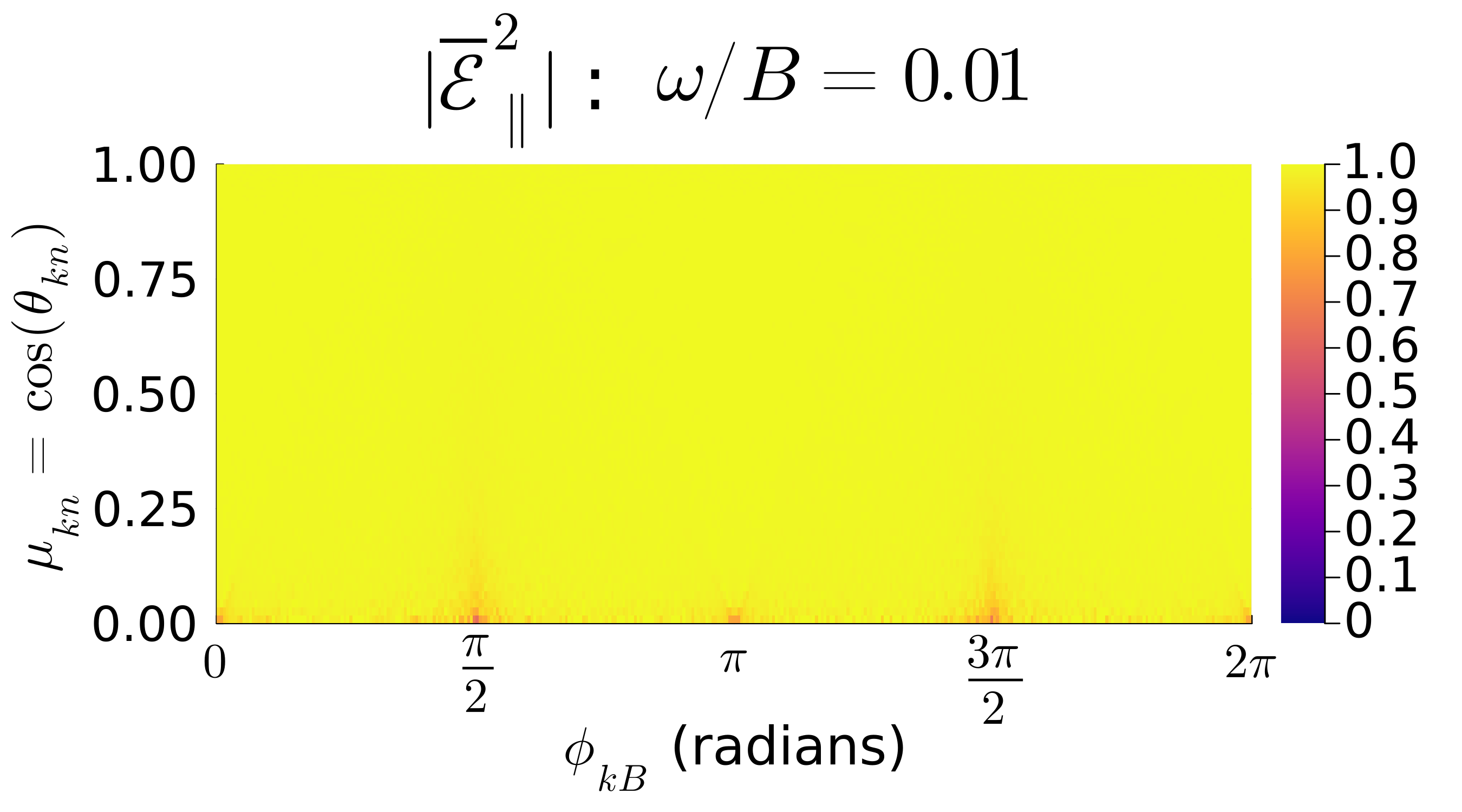} &
    \includegraphics[width=0.333\textwidth]{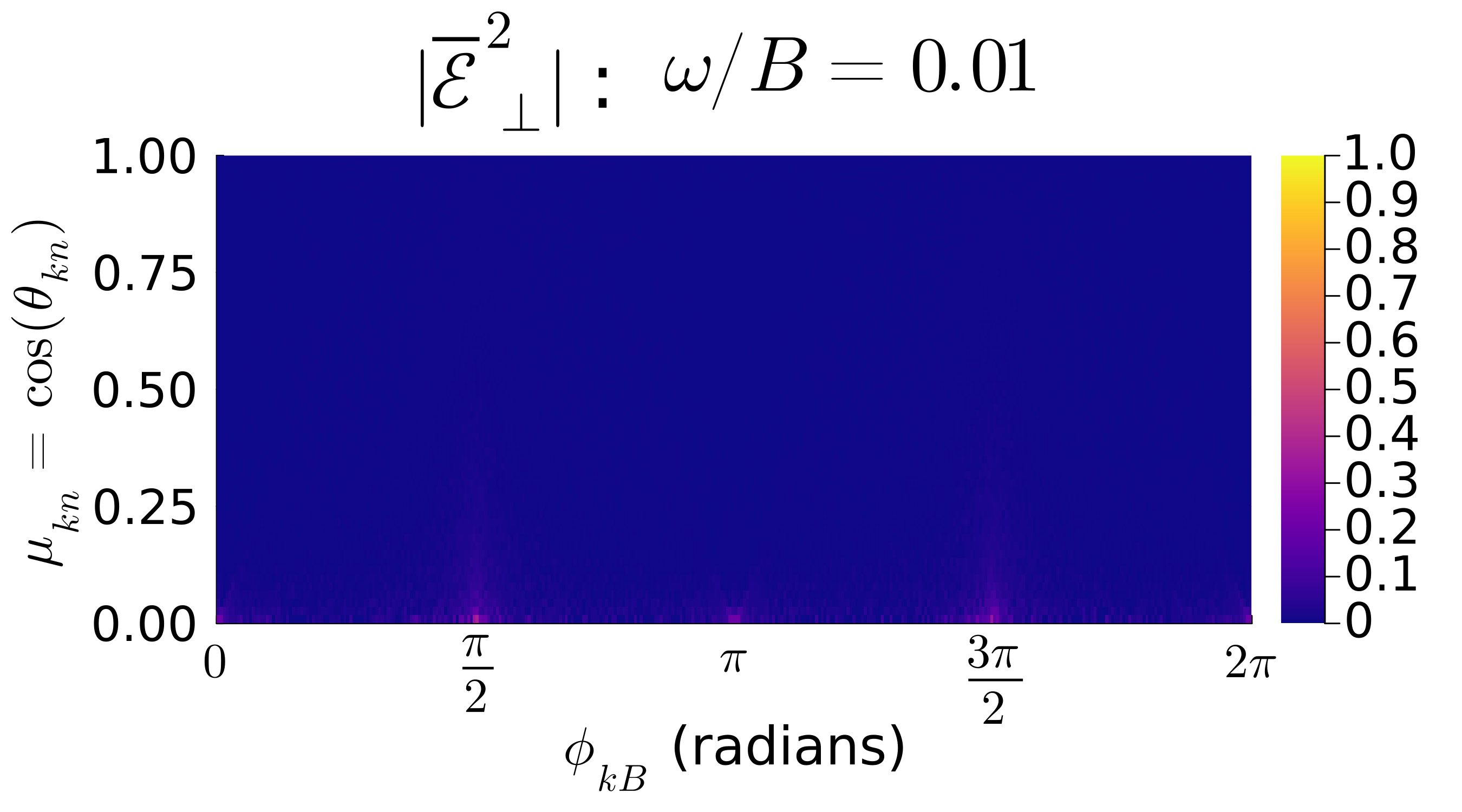} \\
    \includegraphics[width=0.333\textwidth]{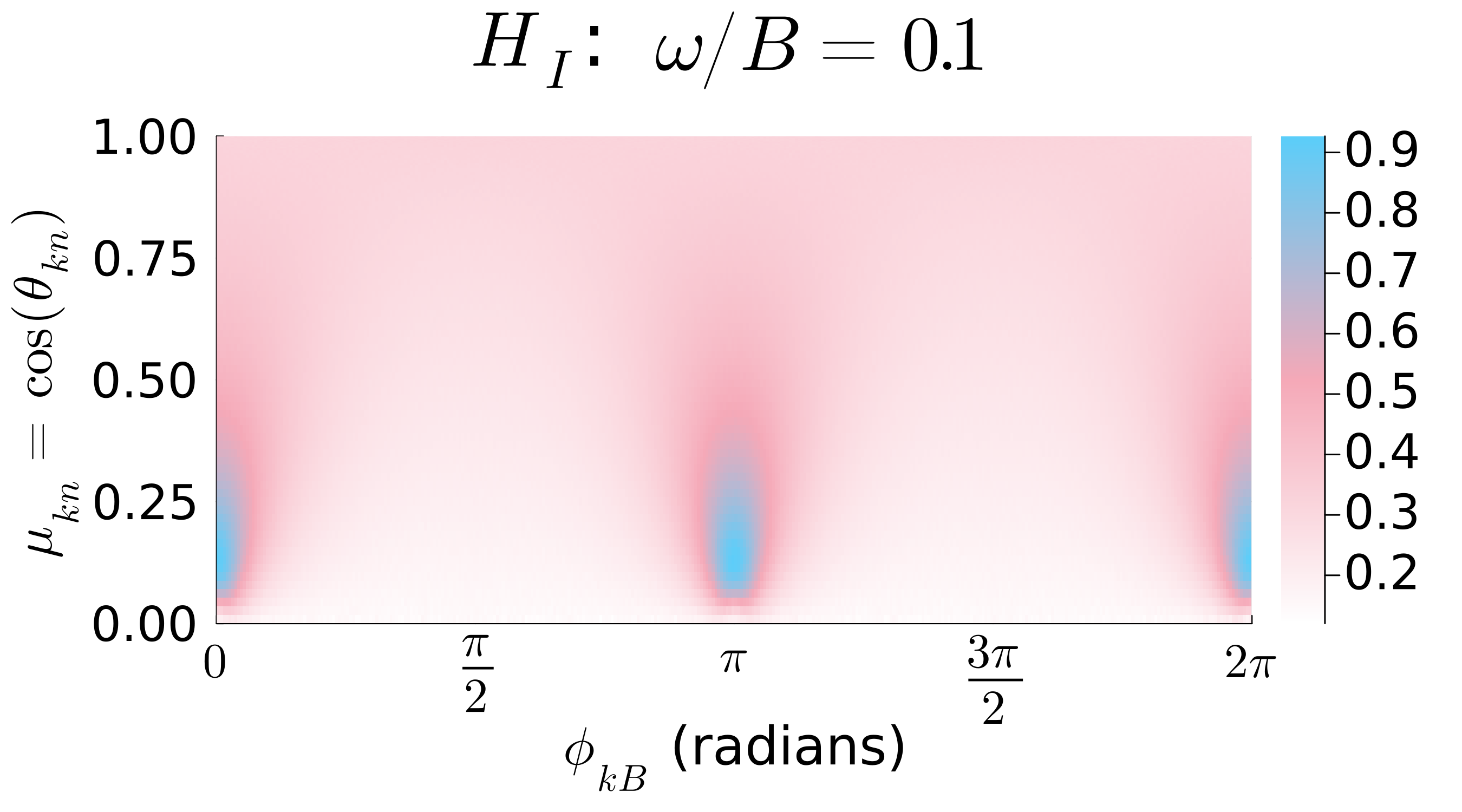} &
    \includegraphics[width=0.333\textwidth]{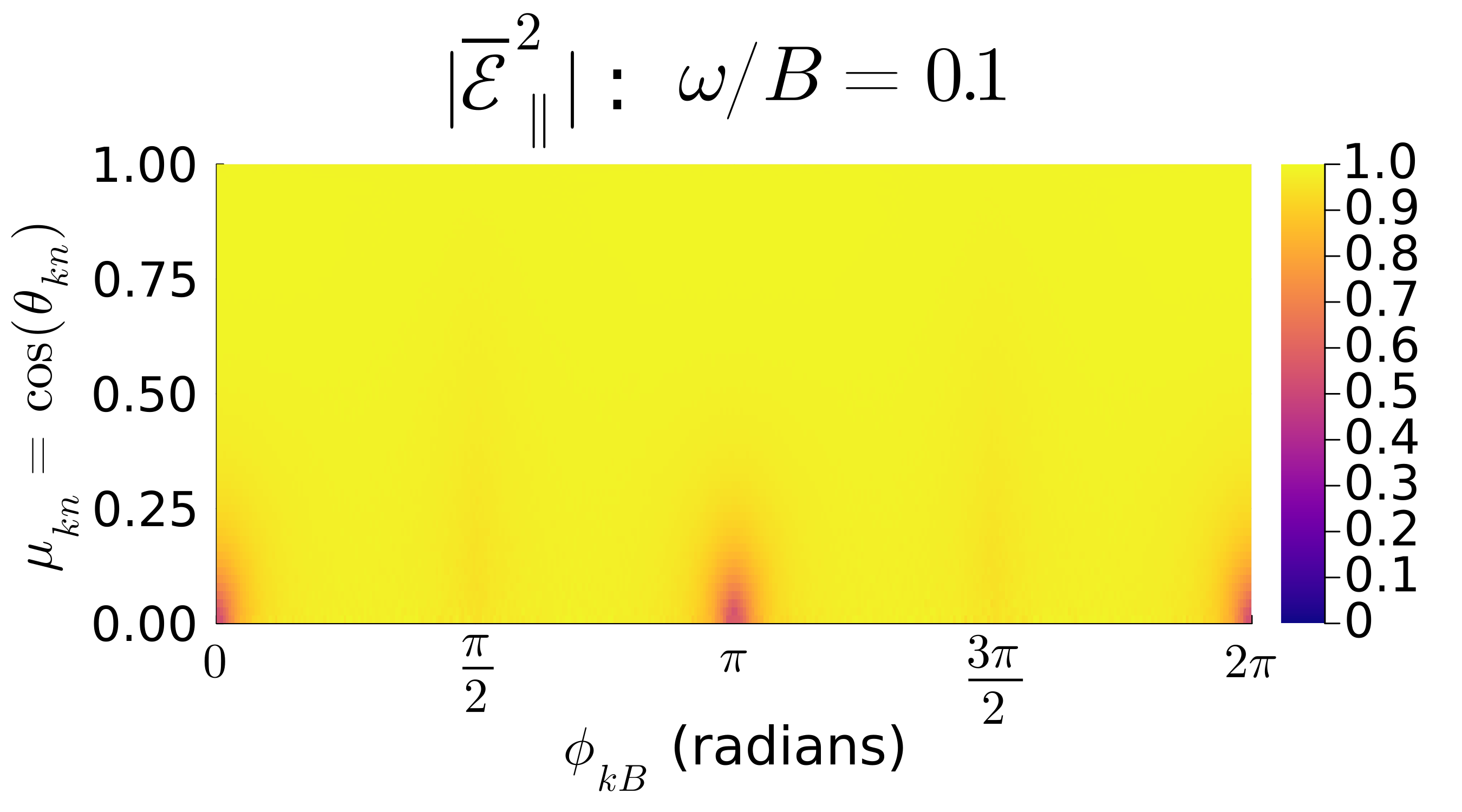} &
    \includegraphics[width=0.333\textwidth]{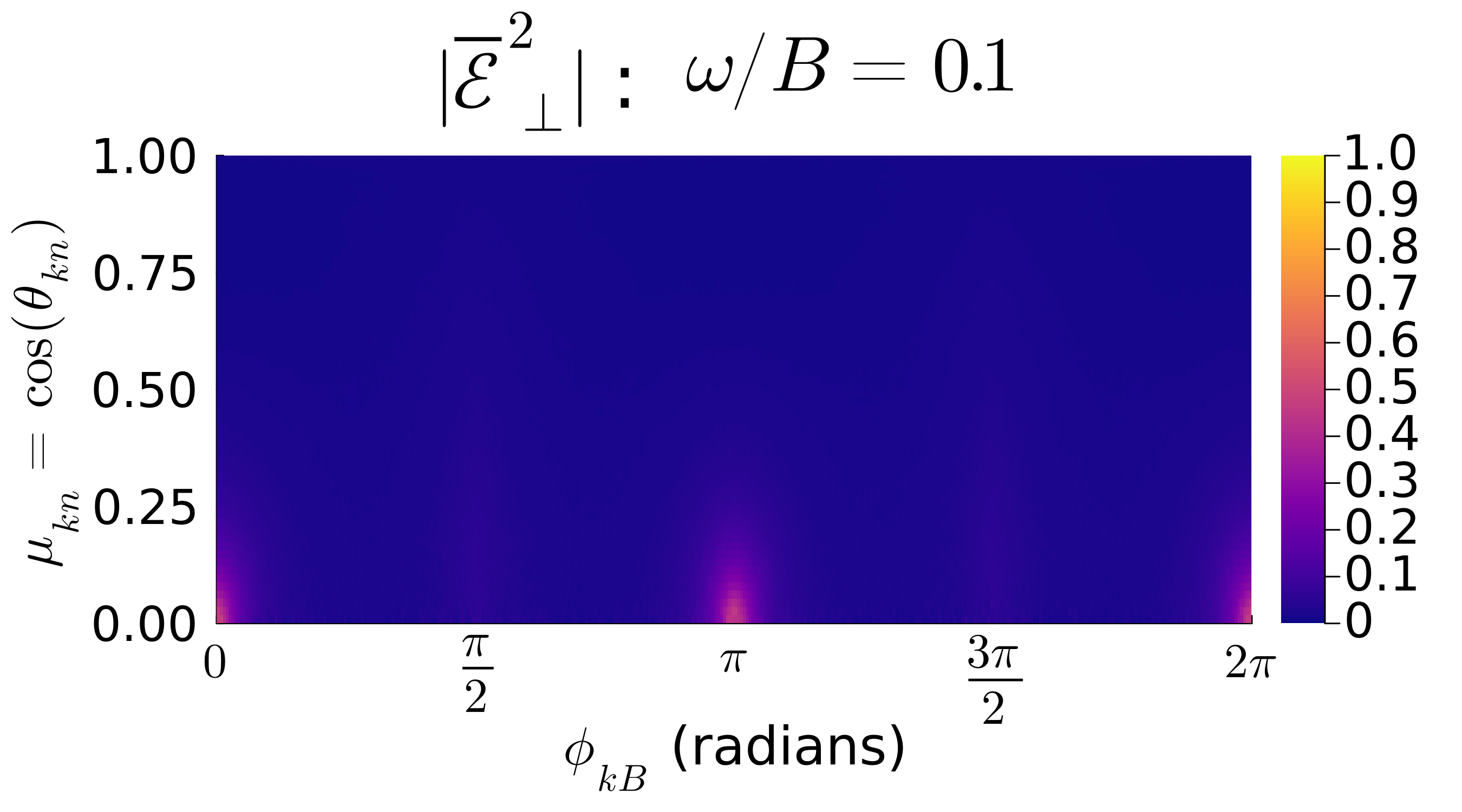} \\
    \includegraphics[width=0.333\textwidth]{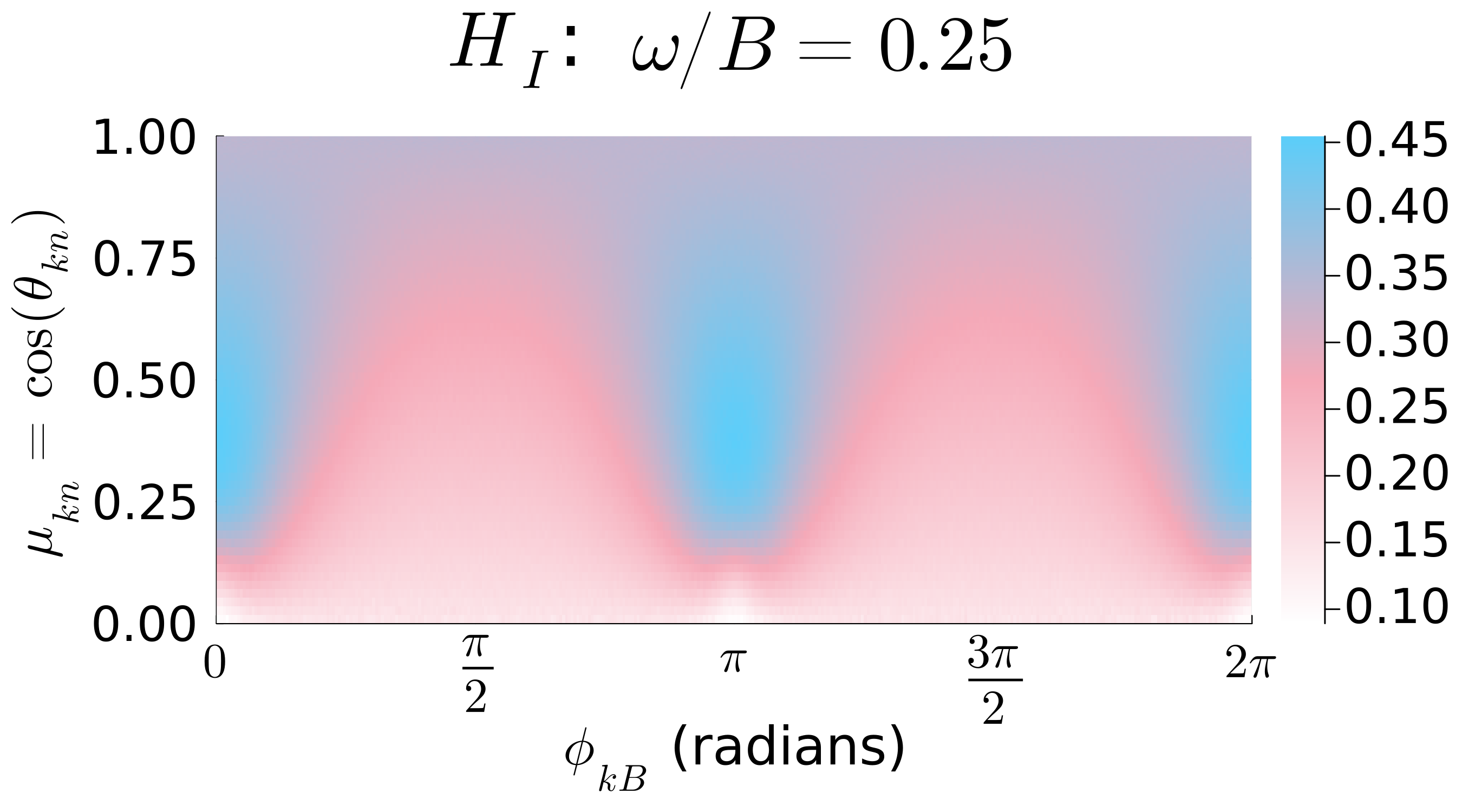} &
    \includegraphics[width=0.333\textwidth]{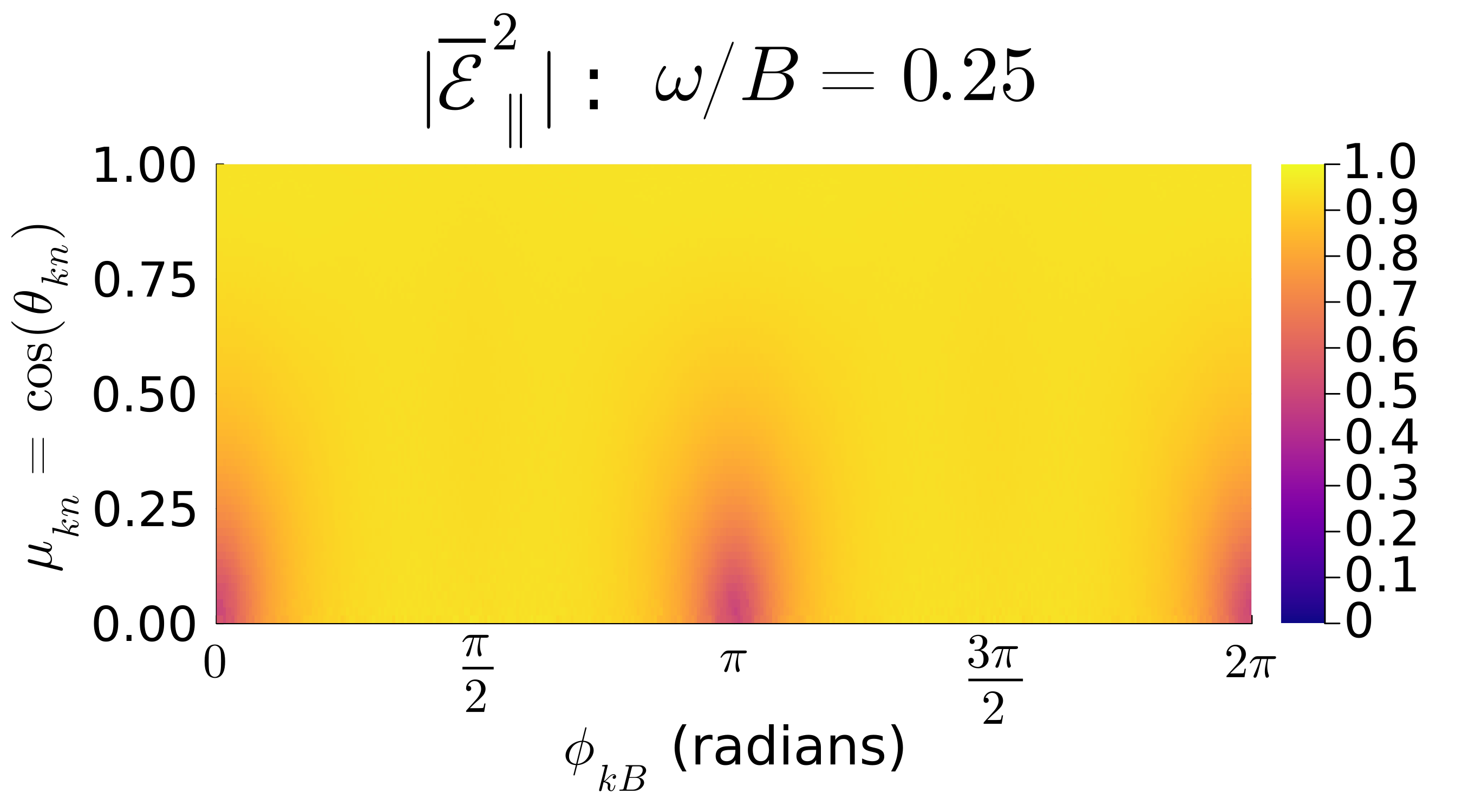} &
    \includegraphics[width=0.333\textwidth]{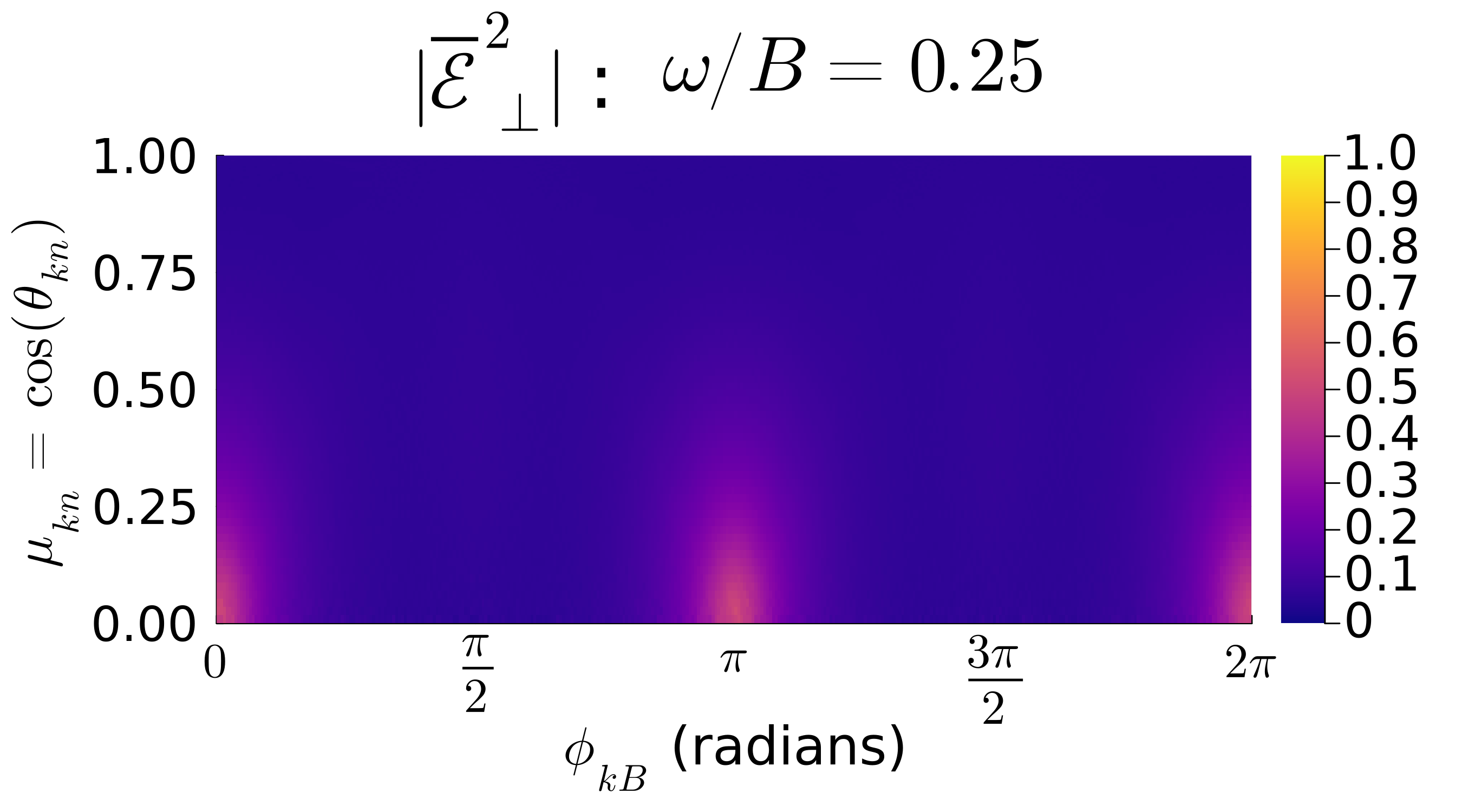} \\
    \includegraphics[width=0.333\textwidth]{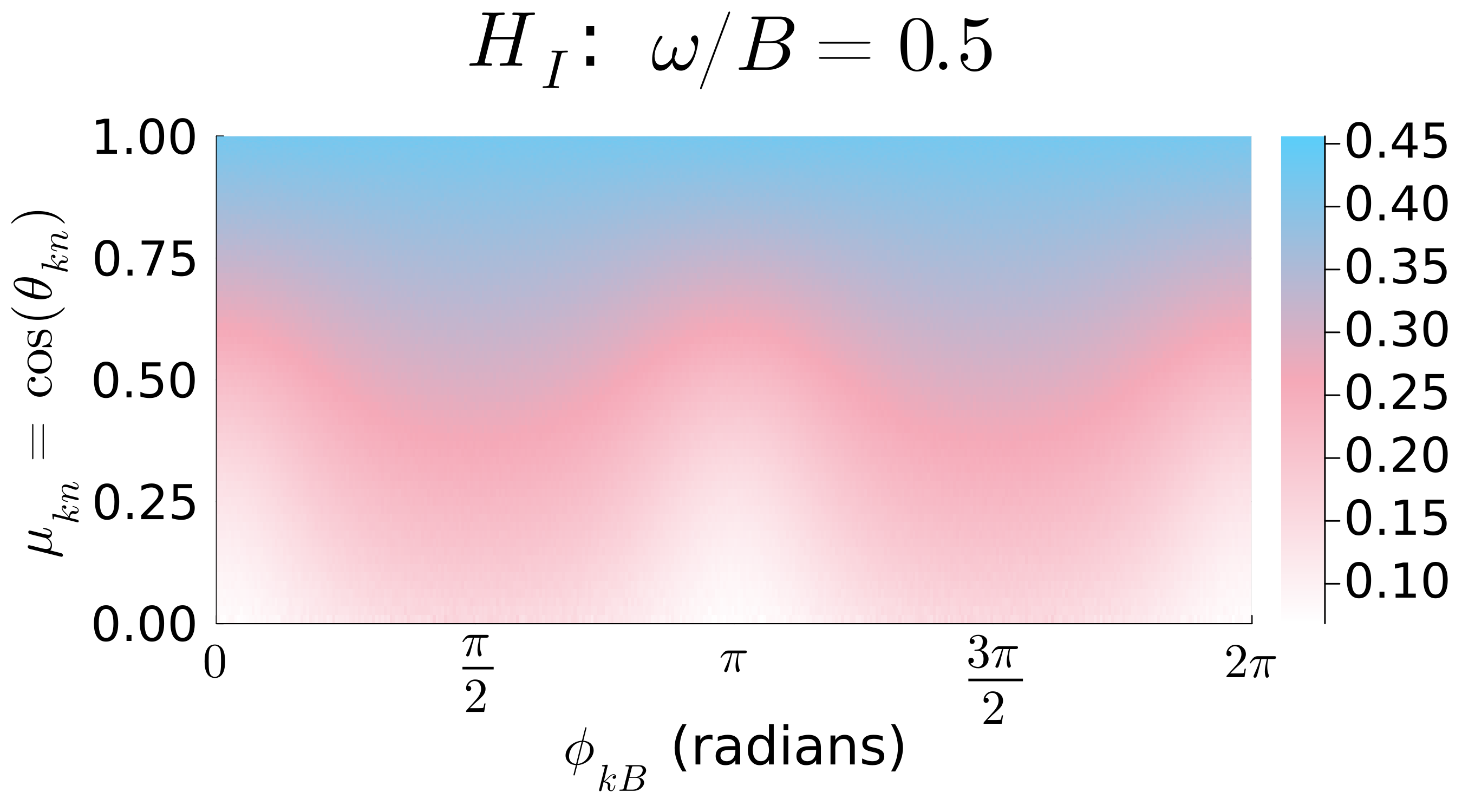} &
    \includegraphics[width=0.333\textwidth]{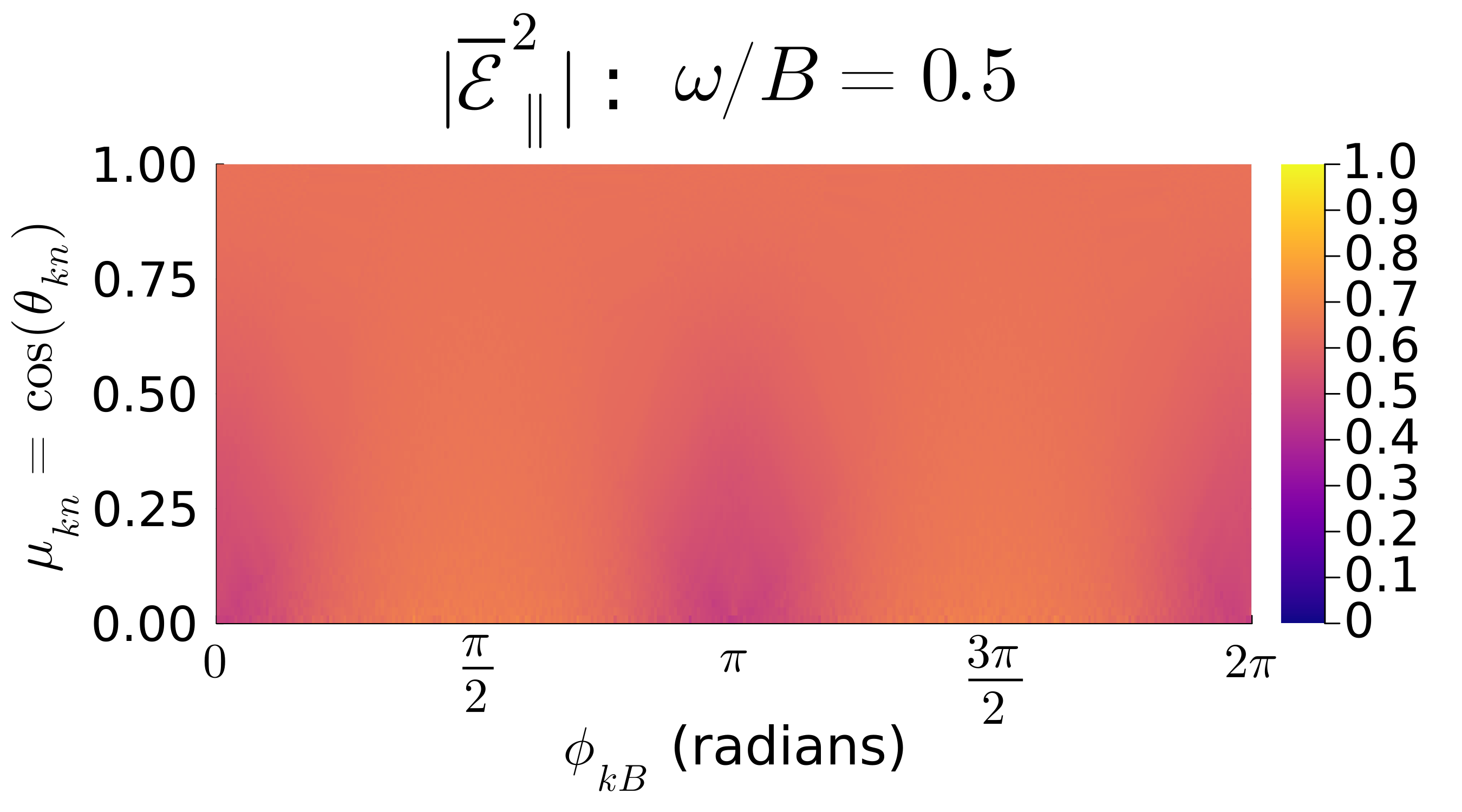} &
    \includegraphics[width=0.333\textwidth]{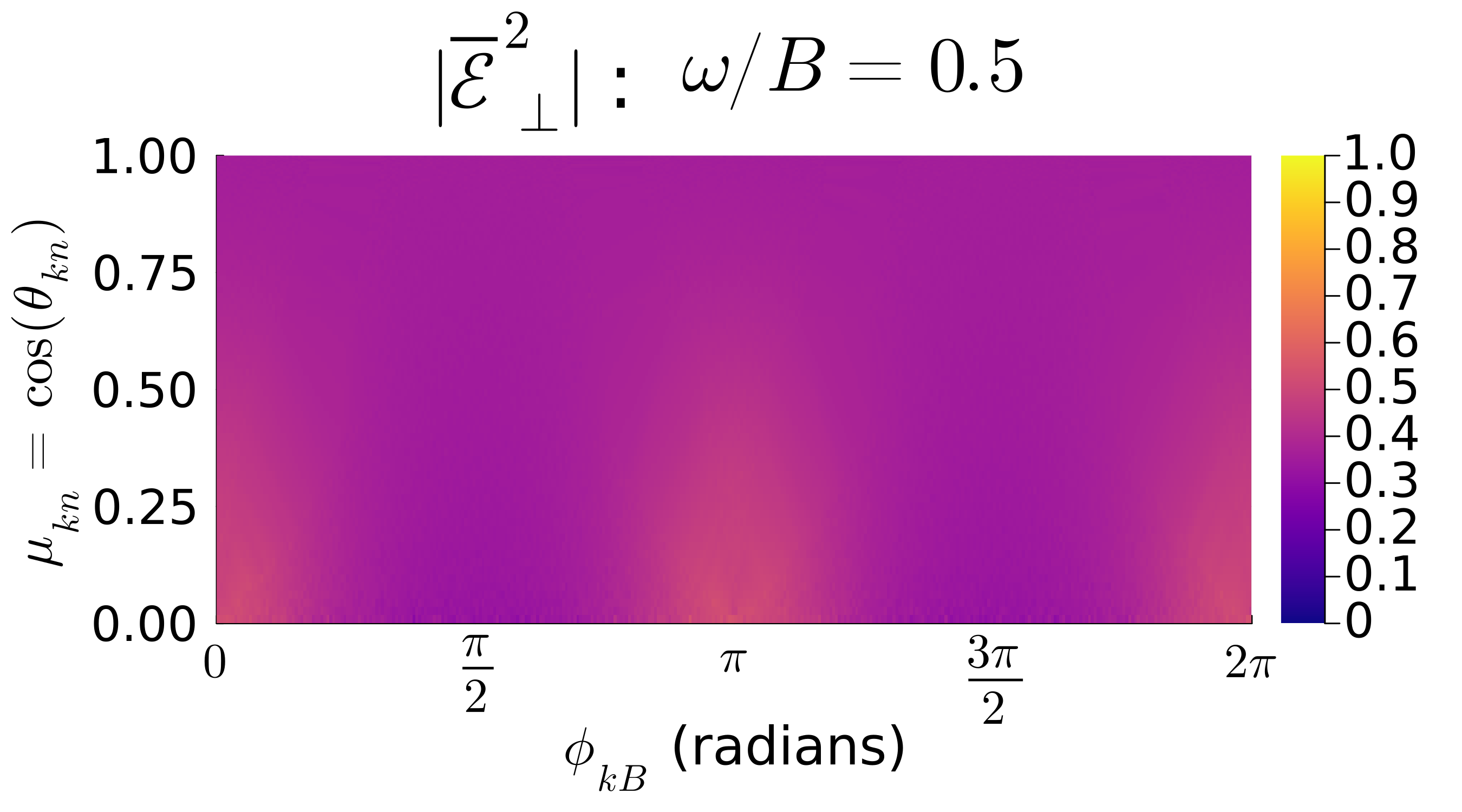} \\
    \includegraphics[width=0.333\textwidth]{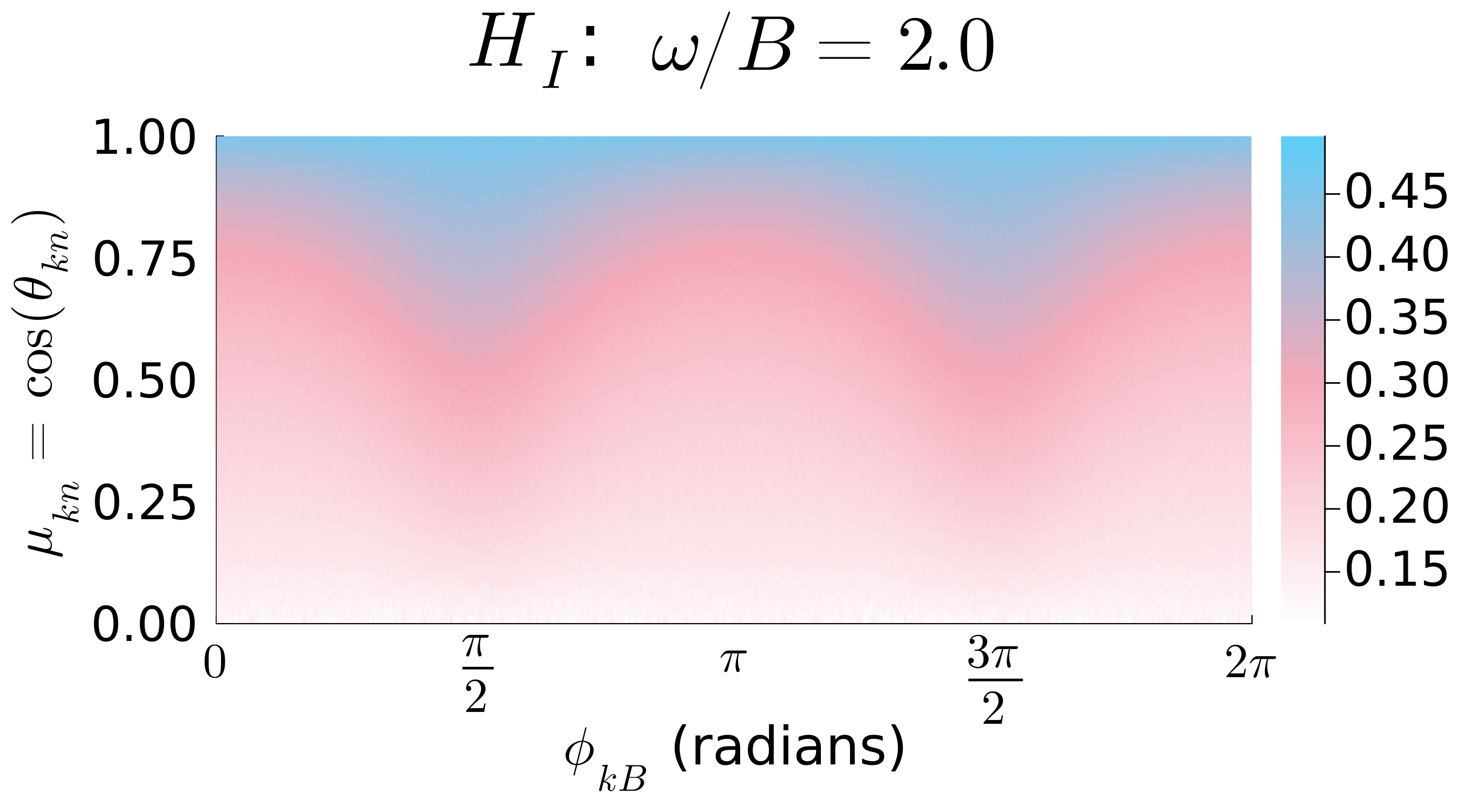} &
    \includegraphics[width=0.333\textwidth]{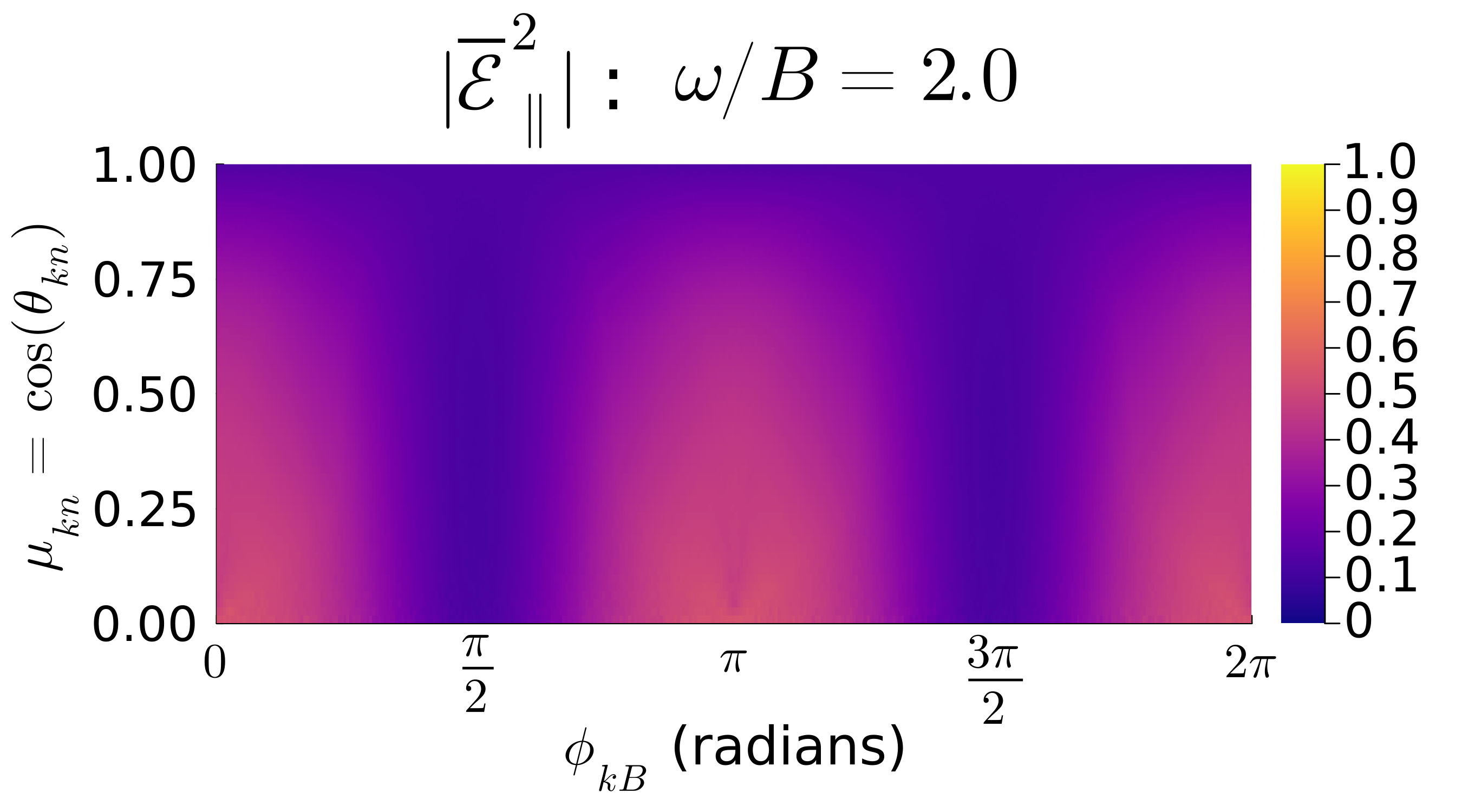} &
    \includegraphics[width=0.333\textwidth]{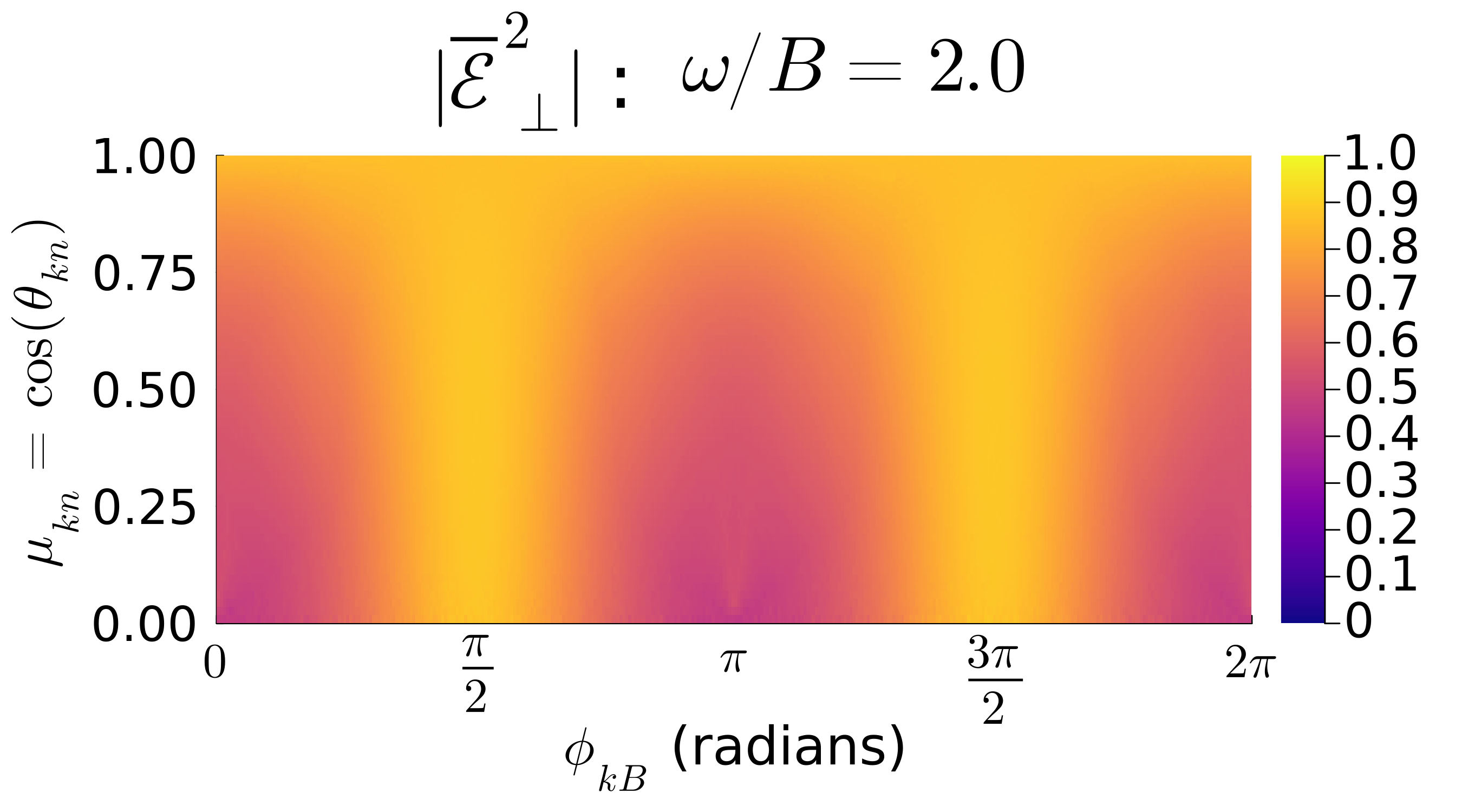} \\
    \includegraphics[width=0.333\textwidth]{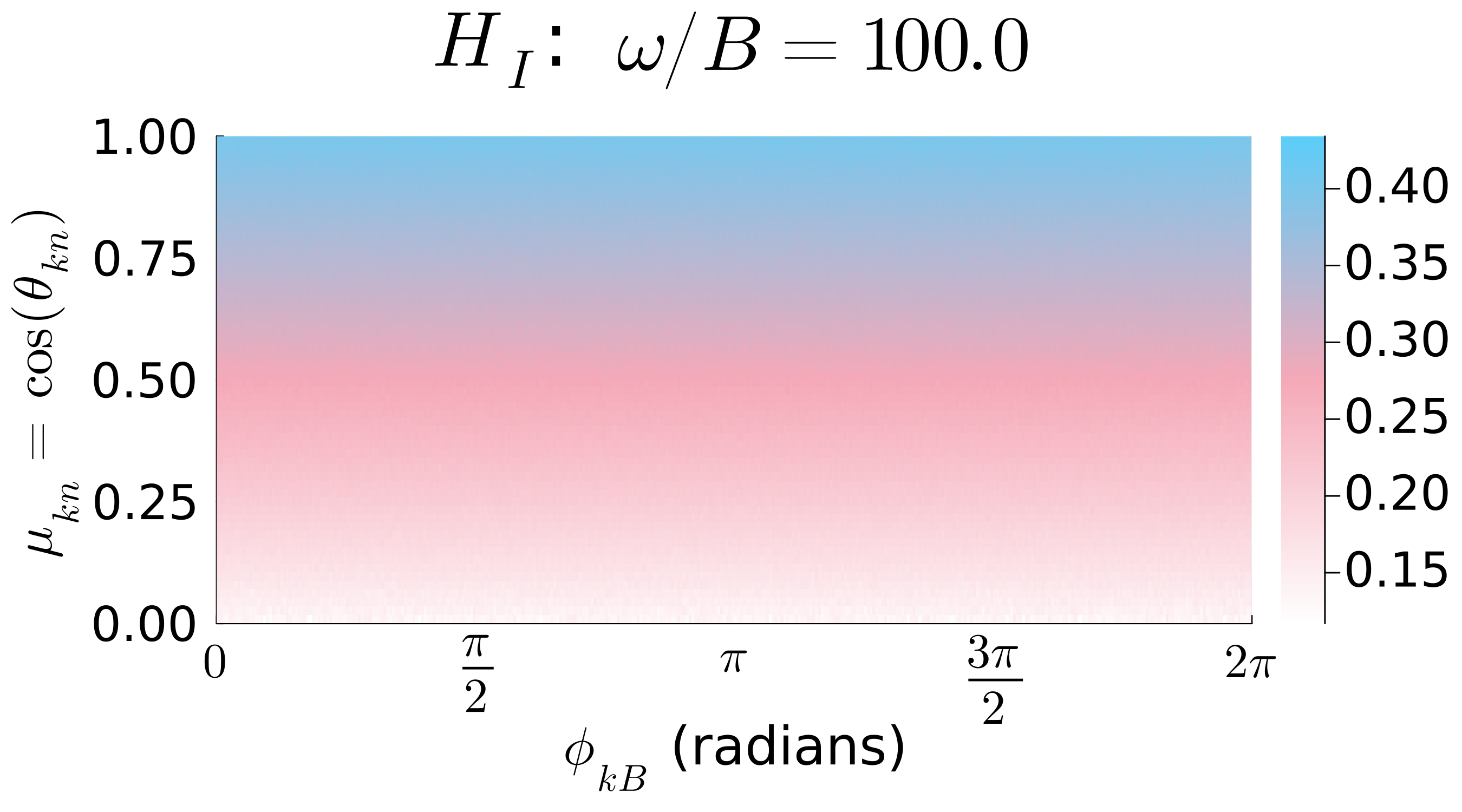} &
    \includegraphics[width=0.333\textwidth]{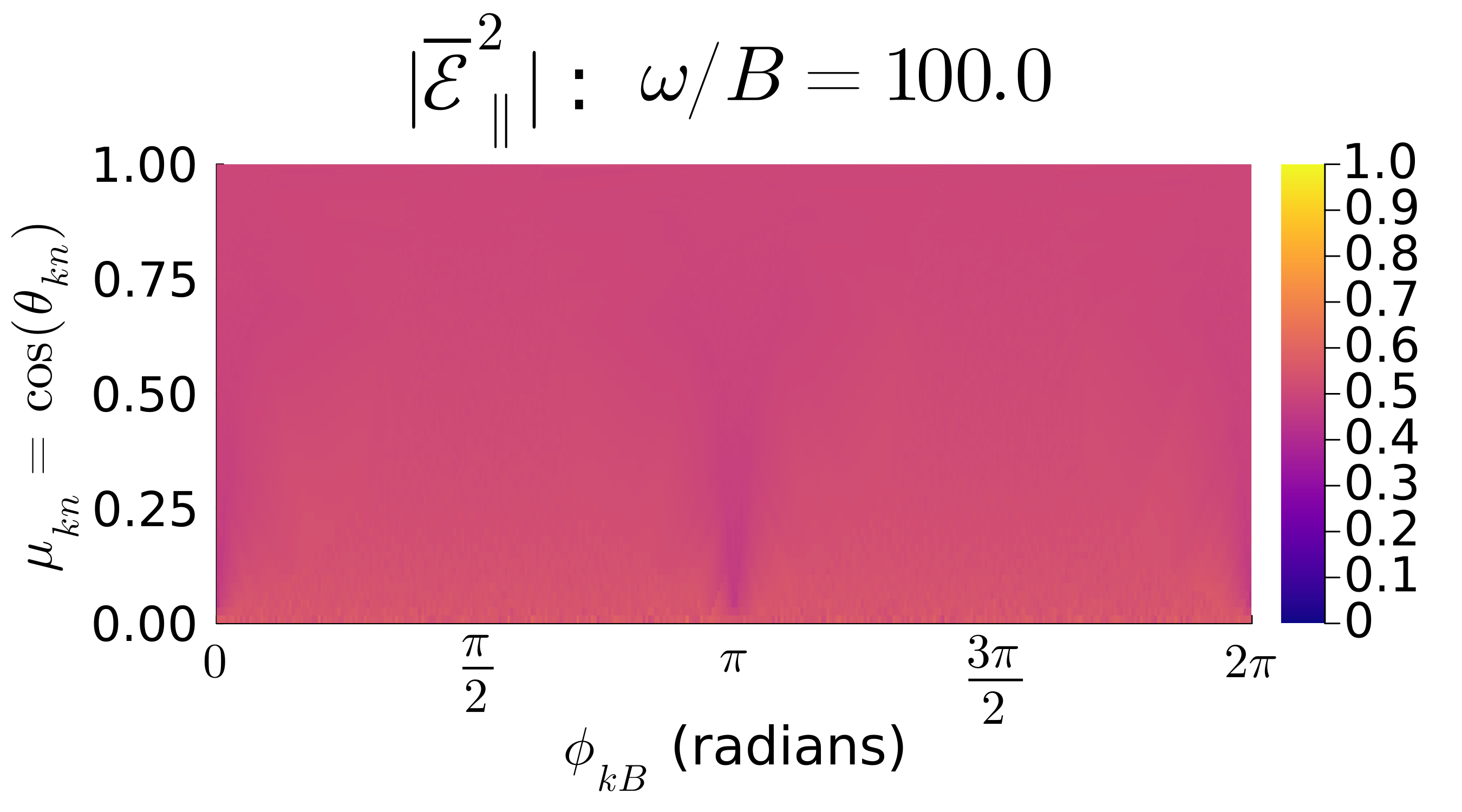} &
    \includegraphics[width=0.333\textwidth]{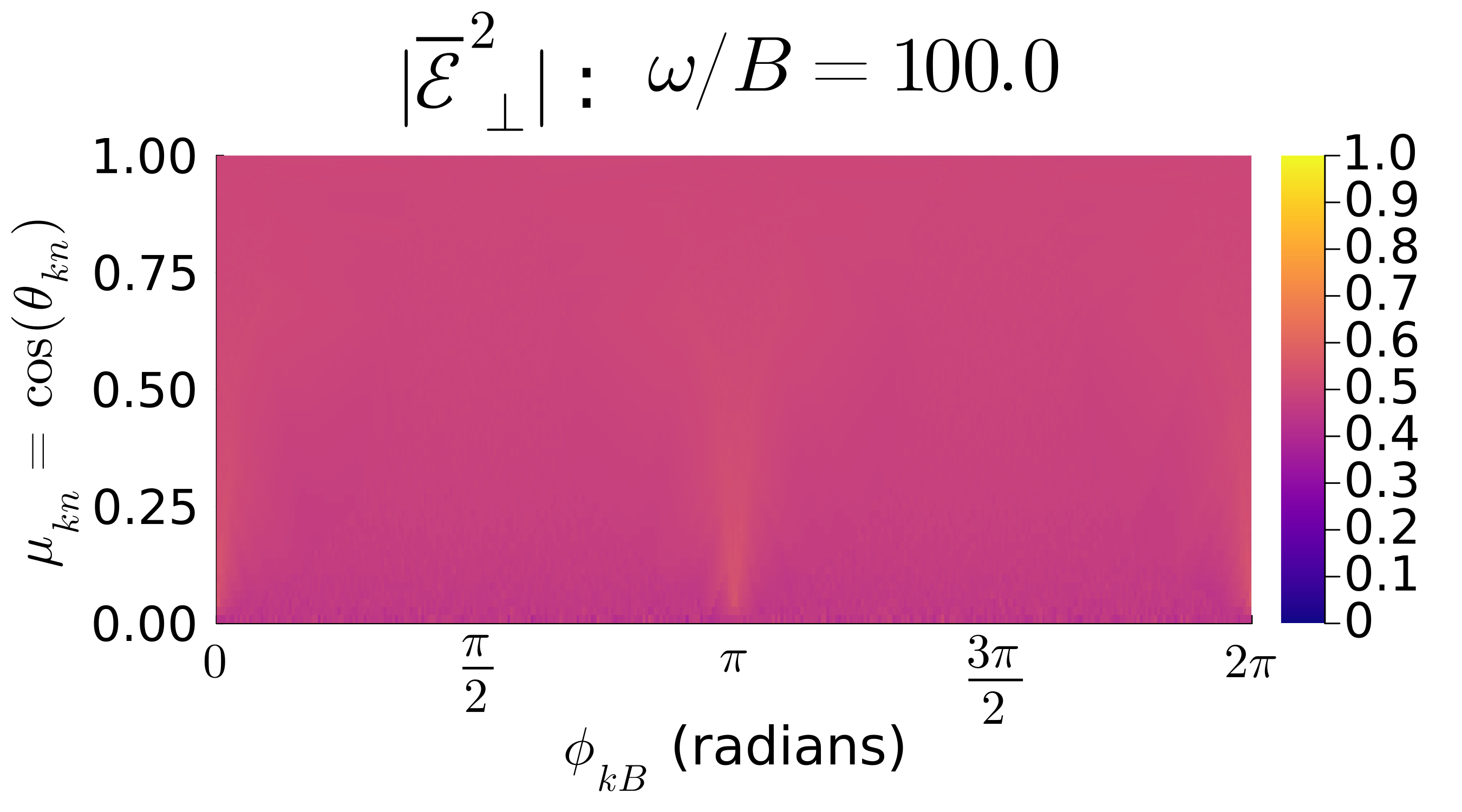} \\
  \end{tabular}
  \caption{Two-dimensional anisotropy angular distributions (for a slab where $\boldsymbol{\hat{n}}_{\cal S} \cdot \boldsymbol{\hat{B}} = 0$) as a function of the cosine of the zenith angle, $\mu_{kn} = \cos\theta_{kn}$, and the azimuthal angle $\phi_{kB}$, of the intensity (Eq.~(\ref{eq:emergent_IQUV}), left column), and the averaged electric field components $|\bar{\cal E}^2_{\parallel}|$ (middle column) and $|\bar{\cal E}^2_{\perp}|$ (right column), obtained in {\sl MAGTHOMSCATT}. Rows correspond, from top to bottom, to $\omega/B=0.01$, $0.1$, $0.25$, $0.5$, $2.0$, and $100.0$, spanning the strong-field to effectively non-magnetic scattering regimes discussed in \S\ref{sec:magthomscatt}. Note the varying normalization color scale for each row for $H_I$.}
  \label{fig:MAGTHOMSCATT}
\end{figure}

\subsection{Anisotropic and Polarized Radiation Emergent from the Burst Photosphere -- MAGTHOMSCATT} 
\label{sec:magthomscatt}

To model the transport of polarized radiation within highly optically thick fireballs along magnetic flux tubes, we use the result from a Monte Carlo radiative transfer simulation named {\sl MAGTHOMSCATT}. This code was originally designed by \cite{2021MNRAS.500.5369B} to treat scattering transport of radiation in magnetized NS surface atmospheres. However, it is applicable for any optically thick environment where magnetic Thomson electron scattering dominates. Although
{\sl MAGTHOMSCATT} does not include free-free opacity from electron-ion scattering (deep in the photosphere) where absorption might be important, the scattering-only treatment is appropriate for the outer layers of the burst for the photon energies under consideration.  Recent developments of {\sl MAGTHOMSCATT} are described in  \cite{2022ApJ...928...82H, 2025ApJ...992..188D}. In the following, we recall the key elements of the simulation.

{\sl MAGTHOMSCATT} employs an electric field vector formalism, where the complex electric field vector $\boldsymbol{\mathcal{E}}$ of each photon is tracked during its transport.  As a result, the Stokes vector $\textbf{S} = (I, Q, U, V)$ can easily be calculated for any reference coordinate. Here, $I$ denotes the radiation intensity, $Q$ and $U$ are related to linear polarization degree (PD) and angle,
and $V$ denotes the circular polarization. Therefore, the full polarization information,  including both linear and circular polarizations, as well as the interplay between them, can be fully captured. Such versatility distinguishes {\sl MAGTHOMSCATT} from previous studies that relied on tracking Stokes parameter information \citep[see, e.g.,][]{Whitney1991} or solving the radiative transfer equation for two normal modes \citep[see, e.g.,][]{Ho2001, Ozel2001}. The code has been robustly validated by various direct comparisons with results from benchmark calculations in the literature, namely, \cite{Sunyaev-1985-AandA, Chandrasekhar-1960-book} in the absence of magnetic fields, and \cite{Whitney1991} in white dwarf magnetic field domains; see \cite{2021MNRAS.500.5369B, 2025ApJ...992..188D} for details.

The fireball surface can be modeled as a collection of emission locales, each of which can be represented locally by a slab, whose normal vector is perpendicular to the local magnetic field direction, $\theta_{\rm B} = \arccos(\hat{\boldsymbol{n}}_{\cal S} \cdot \hat{\boldsymbol{B}}) = 90^{\circ}$, with $\hat{\boldsymbol{n}}_{\cal S}$ being the unit normal vector of the slab and $\hat{\boldsymbol{B}} =\boldsymbol{B} / |\boldsymbol{B} | $  the unit magnetic field vector. 
In the simulation, photons are injected at the bottom of the slab,  and then undergo magnetic Thomson scatterings with electrons, {whose distribution is assumed to be uniform and cold ($\Theta \ll 1$)}. Only photons emerging from the top of the slab are recorded, while those exiting the slab from the bottom are discarded;  see Figure 1 of \cite{2021MNRAS.500.5369B} for an illustration of photons' trajectories within a slab. Each slab simulation in {\sl MAGTHOMSCATT} is performed for a given frequency ratio $\omega / B$.

The anisotropy and polarization of photons at injection are generated according to the so-called anisotropic and polarized (AP) injection protocol, devised by \cite{2021MNRAS.500.5369B}  and further refined in \cite{2025ApJ...992..188D}. In particular,  photons are injected using the accept-reject method according to the following distribution:
\begin{equation}
	I(\mu_0) \; =\; \frac{A_{\omega}(\mu_0)}{\Lambda_{\omega} \sigma(\omega, \mu_0)}
	\quad ,\quad
	\Lambda_{\omega} \; =\; 
	\int_{-1}^{1}\frac{A_{\omega}(\mu_0)}{\sigma(\omega, \mu_0)} d\mu_0 \quad , \quad
	A_{\omega}(\mu_0)  \; = \; \frac{3}{2} \frac{1+ \mathcal{A}(\omega) \mu_0^2}{3+ \mathcal{A}(\omega)}  \quad .
	\label{eq:I_mui}
\end{equation}
Here, $\mu_0 = \hat{\boldsymbol{k}}_0 \cdot  \hat{\boldsymbol{B}}$ is the cosine of the angle made by the photon propagation direction at injection ($\hat{\boldsymbol{k}}_0$) and the local magnetic field direction ($\hat{\boldsymbol{B}}$), and $\sigma(\omega, \mu_0)$ is the total magnetic Thomson scattering cross section. The latter is given by Eq.~(13) in \citet{2021MNRAS.500.5369B}. The initial polarization properties of each photon, {which can be decomposed into $\parallel$ and $\perp$ components (similar to those defined Eq.~(\ref{eq:epar_eperp}) but with the propagation vector $\boldsymbol{k}_0$ at injection)}, are determined by its injected electric field vector, $\boldsymbol{\mathcal{E}}_0$:
\begin{equation}
	\boldsymbol{\mathcal{E}}_0  \; =\; \mathcal{E}_{\perp,0} \hat{\boldsymbol\varepsilon}_{\perp,0} + \mathcal{E}_{\parallel,0} \hat{\boldsymbol\varepsilon}_{\parallel,0}
	\quad  
	\hbox{with}\quad
	\hat{\boldsymbol\varepsilon}_{\perp,0} \; =\; \frac{ \hat{\boldsymbol{B}} \times \hat{\boldsymbol{k}}_0}{|\hat{\boldsymbol{B}} \times \hat{\boldsymbol{k}}_0|}  \quad , \quad  \hat{\boldsymbol\varepsilon}_{\parallel,0}  = \hat{\boldsymbol\varepsilon}_{\perp,0} \times \hat{\boldsymbol{k}}_0 \ , \nonumber
	\label{eq:E_sph-system}
\end{equation}
where $\hat{\boldsymbol\varepsilon}_{\parallel,0}$ and  $\hat{\boldsymbol\varepsilon}_{\perp,0}$, respectively, are parallel and perpendicular to the $\hat{\boldsymbol{k}}_0-\hat{\boldsymbol{B}}$ plane.  The amplitudes $\mathcal{E}_{\parallel,0}$  and $\mathcal{E}_{\perp,0}$ are given by:
\begin{equation}
	\mathcal{E}_{\parallel,0} \; =\; \sqrt{\frac{\Pi_{\omega} + \hat{Q}_{\omega}}{2\Pi_{\omega} }}
	\quad , \quad \mathcal{E}_{\perp,0} \; =\; \frac{ i \hat{V}_{\omega}}{2\Pi_{\omega}	\mathcal{E}_{\parallel,0}} \quad , \quad \Pi_{\omega}  = \sqrt{\hat{Q}^2_{\omega} + \hat{V}^2_{\omega}} \quad ,
	\label{eq:E_theta_phi}
\end{equation}
The two Stokes parameters $\hat{Q}_{\omega}$ and $\hat{V}_{\omega}$ in the above two equations depend on photon frequency ($\omega$) and the direction with respect to the magnetic field, characterized by $\mu_0$:
\begin{equation}
	\hat{Q}_{\omega}(\mu_0) \; =\; \frac{\mathcal{A}(\omega)[\mu_0^2 - 1]}{1+ \mathcal{A}(\omega)\mu_0^2} 
	\quad , \quad 
	\hat{V}_{\omega}(\mu_0) \; =\; \frac{2\mathcal{C}(\omega)\mu_0}{1+ \mathcal{A}(\omega)\mu_0^2} 
	\quad .
	\label{eq:Stokes-inj}
\end{equation}
The coefficients $\mathcal{A}(\omega)$ and $\mathcal{C} (\omega)$ in the expressions of $I(\mu_0)$, $\hat{Q}_{\omega}(\mu_0) $, and $\hat{V}_{\omega}(\mu_0)$ pertain to the anisotropy and circular polarization and are given by Eqs.~(35) and (36) in \cite{Baring-ApJ-2025}.

The final anisotropy and polarization distributions of photons upon emerging from the upper boundary of the slab depend on its thickness, which is defined via an effective optical depth parameter $\tau_{\rm eff}$; see Eq.~(5) in \cite{2022ApJ...928...82H}. For optically thick environs, the scattering saturates, and the anisotropy and polarization characteristics remain unchanged with the increase
of  $\tau_{\rm eff}$. Thus, the optimal value of $\tau_{\rm eff}$ corresponds to the minimum value of $\tau_{\rm eff}$ at which such saturation is achieved; {see Table 1 in \cite{2025ApJ...992..188D} for the values of $\tau_{\rm eff}$}. This value generally depends on the direction of the magnetic field with the slab normal as well as the ratio of photon frequency ($\omega$) to the electron cyclotron frequency ($B$). The AP injection protocol, described in Eqs.~(\ref{eq:I_mui})--(\ref{eq:Stokes-inj}), was devised to allow for convergence at substantially lower  $\tau_{\rm eff}$, hence much shorter runtime, than the isotropic and unpolarized (IU) counterpart. The runtime ratios between AP and IU at different frequency ratios and $\theta_{\rm B}$ values are presented in Table~3 in \cite{Baring-ApJ-2025}.

For each frequency ratio $\omega / B$, the four Stokes parameters for the photons escaping the slab are collected in $n_{\theta} = 90$ zenith angle bins, with $\arccos{(\boldsymbol{\hat{k}} \cdot \boldsymbol{\hat{n}}_{\cal S})}  = \theta_{kn} \in [0, 90^{\circ}]$, and $n_{\phi} = 360$ azimuth angle bins, with $ \arctan\!2{[\boldsymbol{\hat{k}} \cdot (\boldsymbol{\hat{n}}_{\cal S} \times \boldsymbol{\hat{B}})}/(\boldsymbol{\hat{k}} \cdot \boldsymbol{\hat{B}})] = \phi_{kB} \in[0, 360^{\circ}]$. The normalized intensity in each $j \equiv (\theta_{kn}, \phi_{kB})$ bin reads
\begin{equation}
    H_{I,j} = \frac{N_j}{\cos\theta_{kn} \sin\theta_{kn} \Delta\theta_{kn} \Delta\phi_{kB} N_{\rm rec}} \quad , 
    \label{eq:emergent_IQUV}
\end{equation}
where $N_j$ is the number of photons collected in bin $j$, $N_{\rm rec}$ is the total number of photons, and $\Delta\theta_{kn} = \Delta \phi_{kB} = 1 ^{\circ}$. The intensity in Eq.~(\ref{eq:emergent_IQUV}) is normalized such that the integration of photon flux, $H_{I,j}\cos{{\theta_{kn}}}$, over the accessible solid angle, i.e., the upper hemisphere, is unity. 

We considered $n_{\omega} = 58$ frequency ratio values, ranging between $10^{-3}\leq \omega /B \leq 300$. The two-dimensional angular distributions of the intensity of photons emerging from the upper boundary of the slab are displayed in the left column of Figure~\ref{fig:MAGTHOMSCATT} for six selected frequency ratios, $\omega / B = 0.01, 0.1, 0.25, 0.5, 2.0, 100.0$. In the strong magnetic field domain, where $\omega / B \ll 1$, the scattering cross sections are strongly reduced along the magnetic field direction, $\sigma \propto \sigma_T(\omega / B)^2$, with $\sigma_T$ being the non-magnetic Thomson cross section. Thus, the emergent radiation is strongly beamed when $\mu_{kn} = 0$ (i.e., $\theta_{kn} = \theta_B = 90^{\circ}$) and $\phi_{kB} = 0, \pi, 2\pi$. As $\omega /B$ increases--- or, equivalently, as $B$ decreases for a fixed photon energy---the beaming is diminished, as shown in Figure~\ref{fig:MAGTHOMSCATT}. At $\omega / B = 0.5$, the intensity peak becomes less pronounced because photons traveling perpendicular to the slab's normal are less likely to escape. In cases where $\omega / B \gg 1$, the emergent characteristics approximate those of a non-magnetic field scenario \citep{Chandrasekhar-1960-book}.  Using $\omega / B = 1000$ and a $\theta_{kn}$ bin size of $0.0625^{\circ}$, {\sl MAGTHOMSCATT} reproduces the non-magnetic intensity and polarization results in \cite{Chandrasekhar-1960-book} and \cite{Sunyaev-1985-AandA}; see Figure~9 in \cite{2025ApJ...992..188D}. In such conditions, the influence of the magnetic field on photon scattering is negligible, and the radiation anisotropy shows almost no variation with the azimuthal angle, as illustrated in the lower left panel. We note that the anisotropy, and also polarization (which will be discussed below), become invariant for $\omega / B \geq 300 $. Similarly, for the strong-field regime ($\omega / B \ll 1$), as $\omega / B$ decreases, the height of the intensity peak increases, while the width of the peak becomes smaller, i.e., more collimated radiation \citep{2022ApJ...928...82H, 2025ApJ...992..188D}. As a result, the change in the anisotropy is negligible for $\omega /B \leq 0.001$.

\subsection{Vacuum Birefringence} 
\label{subsec:vacuum_birefringence}

VB is a QED effect in which a strongly magnetized vacuum behaves as a birefringent medium, acquiring different refractive indices for the two photon polarization normal modes, namely, the parallel ($\parallel$) and perpendicular ($\perp$) modes \citep{Adler1970, 1971AnPhy..67..599A}.  The $\parallel$ and $\perp$ modes, respectively, correspond to the linear polarization state where the electric field vector $\boldsymbol{\mathcal{E}}$ is in and perpendicular to the plane of the photon propagation vector $\hat{\boldsymbol{k}}$ and the local magnetic field vector $\boldsymbol{B}$. The corresponding polarization coordinate basis is given as in Eq.~(\ref{eq:epar_eperp}). Action of VB in the magnetosphere of magnetars is suggested by the observed strong X-ray polarization in 1E~1547.0-5408 \citep{2025arXiv250919446S,2026arXiv260115452T}.

The electric field vector can be decomposed in terms of the $\parallel$ and $\perp$ components, $ {\cal E}_{\parallel}  =\boldsymbol{\mathcal{E}} \cdot \hat{\boldsymbol{\varepsilon}}_{\parallel}$  and  ${\cal E}_{\perp}  = \boldsymbol{\mathcal{E}} \cdot \hat{\boldsymbol{\varepsilon}}_{\perp}$.
The two vectors $\hat{\boldsymbol{\varepsilon}}_{\parallel}$ and  $\hat{\boldsymbol{\varepsilon}}_{\perp}$ in Eq.~(\ref{eq:epar_eperp}) rotate with variations in $\boldsymbol{B}$ and $\hat{\boldsymbol{k}}$ as the photon propagates through the magnetosphere, while the two amplitudes ${\cal E}_{\parallel}$ and  ${\cal E}_{\perp}$ evolve as \citep{VanAdelsberg2006}:
\begin{equation}
	i\begin{pmatrix}
		\dover{d{\cal E}_{\parallel}}{dl_s} \\[4pt]
		\dover{d{\cal E}_{\perp}}{dl_s}
	\end{pmatrix}
	\; \approx \;
	\begin{pmatrix}
		\dover{-\omega \Delta n}{2c} & i\dover{d\chi}{dl_s} \\[6pt]
		-i\dover{d\chi}{dl_s}  & \dover{\omega \Delta n}{2c}
	\end{pmatrix}
	\begin{pmatrix}
		{\cal E}_{\parallel} \\[4pt]
		{\cal E}_{\perp} \\
	\end{pmatrix}
	\quad .
	\label{eq:dE_ds}
\end{equation}
In the above equation, $\Delta n = n_{\parallel} - n_{\perp} = \alpha_{f}/(30\pi) B^2\sin^2 \theta_{kB}$ is the difference in the refractive indices\footnote{We do not include the effects of dichroic magnetic lensing \citep{1999MNRAS.306..333S}, whose distortions on near-surface $\parallel$-mode trajectories are important for $B~\gtrsim~0.1\times~6\pi/\alpha_f~\sim~300$. Likewise, time-delays on the IRs associated with slower than $c$ propagation of the $\parallel$ mode through regions of the magnetosphere are also not included since it is negligible for most trajectories, except possibly those which curve around and back at low altitudes (e.g., last row of Figure~\ref{fig:archetypes}). This correction as an inter-mode delay is of order $\lesssim 0.3\, \mu$s for typical fields of surface $10^{14}-10^{15}$ G \citep[see also,][]{2026arXiv260517048P}.} between the two polarization modes \citep{1997JPhA...30.6485H} to leading order. Here, $\theta_{kB}$ is the angle between {the directions of propagation $\hat{\boldsymbol{k}}$ and the local external field $\hat{\boldsymbol{B}}$.  Also, the polarization angle $\chi$ is defined as in Eq.~(\ref{eq:cos2chi_from_a2}), and couples tightly to the $\hat{\boldsymbol{k}}-\hat{\boldsymbol{B}}$ plane as it rotates along a photon trajectory \citep{Dinh-2026-ApJ}.}  In the adiabatic regime, where $\omega \Delta n / (2c) \gg d\chi/dl_s$, with $l_s$ being the distance along the propagation direction, the off-diagonal terms are negligible. As a result, the two polarization modes decouple, and their complex amplitudes evolve independently:
\begin{equation}
	{\cal E}_{\parallel} \; = \;  {\cal E}_{\parallel, \rm e}\exp{\left(\frac{i}{2} \Delta \phi \right)} 
	\quad ,\quad  
	{\cal E}_{\perp} \; = \; {\cal E}_{\perp, \rm e}\exp{\left(-\frac{i}{2} \Delta \phi\right)} 
    \quad \hbox{for} \quad
    \Delta\phi \; =\; {{\omega}\over {c}} \int_0^{l_s} \Delta n \, dl_s' \quad ,
 \label{eq:EX_rc}
\end{equation}
where ${\cal E}_{\parallel, \rm e}$ and ${\cal E}_{\perp, \rm e}$ are the mode amplitudes at the emission point, and $\Delta\phi$ is the phase shift between the two modes after the photon travels a distance $l_s$. Neglecting this phase shift, from Eq.~(\ref{eq:EX_rc}), VB preserves the amplitudes of the two polarization modes. Accordingly, the electric field vector rotates around the wave vector with the magnetic field vector as the photon propagates. While the influence of the phase shift $\Delta \phi$ on the circular polarization is important, its impact on the linear polarization properties is {mostly} negligible, {particularly away from the electron cyclotron resonance} \citep{Dinh-2026-ApJ}.  Since we only consider the linear polarization properties, the phase shift $\Delta \phi$ is set to zero for simplicity. 

The evolution equations given by Eq.~(\ref{eq:EX_rc}) are valid when $r < r_{\rm PL}$, where $r_{\rm PL}$ is the so-called polarization-limiting radius for radially propagating photons. For a flat spacetime static vacuum magnetic dipole,  $r_{\rm PL} (\theta) $ is given by \citep{Heyl2002}:
\begin{equation}
	\frac{r_{\rm PL}}{r_\star} \; \approx 230 \,\biggl( \dover{m_e c^2 \omega}{20\, \hbox{keV}}  \biggr)^{1/5} 
    \biggl( \dover{B_p^{\rm flat}\, \sin\theta}{10} \biggr)^{2/5} \quad ,
 \label{eq:r_rc2}
\end{equation}
where $\theta$ is the magnetic colatitude. 

At any altitude during photon propagation, the Stokes parameters are calculated using Cartesian coordinates with screen basis vectors and can be expressed as
\begin{eqnarray}
	I & = &  |{\cal E}_{\parallel,\rm e}|^2 + |{\cal E}_{\perp,\rm e}|^2  \quad ,    \nonumber  \\[3pt] 
    Q & = & (|{\cal E}_{\parallel,\rm e}|^2 - |{\cal E}_{\perp,\rm e}|^2) \cos2\chi - 2\left[\Re({\cal E}_{\parallel,\rm e}{\cal E}^{*}_{\perp,\rm e})\cos\Delta\phi -\Im({\cal E}_{\parallel,\rm e}{\cal E}^{*}_{\perp,\rm e})\sin\Delta \phi\right]\sin 2\chi \nonumber  \\
    &\approx&  (|{\cal E}_{\parallel,\rm e}|^2 - |{\cal E}_{\perp,\rm e}|^2) \cos2\chi \quad , \nonumber  \\[-5.5pt]
 \label{eq:IQUV_rec} \\[-5.5pt]
	U & = & (|{\cal E}_{\parallel,\rm e}|^2 - |{\cal E}_{\perp,\rm e}|^2) \sin2\chi
	   + 2\left[\Re({\cal E}_{\parallel,\rm e}{\cal E}^{*}_{\perp,\rm e})\cos\Delta\phi -\Im({\cal E}_{\parallel,\rm e}{\cal E}^{*}_{\perp,\rm e})\sin\Delta \phi\right]\cos 2\chi \nonumber  \\
    &\approx&  (|{\cal E}_{\parallel,\rm e}|^2 - |{\cal E}_{\perp,\rm e}|^2) \sin2\chi\quad , \nonumber\\[3pt]
	V & = & -2\left[\Re{({\cal E}_{\parallel,\rm e}{\cal E}^{*}_{\perp,\rm e})\sin\Delta\phi} + \Im{({\cal E}_{\parallel,\rm e}{\cal E}^{*}_{\perp,\rm e})}\cos\Delta\phi \right]  \quad . \nonumber
 \end{eqnarray}
When $r<r_{\rm PL}$, the photon polarization vector rotates adiabatically with the magnetic field direction. Therefore, when VB is on, $\chi$ is calculated with the physical polarization vector ${\hat{\boldsymbol{\varepsilon}}_{\parallel,\perp}}$ determined by the magnetic field vector at the polarization-limiting surface. With the Stokes parameters obtained from the local slab simulations presented in \S\ref{sec:magthomscatt}, we {obtained the averaged electric field components in each $(\theta_{kn}, \phi_{kB})$ bin, $|\bar{\cal E}_{\parallel (\perp)}|^2 = \frac{1}{N_j}\sum_{i =1}^{N_j}|{\cal E}_{\parallel (\perp),\rm e}|^2$, with $N_j$ being the number of photons collected in the bin; see Appendix~\ref{sec:slab}.}
 
 The angular distributions of $|\bar{\cal E}_{\parallel}|^2$ and $|\bar{\cal E}_{\perp}|^2$ are shown in the middle and right columns of Figure~\ref{fig:MAGTHOMSCATT}. As shown in the panels in the top three rows, when $\omega /B \ll 1$, the emergent radiation is highly linearly polarized, with $|\bar{\cal E}_{\parallel}|^2 \sim 1$ and $|\bar{\cal E}_{\perp}|^2 \sim 0$. This is due to the dominance of the $\parallel \rightarrow \parallel$ component, ${\rm d} \sigma_{\parallel \rightarrow\parallel} / {\rm d}\Omega_f$,   in the scattering cross sections in most directions; see Eq.~(B9) in \cite{2021MNRAS.500.5369B} for details. Along the magnetic field direction, within the so-called magnetic scattering cone \citep{Baring-ApJ-2025}, the cross sections of photons in both polarization modes, $\sigma_{\perp}$ and $\sigma_{\parallel}$, are of the same order of magnitude and proportional to $\sigma_T (\omega /B )^2$. Therefore, there is a mixing of both polarization modes, as is more evident in the panels in the cases for $\omega / B  = 0.1, 0.25$.

\begin{figure*}[t]
  \centering
  \normalsize

  \noindent
  \begin{minipage}[t]{0.155\textwidth}
    \centering
    $\boldsymbol{\hat{k}\cdot \hat{n}_{\cal S}}$
  \end{minipage}\hfill
  \begin{minipage}[t]{0.155\textwidth}
    \centering
    (VB~off) $\cos 2\chi$
  \end{minipage}\hfill
  \begin{minipage}[t]{0.155\textwidth}
    \centering
    (VB~on) 0.1~keV, $\cos 2\chi$
  \end{minipage}\hfill
  \begin{minipage}[t]{0.155\textwidth}
    \centering
    (VB~on) 1.0~keV, $\cos 2\chi$
  \end{minipage}\hfill
  \begin{minipage}[t]{0.155\textwidth}
    \centering
    (VB~on) 10~keV, $\cos 2\chi$
  \end{minipage}\hfill
  \begin{minipage}[t]{0.155\textwidth}
    \centering
    (VB~on) 100~keV, $\cos 2\chi$
  \end{minipage}

  \vspace{0.3em}
  \hrule
  \vspace{0.5em}

  \noindent
  \begin{minipage}[t]{0.155\textwidth}
    \centering
    \includegraphics[width=\linewidth]{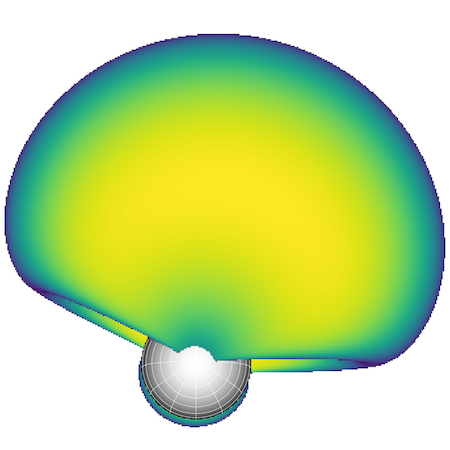}
  \end{minipage}\hfill
  \begin{minipage}[t]{0.155\textwidth}
    \centering
    \includegraphics[width=\linewidth]{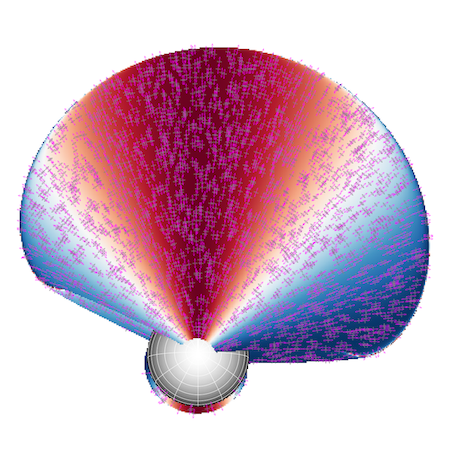}
  \end{minipage}\hfill
  \begin{minipage}[t]{0.155\textwidth}
    \centering
    \includegraphics[width=\linewidth]{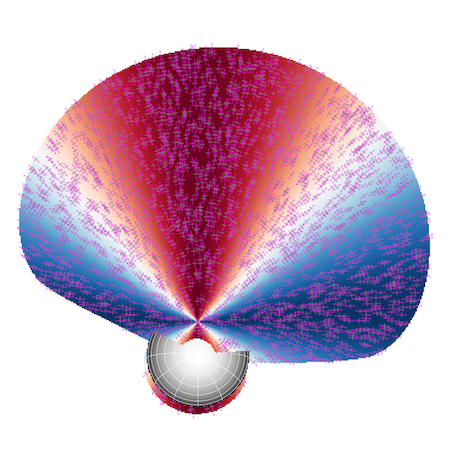}
  \end{minipage}\hfill
  \begin{minipage}[t]{0.155\textwidth}
    \centering
    \includegraphics[width=\linewidth]{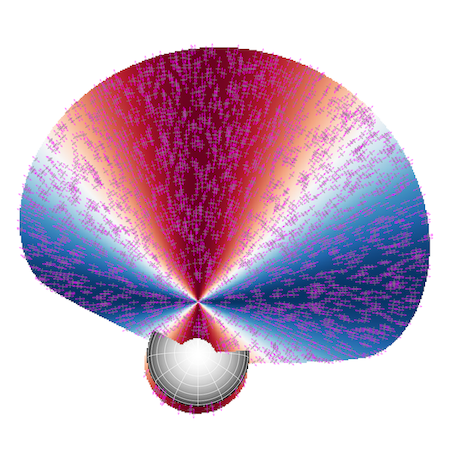}
  \end{minipage}\hfill
  \begin{minipage}[t]{0.155\textwidth}
    \centering
    \includegraphics[width=\linewidth]{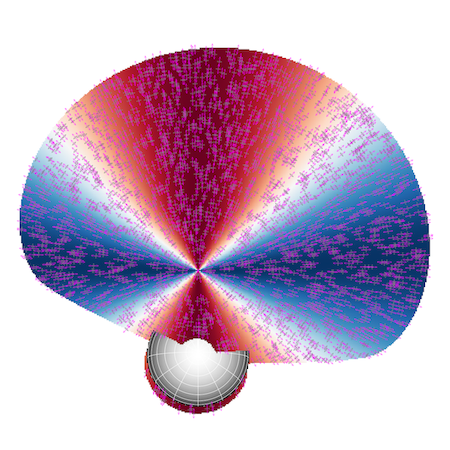}
  \end{minipage}\hfill
  \begin{minipage}[t]{0.155\textwidth}
    \centering
    \includegraphics[width=\linewidth]{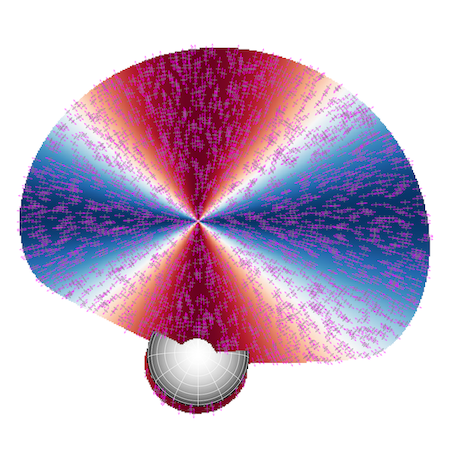}
  \end{minipage}

  \vspace{0.3em}
  \hrule
  \vspace{0.5em}

  \noindent
  \begin{minipage}[t]{0.155\textwidth}
    \centering
    \includegraphics[width=\linewidth]{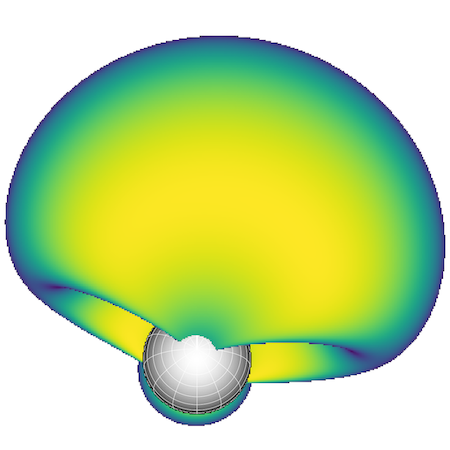}
  \end{minipage}\hfill
  \begin{minipage}[t]{0.155\textwidth}
    \centering
    \includegraphics[width=\linewidth]{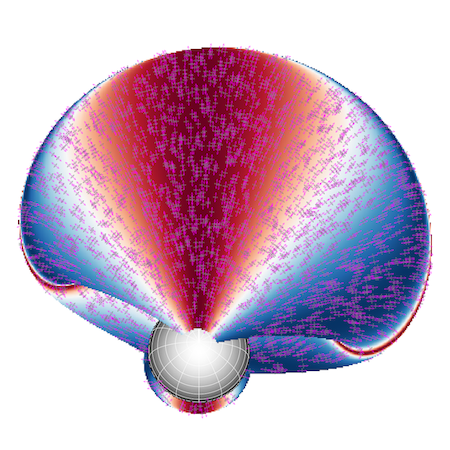}
  \end{minipage}\hfill
  \begin{minipage}[t]{0.155\textwidth}
    \centering
    \includegraphics[width=\linewidth]{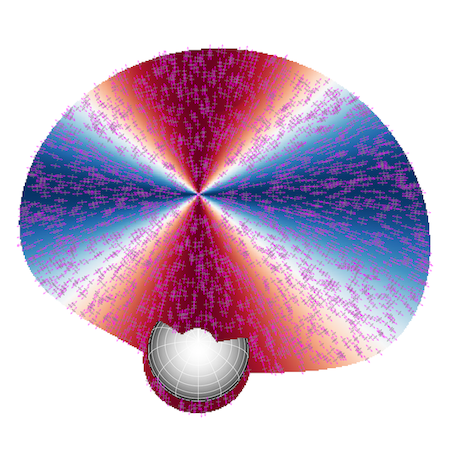}
  \end{minipage}\hfill
  \begin{minipage}[t]{0.155\textwidth}
    \centering
    \includegraphics[width=\linewidth]{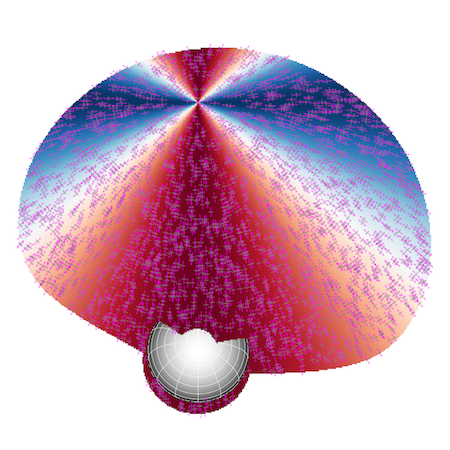}
  \end{minipage}\hfill
  \begin{minipage}[t]{0.155\textwidth}
    \centering
    \includegraphics[width=\linewidth]{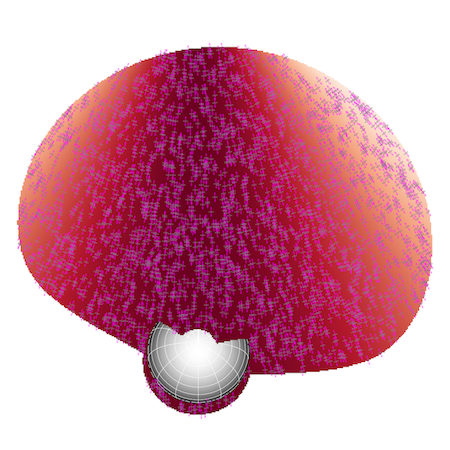}
  \end{minipage}\hfill
  \begin{minipage}[t]{0.155\textwidth}
    \centering
    \includegraphics[width=\linewidth]{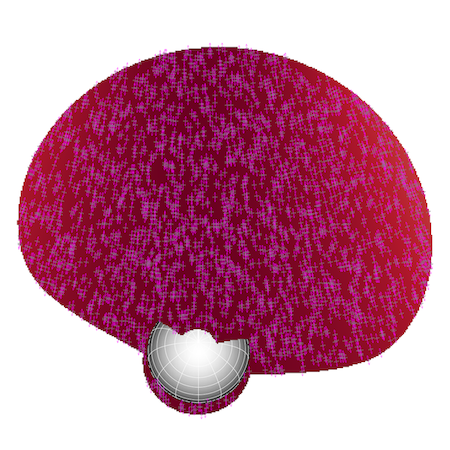}
  \end{minipage}

  \vspace{0.3em}
  \hrule
  \vspace{0.5em}

  \noindent
  \begin{minipage}[t]{0.155\textwidth}
    \centering
    \includegraphics[width=\linewidth]{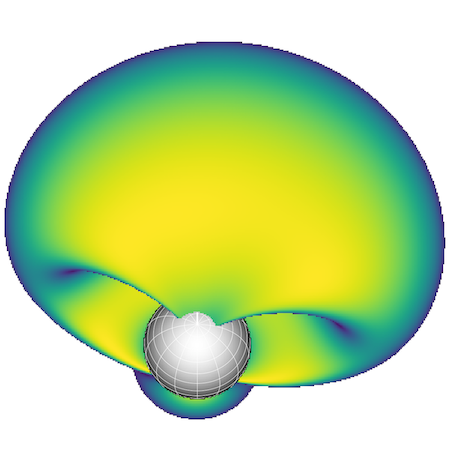}
  \end{minipage}\hfill
  \begin{minipage}[t]{0.155\textwidth}
    \centering
    \includegraphics[width=\linewidth]{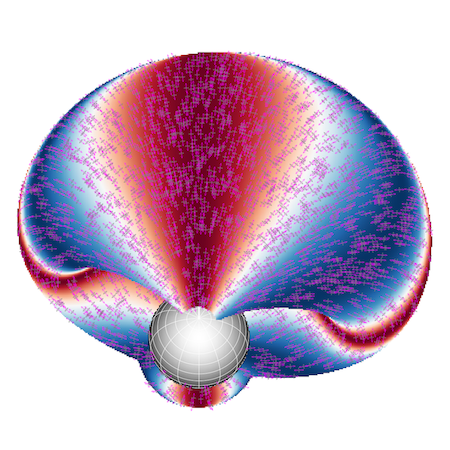}
  \end{minipage}\hfill
  \begin{minipage}[t]{0.155\textwidth}
    \centering
    \includegraphics[width=\linewidth]{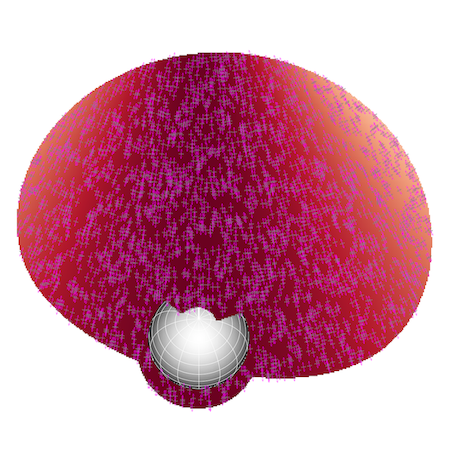}
  \end{minipage}\hfill
  \begin{minipage}[t]{0.155\textwidth}
    \centering
    \includegraphics[width=\linewidth]{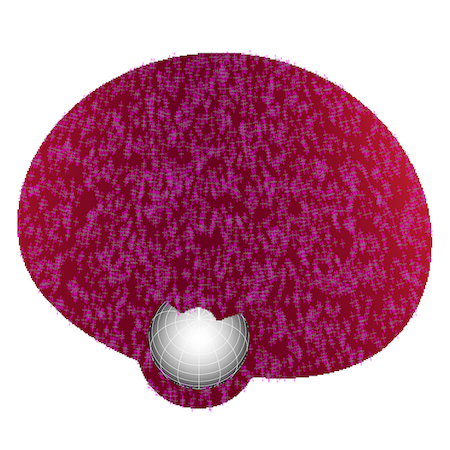}
  \end{minipage}\hfill
  \begin{minipage}[t]{0.155\textwidth}
    \centering
    \includegraphics[width=\linewidth]{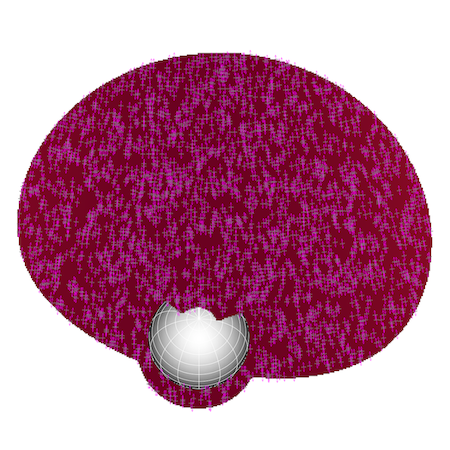}
  \end{minipage}\hfill
  \begin{minipage}[t]{0.155\textwidth}
    \centering
    \includegraphics[width=\linewidth]{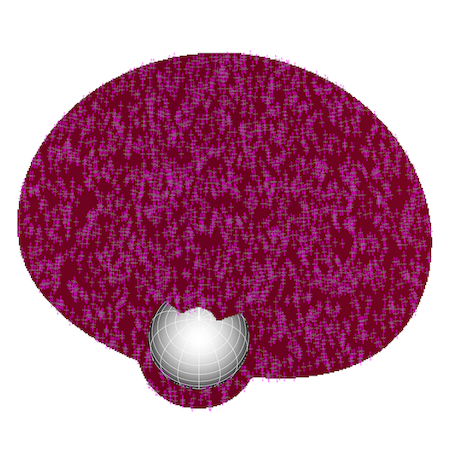}

  \end{minipage}

    \raggedright
  \begin{minipage}[t]{0.3\textwidth}
    \centering
    \includegraphics[width=\linewidth]{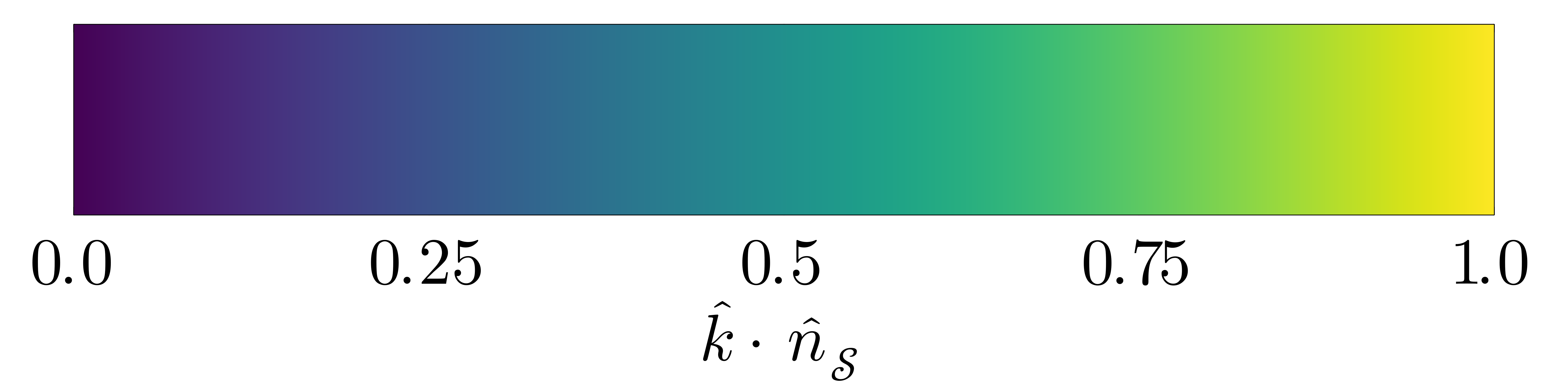}
  \end{minipage}
  \begin{minipage}[t]{0.3\textwidth}
    \centering
    \includegraphics[width=\linewidth]{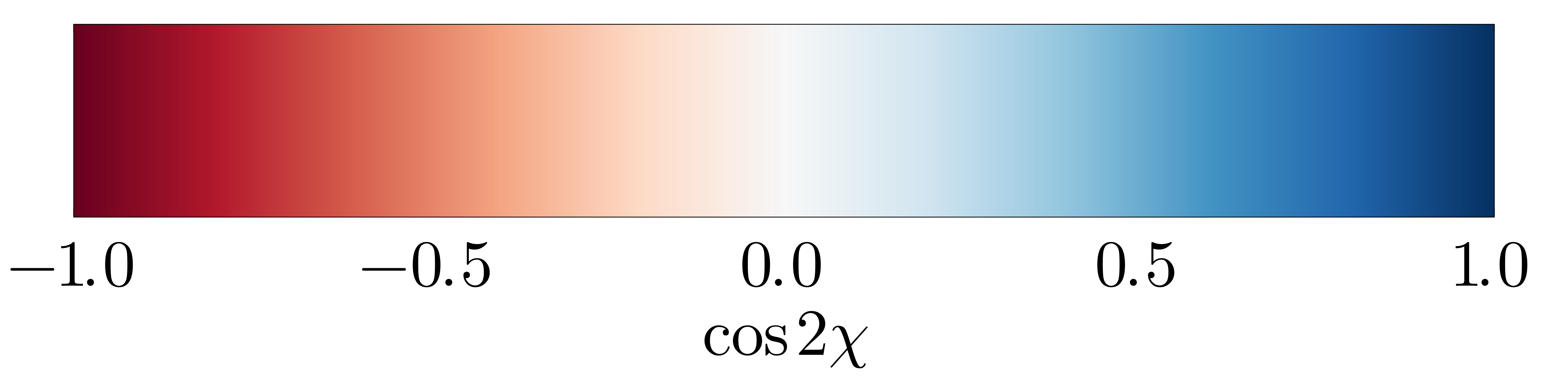}
  \end{minipage}

  \caption{Energy-dependent polarization angle transport with and without VB for three near pole-on geometries (rows) for $B = 10$, $M = 1.7 \,M_\odot$ and $r_\star = 12$~km with viewing angles of $\theta_v = \{ 5^\circ, 15^\circ, 30^\circ \}$ (top to bottom). In each row, the first column shows $\boldsymbol{\hat{k}\cdot\hat{n}_{\cal S}}$ and the second shows $\cos 2\chi$ with VB disabled. The remaining columns show $\cos 2\chi$ with VB enabled at 0.1, 1, 10, and 100~keV. Comparing columns within each row demonstrates how adiabatic VB reorganizes polarization angle structure in a strongly energy-dependent manner for polar viewing angles, approaching the VB-off case at low energies and on-axis views. The color scale indicates $\cos 2\chi$ at the IP, along with a random sample of overlaid polarization angle vectors. As the VB polarization recoupling radius becomes large compared to the emission region, the polarization vectors become more aligned (bottom row panels).  }
  \label{fig:VBdep}
\end{figure*}

\subsection{Polarized Burst Luminosity and Impulse Responses}
 \label{sec:imageplane_sum}

We now wish to assign flux tube photon trajectory hits on the IP onto physical fluxes, and integrate over the image for a distant observer for whom the emission is spatially unresolved.  The resolved source's size on the IP establishes the physical projected emitting area. Let $R_{\rm ip}=Xr_\star$ denote the characteristic IP scale. We use either (i) a uniform Cartesian sampling of the IP, in a regular grid of $n_{\rm pix}\times n_{\rm pix}$ rays, or (ii) a radially log polar sampling of a disk of outer radius $R_{\rm ip}$. The log polar scheme is suited for tall flux tubes with tiny footpoints. Based on knowledge of the loop $r_{\rm max}$ from parameters provided prior to runtime, we set the IP Cartesian half-width, or log polar outer radius, to $X r_\star$ ($X>1$), with $R_{\rm ip}=X r_\star$. For the Cartesian scheme, the linear pixel spacing is $\ell \approx 2R_{\rm ip}/n_{\rm pix}$ and its IP areal weight is uniformly $A_i=\ell^2$. For the log polar scheme, the first radial scale is $\rho_1=R_{\rm ip}/n_{\rm pix}$, the wedge azimuthal spacing is $\Delta\varphi=2\pi/n_{\rm pix}$, and the logarithmic radial spacing is $q=(R_{\rm ip}/\rho_1)^{1/(n_{\rm pix}-1)}$.  The sector edges are $\rho_{m,-}=\rho_1 q^{m-1}$ and $\rho_{m,+}=\rho_1 q^m$ for $m\geq1$ and edges $(0,\rho_1)$ for the innermost wedges. The $m$th radial ray is placed at the area-weighted centroid radius $\rho_m=(2/3)(\rho_{m,+}^3-\rho_{m,-}^3)/(\rho_{m,+}^2-\rho_{m,-}^2)$, with $\rho_0=2\rho_1/3$, and at azimuth $\varphi_n=(n+1/2)\Delta\varphi$ for $n\in\mathbb{N}$. The log polar IP area weight is then $A_{mn}=(\rho_{m,+}^2-\rho_{m,-}^2)\Delta\varphi/2$.

For emission from the flux tube, we assume a BB with dimensionless temperature $\Theta = k_{\rm B} T/(m_e c^2)$. The spectral intensity expressed in dimensionless energy $\omega$ is
\begin{equation}
    \frac{d{\cal S}}{dt dA d\Omega d\omega} \equiv {\cal S}_{\omega} = \frac{m_e c^3}{4 \pi^3 \lambar_c^3} \frac{\omega^3}{\exp(\omega/\Theta)-1} .
\end{equation}
The local temperature in the burst photosphere frame is supplied by Eq.~(\ref{eq:BT_relation}), and the {\sl MAGTHOMSCATT} atmospheres provide $H_I$, $\bar {\cal E}_{\parallel}^2$, and $\bar {\cal E}_{\perp}^2$ as functions of $\omega_{\rm e}/B$, $\mu_{kn}$, and $\phi_{kB}$. The $H_I$ are normalized such that,
\begin{equation}
\int_{0}^{2\pi} \int_0^1 \mu_{kn} H_I \left(\frac{\omega_e}{B}, \mu_{kn},\phi_{kB} \right) d\mu_{kn} d\phi_{kB} = 1
\label{eq:Hnorm}
\end{equation}
for all energies with $\bar{\cal E}_\perp^2 + \bar{\cal E}_\parallel^2 = 1$ for all energies and angles.

The local emitted intensities are taken to be
\begin{align}
{\cal I}_{\rm e,\perp}
&=
\pi {\cal S}_{\omega}
H_I\!\left(\frac{\omega_{\rm e}}{B},\mu_{kn},\phi_{kB}\right)
\bar {\cal E}_{\perp}^2\,f_{\perp}(\omega_{\rm e}) ,
\\
{\cal I}_{\rm e,\parallel}
&=
\pi {\cal S}_{\omega}
H_I\!\left(\frac{\omega_{\rm e}}{B},\mu_{kn},\phi_{kB}\right)
\bar {\cal E}_{\parallel}^2\,f_{\parallel}(\omega_{\rm e}) ,
\label{eq:I}
\end{align}
so that
\begin{equation}
{\cal I}_{\rm e}={\cal I}_{\rm e,\perp}+{\cal I}_{\rm e,\parallel} .
\label{eq:Ie_total}
\end{equation}
The factor of $\pi$ in front accounts for the normalization convention of Eq.~(\ref{eq:Hnorm}) \citep[e.g.,][]{Chandrasekhar-1960-book,2018A&A...615A..50N}. The $f_{\perp}$ and $f_{\parallel}$ encode the chosen photon splitting prescription for ray opacities (\S\ref{sec:splitting}). In the $\perp\rightarrow\parallel\parallel$ case,
\begin{equation}
f_{\perp}=\exp[-\tau_{\perp\rightarrow\parallel\parallel}(\omega_{\rm e})]
\quad , \quad
f_{\parallel}=1 ,
\end{equation}
whereas in the effective all-mode option both factors are $f_{\perp}=f_{\parallel}=\exp[-\tau_{\rm ave}(\omega_{\rm e})]$. When splitting is disabled, $f_{\perp}=f_{\parallel}=1$. 

By Liouville's theorem and conservation of photon occupation number (neglecting in-flight splitting daughter photons), the quantity ${\cal I}_{\omega}/\omega^3$ is invariant along trajectories\footnote{Because photon splitting operates near or induces spectral cutoffs, split daughter photons will generally be a modest contribution to spectra below cutoffs. We defer a full splitting cascade treatment to future work.}. We define $\delta = \omega_{\infty}/\omega_{\rm e} \approx \omega_{\rm ip}/\omega_{\rm e} $ where $\omega_{\rm e}$ is the energy for the emitting frame, $\delta  \equiv 1/(1+z) = (u_\mu k^{\mu})_\infty /(u_\mu k^{\mu})_e \approx \sqrt{-g_{tt}}$. 
Thus,
\begin{equation}
{\cal I}_{\rm ip} (x_o, y_o, \omega_{\rm ip}, t) = \delta^3 {\cal I}_{\rm e} \left[ \boldsymbol{\vec{r}} ; \omega_{\rm ip}/\delta, t - t_{\rm dist}(x_o, y_o) \right]  ,
\label{eq:calIip}
\end{equation}
where ${\cal I}_{\rm e}$ is the specific intensity at position vector $\boldsymbol{\vec{r}}$ along a flux tube, and $t_{\rm dist}$ is the time of arrival at the IP for the curved trajectory, and $\{x_o, y_o\}$ are the IP ray positions. The value of $t - t_{\rm dist}(x_o, y_o)$ establishes $\Delta t_{\rm obs}$ for the IRs.

The differential flux at the IP is $dF_{\rm ip} =  {\cal I}_{\rm ip} d\Omega_{\rm ip} = {\cal I}_{\rm ip} dx_o dy_o/d_{\rm ip}^2 \approx {\cal I}_{i,\rm ip} A_i/d_{\rm ip}^2$. The total flux is a sum over the IP of all rays for each energy, i.e.
\begin{equation}
F_{\rm ip} (\omega_{\rm ip}, t) \approx \sum_{i\in \rm hits} {\cal I}_{i, \rm ip}  \Delta \Omega_{i,\rm ip} \approx  \frac{1}{d_{\rm ip}^2} \sum_{i\in \rm hits} A_i \delta_i^3 {\cal I}_{e,i}.
\end{equation}
We define the isotropic-equivalent Stokes luminosity vector to be $\boldsymbol{L} \equiv 4\pi d_{\rm ip}^2 \boldsymbol{F}_{\rm ip}$. From this construction, luminosities are independent of $d_{\rm ip}$. We discretize energy and time into bins $(\omega_{{\rm ip}, j}, t_k)$
\begin{align}
L_{I,jk} &\approx 4 \pi\sum_{i\in \rm hits} A_i \left({\cal I}_{{\rm e},\parallel,i}+{\cal I}_{{\rm e},\perp,i}\right)\delta_i^3 ,\\
L_{Q,jk} &\approx 4 \pi \sum_{i\in \rm hits} A_i  \left({\cal I}_{{\rm e},\parallel,i}-{\cal I}_{{\rm e},\perp,i}\right)\delta_i^3\cos 2\chi_i ,\\
L_{U,jk} &\approx 4 \pi \sum_{i\in \rm hits} A_i  \left({\cal I}_{{\rm e},\parallel,i}-{\cal I}_{{\rm e},\perp,i}\right)\delta_i^3\sin 2\chi_i .
\label{eq:stokes_sum}
\end{align}
where the arguments of ${\cal I}_e$ are those in Eq.~(\ref{eq:calIip}). From these, luminosity energy-time IRs are computed.

The polarization angle entering the Stokes sums is treated differently in the VB-off and VB-on cases. For VB off, we use the transported electric-field components at the emission point, so that $\cos2\chi$ and $\sin2\chi$ follow directly from Eq.~(\ref{eq:cos2chi_from_a2}). For VB on, we store $\cos2\chi$ and $\sin2\chi$ at sampled energies corresponding to the freeze-out angle at the appropriate polarization-limiting surface. For an arbitrary observed energy $\omega_{{\rm ip},j}$, we linearly interpolate $\cos2\chi$ and $\sin2\chi$ in $\log \omega_{\rm ip}$ between the tabulated values and then renormalize the pair to unit length before constructing Stokes parameters.

Time or energy integrated luminosities are also readily constructed, $\boldsymbol{L}_j = \sum_k \boldsymbol{L}_{jk} \Delta t_k $ or $\boldsymbol{L}_k = \sum_j \boldsymbol{L}_{jk} \Delta \omega_{{\rm ip},j} $, respectively. Likewise, energy or time integrated PD are computed using these forms.

\newcommand{\vcentered}[1]{\raisebox{0.0525\linewidth}[0pt][0pt]{#1}}

\newcommand{\colorbarheight}{0.25\textwidth}
\newcommand{\colorbargap}{-0.12\textwidth}

\begin{figure}[t]
  \setlength{\tabcolsep}{0pt}
  \renewcommand{\arraystretch}{0.0} 
    \centering

    \makebox[\textwidth][c]{
    \llap{
      \raisebox{-.5\height}{
        \includegraphics[
          height=\colorbarheight,
          keepaspectratio
        ]{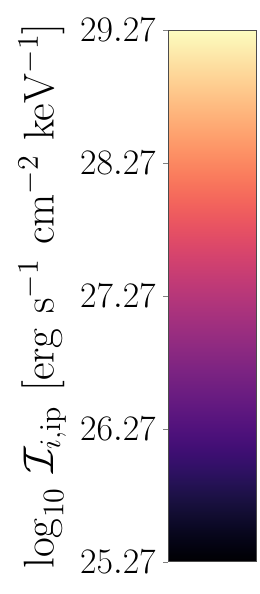}
      }
      \hspace{\colorbargap}
    }
    
  \begin{tabular}[c]{l|ccccccc}
  $\theta_v$ & $\phi_0 = 0^\circ$  & $\phi_0 = 30^\circ$ & $\phi_0 = 60^\circ$ & $\phi_0 = 90^\circ$ & $\phi_0 = 120^\circ$ & $\phi_0 = 150^\circ$ & $\phi_0 = 180^\circ$ \\
  \hline
   \vcentered{$6^\circ$} & \includegraphics[width=0.095\linewidth]{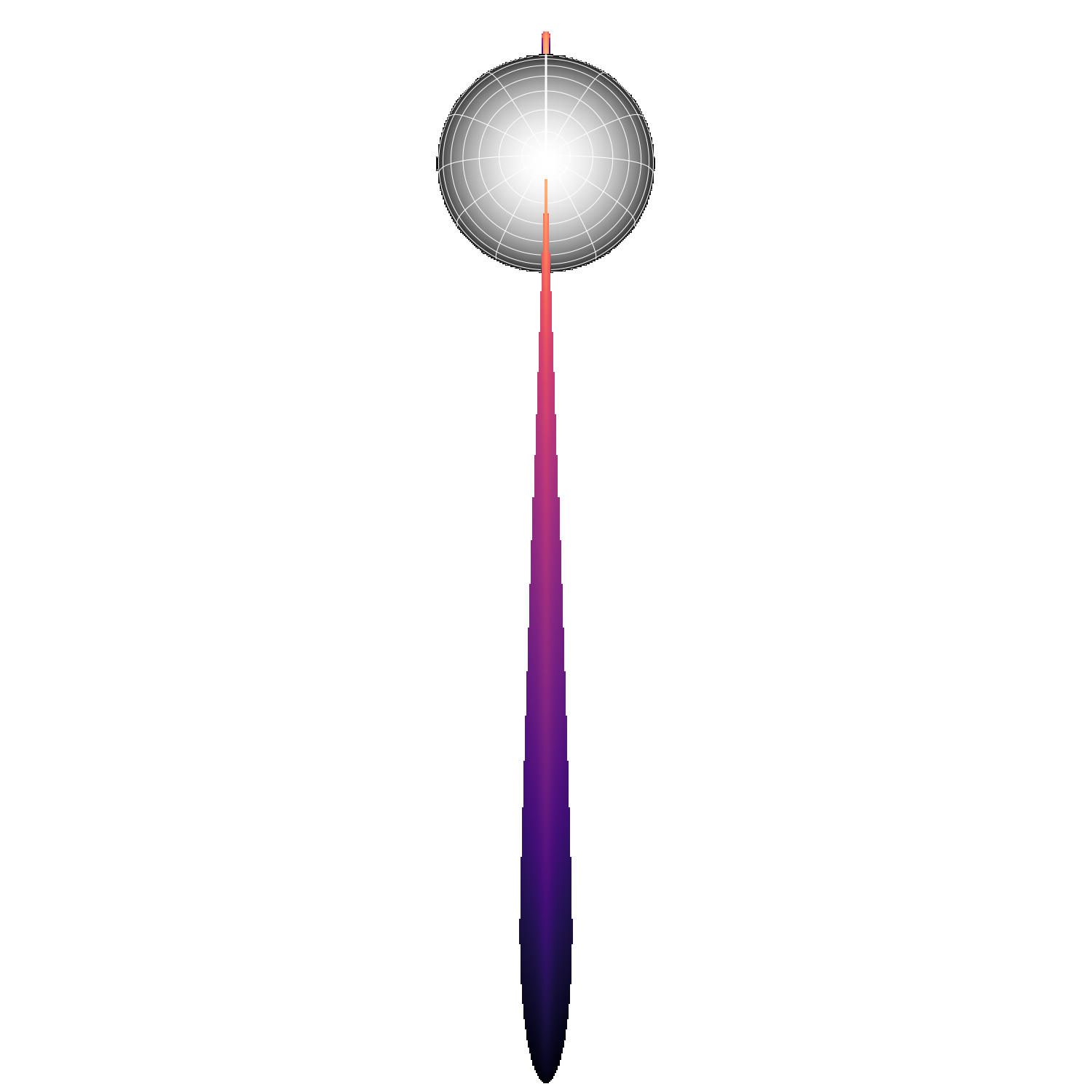} &
    \includegraphics[width=0.095\linewidth]{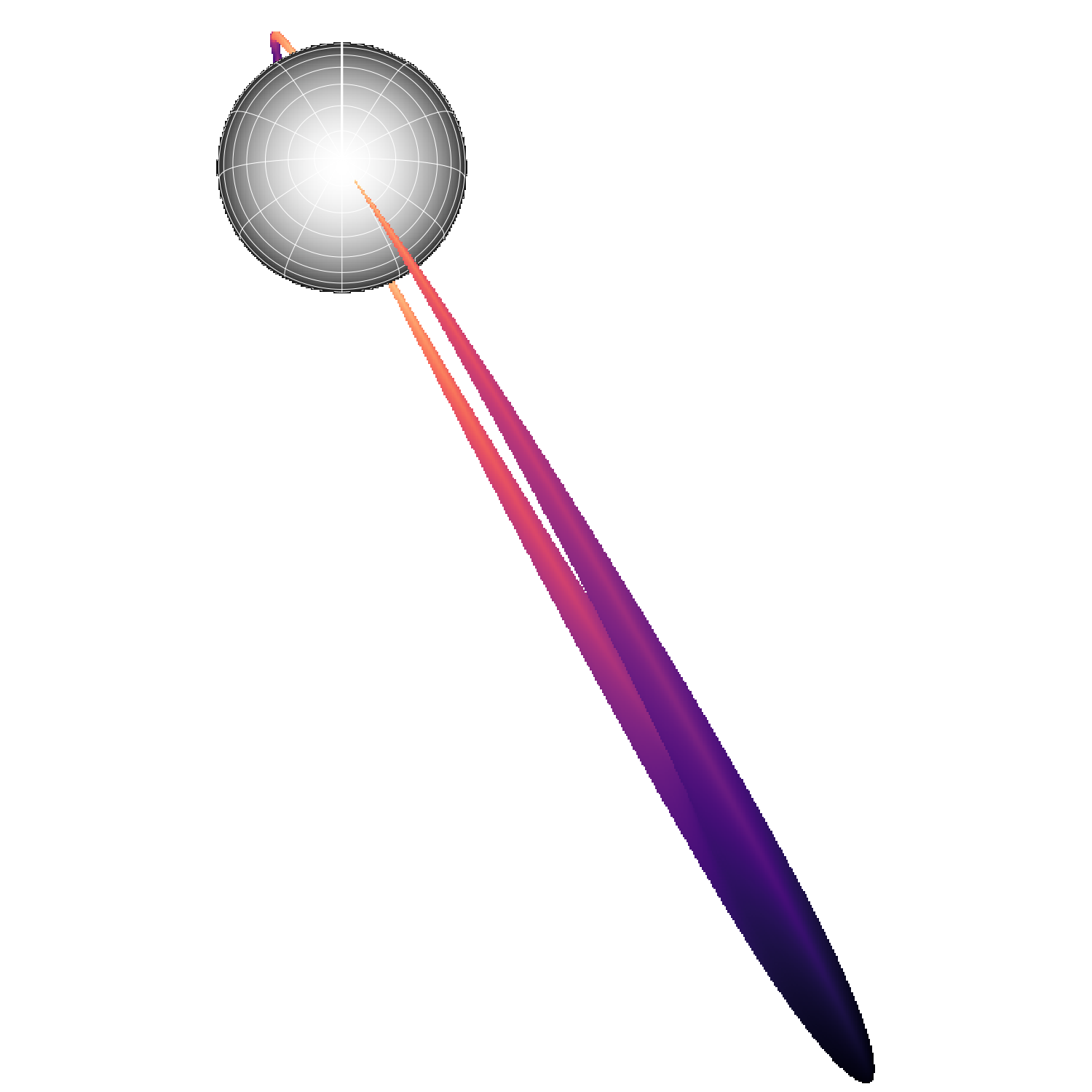} &
    \includegraphics[width=0.095\linewidth]{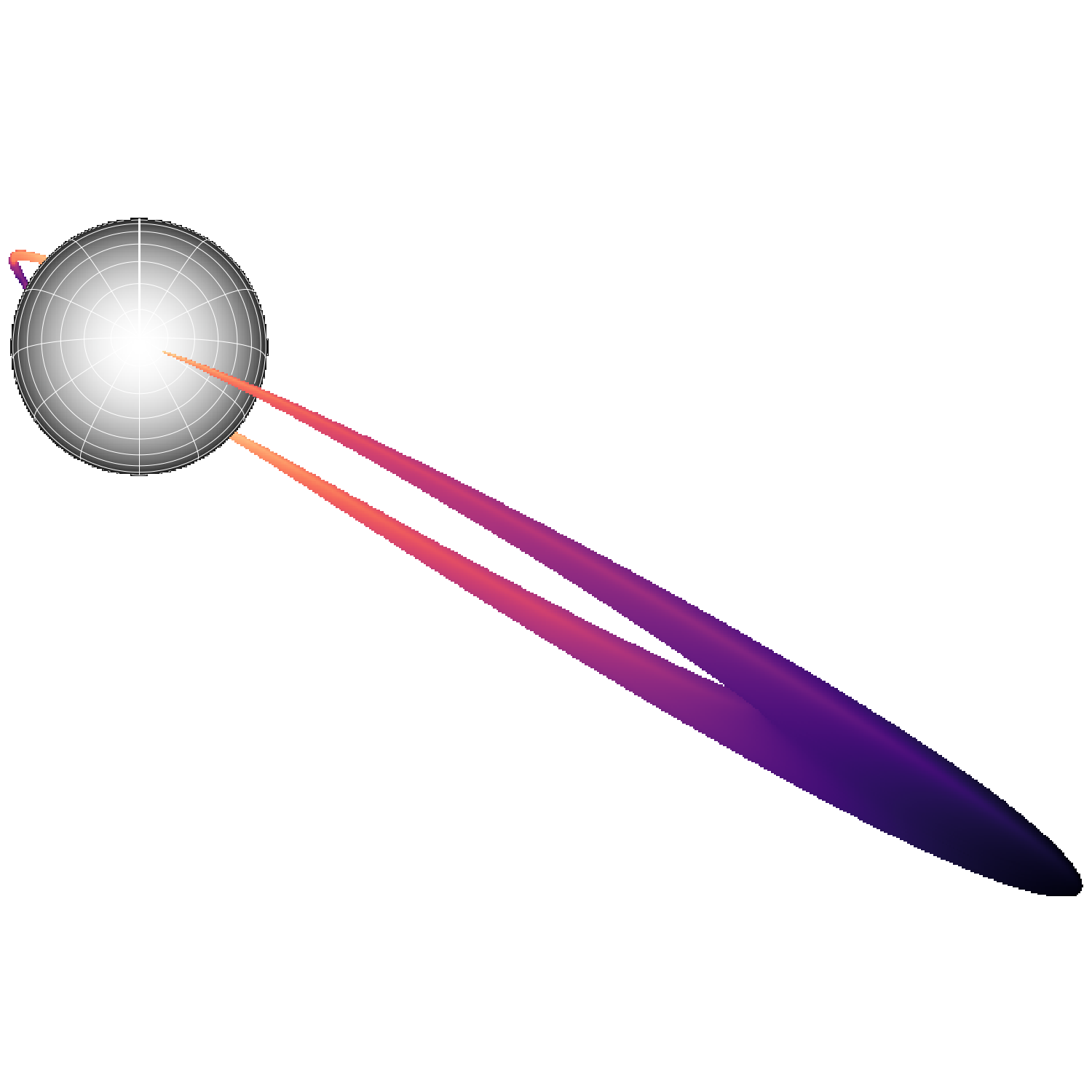} &
    \includegraphics[width=0.095\linewidth]{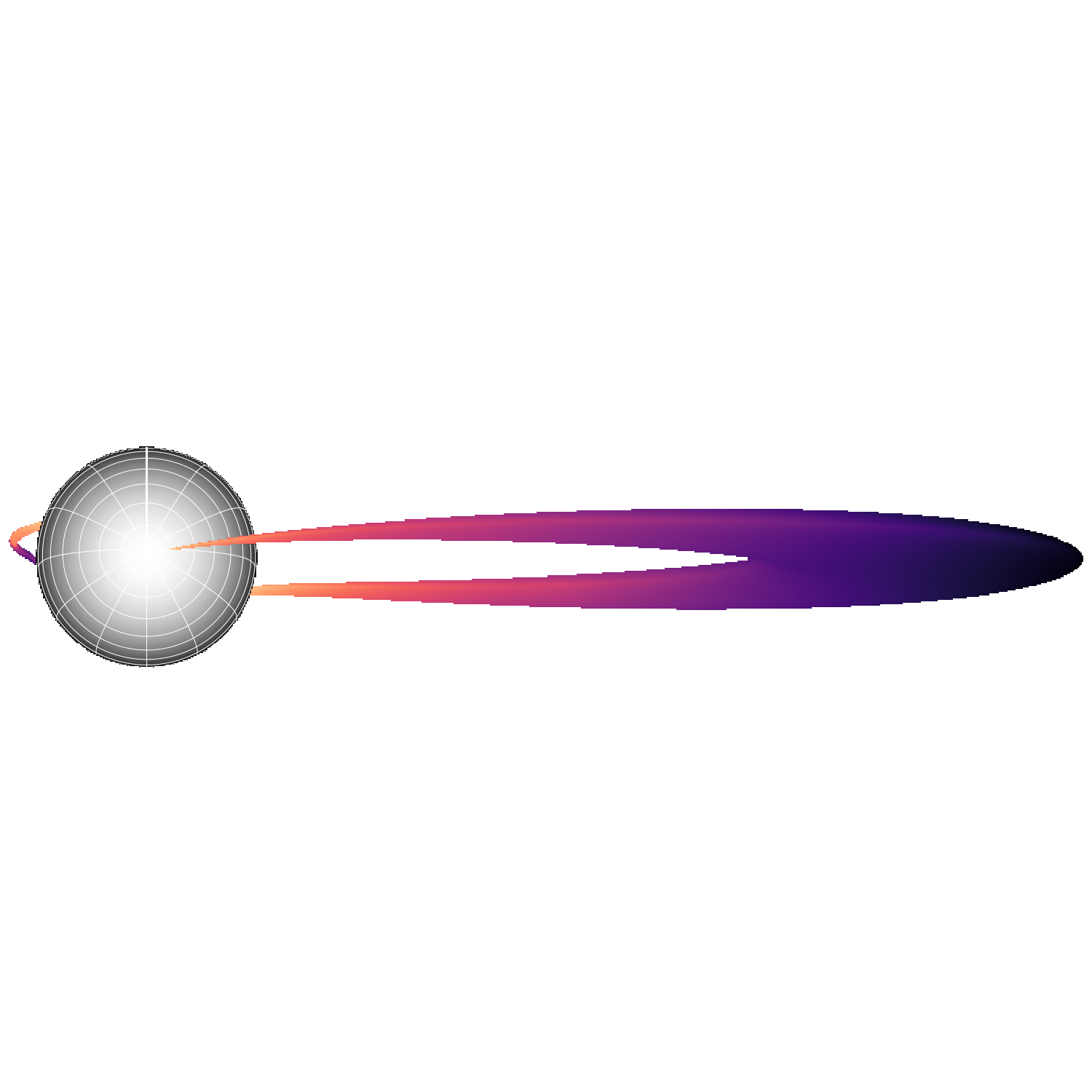} &
    \includegraphics[width=0.095\linewidth]{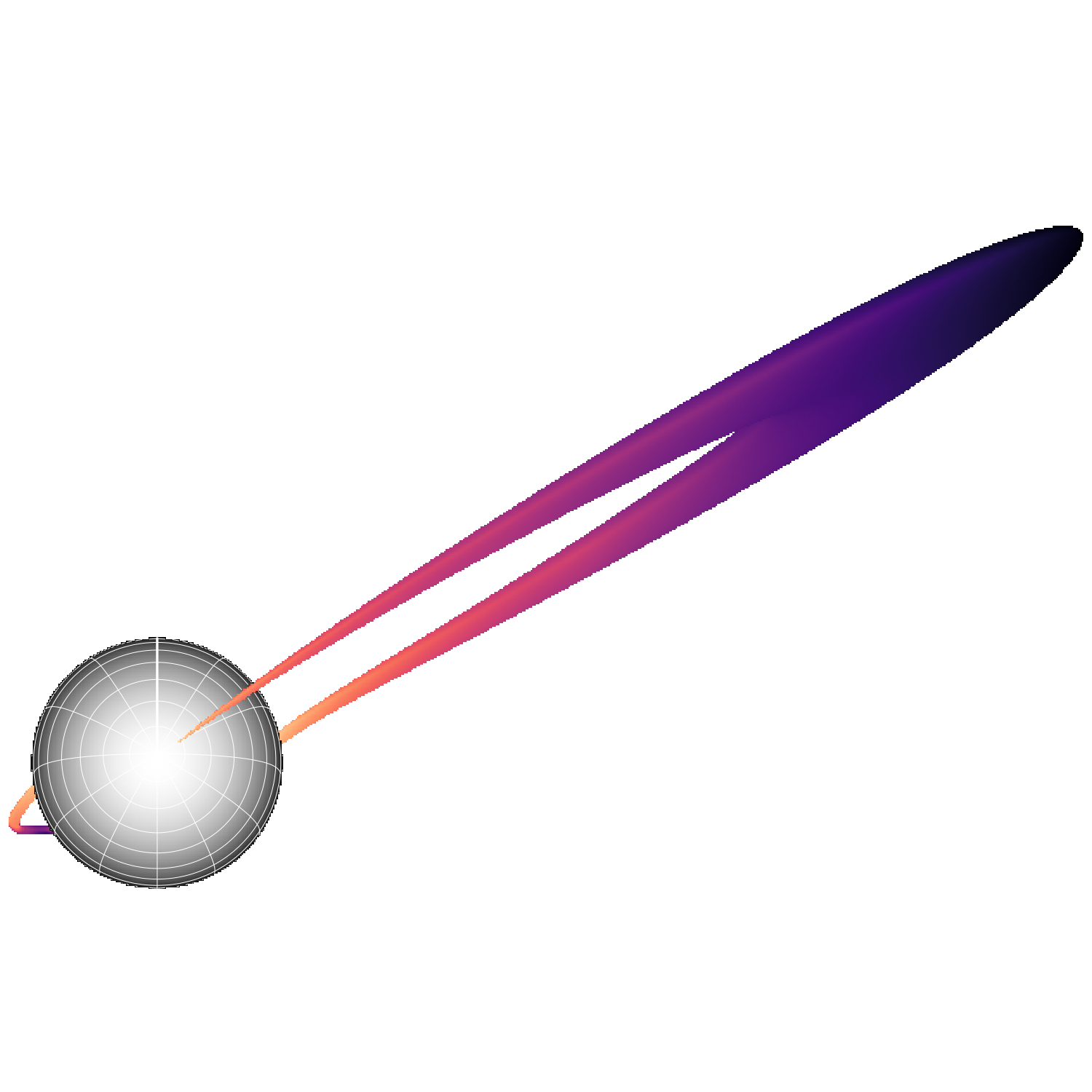} &
    \includegraphics[width=0.095\linewidth]{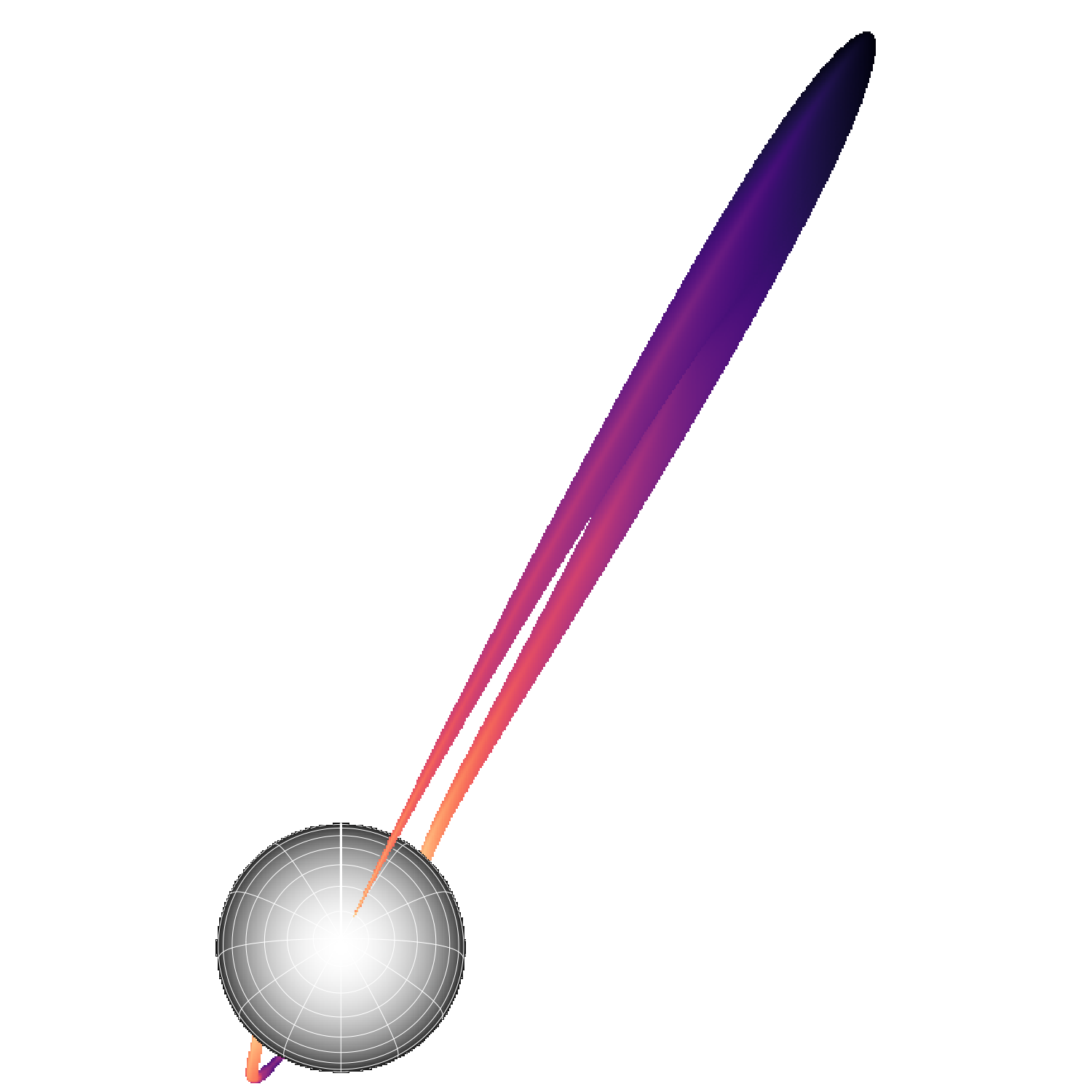} &
    \includegraphics[width=0.095\linewidth]{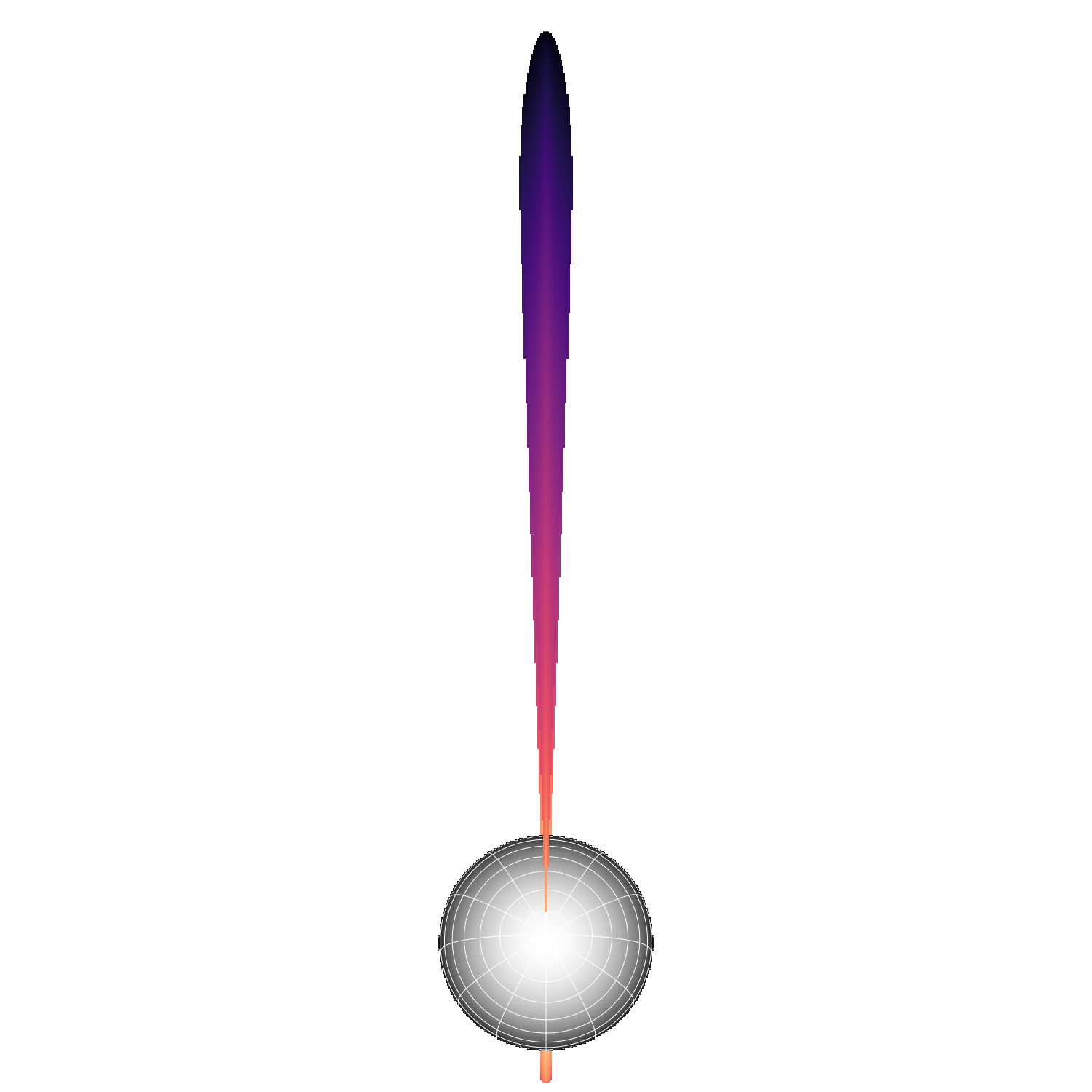}  \\

    \vcentered{$20^\circ$} & \includegraphics[width=0.095\linewidth]{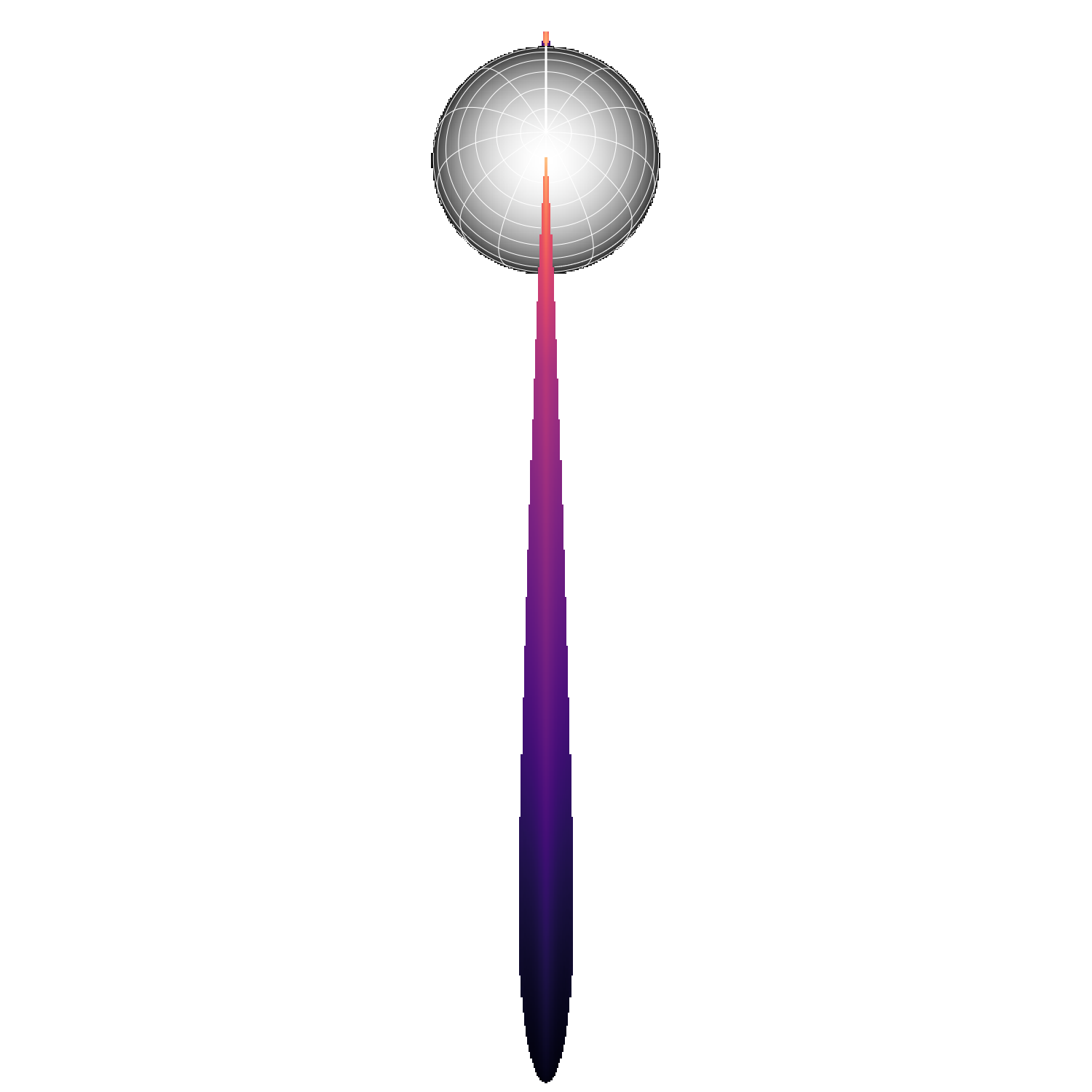} &
    \includegraphics[width=0.095\linewidth]{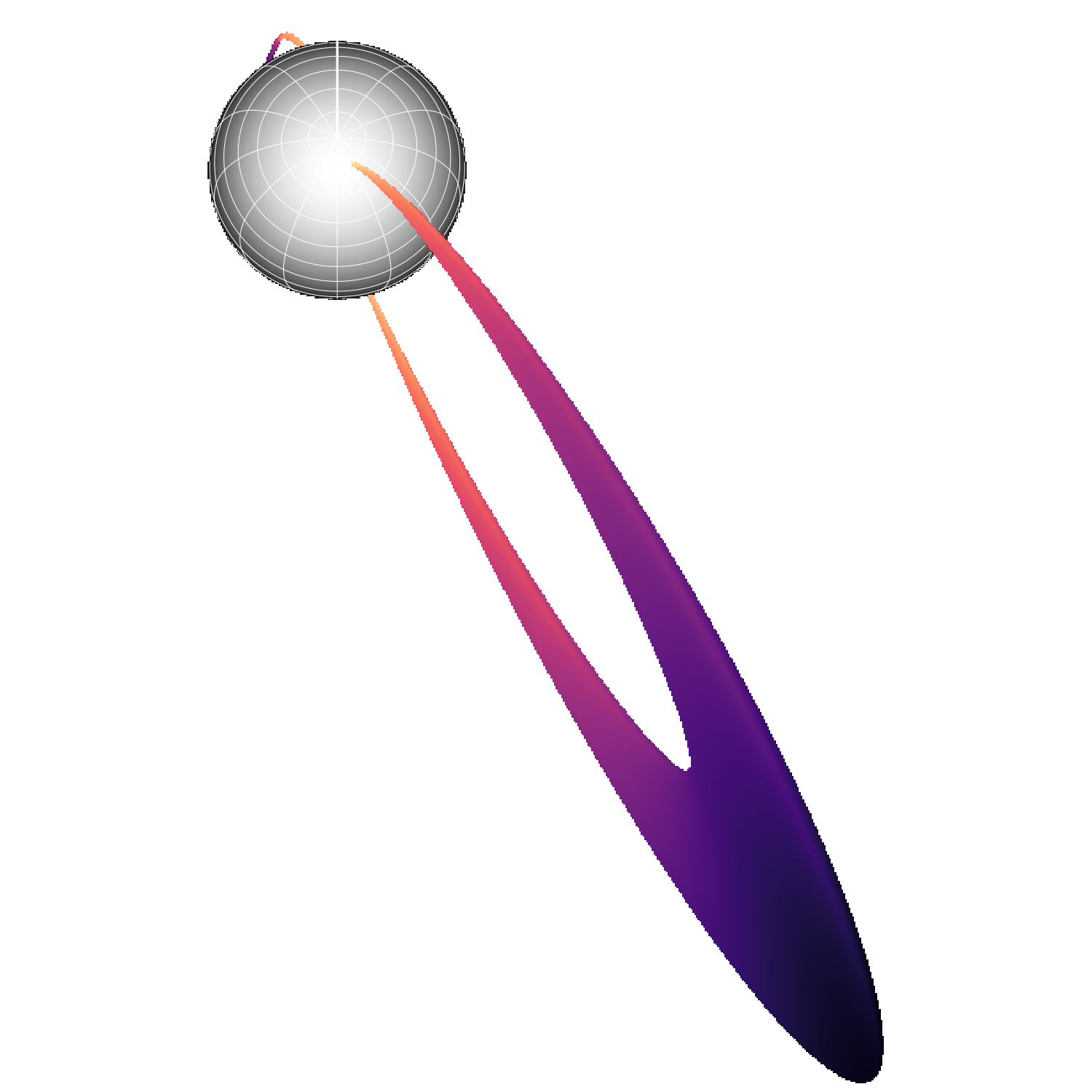} &
    \includegraphics[width=0.095\linewidth]{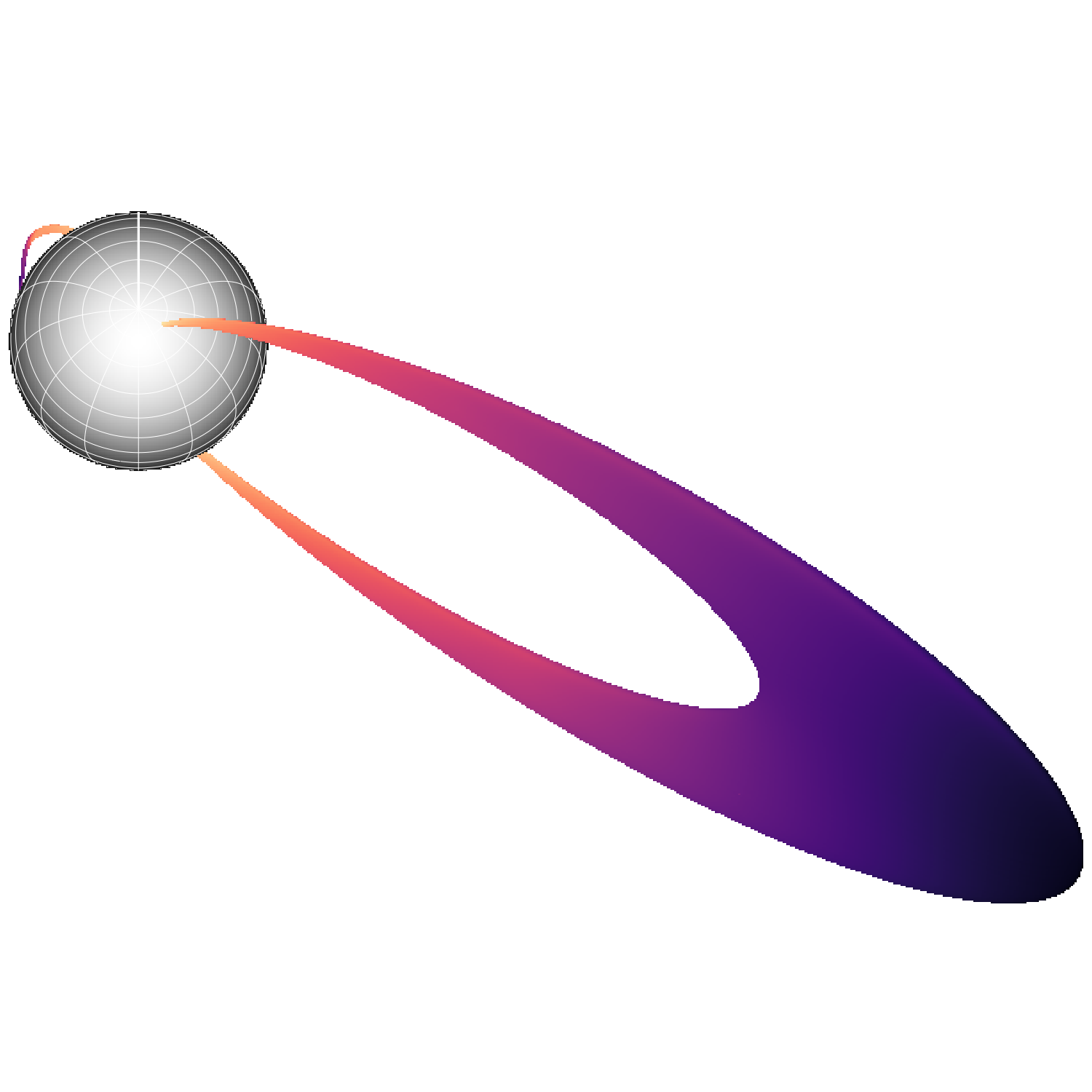} &
    \includegraphics[width=0.095\linewidth]{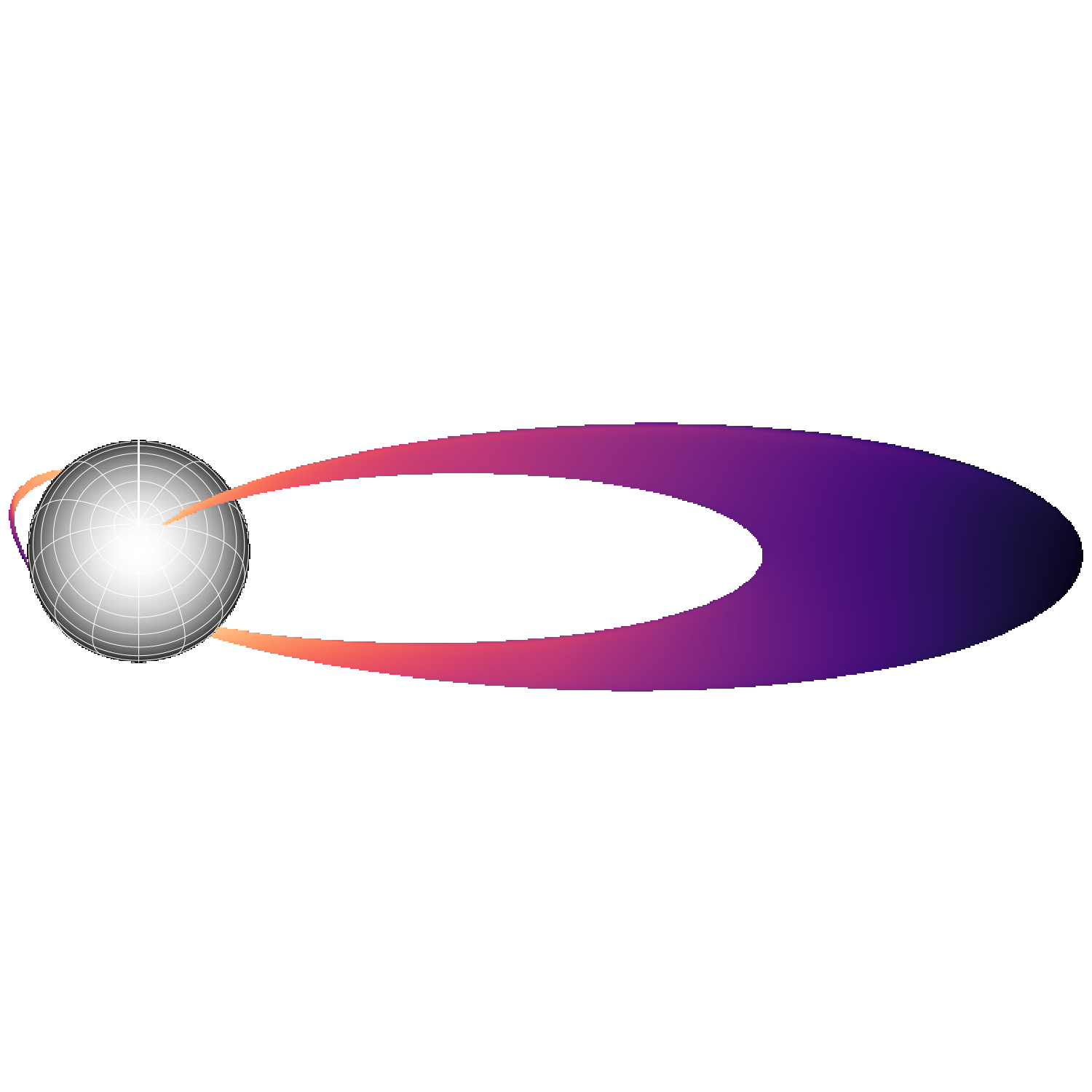} &
    \includegraphics[width=0.095\linewidth]{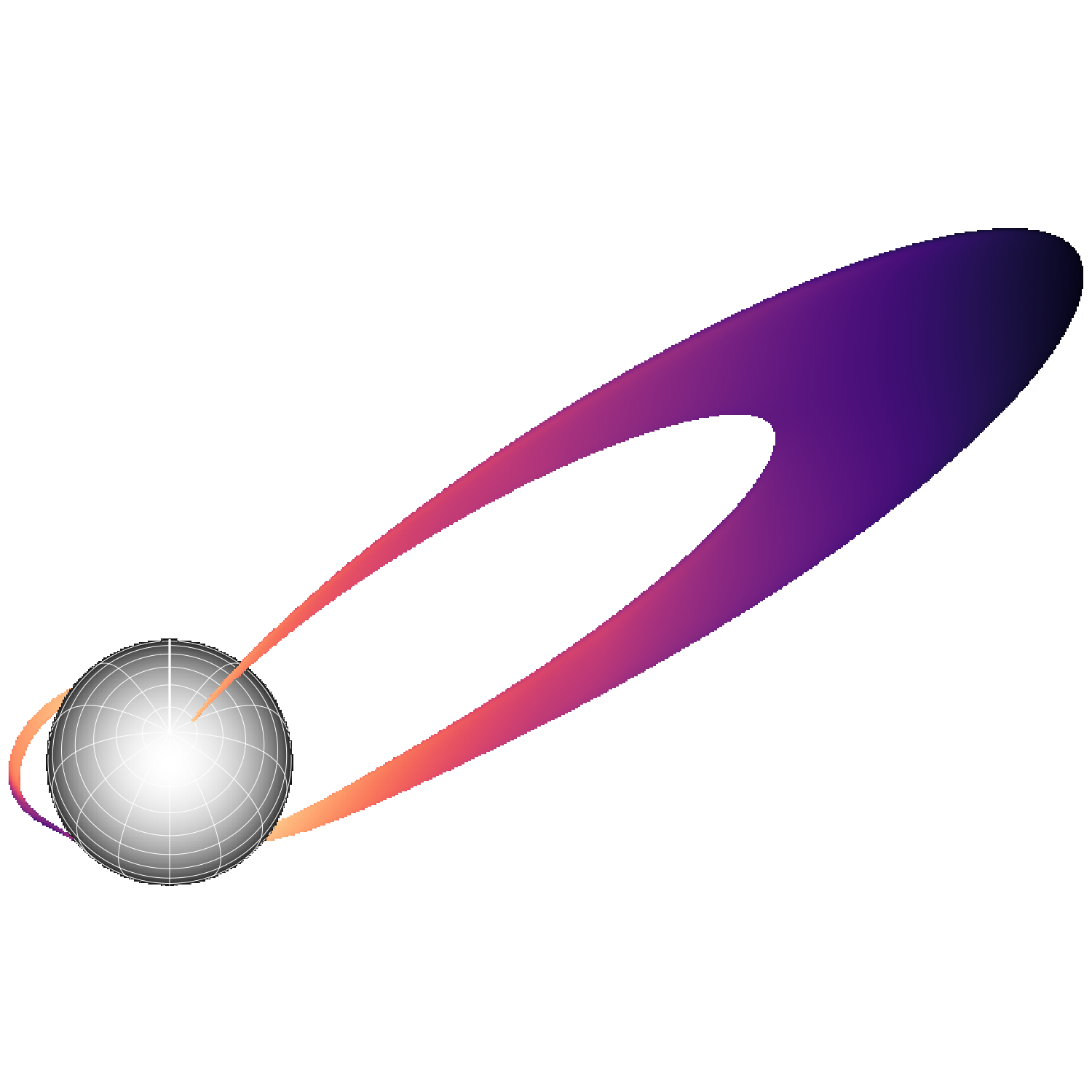} &
    \includegraphics[width=0.095\linewidth]{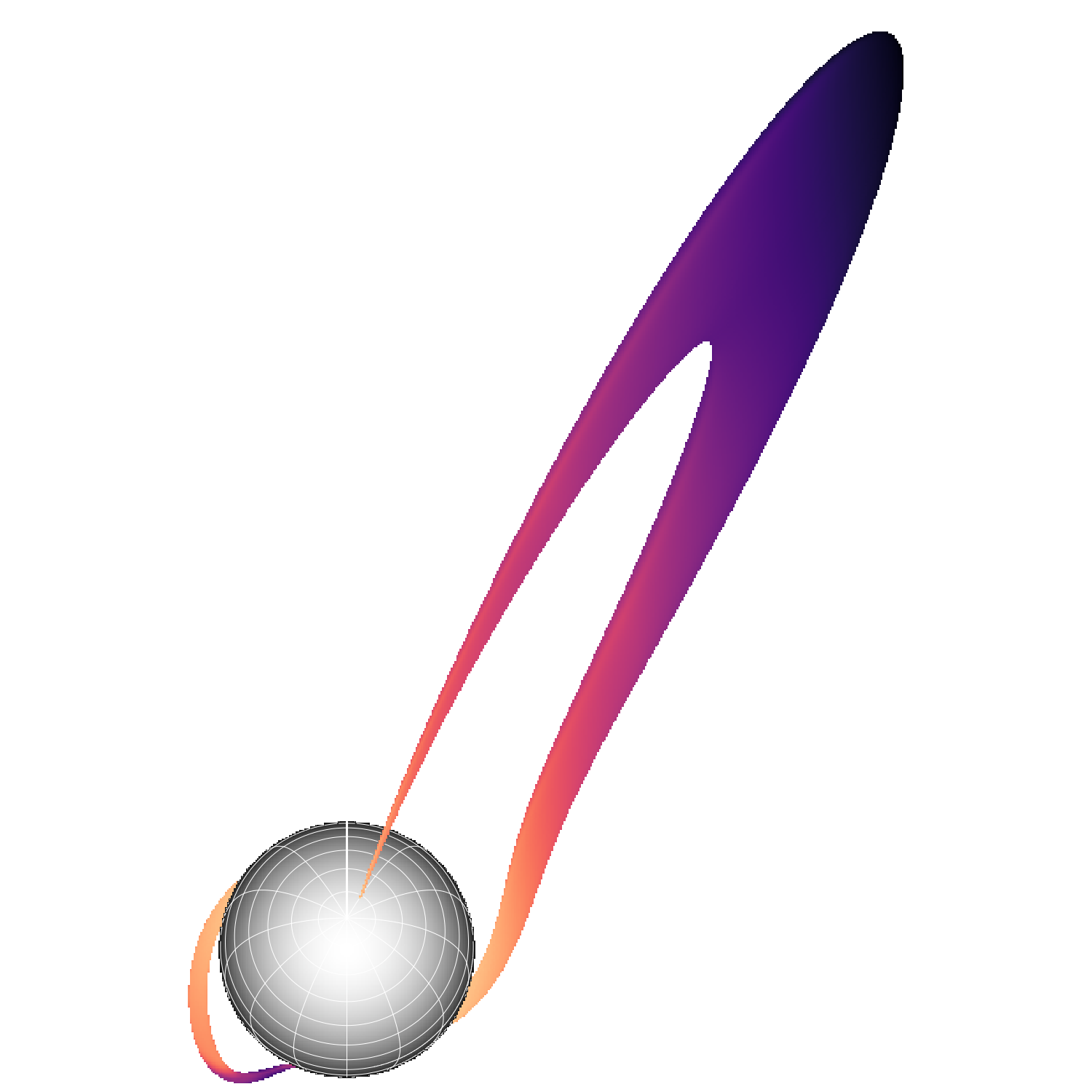} &
    \includegraphics[width=0.095\linewidth]{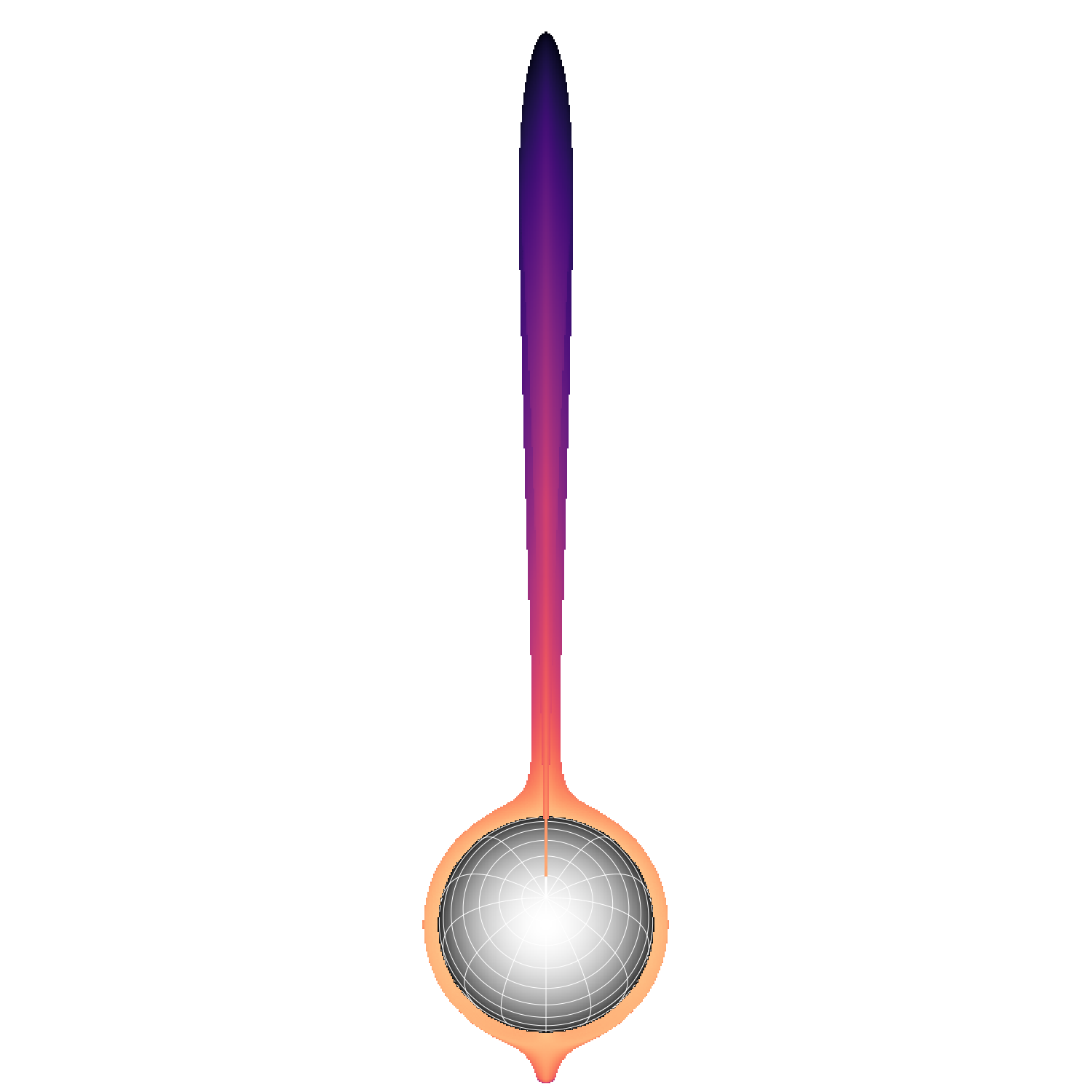}  \\

    \vcentered{$34^\circ$} & \includegraphics[width=0.095\linewidth]{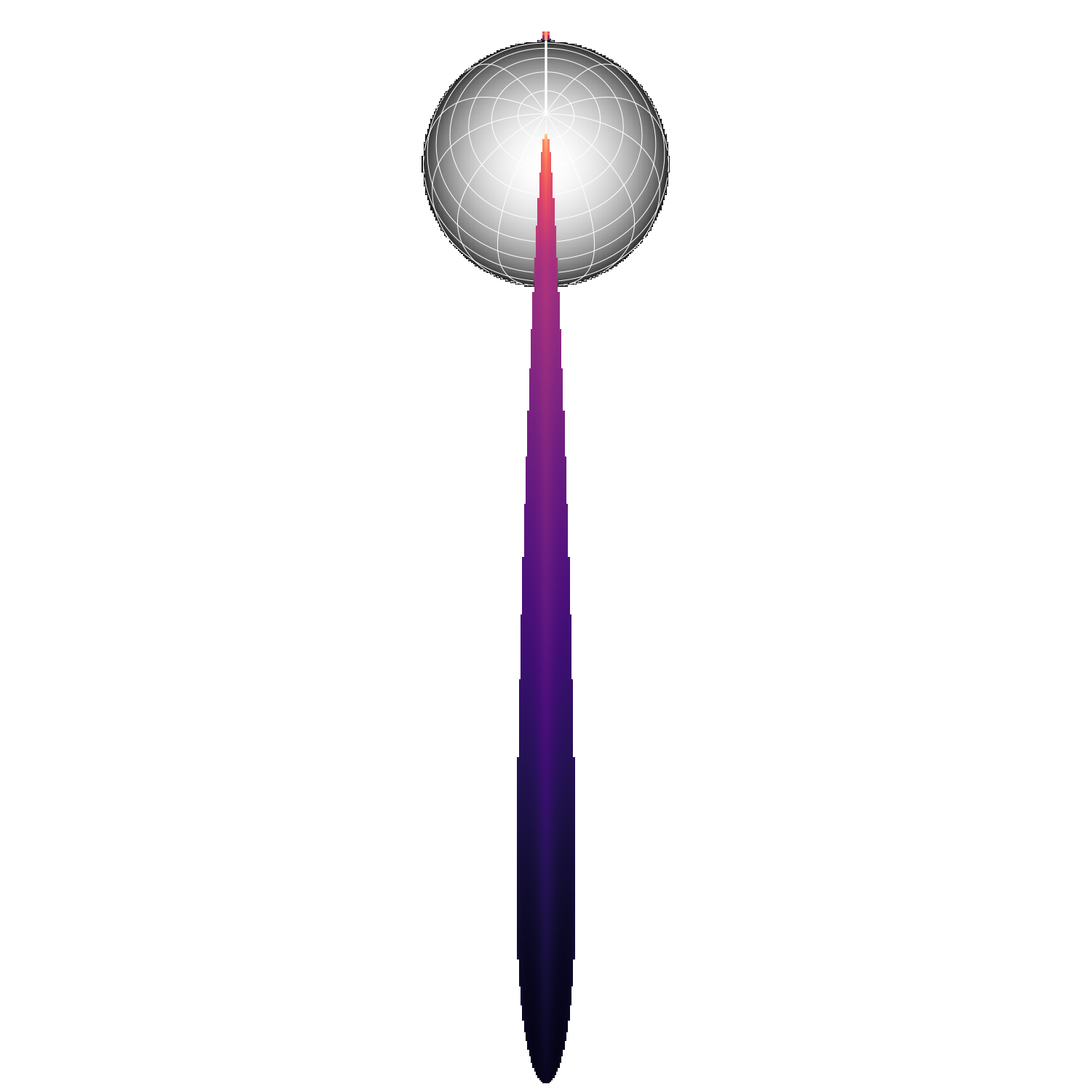} &
    \includegraphics[width=0.095\linewidth]{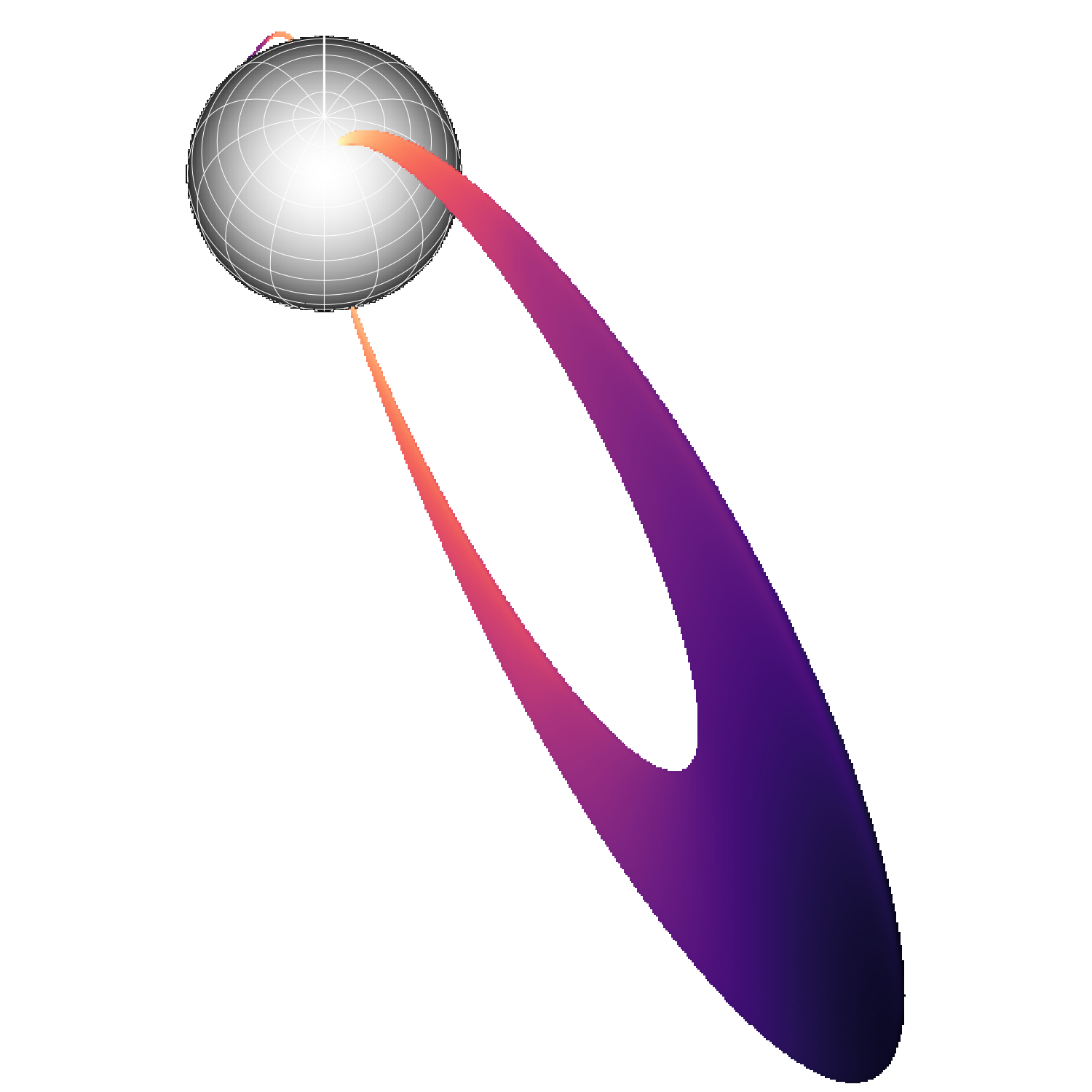} &
    \includegraphics[width=0.095\linewidth]{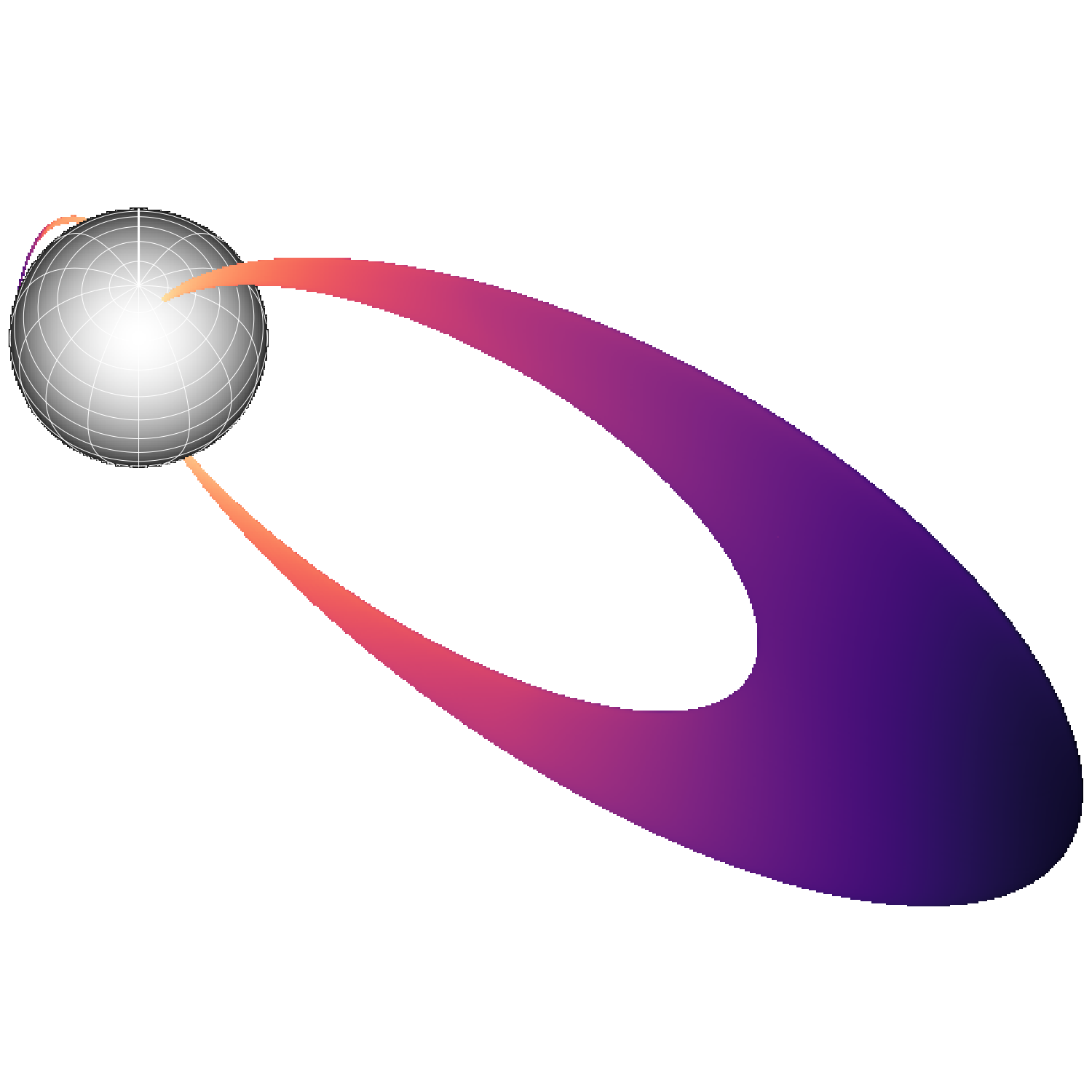} &
    \includegraphics[width=0.095\linewidth]{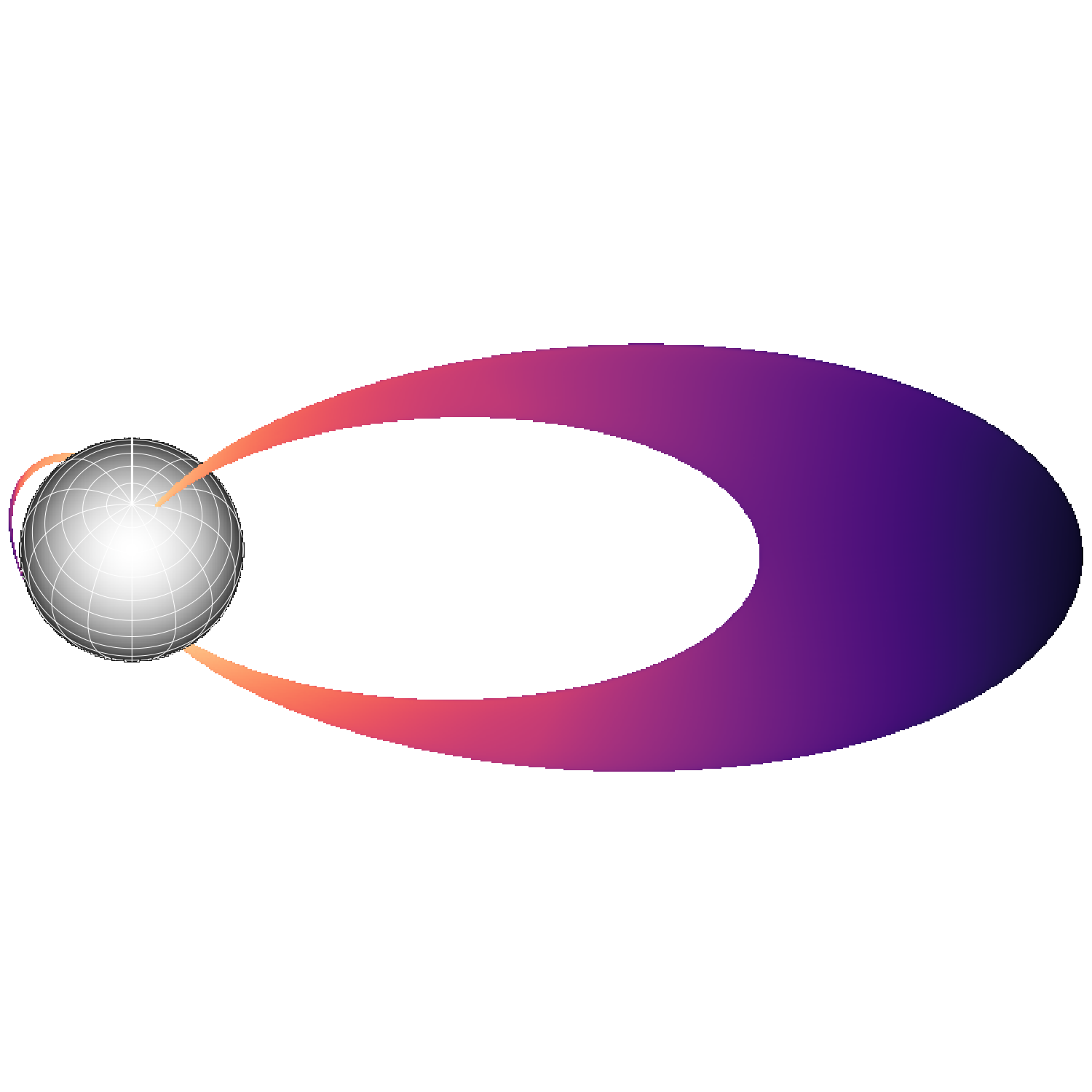} &
    \includegraphics[width=0.095\linewidth]{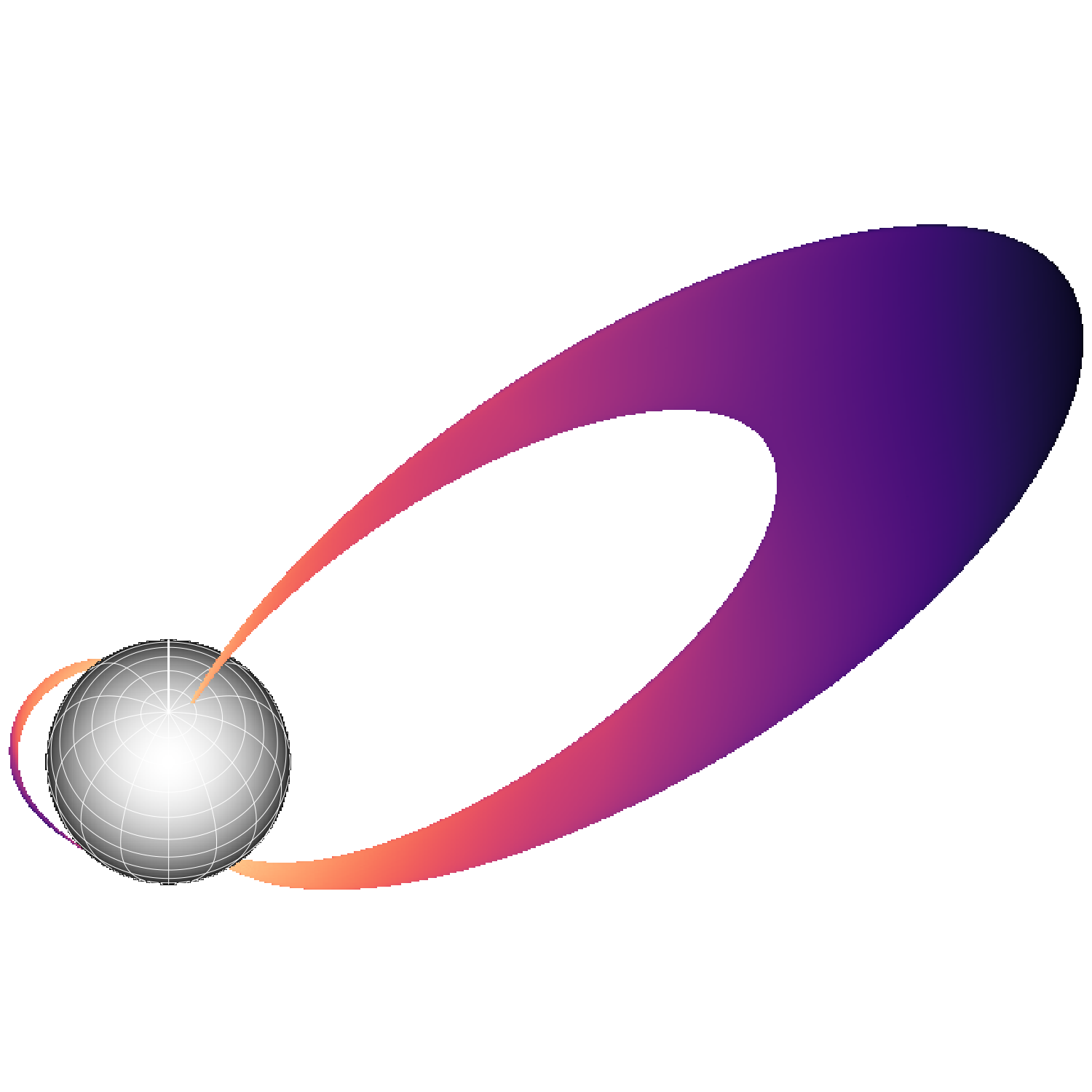} &
    \includegraphics[width=0.095\linewidth]{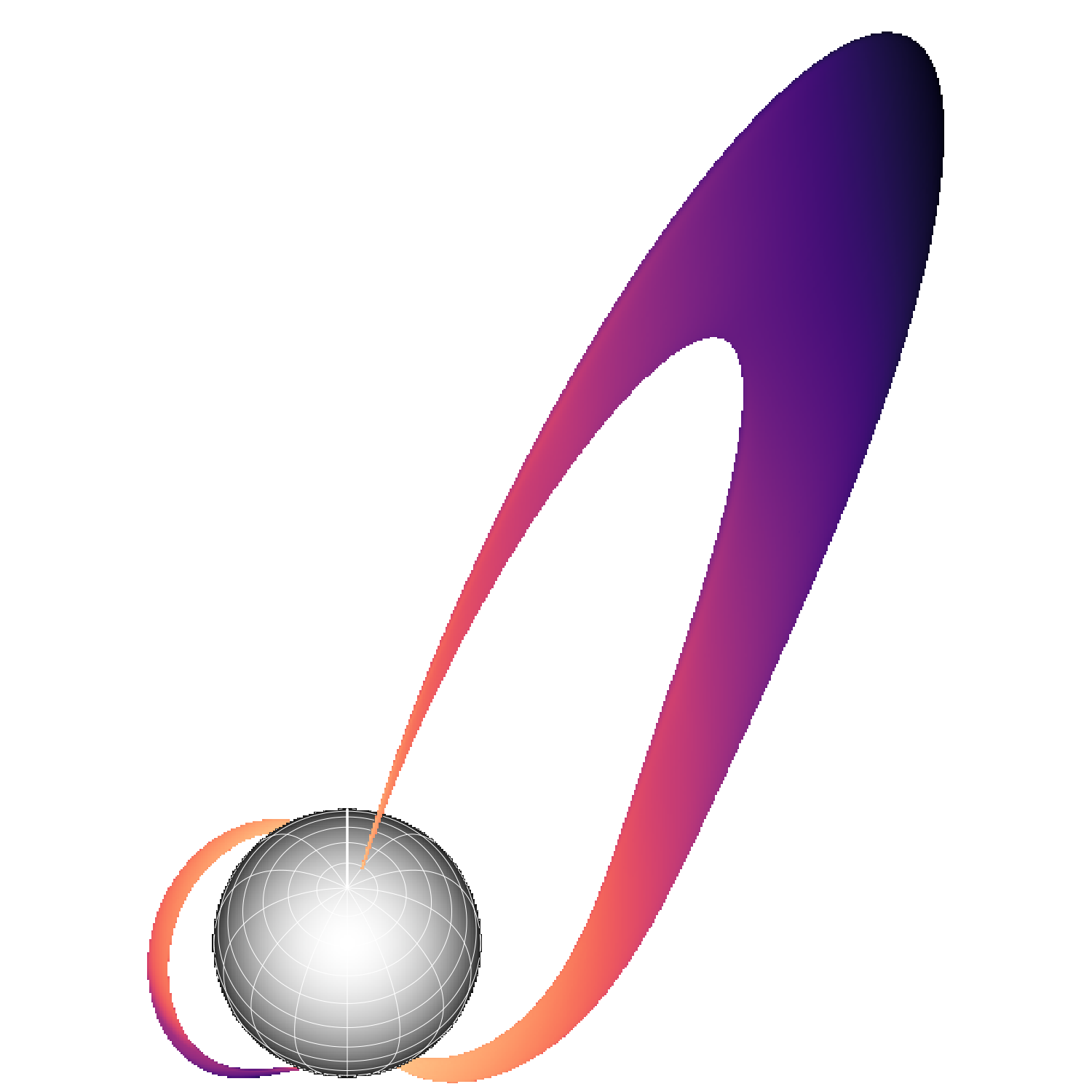} &
    \includegraphics[width=0.095\linewidth]{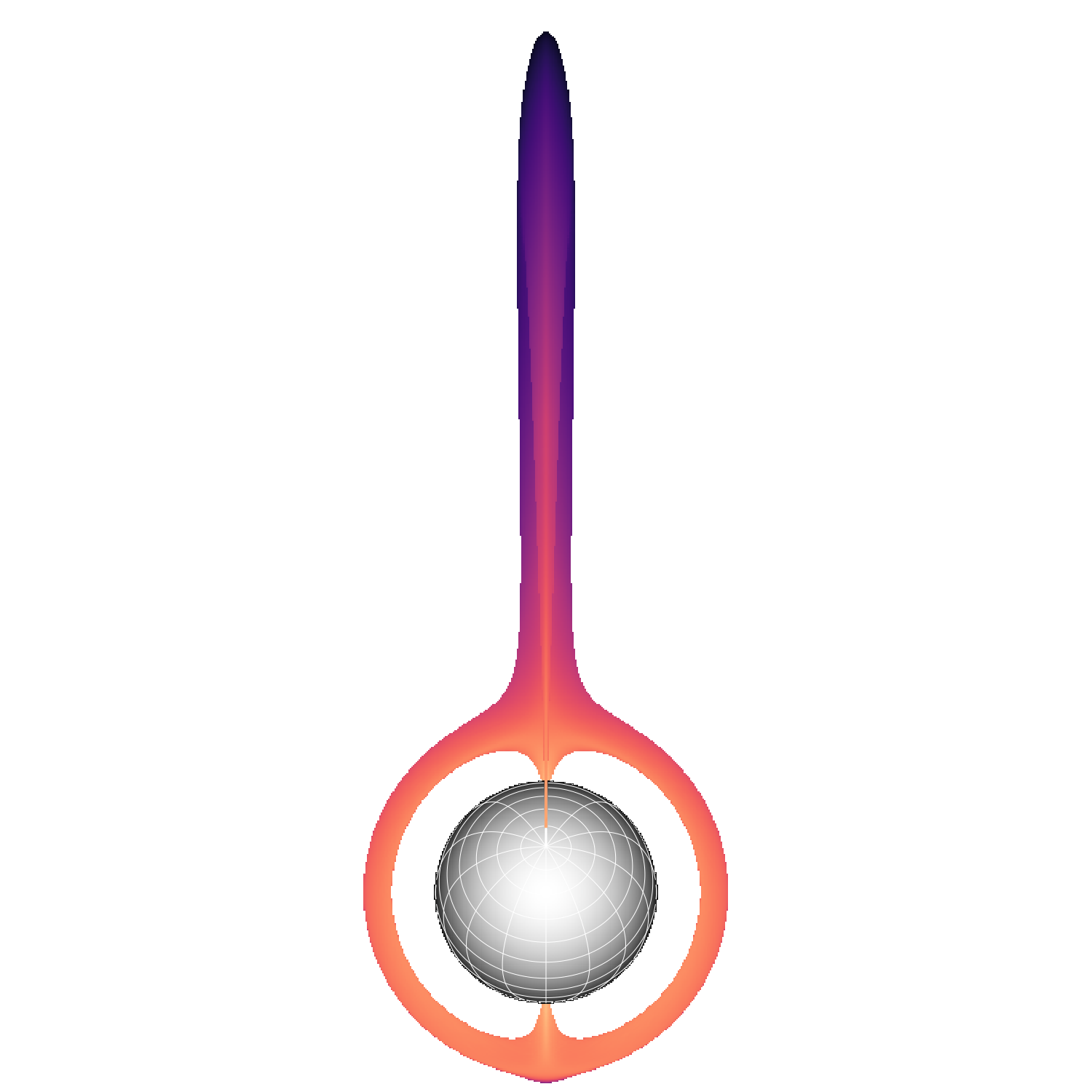}  \\

    \vcentered{$48^\circ$} & \includegraphics[width=0.095\linewidth]{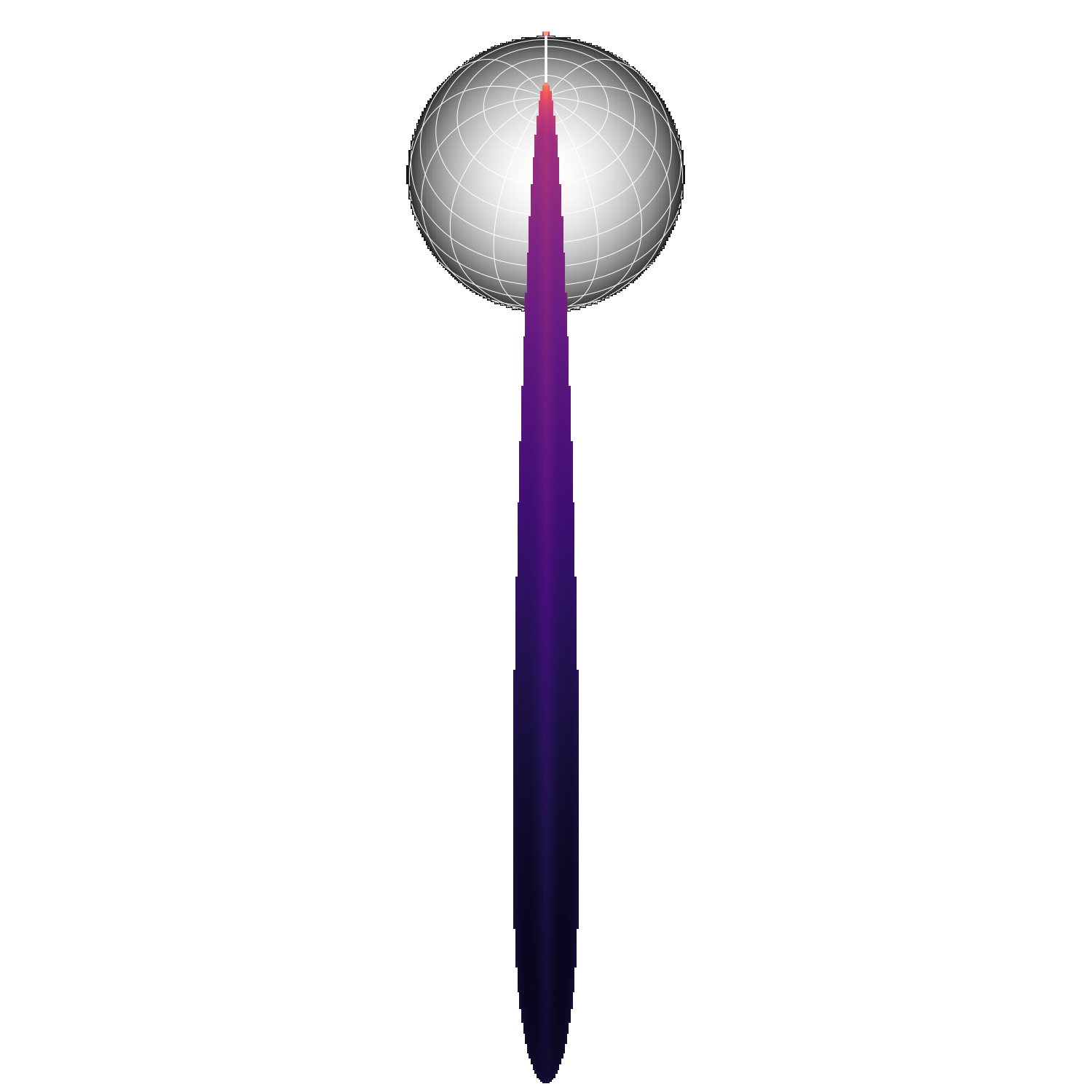} &
    \includegraphics[width=0.095\linewidth]{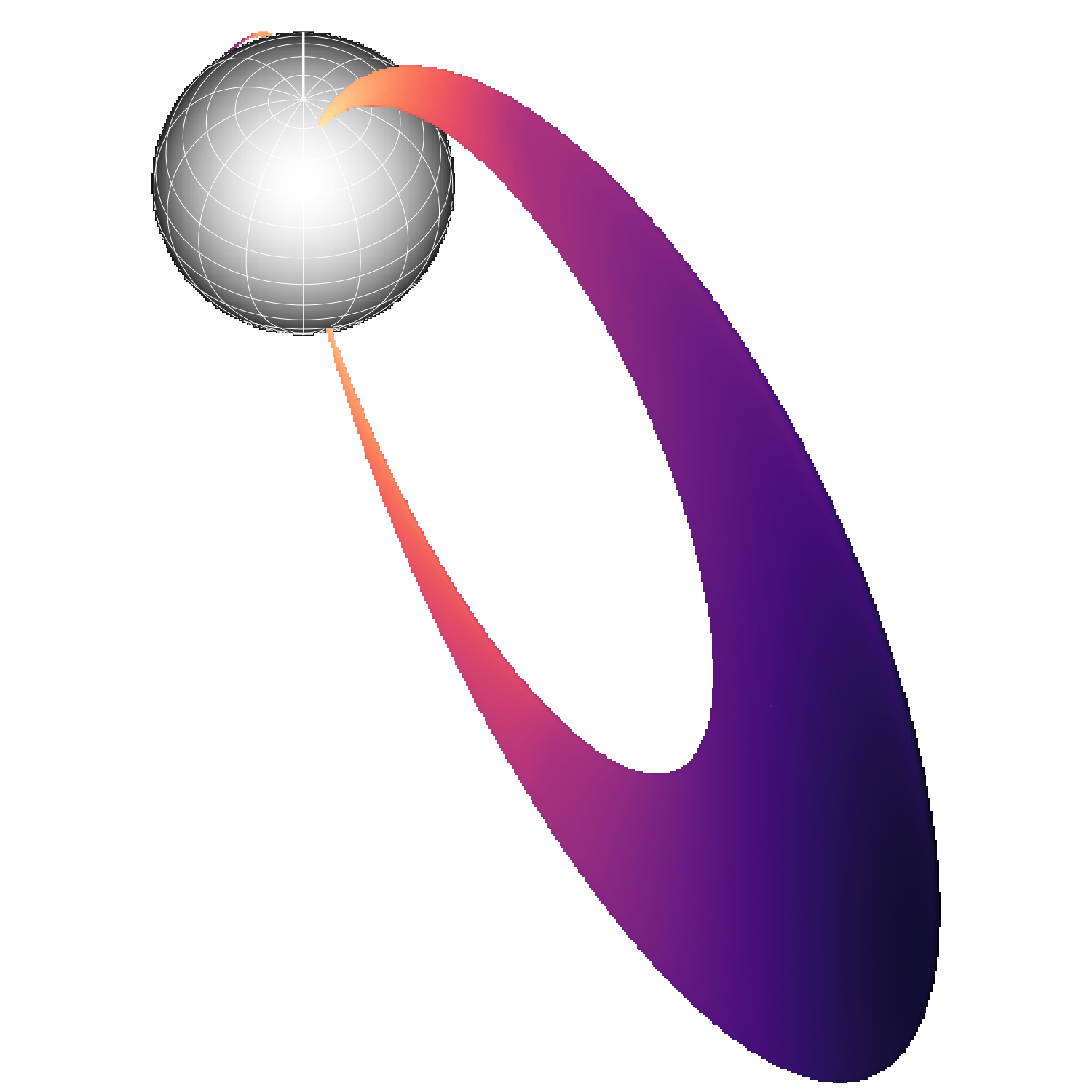} &
    \includegraphics[width=0.095\linewidth]{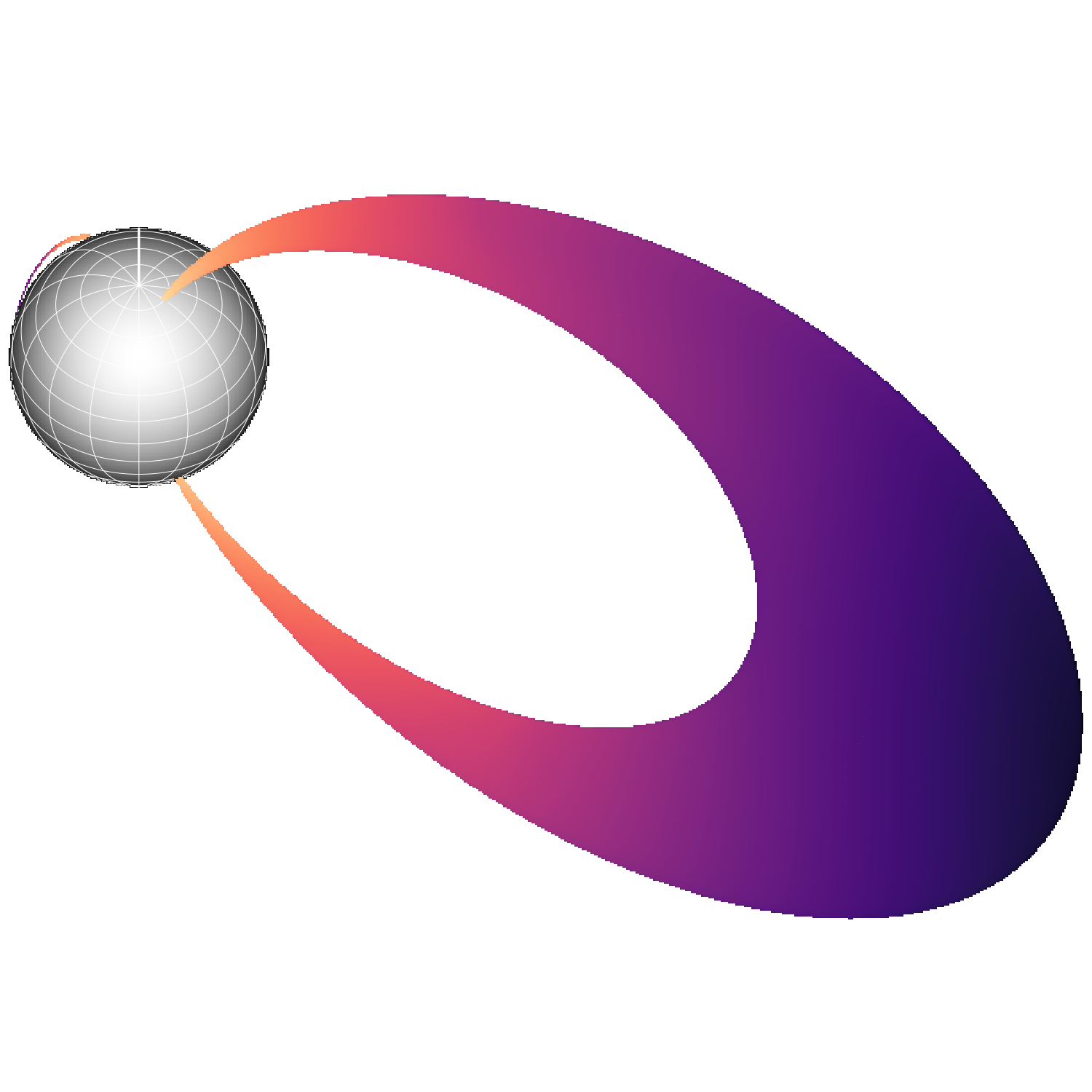} &
    \includegraphics[width=0.095\linewidth]{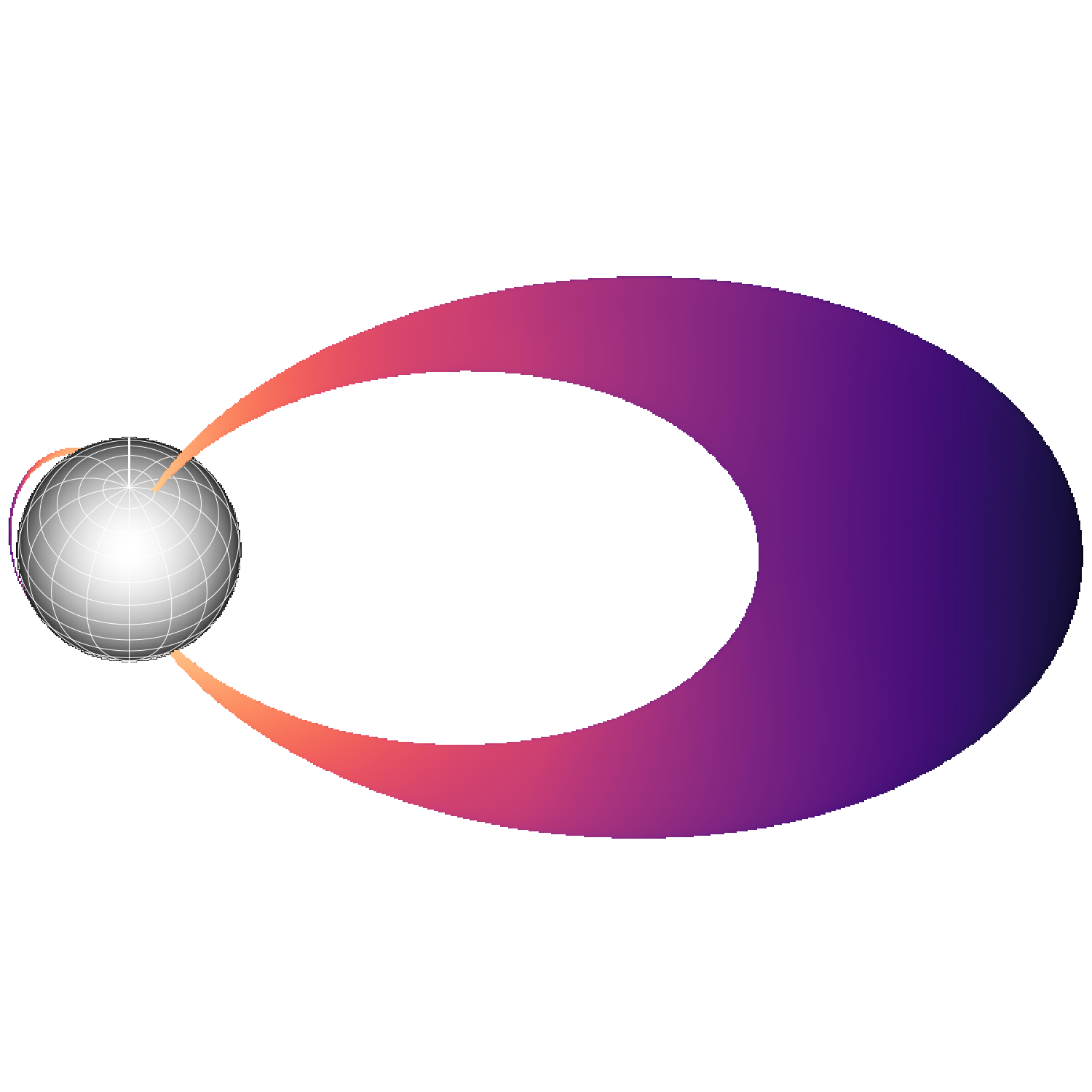} &
    \includegraphics[width=0.095\linewidth]{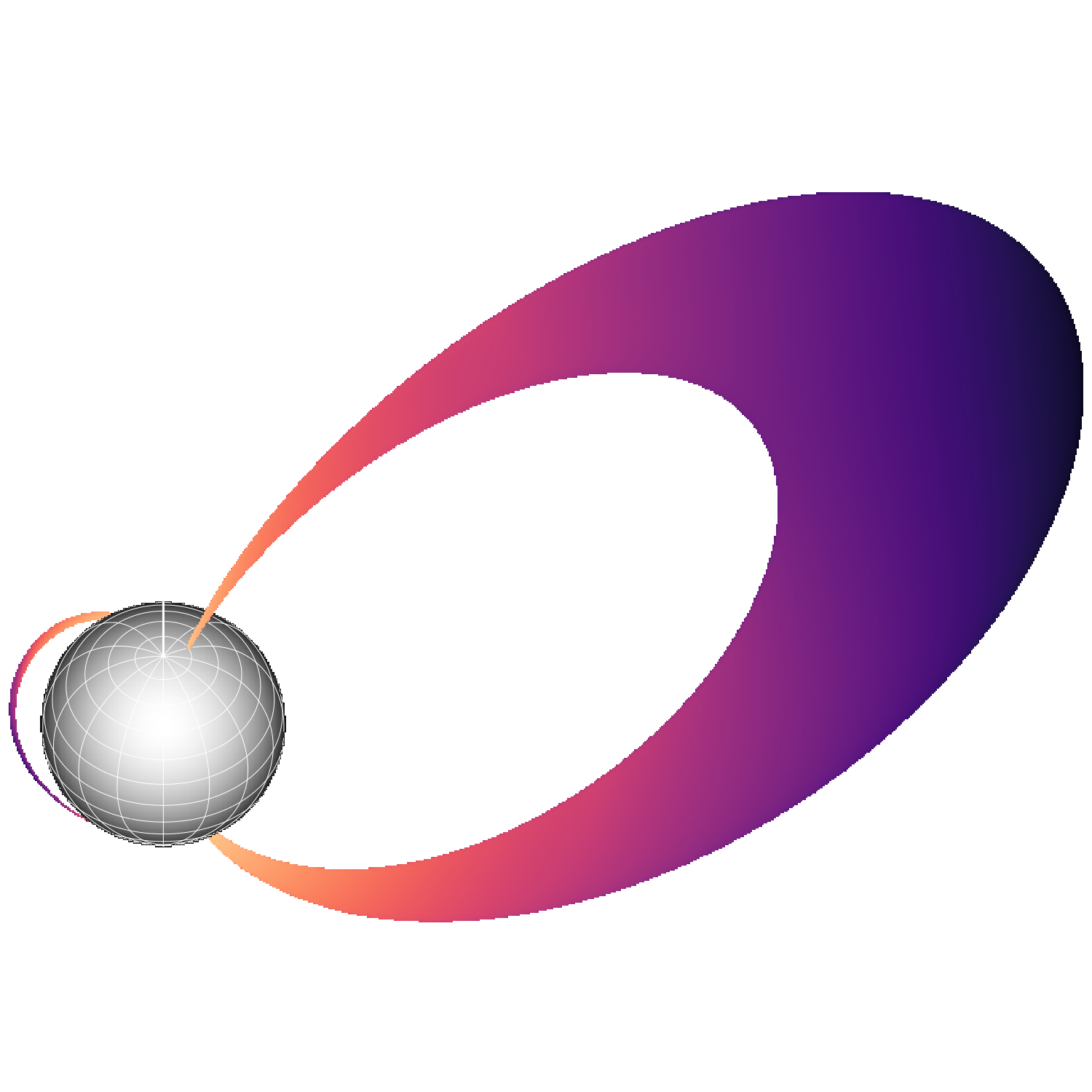} &
    \includegraphics[width=0.095\linewidth]{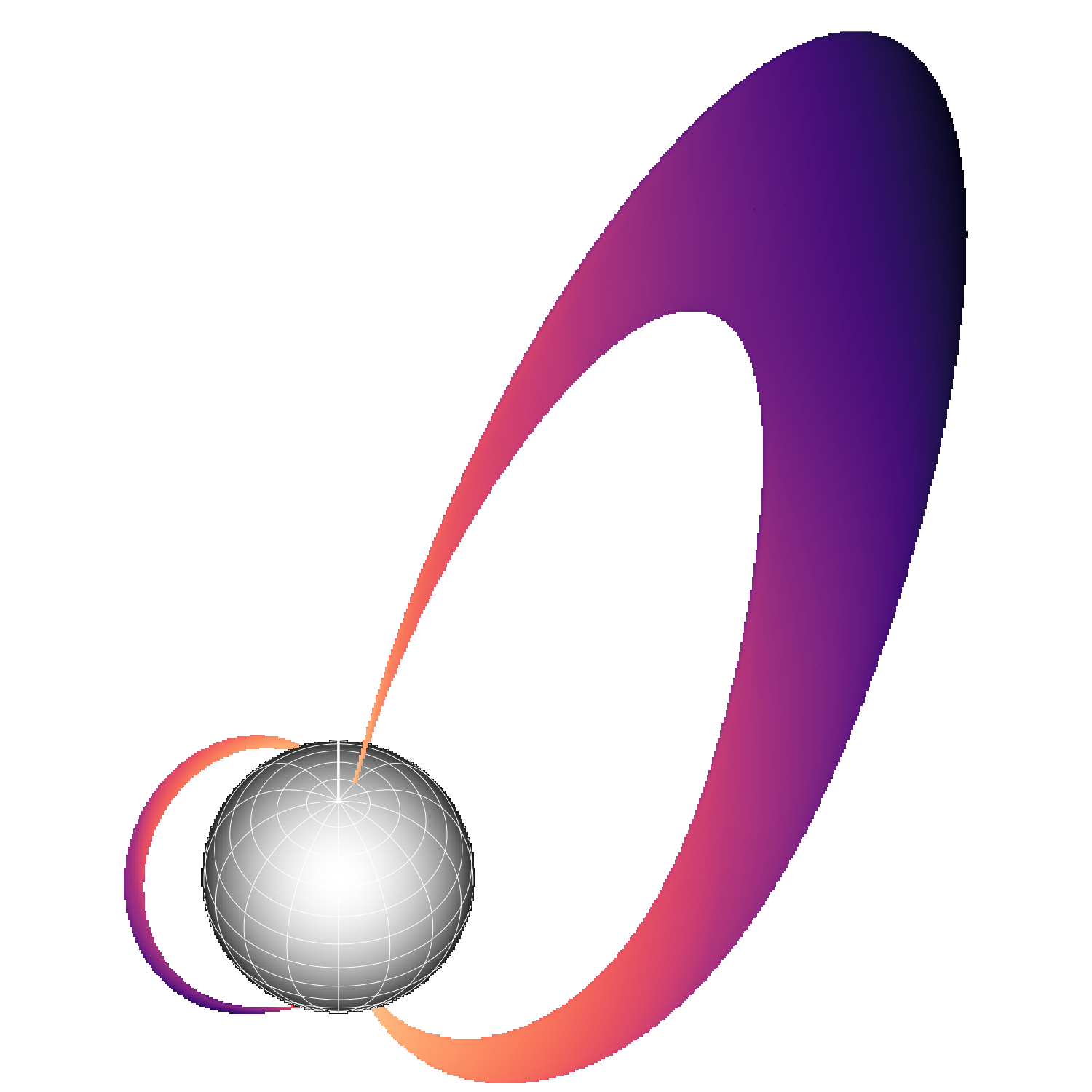} &
    \includegraphics[width=0.095\linewidth]{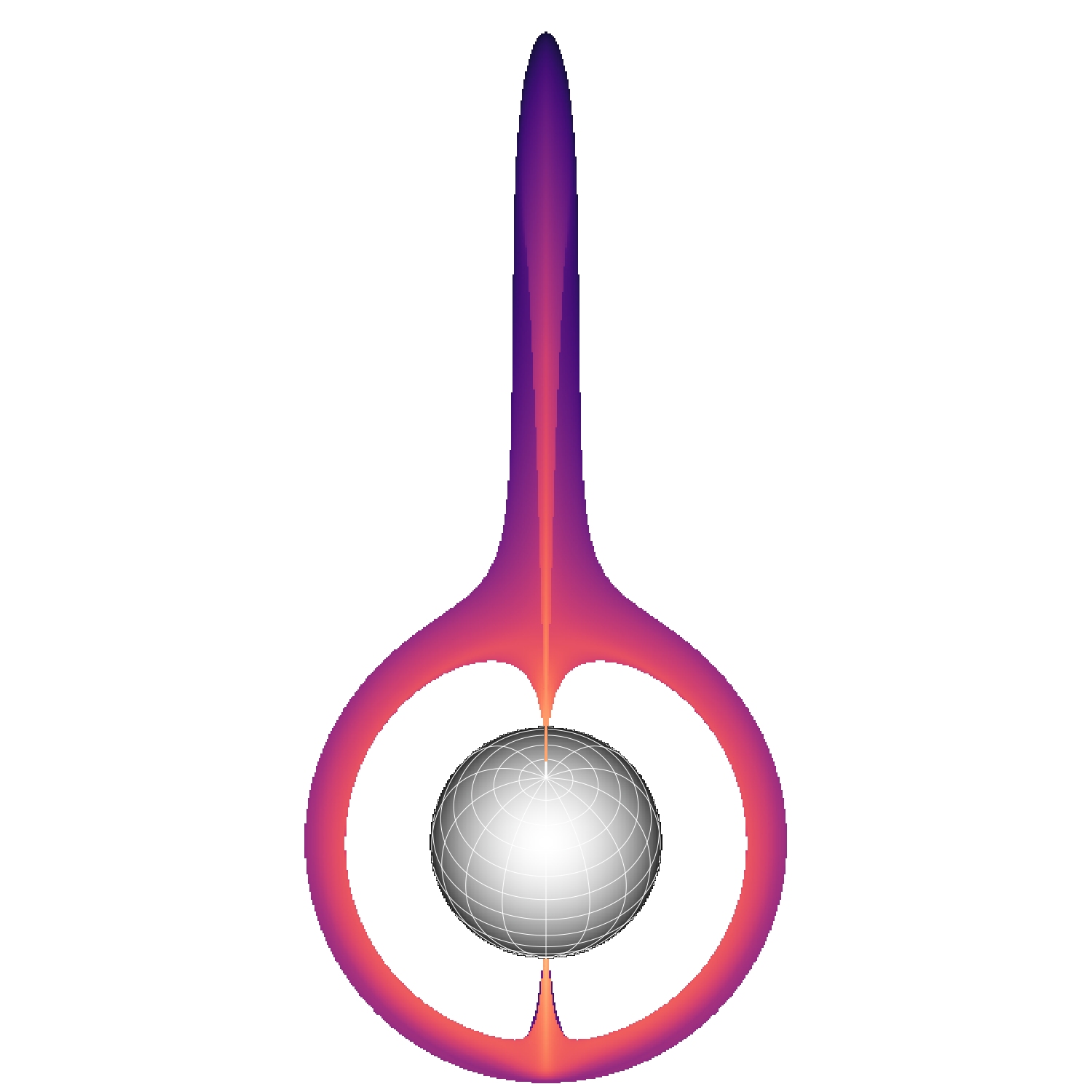}  \\

    \vcentered{$62^\circ$} &\includegraphics[width=0.095\linewidth]{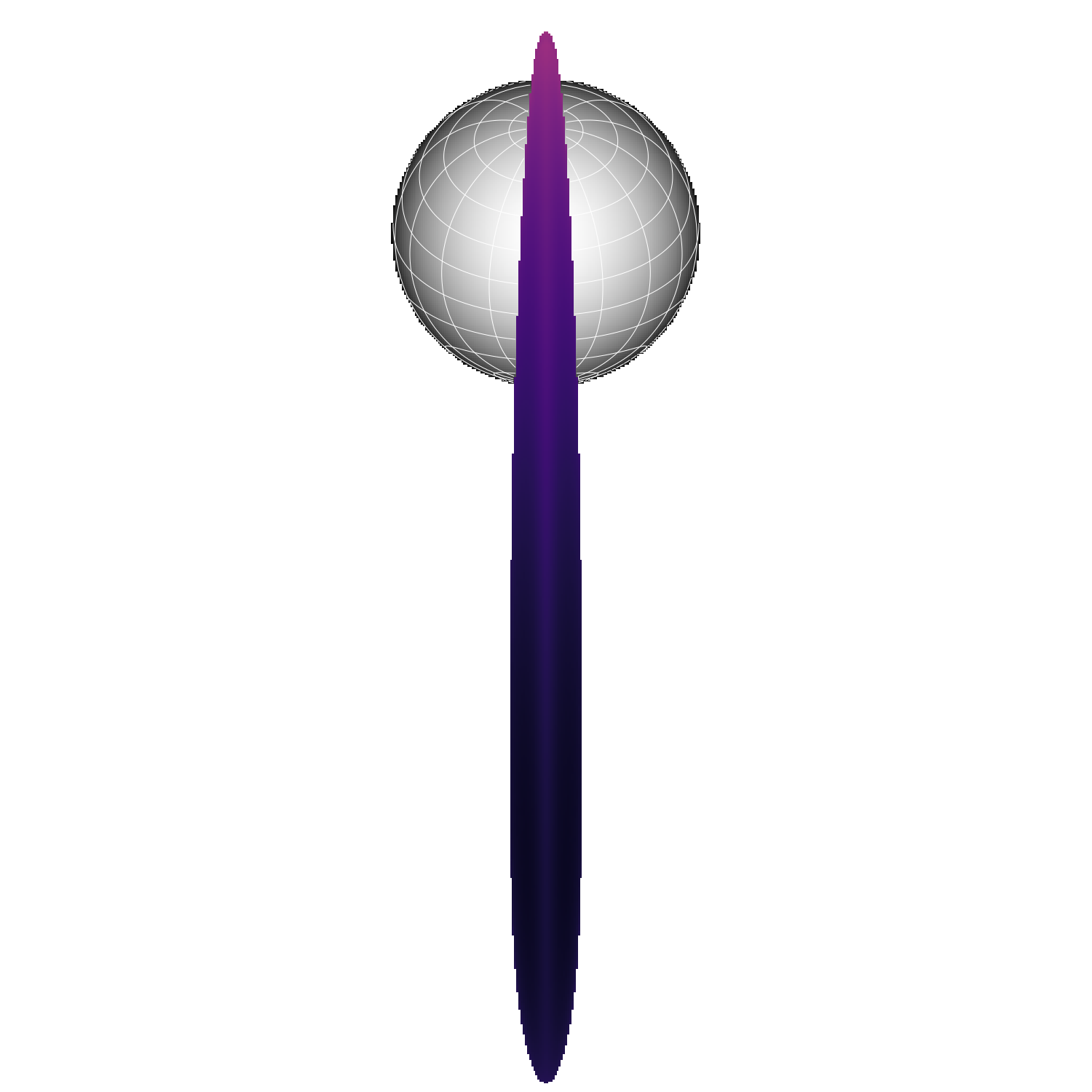} &
    \includegraphics[width=0.095\linewidth]{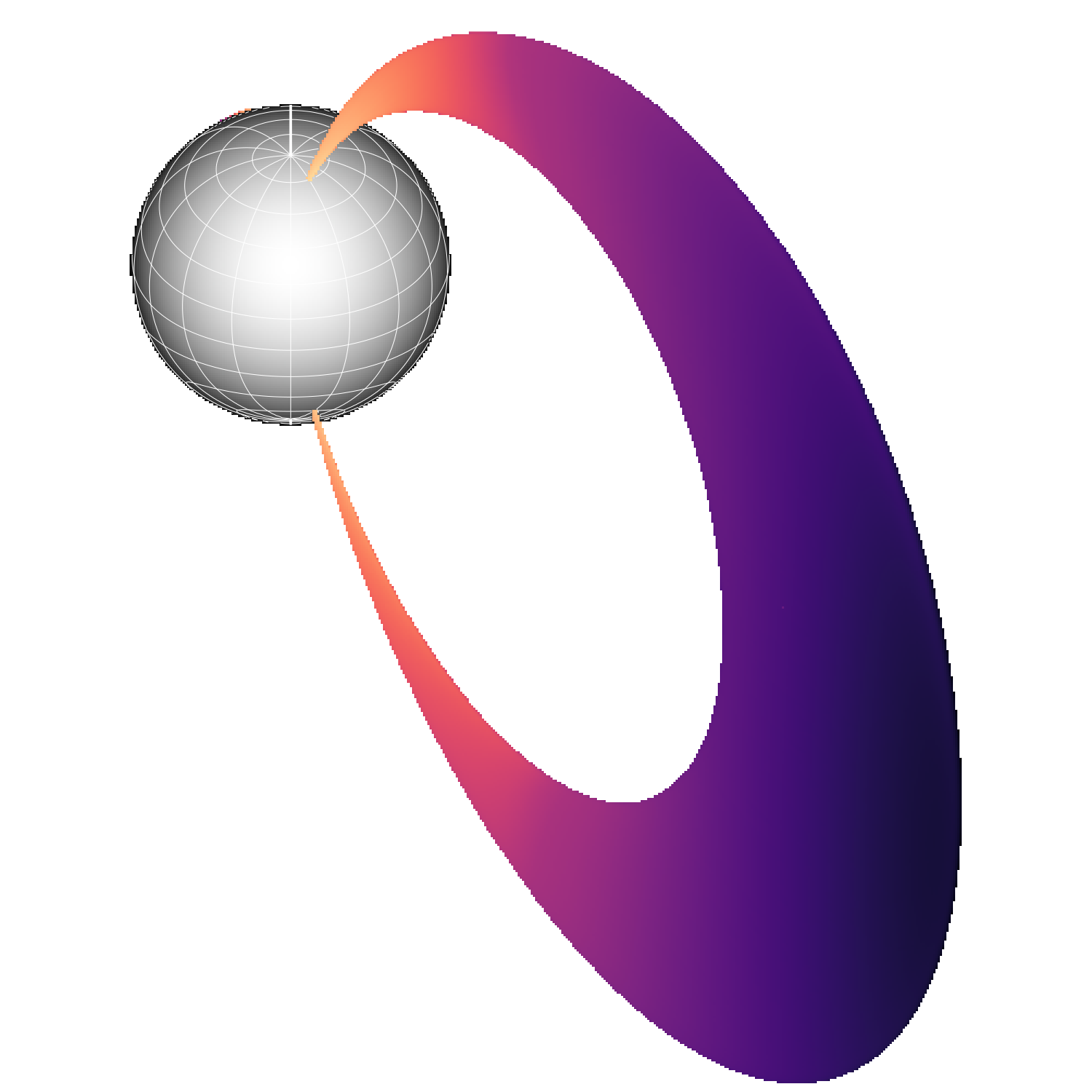} &
    \includegraphics[width=0.095\linewidth]{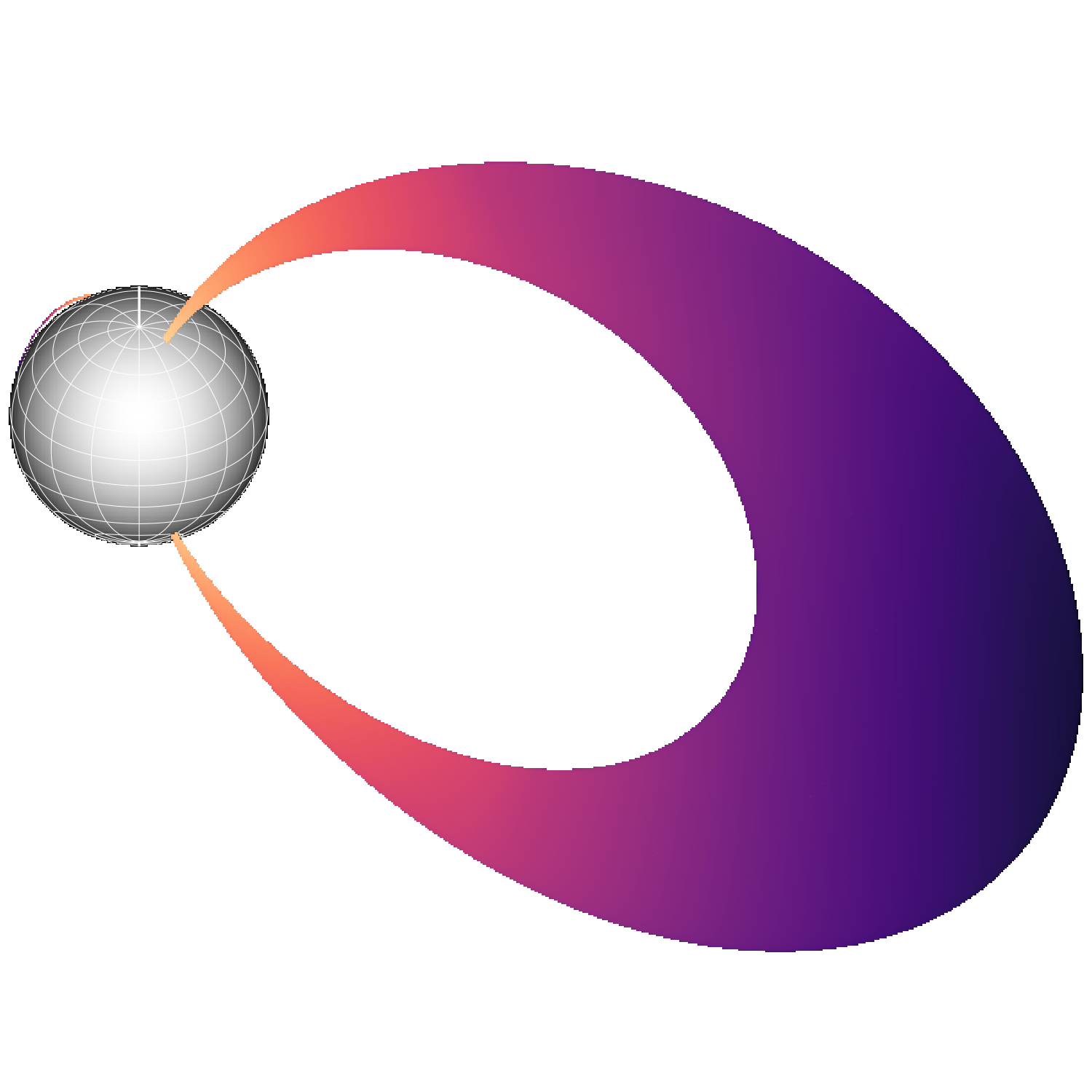} &
    \includegraphics[width=0.095\linewidth]{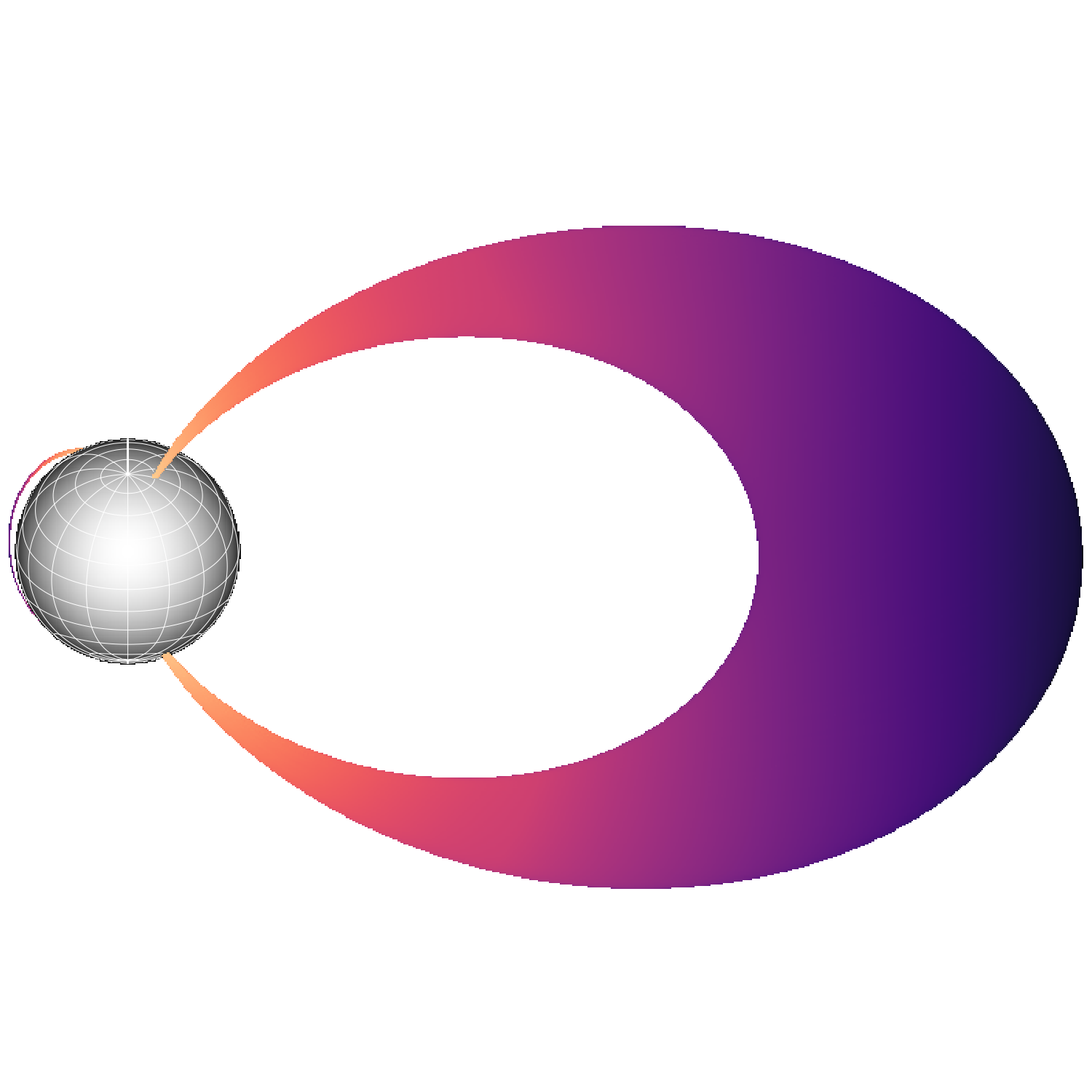} &
    \includegraphics[width=0.095\linewidth]{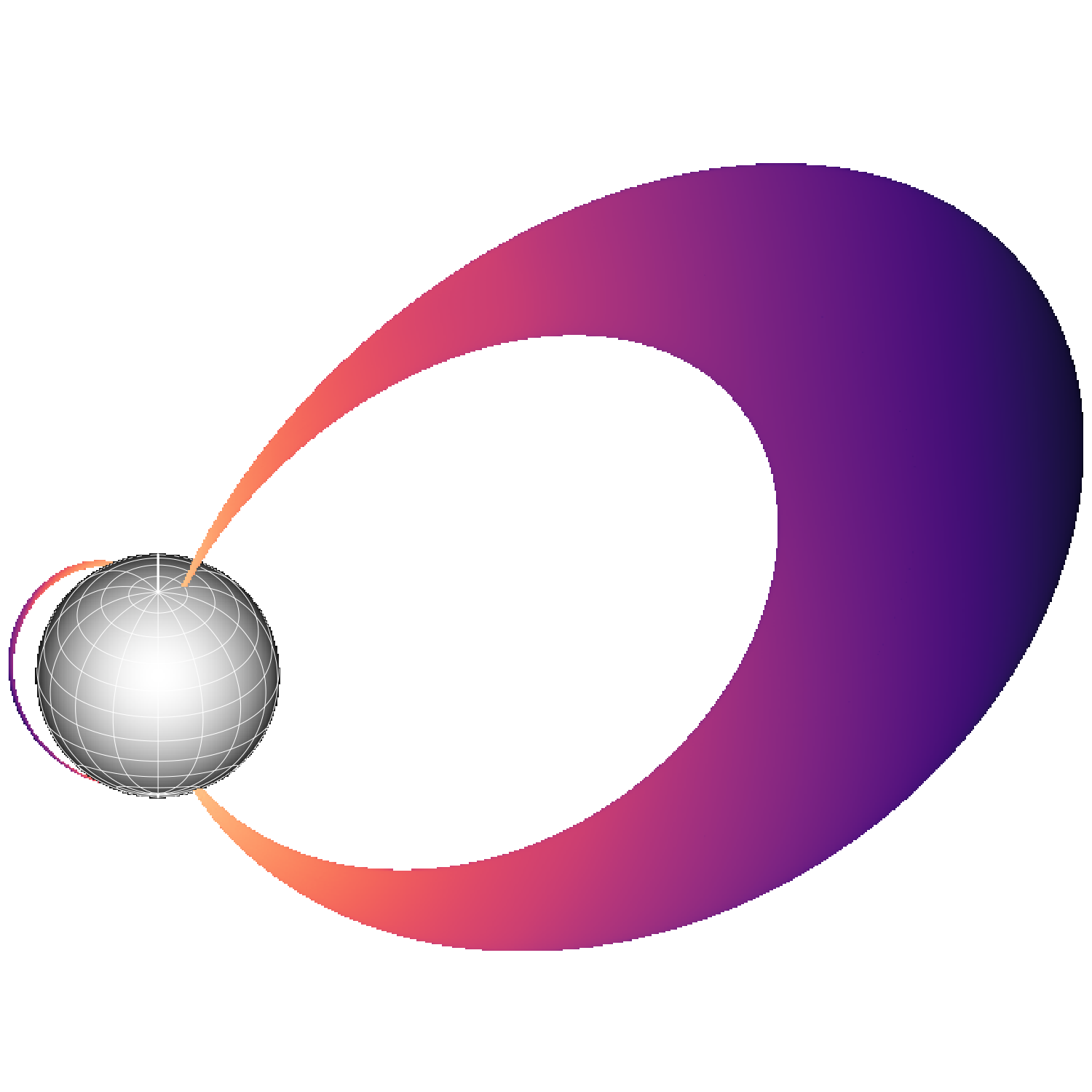} &
    \includegraphics[width=0.095\linewidth]{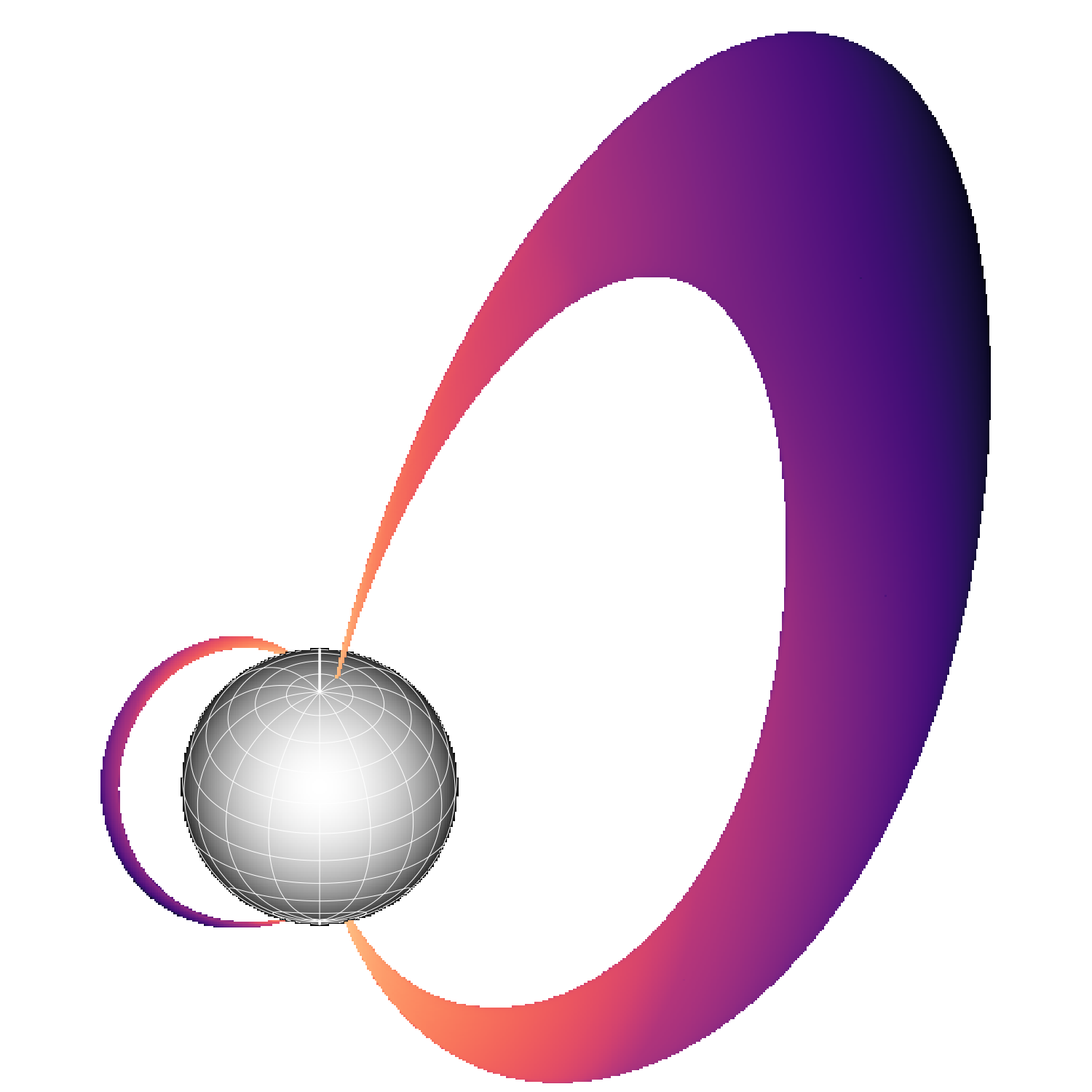} &
    \includegraphics[width=0.095\linewidth]{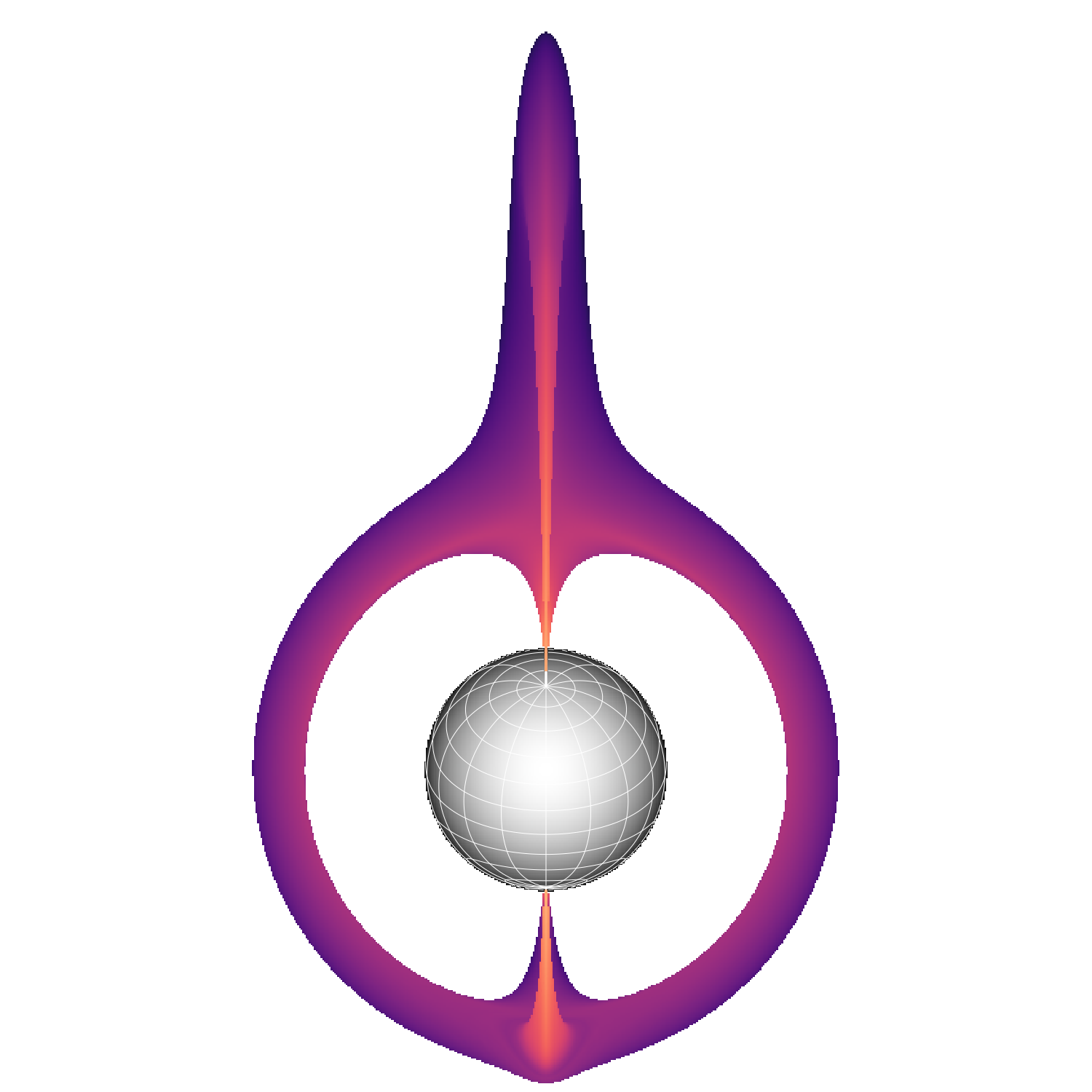}  \\

    \vcentered{$76^\circ$} &\includegraphics[width=0.095\linewidth]{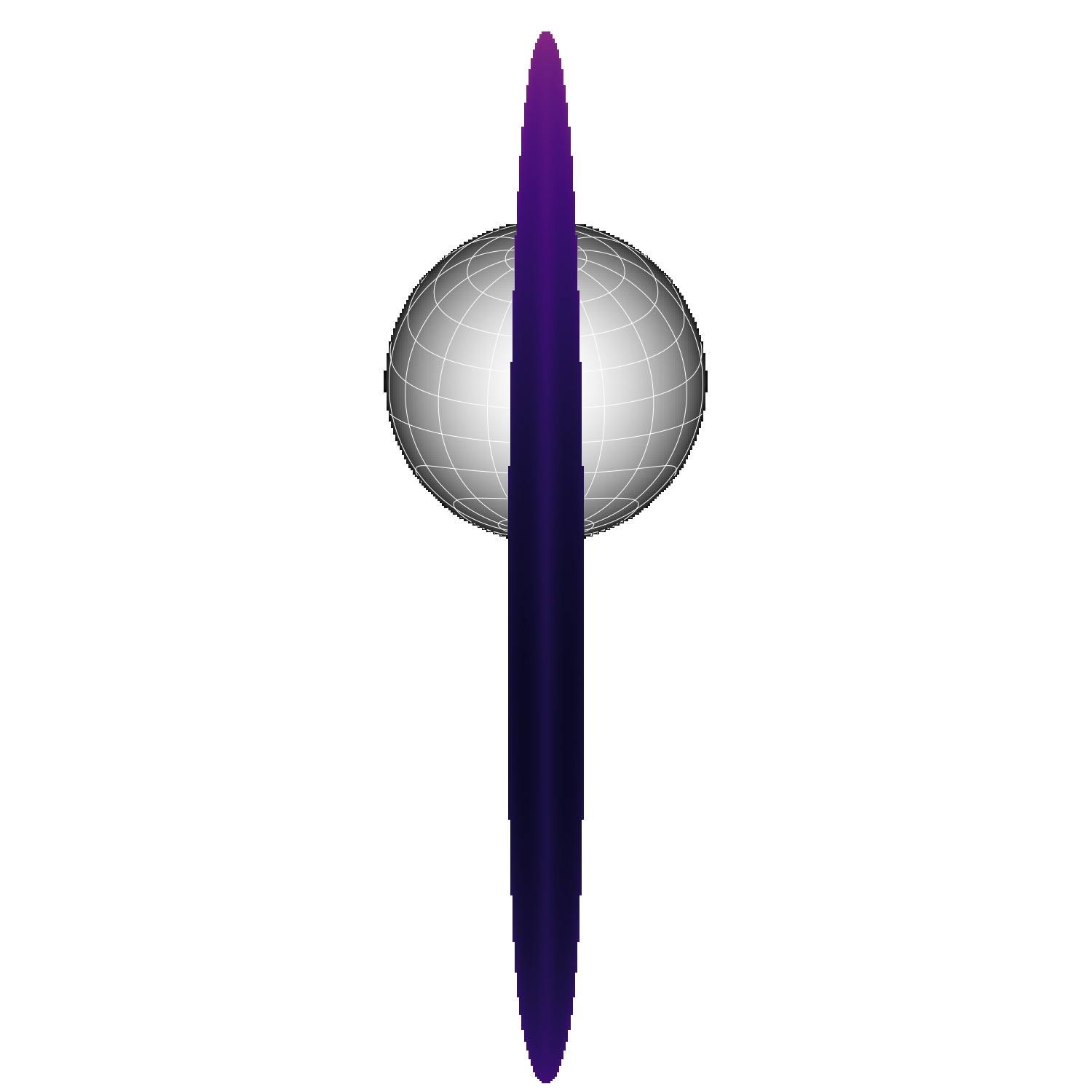} &
    \includegraphics[width=0.095\linewidth]{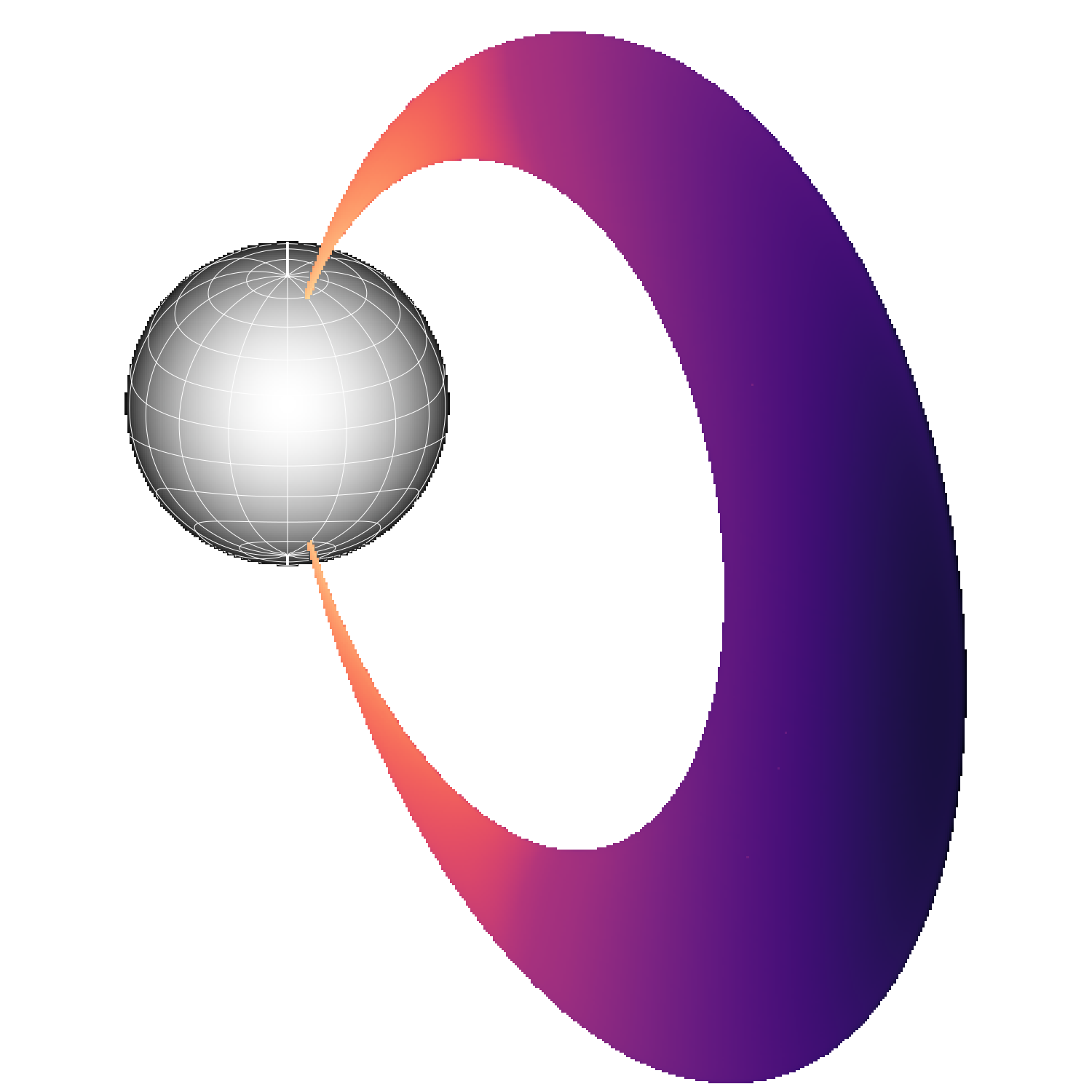} &
    \includegraphics[width=0.095\linewidth]{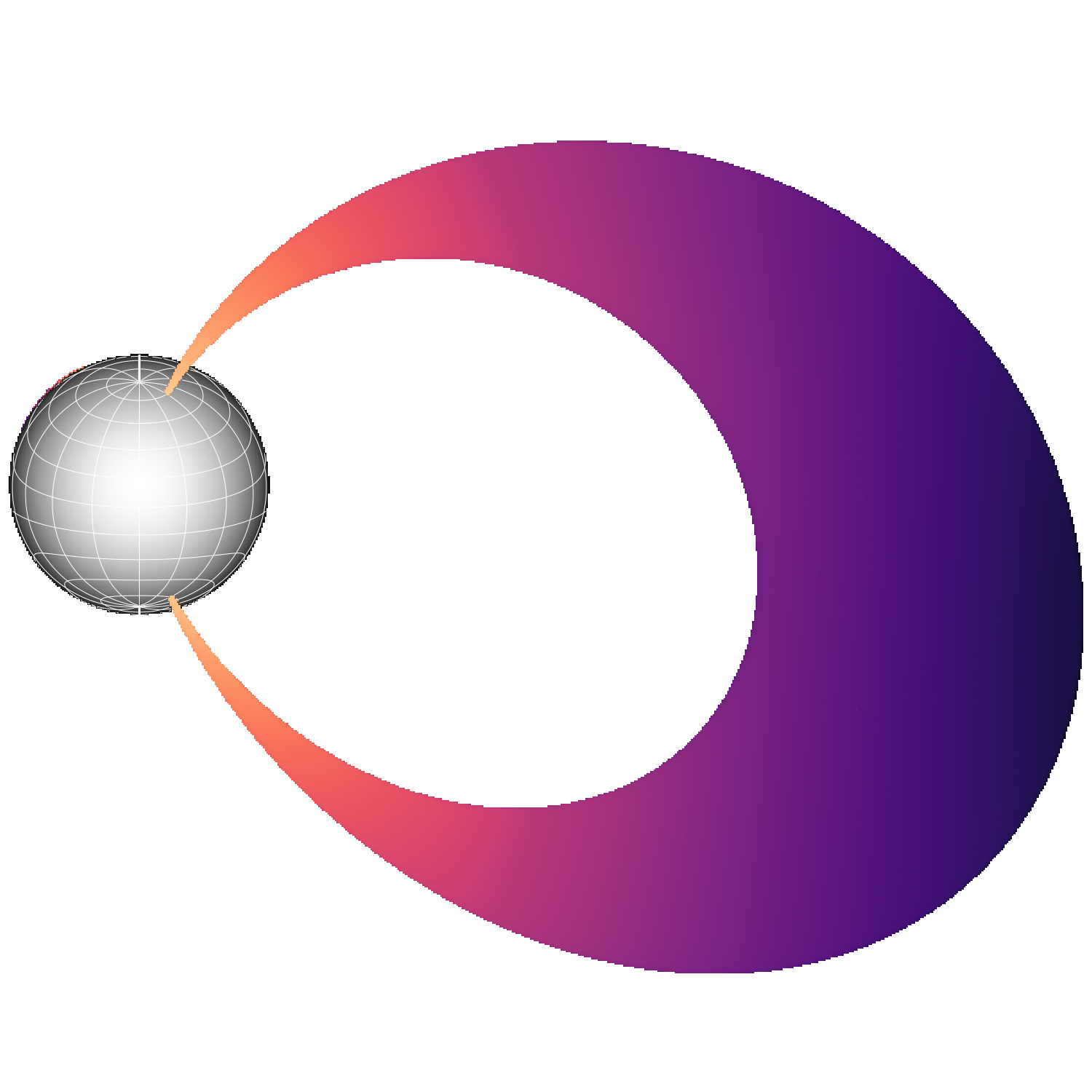} &
    \includegraphics[width=0.095\linewidth]{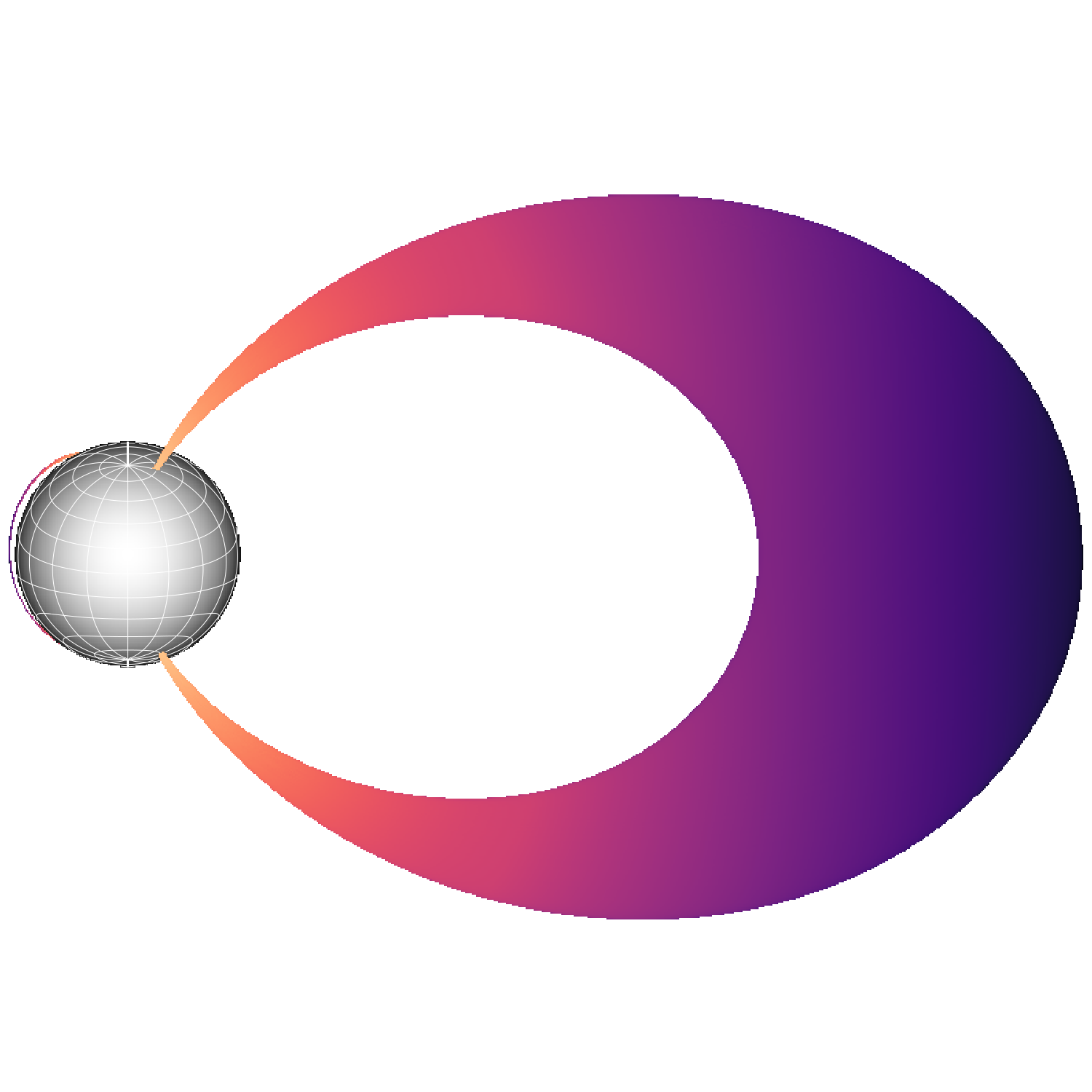} &
    \includegraphics[width=0.095\linewidth]{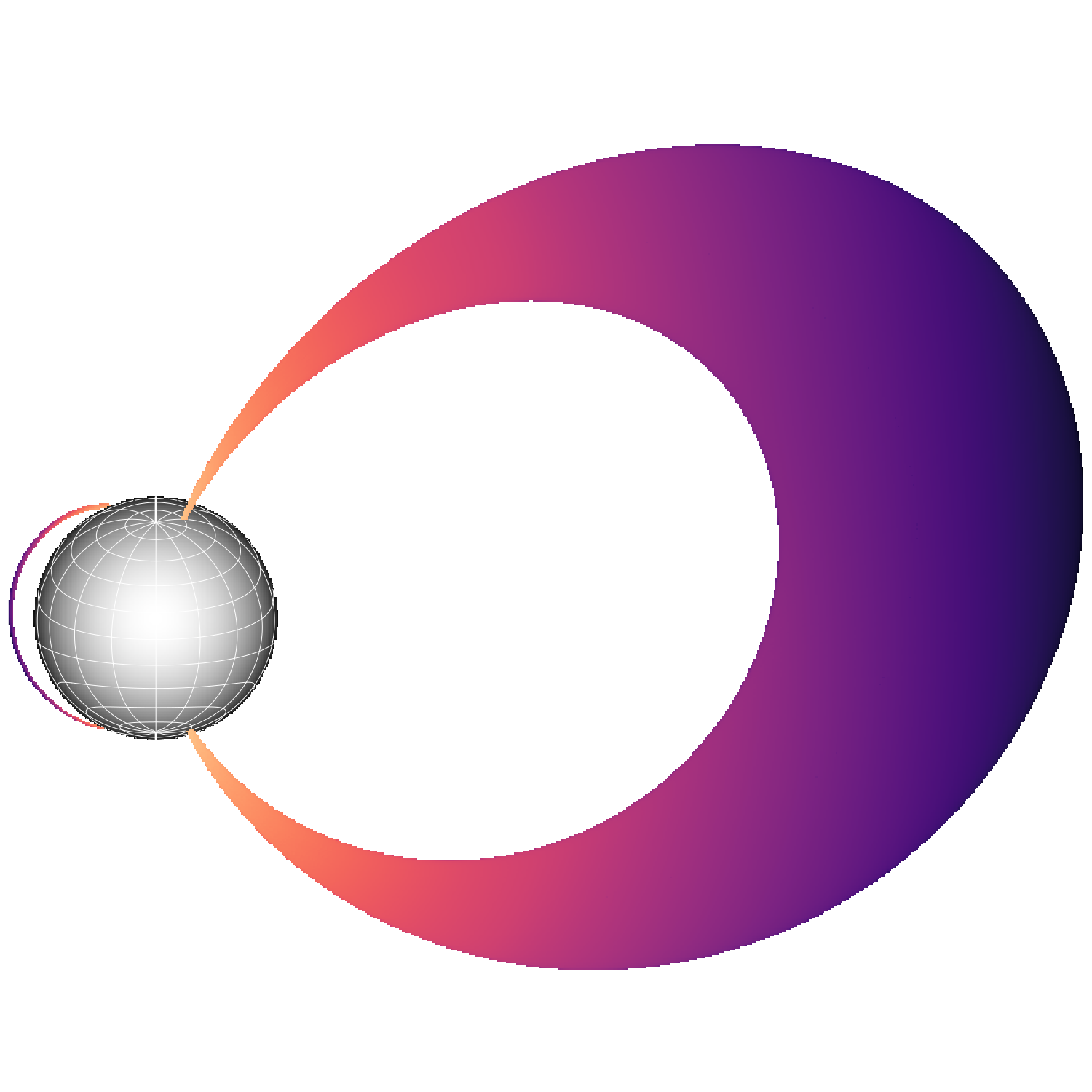} &
    \includegraphics[width=0.095\linewidth]{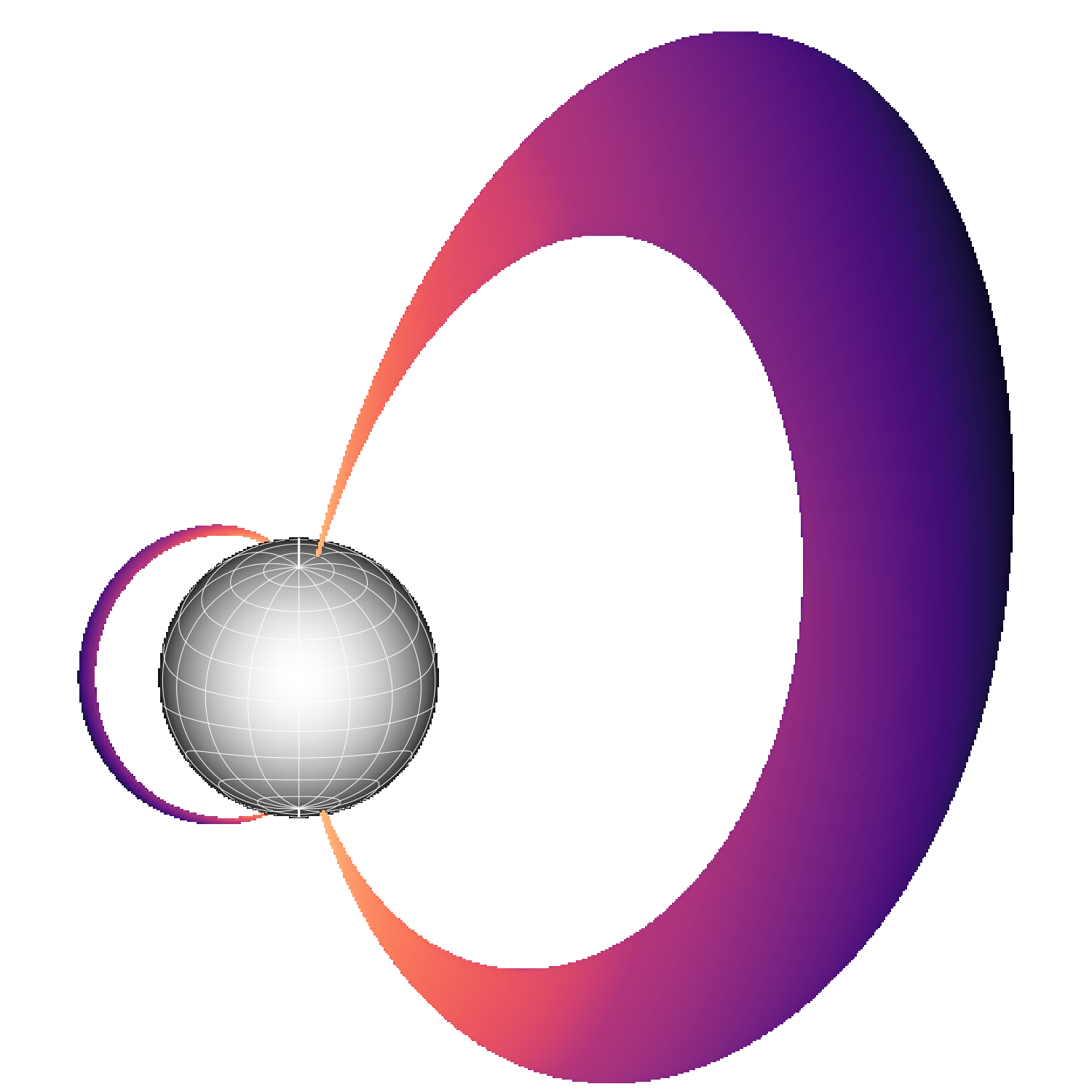} &
    \includegraphics[width=0.095\linewidth]{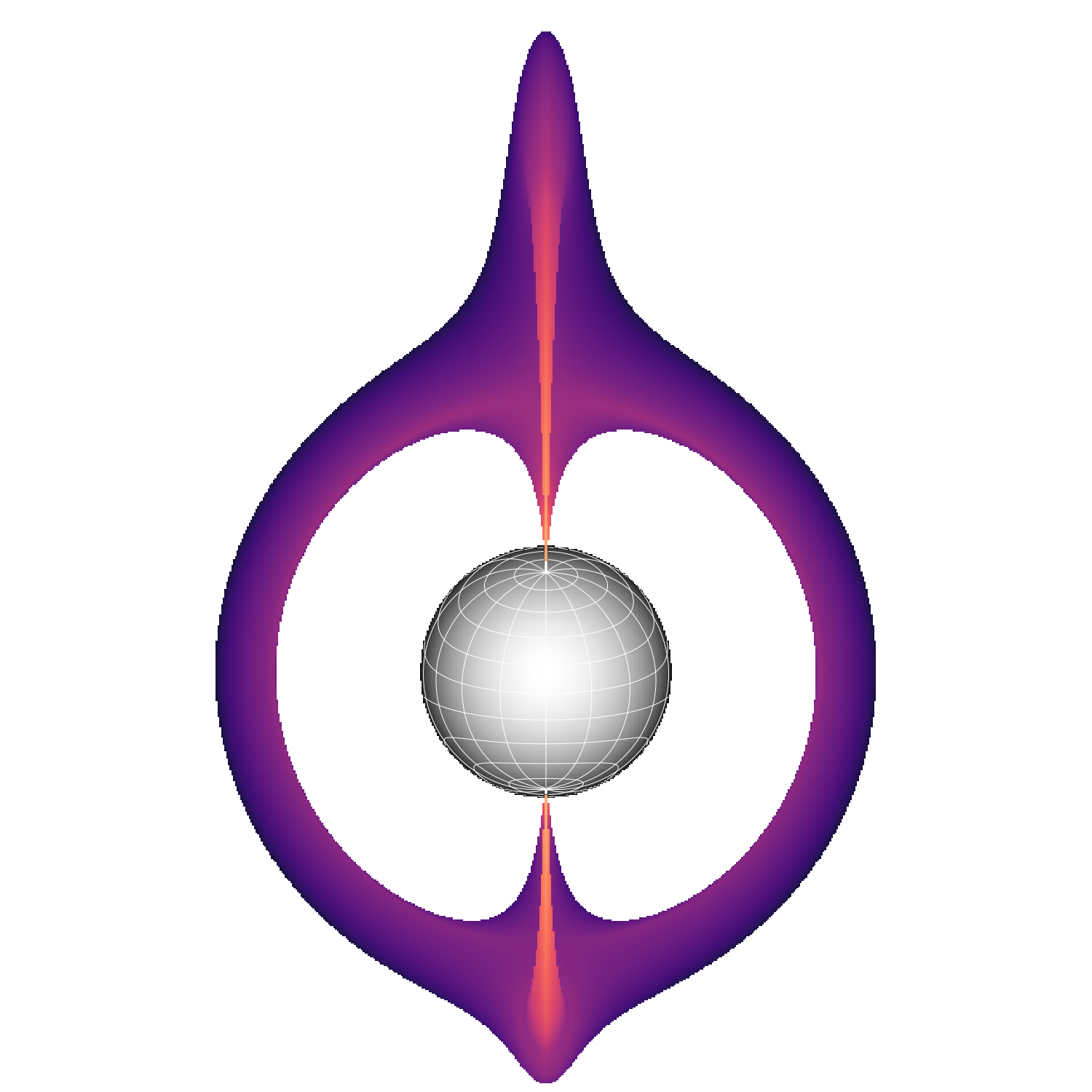}  \\

    \vcentered{$90^\circ$} &\includegraphics[width=0.095\linewidth]{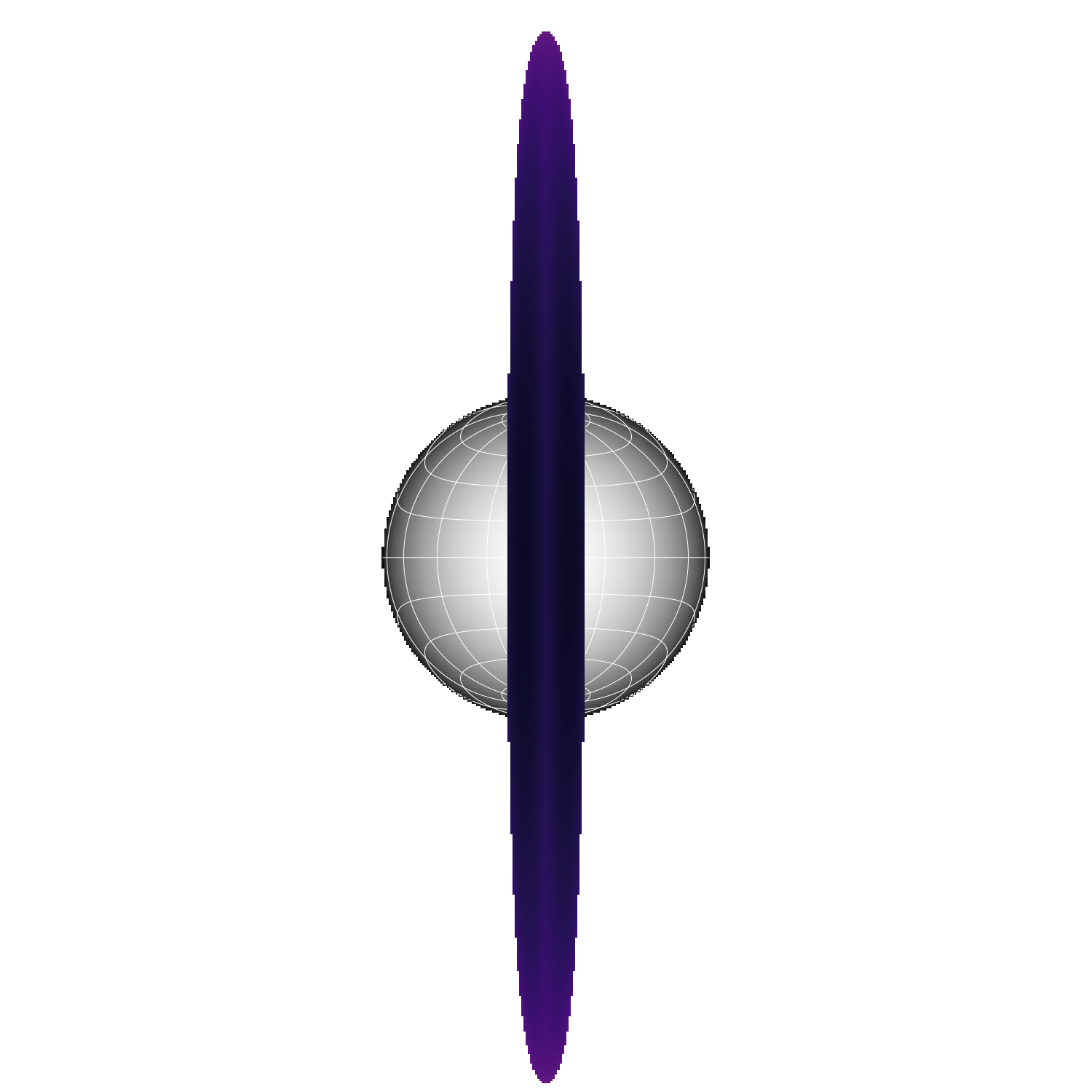} &
    \includegraphics[width=0.095\linewidth]{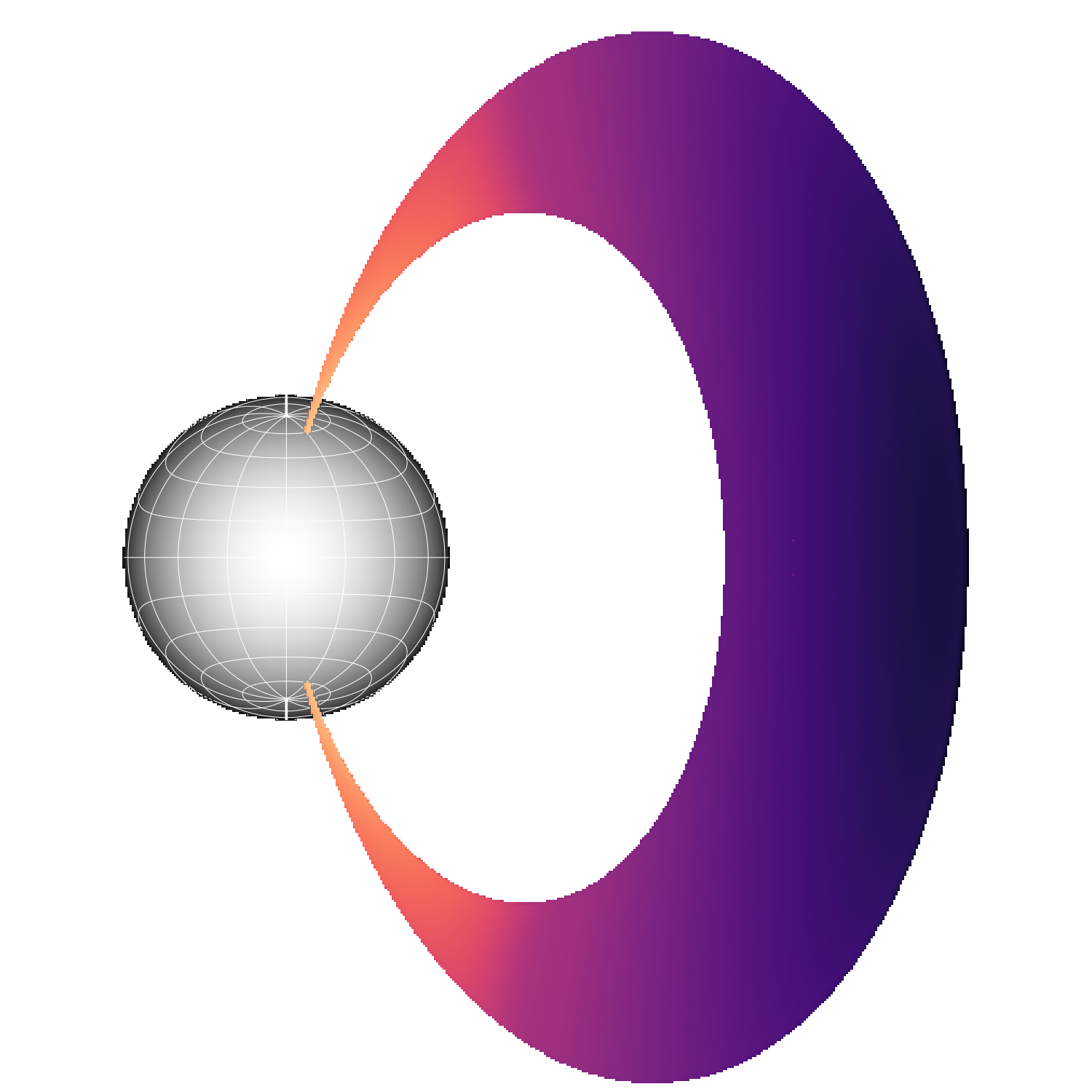} &
    \includegraphics[width=0.095\linewidth]{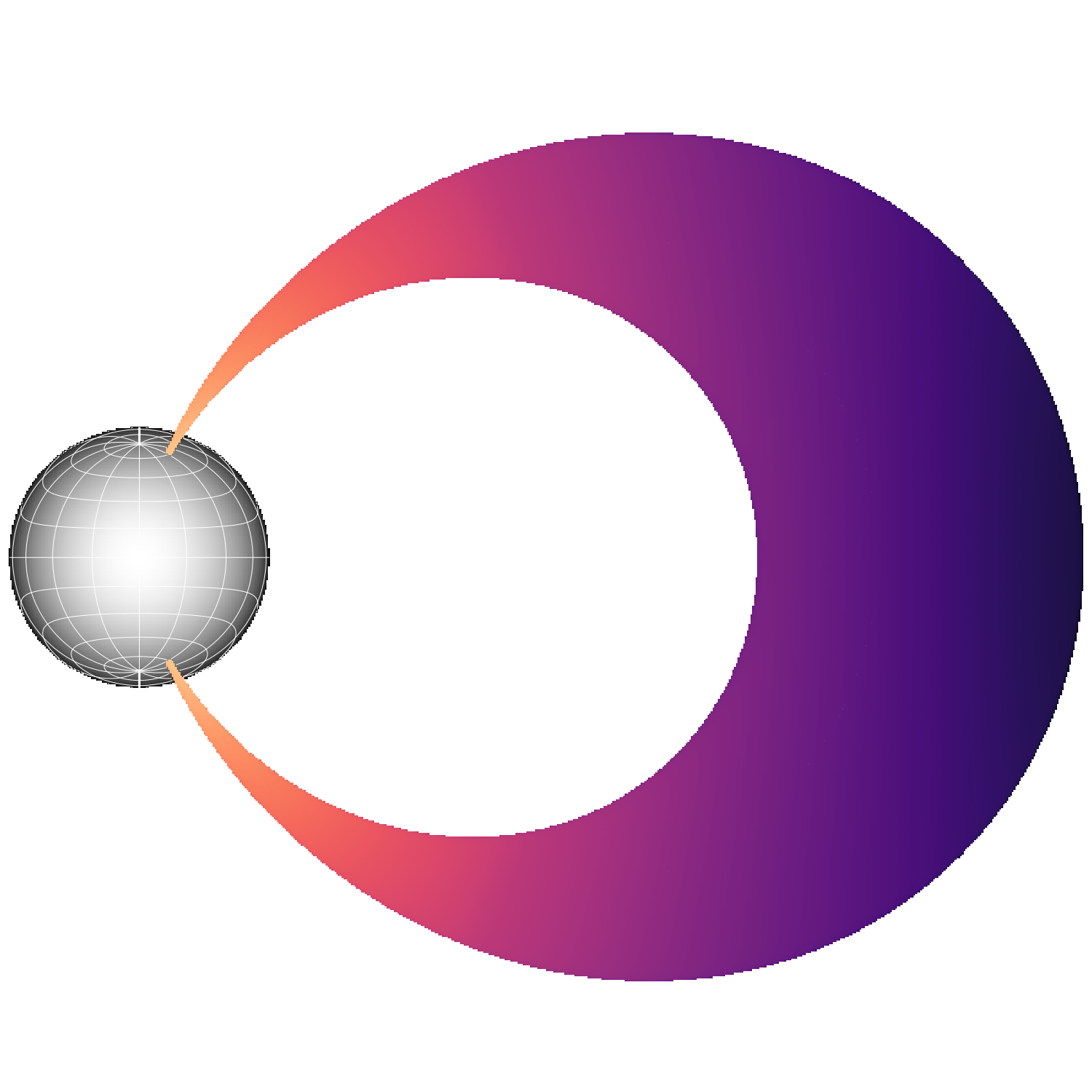} &
    \includegraphics[width=0.095\linewidth]{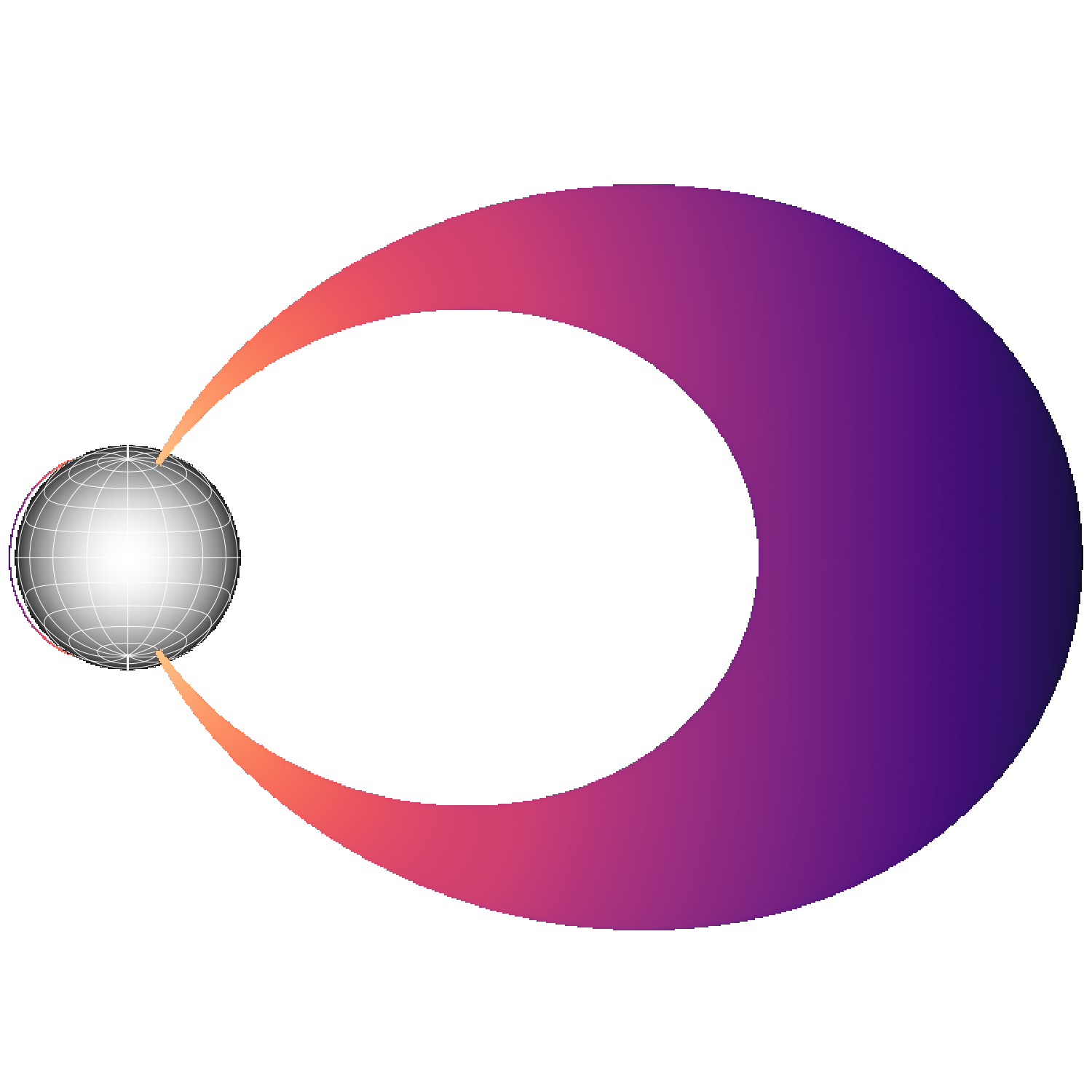} &
    \includegraphics[width=0.095\linewidth]{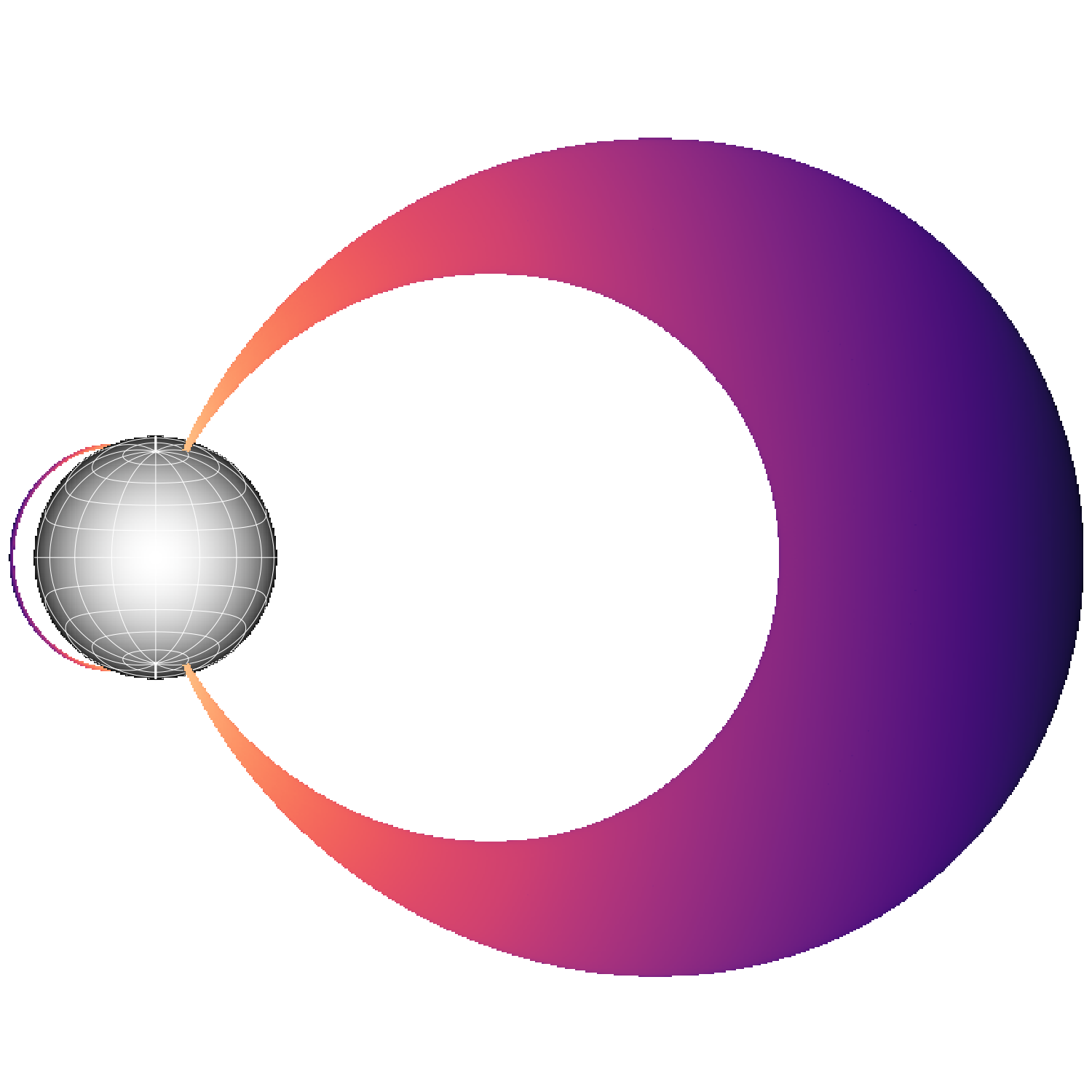} &
    \includegraphics[width=0.095\linewidth]{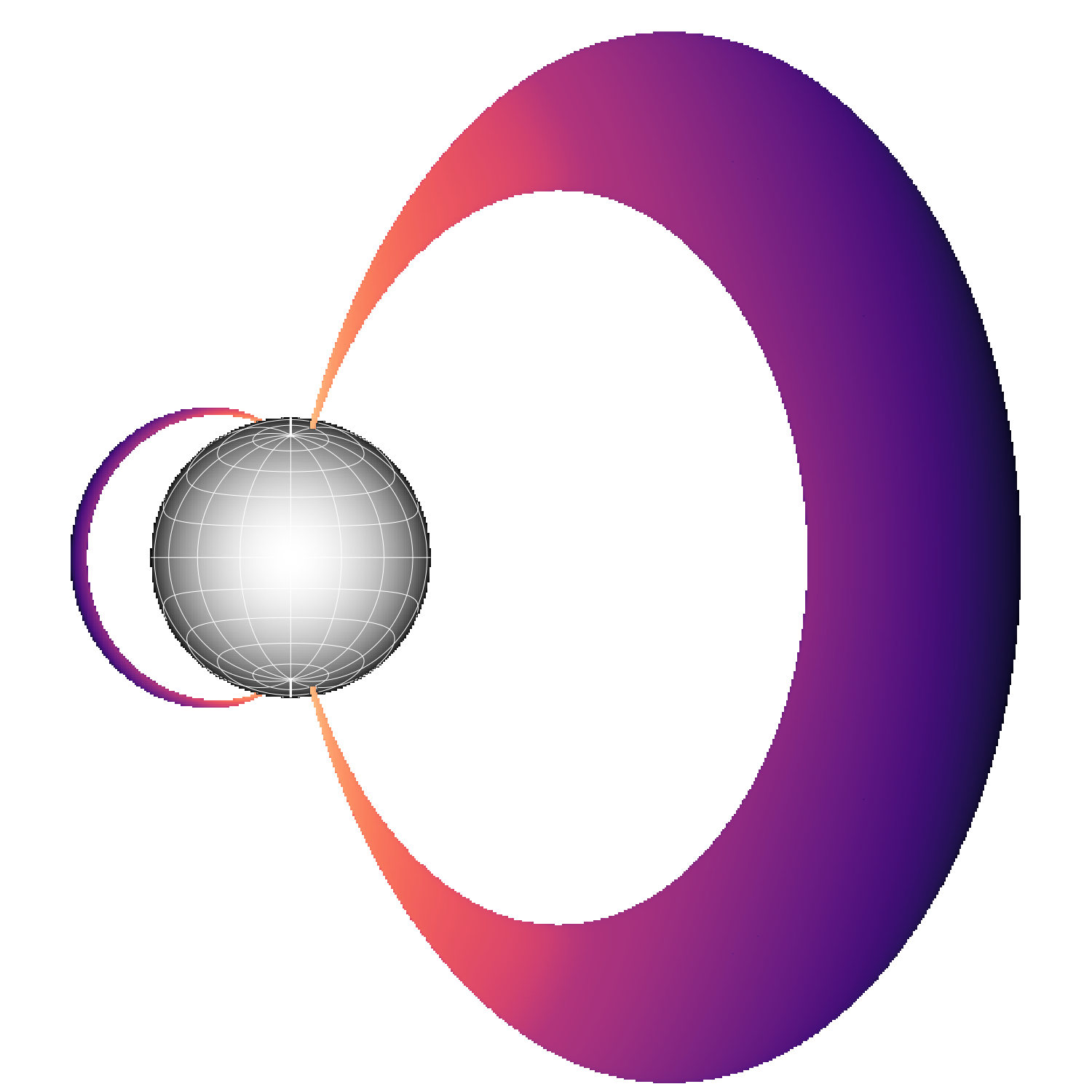} &
    \includegraphics[width=0.095\linewidth]{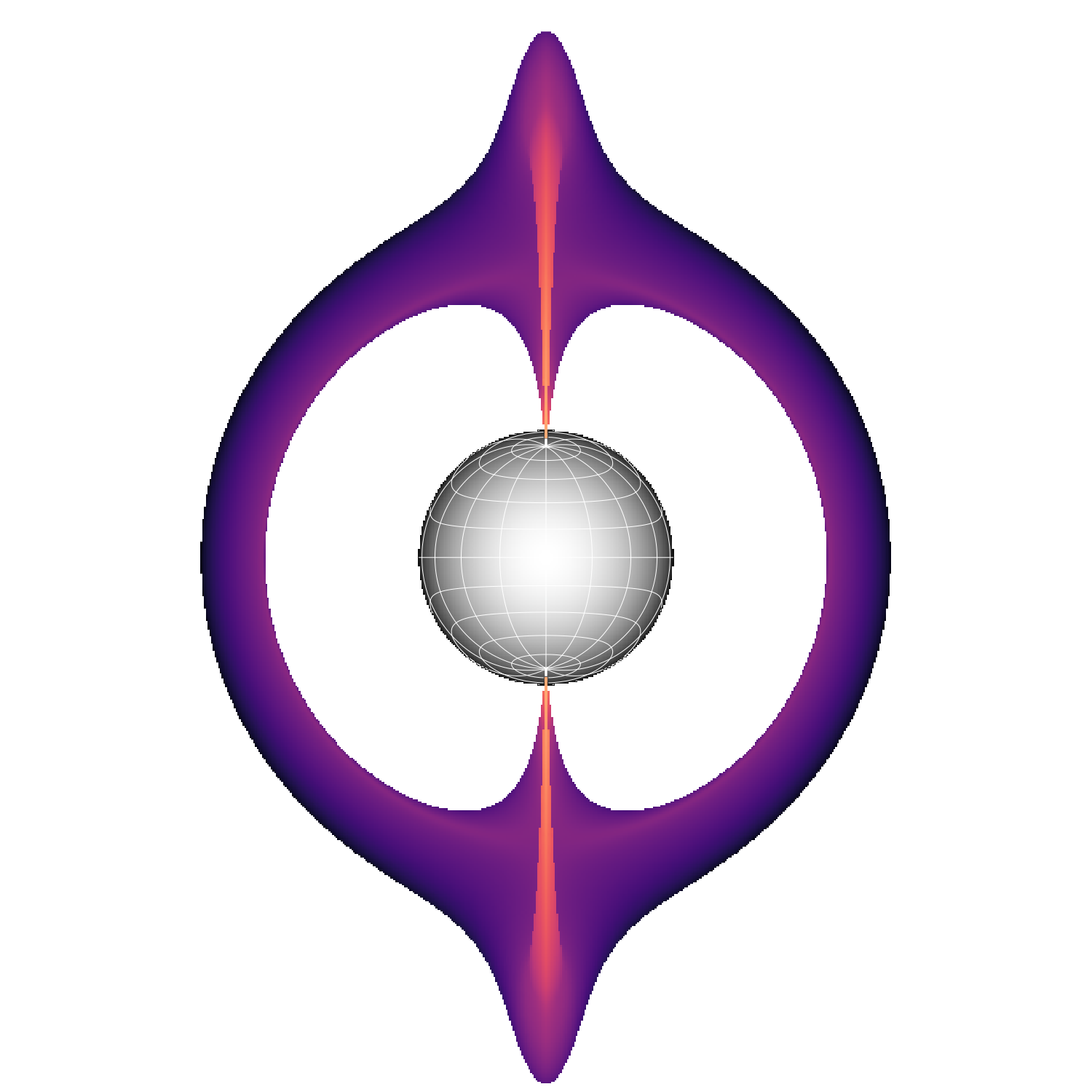}  \\
  \end{tabular}
  }
  \centering
      \includegraphics[width=0.98\textwidth]{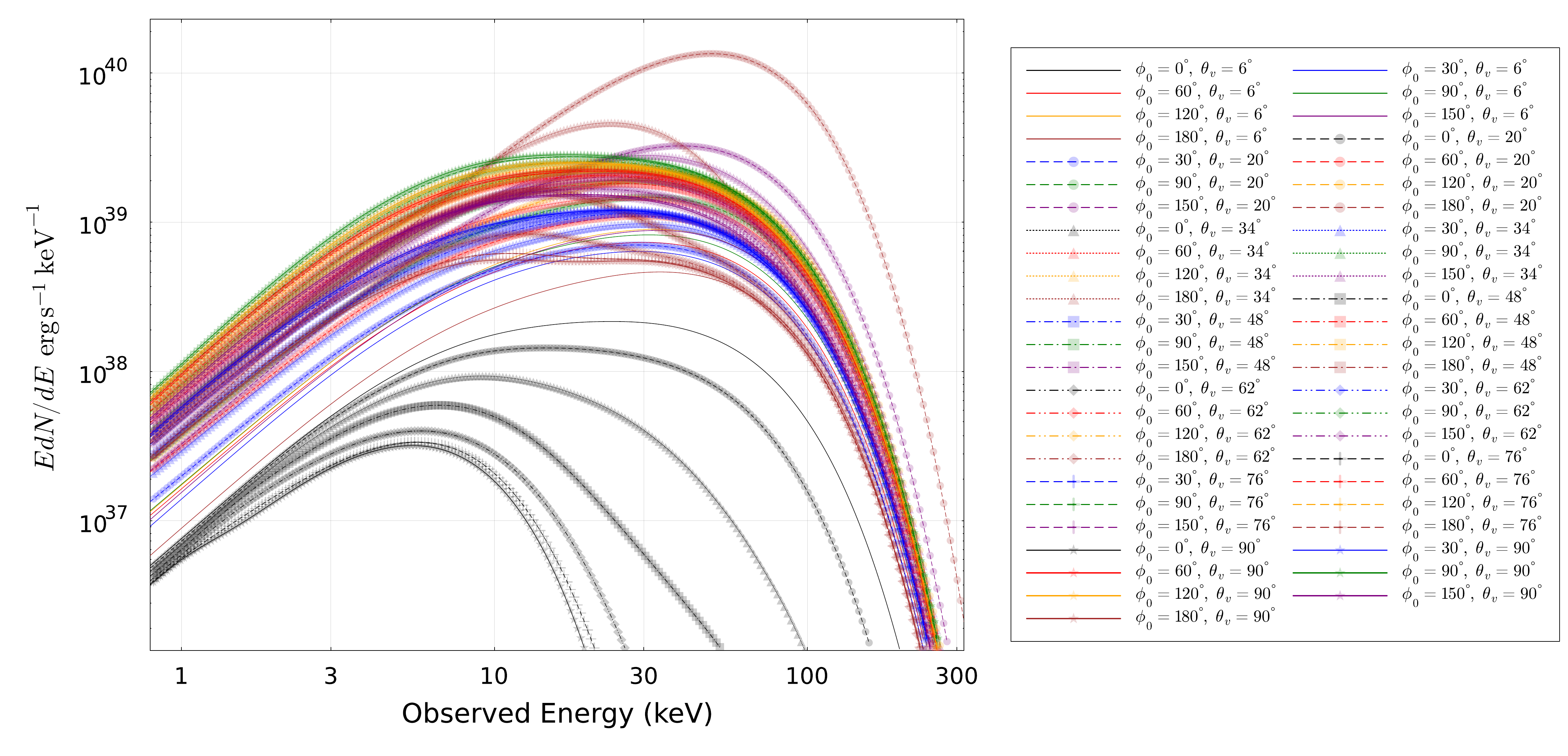} 
      \vspace{-4mm}
  \caption{{\bf Top:} Suite of panels at $12$~keV varying $\theta_v \in \{6^\circ,...,90^\circ\}$ and $\phi_0 \in \{0^\circ,...,180^\circ \}$ with  $\theta_0 = 18^\circ$, $\Delta \theta_0 = \Delta\phi_0 = 2^\circ$, $M=2.0 \,M_\odot$, $r_\star = 12$~km, $B_p^{\rm flat} = 10$, $\alpha = 1/3$, $m_e c^2 \Theta_{\rm max} = 30$~keV, $\eta = 0$, and splitting $\perp\rightarrow \parallel\parallel$ only. {\bf Bottom:} Attendant time-integrated snapshot spectra for each case. This demonstrates that geometric changes can lead to strong spectral and apparent luminosity variations. There is strong geometric and visibility enhancement of luminosity of bursts behind the NS, compared to those in the foreground that can occult high luminosity low altitude regions of the burst. The high luminosity outlier $\theta_v = 20^\circ$ and $\phi_0 = 180^\circ$ is a result of surface grazing trajectories near the back and lensed polar footpoints that are strongly beamed along $\boldsymbol{B}$.  }
  \label{fig:7x7panels}
\end{figure}

\section{Results}
\label{sec:results}

There are eleven parameters associated with the models presented in this work, see Table~\ref{tab:model_params}, plus the choice of splitting attenuation ($\perp\rightarrow \parallel\parallel$, or all three modes) and the switch for VB on or off. We cannot do justice in this work to exploring the full set of parameters. Here, we explore a limited set of instructive cases and highlight qualitative features that might arise from different physical situations.

\begin{figure}[t]
\centering
\includegraphics[width = 0.8\textwidth]{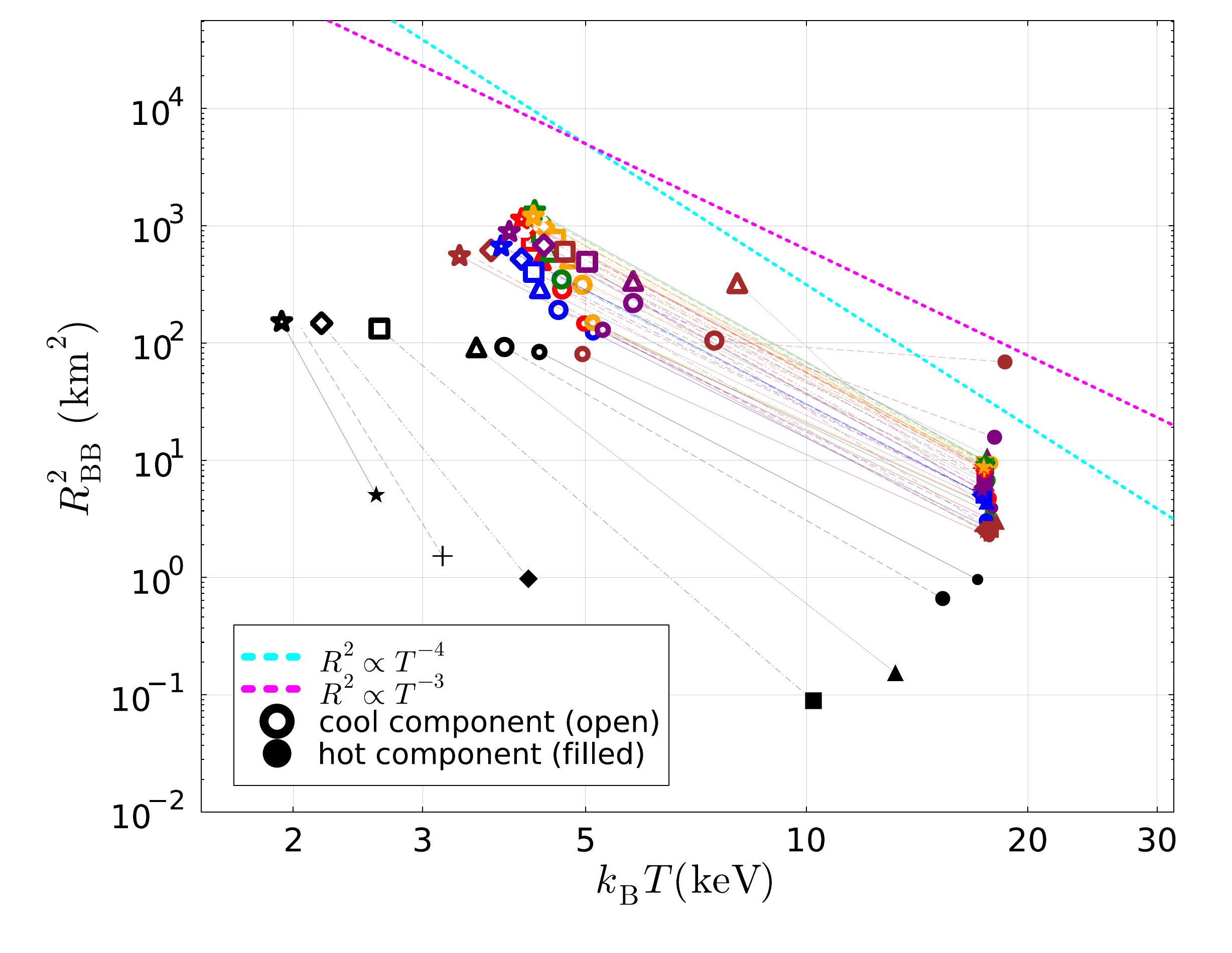} 
\caption{Two-BB phenomenological decomposition of the synthetic spectra presented in Figure~\ref{fig:7x7panels}. The high temperature component has dissimilar dispersion of its hot-BB temperature than the cool component. This is a natural outcome of atmosphere anisotropy, projection and self-occultation for the fixed temperature models in Figure~\ref{fig:7x7panels}. }
\label{fig:7x7twoBB}
\end{figure}

\subsection{Spectra, Geometry and Reproducing Two-Blackbody Model Phenomenology}

To isolate geometric influences on the burst spectrum, let us first consider only varying viewing angle $\theta_v$ and $\phi_0$. In Figure~\ref{fig:7x7panels}, we depict a grid of IP reconstructions at $12$~keV for different viewing angles $\theta_v$ and burst footpoint placement in azimuth, fixing $\theta_0 = 18^\circ$, $\Delta \theta_0 = \Delta\phi_0 = 2^\circ$, $M=2.0 \,M_\odot$, $r_\star = 12$~km, $B_p^{\rm flat} = 10$, $\alpha = 1/3$, $m_e c^2 \Theta_{\rm max} = 30$~keV, $\eta = 0$, and splitting only $\perp\rightarrow \parallel\parallel$. Additionally, all snapshot spectra are shown for each case in the bottom panel. There is more than an order of magnitude variability in observed luminosities merely by varying these geometric parameters. Four factors are at play in this observed variation: (1) the apparent area of the burst, (2) self-occultation of luminous, higher-temperature regions near the footpoints, (3) visibility enhancement and magnification by lensing, and (4) energy-dependent anisotropy from polarized radiative transfer in {\sl MAGTHOMSCATT}. Thus, strong luminosity and spectral variations can result from geometric factors alone rather than intrinsic burst spectral energy injection or size, although in a magnetar burst storm, both factors should be in play (e.g., earthquake-like power-law distribution in energy injection) for a single active burst location on the surface. 

In Figure~\ref{fig:7x7twoBB}, we depict the results of fitting phenomenological two-BB models to our synthetic spectra presented in Figure~\ref{fig:7x7panels}. The parameters are chosen to resemble observed magnetar burst luminosities and spectral extent. From these two-BB fits, the hot BB temperature is a reasonable lower limit for $m_e c^2 \Theta_{\rm max}$, given gravitational redshift and multicolor spatial extent of the footpoint. Meanwhile, larger scatter results in the cool BB $R^2$ and $k_{\rm B} T_{\rm cool}$ owing to geometric projection effects and surface anisotropy. Lines of $R^2 \propto k_{\rm B}T^{-3}$ and $\propto k_{\rm B}T^{-4}$ are overlaid. For the adopted value of $\alpha = 1/3$, our scheme Eq.~(\ref{eq:BT_relation}) reproduces the observed trend reasonably well, as discussed in \S\ref{sec:intro} and \S\ref{sec:tempspec}. It is also interesting to note that with $\alpha = 1/3$, the trend is somewhat steeper than $R^2 \propto k_{\rm B}T^{-3}$, possibly the result of self-occultation of hotter regions. This effect is most extreme for the black points, which sample $\phi_0 = 0^\circ$.

The choice of parameters for Figure~\ref{fig:7x7panels} is deliberate to broadly match the two-BB area and temperatures reported in magnetar SBs. We find that bursts closer to poles have greater spectral breadth, a fact captured in the two-BB fit temperature distinction of hot and cool temperature components. The apparent areas of the two-BB fits are also informative, with small hot components $R^2$ indicative of limited and narrow quasi-polar footpoints. Meanwhile, large bursts extending in azimuth or also to the magnetic equator can phenomenologically appear with a more narrow spectral extent, while with overall higher $R^2$ and luminosity.  The sequence of flux-dependent two-BB fit reported in Figure~6 of \citet{2014ApJ...785...52Y} thus can be accounted for in our framework, with most bursts analyzed in that work compatible with footpoint angular sizes of only a few degrees, $m_e c^2 \Theta_{\rm max} \sim 30-50$~keV, and footpoint colatitude of $\sim 10^\circ-20^\circ$. In contrast, bursts with footpoint loci that span larger bubbles with more equatorial footpoints have a murkier distinction between hot and cool components but larger apparent areas. This phenomenology is also visible for higher $R^2$ bursts in Figure 6 of \citet{2014ApJ...785...52Y}. Finally, bursts with even narrower footpoints closer to the pole, $\Delta \theta_0 \sim \Delta \phi_0 \lesssim 1^\circ$  (corresponding to $\sim 0.01-0.1$ km$^2$ active regions) attain larger spectral separation between hot and cool components and are similar to the lower flux bursts in \citet{2014ApJ...785...52Y}.

\subsection{Impulse Responses and Burst Polarization Diagnostics}

Let us now consider detailed and diagnostic polarized energy-time impulse responses for illustrative burst geometries. In Figures~\ref{fig:ResultsFig1}--\ref{fig:ResultsFig10}, we present a suite of model outputs for different parameter choices (Table~\ref{tab:model_params}) fixing $B_p^{\rm flat}=10$, $r_\star = 12$~km. Each figure contains the same set of diagnostic panels: top-row IP intensity maps at 4, 12, 36, and 108~keV, middle-row 3D IRs for Stokes-$I$ and polarization ($Q/I$, $U/I$) with VB off/on with the same color map as the intensity maps, and bottom-row $Q/I$--$U/I$ tracks with the time-integrated spectrum and PD. Note that the delays and features (particularly for secondary lensed components) in the IR not only involve R{\o}mer geometric contributions but also Shapiro delay \citep{1964PhRvL..13..789S}.

Figure~\ref{fig:ResultsFig1} encapsulates the complexity of the energy-time structure that might arise from a spatially extended burst with activation regions roughly azimuthally extended near a pole when viewed behind the star, a case highlighted in Figure~\ref{fig:archetypes} and also Figure~\ref{fig:VBdep} with different viewing angles. From the top panels of Figure~\ref{fig:ResultsFig1} and the Stokes-$I$ IR, it is clear that multiple features coexist with temporal caustics -- the direct outer and inner parts of the flux tube, a lensed image, and a gap associated with the footpoints at $\sim 50-70~\mu$s. More temporal and polarization structure in the VB-off case is clearly visible at higher energies, where these multiple components are less smeared. Given the spatially extended nature of the burst, the VB off PD is relatively low at lower energies, only rising at higher energies where there is less spatial cancellation on the IP. 

Figure~\ref{fig:ResultsFig2}, by contrast, is a relatively restricted set of footpoints with a lower $\Theta_{\rm max}$ behind the star for the observer. This case highlights the energy-time gating of the secondary image by the stellar surface. The low temperature also highlights a case where most of the burst is in a regime of $\omega/B \ll1$. This burst is clearly less spectrally extended than other cases, but because of the secondary image and light bending, it has strongly enhanced visibility (``magnified") over flat spacetime.

Figures~\ref{fig:ResultsFig3} and \ref{fig:ResultsFig4} highlight the instructive cases of a large azimuthally symmetric torus snapshot viewed either edge-on or nearly pole-on. Such extended fireball tori may arise in the evaporating portion of a giant flare's pulsating tail. The edge-on case's IR is clearly more temporally extended from the resulting geometric and R{\o}mer geometric+Shapiro delays, with a pronounced temporal caustic for the back side of the torus. Meanwhile, the symmetry of the pole-on situation results in a large cancellation of net polarization, for both VB on or off. Additionally, in the pole-on Figure~\ref{fig:ResultsFig4} situation, there is an abrupt transition in the IR at $\sim 40~\mu$s associated with the truncation of the inner region at the footpoints by the NS surface, with outer soft parts of the spectrum only emerging after this.

Figures~\ref{fig:ResultsFig5} and \ref{fig:ResultsFig6} are tori viewed at an inclined viewing angle at a hotter temperature of $m_e c^2\Theta_{\rm max} = 30$~keV, with a switch in the allowed modes of photon splitting. A prominent temporal caustic is present at $\sim 160~\mu$s and is associated with the far side of the torus (see first column of Figure~\ref{fig:archetypes}). Given the high temperature, the spectrum extends beyond 100~keV in Figures~\ref{fig:ResultsFig5}. In contrast, splitting of both modes curtails the high-energy part of the spectrum in the case of Figure~\ref{fig:ResultsFig6}, resulting in a steeper cutoff and narrower spectral breadth. Split photons will also modestly contribute to the spectrum at lower energies. From the top panel of 108~keV, it is apparent that only a narrow annulus escapes at higher energies owing to splitting.

Figure~\ref{fig:ResultsFig7} highlights a case with an active footpoint closer to the magnetic pole, reaching higher altitudes similar to Figure~\ref{fig:7x7panels}. This results in an IR extending to nearly $500\,\mu$s, and is among the most spectrally extended bursts of the cases shown. The two footpoints, as high-energy cusps, are also clearly discernible in the IR, imprinting a temporal signal commensurate with a NS diameter or half-circumference crossing timescale with Shapiro and R{\o}mer delays. The energy-dependent polarization character is also particularly rich owing to depolarization from some regions of the burst with photon energies passing through $\omega/B \sim 1$ (see Figure~\ref{fig:MAGTHOMSCATT}) at higher altitudes when $B$ drops. The $Q/I$ and $U/I$ IRs contain discontinuities that are preserved even with VB on, unlike some of the other cases, e.g., Figure~\ref{fig:ResultsFig2}.

Figure~\ref{fig:ResultsFig8} highlights a high compactness ($M=2.0 \,M_\odot$), pole-on view that involves a completely detached lensed component in the IR, which is the normally invisible lower portion of the footpoint associated with the visible direct component. The three higher energy components, corresponding to three visible footpoints, all have temporal gaps commensurate with the geometric plus Shapiro delay of geodesics. Here again, the polarization structure is rich, with the VB on and off cases similar in time-integrated behavior, but quite different in $Q/I$ and $U/I$ IRs. Here, the burst spectrum is also spectrally broadened compared to other cases highlighted owing to its size spanning a multitude of temperatures. In contrast, reducing the compactness with $M=1.4 \,M_\odot$ with otherwise identical parameters (Figure~\ref{fig:ResultsFig9}) results in no secondary lensed component or temporal gap in the IR. Shapiro and R{\o}mer delay is also reduced, with two hot regions associated with the footpoints having a mildly shorter timescale compared to the $M=2\,M_\odot$ case. Thus, NS mass and radius can be dramatically encoded in the IR's temporal structure.

Finally, in Figure~\ref{fig:ResultsFig10} we show the instructive case of a loop directly behind the NS viewed edge-on. Because of lensing and magnification, visibility is not curtailed but enhanced with both sides of the loop being visible -- this is a smaller version of cases depicted in Figure~\ref{fig:7x7panels}.

Taken together, SB observables are governed by the interplay of a limited set of geometric, thermodynamic, and radiative-transfer parameters, each leaving a characteristic but not generally unique imprint. The viewing angle $\theta_v$ and fireball shape primarily determine visibility, self-occultation (e.g., Figures~\ref{fig:ResultsFig3}-\ref{fig:ResultsFig5}), and the relative contributions of direct and lensed portions, thereby setting the overall structure of the IRs, and phase dependence. This also principally sets the apparent emitting area. The fireball temperature structure, mainly through $\Theta_{\max}$ and the scaling index $\alpha$, controls the spectral breadth and the relative weighting of compact hot and extended cool emitting zones. The QED transport assumptions, specifically the efficacy of VB and the allowed photon-splitting channels, chiefly govern the energy dependence of the polarization signal and spectral cutoff shape. NS mass and radius enter through gravitational redshift, light bending, secondary-image magnification, and differential time delays. They are therefore expressed most clearly in the morphology of delayed components and discontinuities in the IR, rather than spectroscopically.

\section{Modeling Observed Burst Spectra}
\label{sec:model_data}

We have presented new physically grounded fireball models for magnetar bursts and giant flare tails, accounting for GR lensing, strong-field photosphere polarization and anisotropy, QED photon splitting and VB. Prior SB analyses have largely relied on phenomenological spectral decompositions that are useful descriptively but do not map uniquely onto geometry or microphysics \citep[e.g.,][]{2004ApJ...612..408F,2008ApJ...685.1114I,2012ApJ...749..122V,2014ApJ...785...52Y,2025ApJ...988..282D}. Our models provide a framework for relating burst observables to underlying source geometry and radiative-transfer physics, and potentially to fundamental source parameters. We find bursts with angular footpoint radii of a few degrees $\Delta \phi_0 \sim \Delta \theta_0 \lesssim 5^\circ$ placed at $\theta_0 \lesssim 20^\circ$ magnetic colatitude can reproduce aspects of the empirical two-BB phenomenology of SBs, and an adiabatic description of the fireballs appears broadly consistent with those trends. Lower colatitude quasi-polar fireballs are more spectrally extended, and naturally have greater distinction between hot and cool components in a two-BB phenomenological spectral decomposition. 

Qualitatively new features are present with GR and in 3D. Across the cases examined,  multiple branches (direct + lensed delayed components) are a generic outcome if fireballs are observed at favorable angles, a situation likely realized given a magnetar's rotation and misalignment between spin and magnetic axes. Stellar occultation or temporal gating of the burst are key to the branch morphology and energy-dependent visibility. Quasi-polar, higher-altitude geometries (Figures \ref{fig:7x7panels} and \ref{fig:ResultsFig7}) are among the most spectrally extended cases. In contrast, more localized small bursts retain sharper geometric substructure. Spatially extended bursts, especially those attaining high altitudes, generally have a lower net PD below $10$~keV even if VB is active since photon energies pass through the local cyclotron resonance. Conversely, smaller and more localized bursts, or those with higher temperatures, are predicted to show systematically higher PD. This situation should also translate to the evaporating pulsating tail of magnetar giant flares -- PD may increase with time as the fireball dissipates and becomes increasingly more spatially restricted \citep{2001ApJ...561..980T}. Smaller fireballs preserve sharper energy-time structure or spin-phase dependence. Only small regions can remain fully hidden behind the star for typical compactness (see Appendix~\ref{sec:invis}), and even so, only for limited viewing geometries or spin phases. Hence, complete non-detection by geometric occultation requires compact activation regions with restricted altitude and unfavorable footpoint placement. This rules out models which invoke SBs occurring or emitting at higher altitudes. 

Beyond fitting individual bursts, the present model suggests a set of population-level correlations that may be worth testing observationally: our calculations predict that multi-peaked high-energy IRs and detached delayed branches should preferentially appear at rotational phases where a substantial fraction of the emitting volume is situated on the far side of the NS relative to the observer. Likewise, temporal-caustic signatures should not appear randomly in phase, but cluster in geometries with strong lensing magnification (see Figure~\ref{fig:7x7panels}). Such phase-conditioned statistics possibly provide a promising route for testing or falsifying simplified burst geometries. Empirical correlations observed in spectroscopic studies of bursts \citep[e.g.,][]{2008ApJ...685.1114I,2014ApJ...785...52Y,2025ApJ...988..282D} might be reproduced by varying a limited set of parameters for a fixed active region on the crust. Geometric viewing changes may also mimic spectral evolution. This implies that phenomenological time-resolved spectral fits may partially fold geometric transfer effects into inferred BB model variability. Meanwhile, unusual bursts might suggest different active locales on the surface. Varying crust locales might also explain why isolated SB storms of magnetars can exhibit different aggregate spectral character \citep[e.g., for SGR~1935+2154,][]{2023ApJ...950..121R,2024ApJ...969...38R,2025ApJ...989...63H,2025ApJS..276...60R}. At the same time, the signatures identified here might not be generally unique, and robust source constraints will likely require bright bursts or ensembles of bursts together with possible extensions of the models presented here. For energetic bursts with higher temperatures, all-mode photon splitting may be noticed from spectra alone if steep spectral cutoffs are observed above $\sim 50-100$~keV. In contrast, $\perp\rightarrow \parallel \parallel$ splitting attenuation is a minor effect on spectra as most emergent photons from the magnetized atmosphere are in the $\parallel$ mode at high photon energies or low altitudes.

\subsection{Illustrative Application to SGR~1935+2154 - A quasi-polar nature to its FRB-associated burst?}

\begin{figure}[t]
    \centering
        \includegraphics[width=0.47\linewidth]{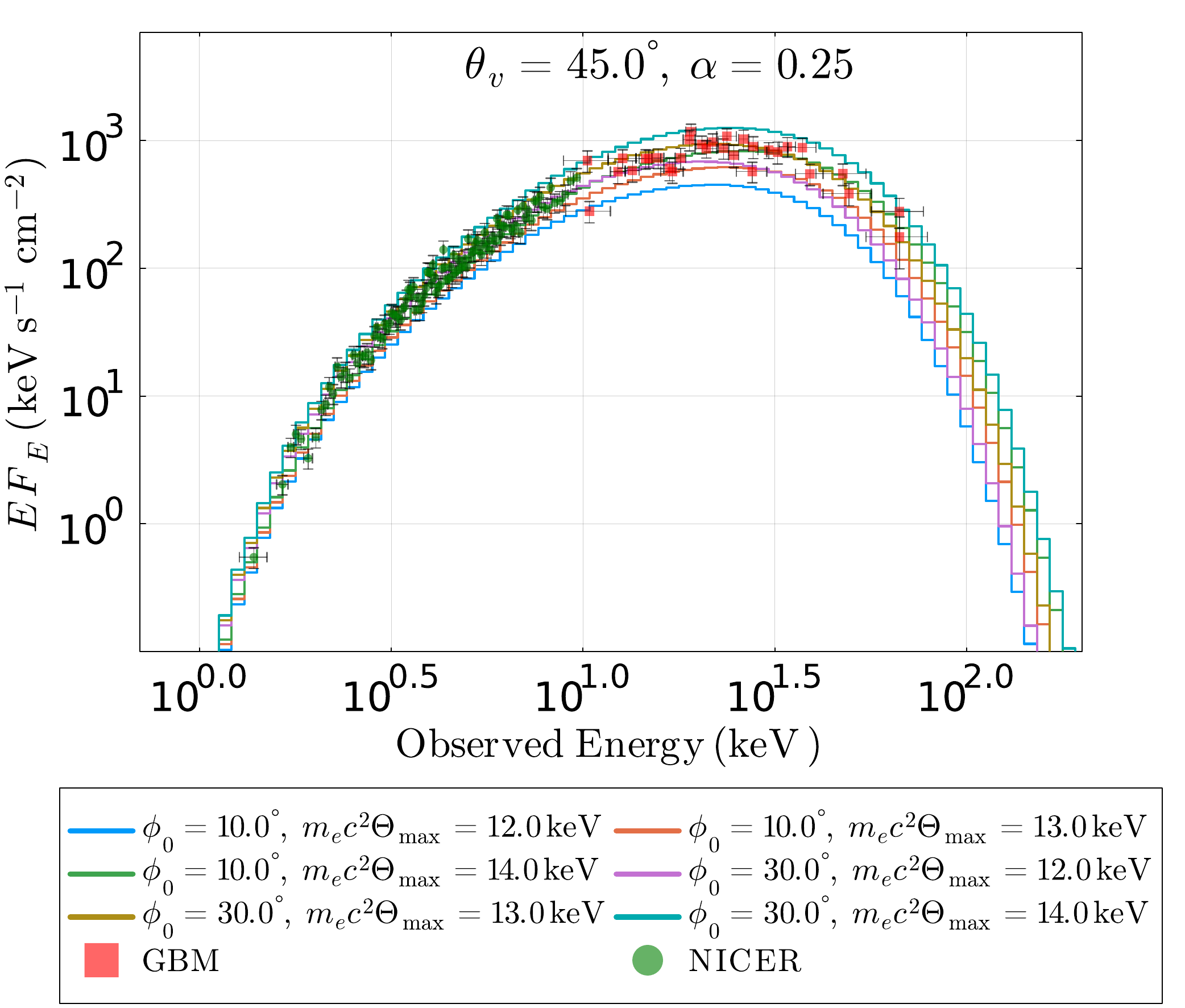} 
        \includegraphics[width=0.47\linewidth]{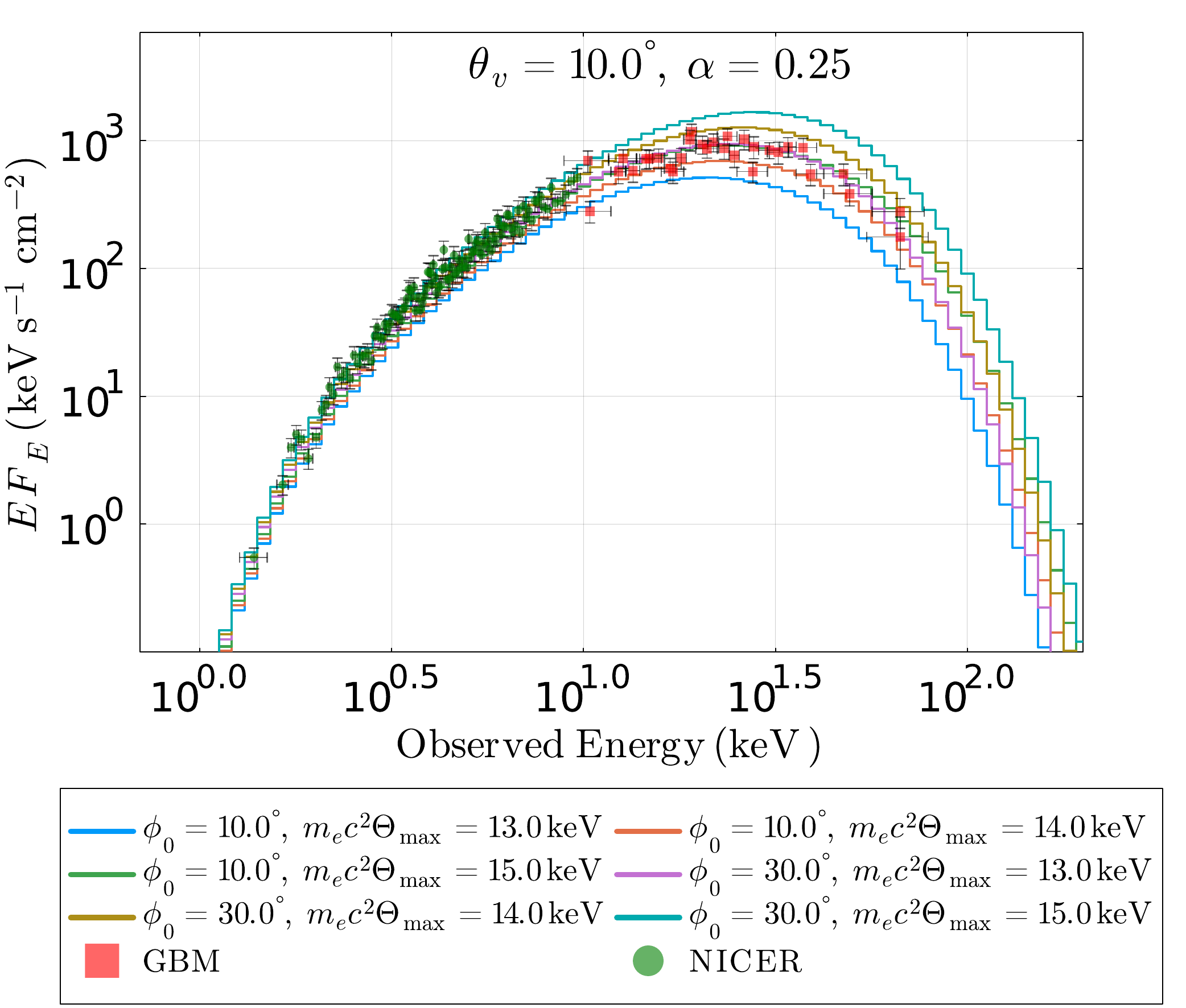}
       \\
        \includegraphics[width = 0.47\linewidth]{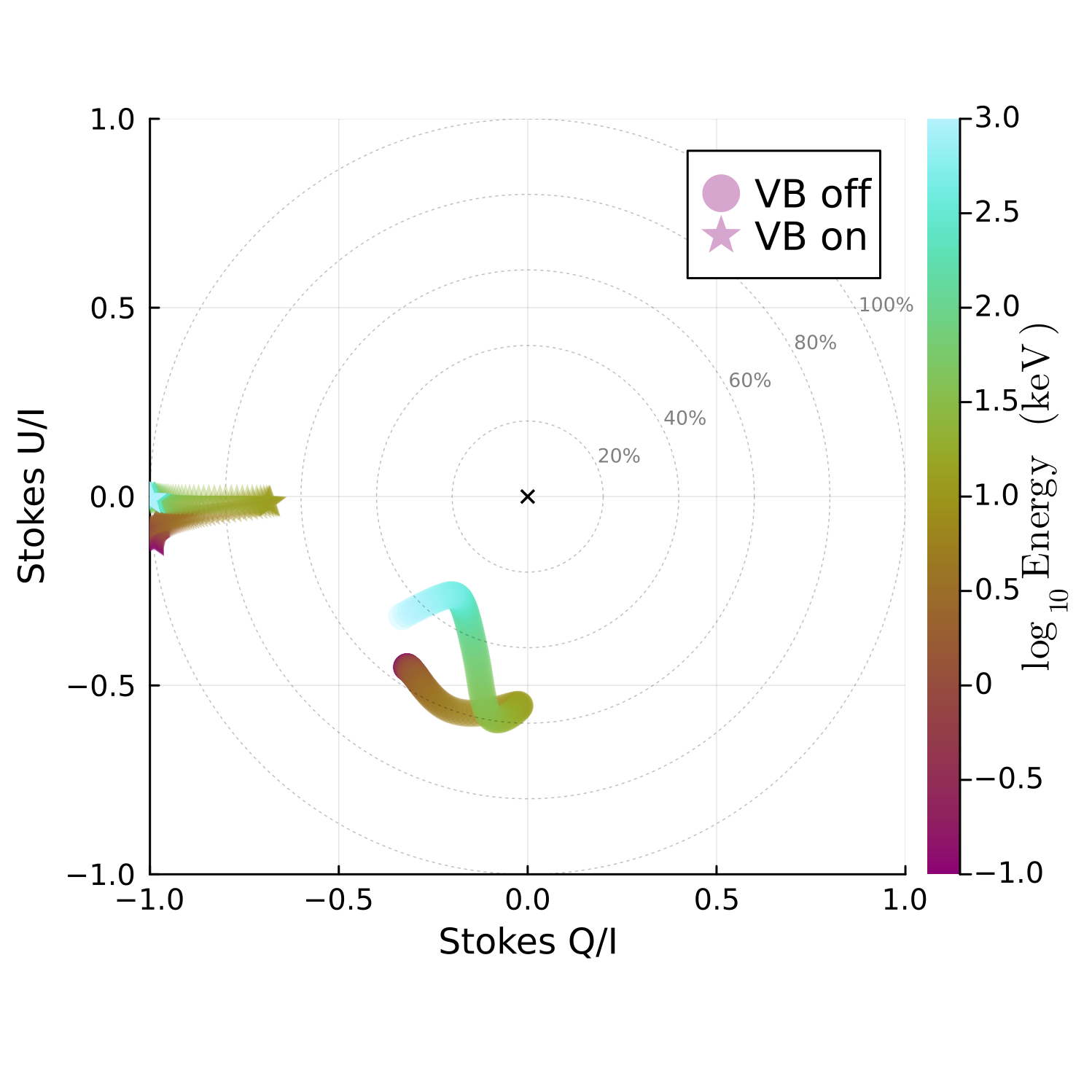}
         \includegraphics[width = 0.47\linewidth]{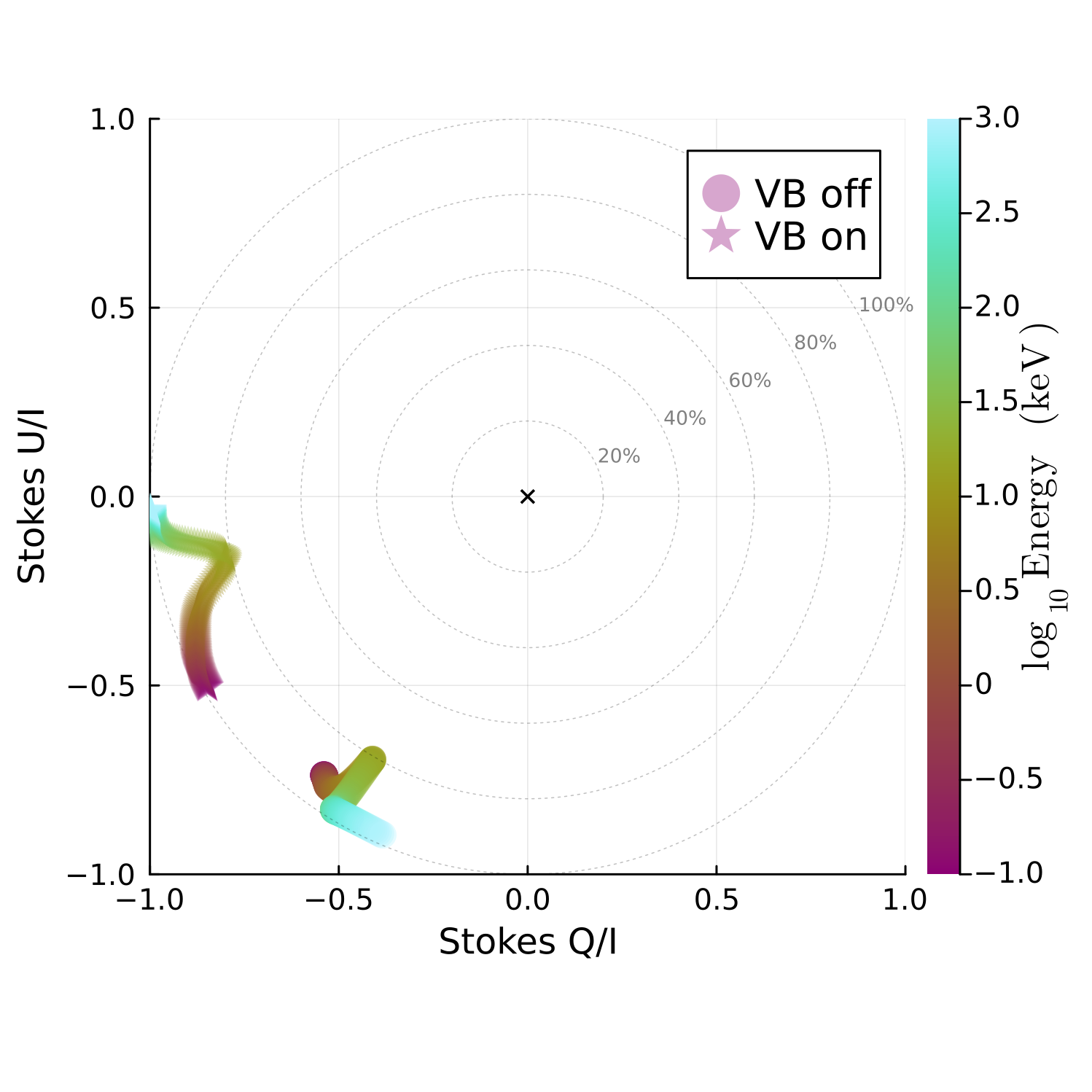}
         \vspace{-0.8cm}
    \caption{Top panels: Comparison between the simulated spectra and that of a SGR~1935+2154 burst observed simultaneously with NICER and Fermi GBM \citep{2021NatAs...5..408Y}. Two viewing angles are considered: $\theta_v = 45^{\circ}$ (top left panel) and $\theta_v = 10^{\circ}$ (top right panel). Bottom panels: $Q-U$ trajectories as a function of energy on a logarithmic scale, shown for VB off (filled circles) and VB on (filled stars), obtained using $\phi_0 = 30^{\circ}$ and $m_ec^2\Theta_{\rm max} = 13$~keV for $\theta_v = 45^{\circ}$ (bottom left panel) and $\theta_v = 10^{\circ}$ (bottom right panel). The PD is indicated by the concentric gray dotted circles.   The fixed parameters are $\theta_0 = 25^\circ$,  $\Delta \theta_0 =3^{\circ}$, $ \Delta\phi_0 = 6^\circ$, $M=1.5 \,M_\odot$, $r_\star = 12$~km, $B_p^{\rm flat} = 10$, $\alpha = 1/4$,  $\eta = 0$, and splitting $\perp\rightarrow \parallel\parallel$ only. See text for details.}
    \label{fig:NICER_Fermi_data}
\end{figure}

\begin{figure}[t]
    \centering
        \includegraphics[width=0.495\textwidth]{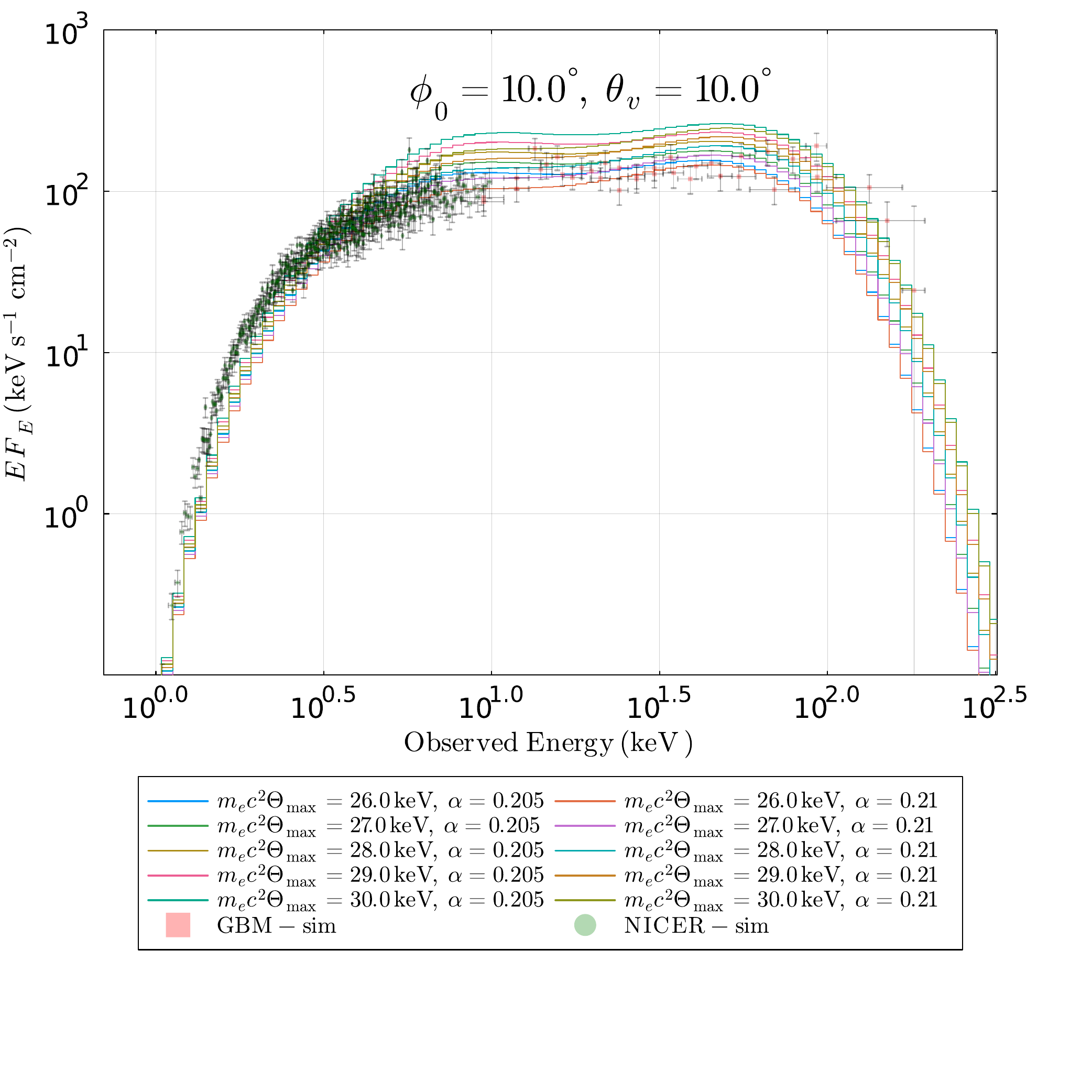}
        \includegraphics[width=0.495\textwidth]{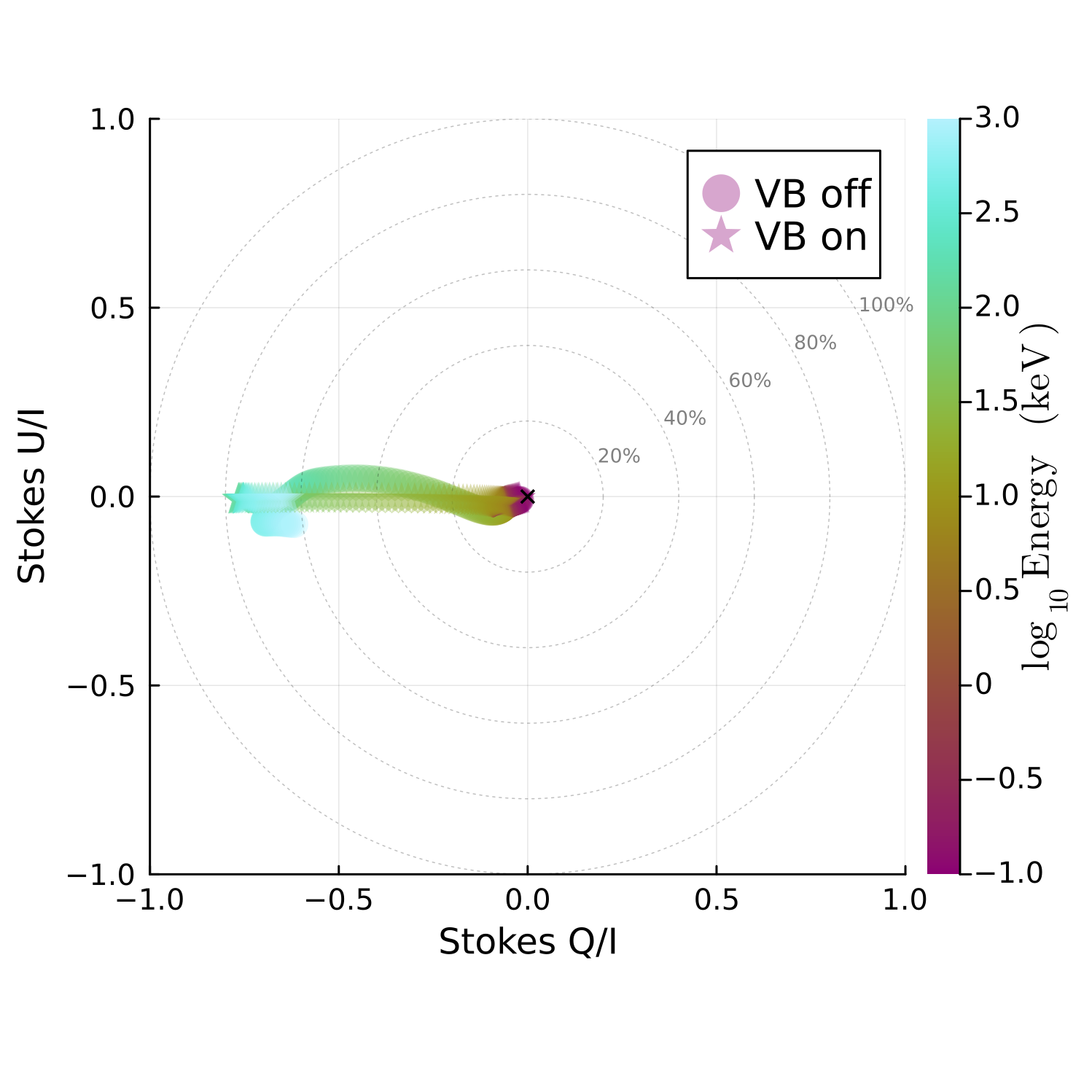}
    \caption{Left panel: Comparison between the model spectra and the simulated NICER+{\it{Fermi}}-GBM data for the FRB-associated burst \citep{2021NatAs...5..408Y}. Right panel: $Q-U$ trajectories as a function of energy on a logarithmic scale, shown for VB off (filled circles) and VB on (filled stars), obtained using $\alpha = 0.21$ and $m_ec^2\Theta_{\rm max} = 29$~keV. The fixed parameters are  $\phi_0 =10^\circ$, $\theta_v  = 10^{\circ}$,   $\theta_0 = 5^\circ$,  $\Delta \theta_0 = \Delta\phi_0 = 0.1^\circ$, $M=1.5 \,M_\odot$, $r_\star = 12$~km, $B_p^{\rm flat} = 10$, $\eta = 0$, and splitting $\perp\rightarrow \parallel\parallel$ only.}
    \label{fig:FRB-burst}
\end{figure}

To demonstrate the applicability of our model, let us compare the snapshot model spectra with burst data for SGR~1935+2154 during its prominent April 2020 activity.  We note that the exposition here solely serves as a proof of concept, rather than an in-depth parameter exploration with statistical rigor. For this purpose, we fix the $B_p^{\rm flat} = 10$, stellar mass and radius to $M = 1.5\,M_\odot$ and $r_\star = 12$~km, respectively, and assume $\eta = 0$. Only photon splitting in the $\perp$ mode, $\perp \rightarrow \parallel\parallel$, is considered. To account for the absorption from the interstellar medium along the line of sight to the source, we adopt a fixed neutral hydrogen column density of $N_H = 2.7\times10^{22}$~cm$^{-2}$ as in \citep{2021NatAs...5..408Y}. The other parameters in Table~\ref{tab:model_params} are manually adjusted to match the observed spectra. The absorbed flux is $F_{E} = \exp \left[-N_H \sigma_{\rm ISM}(\omega)\right] L_{I}/(4\pi d^2)$ where the distance to the source is fixed at $d = 6.6$~kpc\footnote{Distance estimates to SGR 1935+2154 vary widely, from $>1.5$~kpc \citep{2021MNRAS.503.5367B}, $4.4$~kpc \citep{2020ApJ...898L..29M}, $6.6$~kpc \citep{2020ApJ...905...99Z}, to $12.5$~kpc \citep{2018ApJ...852...54K}. This uncertainty largely affects the model normalization (i.e. flux tube cross sectional area) but not spectropolarimetry diagnostics.}, $\sigma_{\rm ISM}$ is 
the ISM photoelectric absorption cross section \citep{Wilms_2000}, and $L_I$ is the time-marginalized snapshot model luminosity. 

In the top left panel of Figure~\ref{fig:NICER_Fermi_data}, we show the comparison between the simulated spectra obtained with two different values of the footpoint longitude, $\phi_0 = \{10^{\circ}, 30^{\circ}\}$, and that of a representative burst observed simultaneously with {\it Fermi}-GBM and NICER \citep{2021NatAs...5..408Y} in the April 2020 burst storm of SGR~1935+2154. The data here has been folded through the instruments' responses. Three different values of temperature are considered, $m_ec^2 \Theta_{\rm max} = \{12, 13, 14\}$~keV. This configuration corresponds to an oblique viewing angle of $\theta_v = 45^{\circ}$.  We also identify an alternative scenario, corresponding to $\theta_v = 10^{\circ}$, that can also reproduce the observed data, as shown in the top right panel. For the latter, we adopt slightly higher temperatures, $m_ec^2 \Theta_{\rm max} = \{13, 14, 15\}$~keV, while keeping the same footpoint longitude values. This demonstrates that the spectroscopy dictates values of $\Theta_{\rm max}$ that are largely insensitive to the viewing direction, except for extreme lensing or self-occultation scenarios (Figures~\ref{fig:7x7panels}-\ref{fig:7x7twoBB}).

In the bottom row of Figure~\ref{fig:NICER_Fermi_data}, we present the polarization information, specifically the $Q-U$ trajectories with VB off (filled circles) and VB on (stars), for $\theta_v = 45^{\circ}$ (left) and $\theta_v = 10^{\circ}$ (right).  Among the six ($\phi_0, m_ec^2\Theta_{\rm max}$) pairs displayed in the top row of the figure, we obtained the $Q-U$ results (bottom row) for a representative case where  ($\phi_0, m_ec^2\Theta_{\rm max}$) = (30$^{\circ}$, 13~keV). While the spectral results obtained at the two viewing angles are almost identical, the polarization characteristics of the two cases are visually distinct. Therefore, tighter constraints can be expected with future spectropolarimetric observations.  Considering the bottom left, with VB-on, there is a slight depolarization at $\sim10-30$~keV owing to these energies passing through the local cyclotron resonance (see also the case in Figure~\ref{fig:ResultsFig7}), which physically arises from the polarization behavior of the scattering cross section employed in {\sl MAGTHOMSCATT}. Such a cyclotronic depolarization arises at similar and higher energies for VB off simulations.  The view of $\theta_v =10^\circ$ with longitude of $\phi_0 = 30^\circ$ is not sufficiently pole-on for the VB-on case to depolarize, since the longitude is on the front side of the star for the observer. The projected pole of the polarizing-limiting surface is on the opposite side of the star -- this behavior can be discerned from the intermediate case in Figure~\ref{fig:VBdep}. The bottom right panel is similar to the cases highlighted Figures~\ref{fig:ResultsFig8}--\ref{fig:ResultsFig9}; it is noticeable that it exhibits high PD even with VB-off, because the flux tube is a coherent linear structure on the IP: there is little depolarization from contributions from different burst regions.  Yet, a much greater variation in Stokes U is observed for VB on than for VB off, along with different ranges for both Q and U; those both augur the possibility of useful diagnostics on the magnetospheric action of VB when sensitive spectropolarimetry is eventually available.

We also illustrate in Figure~\ref{fig:FRB-burst}  (left panel) simulated data for the FRB-associated hard X-ray burst of April 28th, 2020, which was unusually spectrally extended \citep{2020ApJ...898L..29M,2021NatAs...5..378L}. Here, we simulate the reported spectral constraints through NICER and {\it Fermi}-GBM instrument responses precisely as in \citet{2021NatAs...5..408Y}. Delivering acceptable spectral models for this soft gamma-ray burst is more challenging, and requires deviating from $\alpha \sim 1/3$. Specifically, we find that $\alpha \lesssim 1/4$ solutions work best, corresponding approximately to the adiabatic expansion of blackbodies. The relatively large spectral extent along with flat spectrum requires placing the burst footpoint close to magnetic pole, while imposing a narrow activation ($\Delta\theta_0 = \Delta\phi_0 \sim 0.1^\circ$) region to decrease the apparent emitting area so as not to overpredict the observed flux level at higher energies. In this case, the impact of VB on polarization is insignificant, as shown in the right panel, owing to the near-pole on geometry and tallness of the loop.  Moreover, not all of the high-altitude part of the flux can be visible to approximately match the low energy spectrum.  Only pole-on geometries or fine-tuned edge-on geometries with small $\phi_0$ are compatible with the data for minimalist assumptions. This is extremely interesting, as near pole activation is favorable for FRBs from the inner magnetosphere to evade absorption \citep[e.g.,][]{2024MNRAS.529.2180L}, but that pole-on viewing geometries may also be favored in extragalactic FRBs \citep{2025ApJ...982...45B}.  We remark that a polar origin and viewing for this special hard X-ray burst was invoked in the interpretation provided by \cite{2021NatAs...5..408Y}, and naturally leads to increased transparency to attenuation processes such as photon splitting \citep{2019MNRAS.486.3327H}. Further note that because the scattering transfer in {\sl MAGTHOMSCATT} generates a predominance of the $\parallel$ mode of polarization, splitting by the $\perp\to\parallel\parallel$ mode only delivers a minimal attenuation at the highest energies.

In the regime of small $\theta_0$, the assumptions of our model tend to break down, such as the optically thick (throughout) nature of the fireball (see  \S\ref{sec:tempspec}). In particular, for footpoints with $\theta_0 \ll 2^\circ$, the flux tubes approach the light cylinder and so strong distortions and sweepback from plasma loading are to be expected. However with $\theta_0 \sim 5^\circ$, $r_{\rm max}/r_\star \approx 100$ and $r_{\rm max} \Omega/c \sim 10^{-2}$ so the models shown in Figure~\ref{fig:FRB-burst}  are well within the light cylinder for SGR~1935+2154. Moreover, we note that the spectral extension of the April 28th burst corresponds approximately to a temperature range spanning a factor of $\sim 15-30$ (unabsorbed), which for $AT^4 \propto T^4/B \propto T^4 r^3$ conservation along an activated field loop (i.e., $\alpha \sim 1/4$), corresponds to a maximum altitude of $\sim (37-93) r_\star$ for the loop; this is roughly consistent with the estimate just above.

These parameter choices should not be overinterpreted, but rather taken to mean the current model already has flexibility to accommodate such an unusual burst. The value of $\alpha \approx 0.25 \lesssim 1/3$ implies departures from conditions corresponding to the conservation of photon number.  Meanwhile, moving away from the pole requires even smaller $\alpha$ to broaden the spectrum. Alternatively, the burst may also be a composite fireball, with multiple activated regions. A rare global crustal torsional oscillation possibly transpired in SGR~1935+2154's FRB-associated burst \citep{2022ApJ...931...56L,2023ApJ...953...67G}, and thus one might envision multiple burst active regions at amplitude antinodes in the crust. In conclusion, for the FRB-associated high energy burst from SGR~1935+2154, we favor a small activated footpoint close to the magnetic pole (possibly viewed pole-on) provided that our minimalistic assumptions hold: (1) a single rather than composite burst (2) purely thermal fireball rather than additional non-thermal emission components, e.g. unsaturated Comptonization. Ruling out these scenarios requires high quality temporal or spectropolarimetric data for similar future bursts.

\section{Discussion}
\label{sec:discussion}

\subsection{From Impulse Responses to Source Constraints?}

The models presented in this work can calculate not only snapshot broadband spectra and energy-dependent polarization observables, but also energy-time Stokes IRs that encode source geometry, compactness, and activation on the stellar surface.
It is therefore natural to ask how might the model IRs deliver source constraints on magnetars via bursts? Let us assume we are agnostic to the burst temporal injection/driving history. A burst may experience an unspecified transient perturbation $h(t)$ (suitably normalized) which is convolved with the IR, and we do not impose a preferred shape a priori on $h(t)$. The most incisive constraints can be made if $h(t)$ is photon energy and perhaps polarization independent\footnote{This could result from a high frequency crust perturbation during a burst, from a small burst (or hard X-ray emitting footpoints of a larger burst) whose dynamical time is shorter than the R{\o}mer geometric and Shapiro delay timescales of lensed secondary images.}. This approach in black hole systems has a strong track record and rich history for AGN reverberation and X-ray binaries \citep[e.g.,][]{1982ApJ...255..419B,1995MNRAS.272..585C,1999ApJ...514..164R,2019Natur.565..198K,2021iSci...24j2557C,2021Natur.595..657W}. Given the strongly polarized nature of magnetar bursts, much additional information is encoded and accessible in the IRs of Stokes Q and U beyond Stokes I, a facet that is generally not true for black hole systems.

In short, consider the IR convolved with $h(t)$,
\begin{equation}
\boldsymbol{L}^\prime(t, \omega) = \int_{0}^{\infty}d\tau\;\boldsymbol{L}(\tau;\vec{\theta}, \omega) h(t-\tau),
\label{eq:stokes_conv}
\end{equation}
where $\vec{\theta}$ is the parameter set and $\boldsymbol{L}^\prime$ is the observed dataset. The ``cross-spectrum" method \citep{2014A&ARv..22...72U} allows one to isolate IR physics (i.e., $\vec{\theta}$) even when $h(t)$ is unknown. In Fourier space at energy $\omega$, $\boldsymbol{\tilde{g}}(f)=\boldsymbol{\tilde{L}}(f;\vec{\theta})\,\tilde{h}(f)$. For two energies $\omega_i,\omega_j$, $C(\omega_i,\omega_j,f)=\tilde{g}_{\omega_i}(f)\tilde{g}^*_{\omega_j}(f)=\tilde{L}_{\omega_i}(f)\tilde{L}^*_{\omega_j}(f)|\tilde{h}(f)|^2$.
Because $|\tilde{h}(f)|^2$ is non-negative, it alters the amplitude but not the phase, so  $\phi_{ij}(f)=\arg C(\omega_i,\omega_j,f)=\arg \tilde{L}_{\omega_i}(f)-\arg \tilde{L}_{\omega_j}(f)$, with corresponding energy-dependent lag timescale $\tau_{ij}(f)= \phi_{ij}/(2\pi f)$.

For individual bursts, such machinery requires bright bursts with $\gtrsim 10^5$ counts/s \citep[e.g., those seen in SGR~1935+2154][]{2021ApJ...916L...7K} given the $\sim0.1$~ms structures in the IR for potentially detectable energy-dependent temporal signatures. With multiple bursts from the same magnetar, this requirement is less stringent because a subset of source-level fundamental parameters $\vec{\theta}$ is shared, the spin phase of the magnetar is known at the particular burst time, while intrinsic burst parameters are not.  Thus, these nuisance parameters may be stacked or marginalized over in a likelihood framework to yield fundamental source parameters. Computation of accurate IRs requires relatively high IP resolution. This is computationally demanding, at a few minutes per model evaluation for $\sim10^6$ trajectories on a modern laptop, and thus future work likely requires development of surrogate models \citep[e.g.,][]{2025ApJ...991..169O} for accelerated and accurately sampled posteriors in inference. 

\subsection{Caveat Emptor}

In this first work, we have made several approximations, which within our framework, could readily be explored in future work. Our fireball treatment is a static snapshot. Rotation of the magnetar is not captured, so current model applicability is limited to co-adding data over time segments much shorter than the magnetar spin period. A time-dependent prescription for fireball evaporation or cooling \citep[e.g.,][]{2001ApJ...561..980T}, possibly with internal structure, is not yet implemented. We also neglect aberration and Doppler boosting, so relativistically expanding fireballs \citep[e.g., for magnetar giant flares:][]{2021Natur.589..207R} are not yet represented. We have not included any azimuthal temperature gradients in the fireball, besides those induced by a nonzero tilt angle $\eta$. Such temperature gradients perpendicular to the local magnetic field could exist, given the long photon diffusion timescales, although they would only be important for large fireballs such as giant flare pulsating tails. We have not included curved spacetime effects in the calculation of VB and the polarization limiting radius, which might be important for low energies ($\lesssim 1$~keV) and pole-on viewing geometries. Magnetic lensing from VB, relevant for $B\gtrsim 10^{16}$~G, is also not included in the GR geodesic trajectory calculations. We have specialized to a dipole field, although multipolar fields can readily be incorporated in the future at the cost of additional free parameters. We have not included reflection \citep[e.g.,][for a cold plasma nonrelativistic treatment]{1978Natur.271..216L}, Compton back-scattering, or reprocessing components from the NS surface atmosphere  or scattering layers at higher altitudes. If they are beyond the VB polarization recoupling radius, they might have a large effect on polarization. Our treatment of the photosphere scattering cross section is in the magnetic Thomson limit, and is thereby limited to nonrelativistic temperatures and does not account for possible Comptonization reprocessing layers or sheaths from hot electrons \citep[e.g.,][]{2012A&A...545A.120S}, which has not yet been considered in a strongly magnetized context.

\subsection{Outlook}

We find there is substantial physics information concealed in magnetar SBs not captured by phenomenological spectroscopic models historically adopted in the literature. The models in this work, particularly for energy-time information, require thousands of counts in a burst or giant flare pulsating tail and potentially more for meaningful energy-dependent Stokes Q and U inferences. The number of detected magnetar SBs is in the tens of thousands. Sizable archival data are available for CGRO-BATSE, RXTE, INTEGRAL, Konus-Wind, {\it Swift}-BAT, {\it Fermi}-GBM, SVOM, GECAM, Insight-HXMT and other high-energy instruments. Additionally, a significant number of bursts exist in the NICER archive with microsecond time resolution \citep{2025arXiv251212291C}, some with contemporaneous coverage with NuSTAR \citep{2025ApJ...989...63H}. It is conceivable that many of the bright bursts in these populations may already have sufficient count rates to offer interesting constraints on source geometry and physics.  Planned or proposed efforts offer the possibility for high effective area \citep[e.g., STROBE-X, STARBURST, HEX-P, and HEROIX,][]{2019arXiv190303035R,2024HEAD...2140602K,2024FrASS..1157834M,2025HEAD...2210204C} or polarization capability to transients in the relevant hard X-ray to soft gamma-ray range \citep{2022hxga.book...33B}. These polarimeters include IXPE \citep{2022JATIS...8b6002W}, GOSox \citep{10.1117/12.2596186}, Astrosat-CZTI \citep{2015A&A...578A..73V,2022cosp...44.2253C}, THOR-SR \citep{2026APh...17503181S}, XL-Calibur \citep{2021APh...12602529A}, XPoSat/POLIX \citep{2022cosp...44.1854M,2022cosp...44.1853P}, eXTP \citep{2025SCPMA..6819502Z}, ComPol \citep{2023NIMPA104667662C}, EXPO \citep{Soffitta2026EXPO}, POLAR-2 \citep{2022icrc.confE.580D,2026MNRAS.547ag452G}, Daksha-CZTI \citep{2023JATIS...9d8002B}, AMEGO-X \citep{2022JATIS...8d4003C}, VLAST \citep{2024NuScT..35..149P}, GammaTPC \citep{2025arXiv250214841S}, GRAMS \citep{2020APh...114..107A}, LiquidO \citep{2025APh...17203135S}, MASS \citep{2024ExA....57....2Z}, APT \citep{2022icrc.confE.655B,2022icrc.confE.590A}, or possibly COSI \citep{2024icrc.confE.745T}. 

A dedicated mission specifically for magnetar burst science would necessitate a large effective area and $\mu s$ capability (with low deadtime or saturation effects at high count rates) with broadband coverage in the $\sim1-300$ keV band, ideally with polarimetric capability. Data may also be co-added from multiple instruments, for instance those used in future spacecraft navigation networks \citep[e.g.,][]{2023ApJS..266...16L}.

Joint fitting of I, Q/I, U/I and their energy dependence should be substantially less degenerate than Stokes I alone. The models set the stage for inference across multiple bursts. Compactness and geometry may become constraining through coupled observables. Our models predict complex correlations between photon energy, magnification/flux, and rotational phase. Understanding these correlations (and IR functions) will be key to eliminating degeneracies and constraining the fundamental parameters of magnetars (mass, radius, spin axis direction relative to the observer, magnetic obliquity) across many bursts. These parameters will not vary between different bursts from a given magnetar. This will be the subject of future work.

More broadly, inferences on active crustal zones in magnetars can inform on the trigger locales and stresses in magnetars, relate magnetothermal models of field evolution in the crust, and possibly the peculiar conditions necessary for observable FRBs.

\software {Eigen} \citep{eigenweb}, GNU Scientific Library \citep{gough2009gnu}

\begin{acknowledgements}

Z.W. thanks Dan Wilkins, Daniela Huppenkothen, Teruaki Enoto, Paz Beniamini and Sujay Mate for interesting conversations. M.G.B. thanks NASA for support through the grants 80NSSC22K0777, 80NSSC25K7257 and 80NSSC25K0079. The material is based upon work supported by NASA under award number 80GSFC24M0006 and 80NSSC22K1908, and under grants 22-ADAP22-142, 22-TCAN22-0027.  S.G. acknowledges the support of the CNES. This work has made use of the NASA Astrophysics Data System.

\end{acknowledgements}

\appendix
\section{The Geometric Invisibility Region Behind a Neutron Star}
\label{sec:invis}

Let us consider the maximum height above a NS that is invisible to a distant observer situated on the opposite side in Schwarzschild spacetime. The boundary of the invisible region is therefore set by the photon geodesic that just grazes the stellar surface. For a given compactness $\varsigma=2M/r_\star$ the critical ray on the boundary has closest approach $r_{\min}=r_\star$ and impact parameter
\begin{equation}
b_{\star}=\frac{r_\star}{\sqrt{1-\varsigma}} .
\end{equation}
If $\vartheta$ is the polar angle measured from the observer axis, then the boundary of the hidden region is obtained from the accumulated angular deflection of this grazing ray,
\begin{equation}
\vartheta_{\rm invis}(r)=\int_{r_\star}^{\infty}
\frac{b_{\star}\,dr^\prime}{{r^\prime}^2\sqrt{1-\dfrac{b_{\star}^2}{{r^\prime}^2}\left(1-\dfrac{2M}{r^\prime}\right)}}
+
\int_{r_\star}^{r}
\frac{b_{\star}\,dr^\prime}{{r^\prime}^2\sqrt{1-\dfrac{b_{\star}^2}{{r^\prime}^2}\left(1-\dfrac{2M}{r^\prime}\right)}} ,
\end{equation}
with regions satisfying $\vartheta_{\rm invis}(r) \leq \vartheta \leq \pi$ being occulted. Rescaling to $u = r^\prime/r_\star$, the most distant occulted point lies on the ``anti-observer" hemisphere at $\vartheta=\pi$, and its dimensionless radius $R_{\max}^{\rm invis}$ is found by solving the Volterra-type integral equation,
\begin{equation}
\int_{1}^{\infty}
\frac{du}{u^2\sqrt{1-\varsigma + (\varsigma-u)/u^3}}
+
\int_{1}^{R_{\max}^{\rm invis}}
\frac{du}{u^2\sqrt{1-\varsigma + (\varsigma-u)/u^3}}
=\pi .
\label{eq:invis}
\end{equation}
There is no invisible zone for $ \varsigma \geq 0.568$ \citep{2018PhRvD..98d4017S}, a well-known result.  For typical NS parameters ($M= 1.4 \, M_\odot$, $r_\star = 12$~km, i.e., $\varsigma = 0.34$), $R_{\max}^{\rm invis} \approx 1.35 $ (i.e., a maximum height of $0.35 r_\star$ beyond the star) implying only very small flux tubes with near-equatorial surface activation footpoints can entirely evade an observer. To an accuracy of $\lesssim 5\%$ for the numerical solution of Eq.~(\ref{eq:invis}), we find the fit $R_{\max}^{\rm invis} \approx (1+ 0.0008\varsigma^{-9.1})^{0.11}$.

\section{Outputs from Slab Simulations}
\label{sec:slab}
For the local transfer geometry we define the ray-aligned basis at the flux
tube surface,

\begin{equation}
\left\{\hat{\boldsymbol \varepsilon}_{xk},\hat{\boldsymbol \varepsilon}_{yk},\hat{\boldsymbol \varepsilon}_{zk}\right\}
=
\left\{
\left(\frac{\hat{\boldsymbol n}_{\cal S}\times\hat{\boldsymbol k}}{|\hat{\boldsymbol n}_{\cal S}\times\hat{\boldsymbol k}|}\right)\times\hat{\boldsymbol k},
\frac{\hat{\boldsymbol n}_{\cal S}\times\hat{\boldsymbol k}}{|\hat{\boldsymbol n}_{\cal S}\times\hat{\boldsymbol k}|},
\hat{\boldsymbol k}
\right\},
\label{eq:xyzS}
\end{equation}

with $\hat{\boldsymbol \varepsilon}_{xk}$ in the meridional plane spanned by $\hat{\boldsymbol n}_{\cal S}$ and $\hat{\boldsymbol k}$. These are all defined for the stationary local observer, e.g., tetrad frame Eq.~\ref{eq:k_tetrad_components}. The angle $\psi_B$ between $\hat{\boldsymbol{\varepsilon}}_{\parallel}$ and $\hat{\boldsymbol{\varepsilon}}_{xk}$ obeys (for $\hat{\boldsymbol{B}}\cdot\hat{\boldsymbol{n}}_{\cal S}=0$),

\begin{equation}
\tan\psi_B
=
-\frac{\hat{\boldsymbol{n}}_{\cal S}\cdot(\hat{\boldsymbol{B}}\times \hat{\boldsymbol{k}})}{(\hat{\boldsymbol{B}}\cdot\hat{\boldsymbol{k}})(\hat{\boldsymbol{k}}\cdot\hat{\boldsymbol{n}}_{\cal S})}
= -\frac{\tan\phi_{kB}}{\cos\theta_{kn}}.
\label{eq:psiB}
\end{equation}

The Stokes parameters for each photon at emergence can be obtained using the two complex components, ${\cal E}_{xk} = \boldsymbol{\cal E}_{\rm e} \cdot \hat{\varepsilon}_{xk}$ and ${\cal E}_{yk} = \boldsymbol{\cal E}_{\rm e} \cdot \hat{\varepsilon}_{yk}$, of its electric field vector $\boldsymbol{\cal E}_{\rm e}$, as follows:
\begin{equation}
Q \;  = \; {\cal E}_{xk}{\cal E}_{xk}^{*} -{\cal E}_{yk}{\cal E}_{yk}^{*} \quad , \quad
	U \; = \; {\cal E}_{xk}{\cal E}_{yk}^{*} +{\cal E}_{xk}^{*} {\cal E}_{yk}  \quad , \quad V \; =\;  i ( {\cal E}_{xk}{\cal E}_{yk}^{*} -{\cal E}_{xk}^{*} {\cal E}_{yk}) \ . \label{eq:IQU_S}
\end{equation}
The Stokes parameters in each bin $j \equiv (\theta_{kn}, \phi_{kB})$ can be written as
\begin{equation}
    (H_Q, H_U, H_V)_j = \frac{1}{N_j}\sum_{i =1}^{N_j} (Q, U, V)_i \quad, 
\end{equation}
where $N_j$ is the number of photons collected in bin $j$. With the Stokes parameter $H_Q$ obtained from the local slab simulations, we 
approximate the two components $|\bar{\cal E}_{\parallel}|^2$ and $|\bar{\cal E}_{\perp}|^2$ using $|\bar{\cal E}_\parallel|^2 \approx (1 + H_Q/\!\cos2\psi_B)/2$ and $|\bar{\cal E}_\perp|^2 \approx (1 - {H_Q}/\!\cos 2\psi_B)/2$. We note that, with this approximation, the values of  $|\bar{\cal E}_{\parallel, \perp}|^2$ encounter artificial discontinuity when {$\cos2\psi_B =0$}. To regularize this, we replace the $|\bar{\cal E}_{\parallel, \perp}|^2$ values at the ($\theta_{kn}, \phi_{kB}$) bins where {$\cos2\psi_B = 0$} by the interpolated values from the surrounding bins.

\ResultsFigure{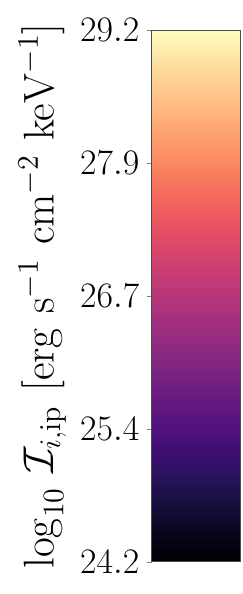}{2026-05-06_1923}{fig:ResultsFig1}{A fiducial partial-arc fireball with parameters as indicated and $\perp$-mode only splitting attenuation. This geometry highlights a rich time-energy structure possible in the IRs. Top row: IP intensity at 4, 12, 36, and 108 keV. Middle row: Stokes-$I$ IR (left) and polarization maps $Q/I$ and $U/I$ with VB off (top-right pair) and on (bottom-right pair). Bottom row: $Q$--$U$ trajectories and energy-dependent PD. This partial arc geometry exhibits coexisting direct and lensed delayed branches with strong chromatic polarization evolution.}

\ResultsFigure{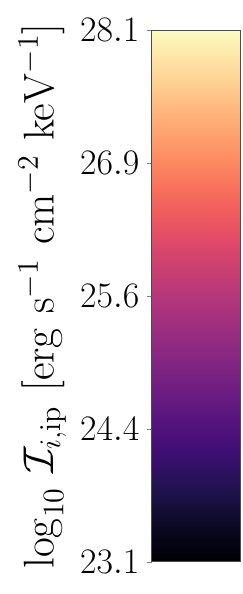}{2026-05-06_2156}{fig:ResultsFig2}{Case of a cool fireball with smaller footpoint locus, similar to some bursts observed by NICER. Relative to Figure~\ref{fig:ResultsFig1}, the smaller emitting area and lower temperature suppress overall luminosity, yielding a shorter and softer IR.}

\ResultsFigure{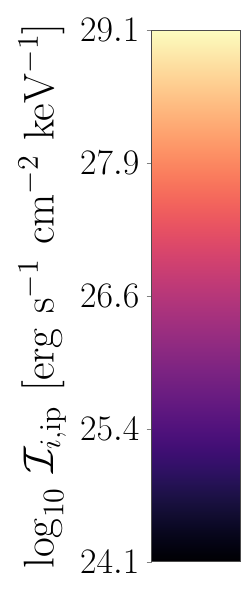}{2026-05-06_2348}{fig:ResultsFig3}{Case of an edge-on axisymmetric fireball.  Edge-on viewing strengthens occultation gating, producing more strongly energy-stratified delayed branches.}

\ResultsFigure{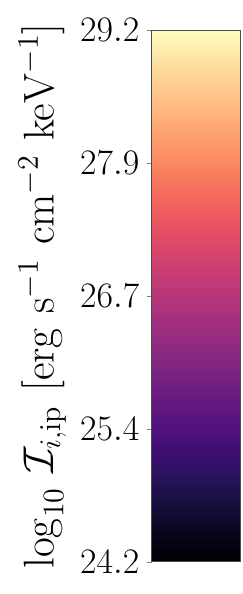}{2026-05-07_1116}{fig:ResultsFig4}{Fireball viewed nearly ($\theta_v = 0.1^\circ$) along the magnetic axis. Relative to edge-on annulus case Figure~\ref{fig:ResultsFig3}, the viewing geometry causes cancellation of net polarization in the VB-off case. Owing to the low altitude of the polarization limiting radius, the VB-on case is also nearly identical to VB-off expectations. The overall time structure in the IR is also much shorter in duration than Figure~\ref{fig:ResultsFig3}.}

\ResultsFigure{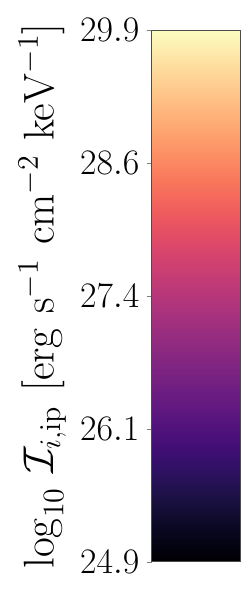}{2026-05-07_1259}{fig:ResultsFig5}{Inclined axisymmetric fireball at higher temperature ($m_e c^2 \Theta_{\rm max} = 30$~keV) with only one mode of splitting. The off-axis viewing angle significantly enhances observed net polarization, especially with VB on. Here the $E^2dN/dE$ approaches $10^{41}$ erg/s, similar to the pulsating tails of intermediate and giant flares.}

\ResultsFigure{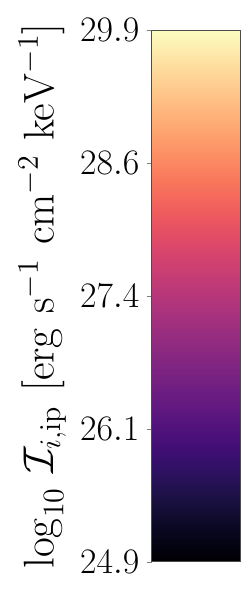}{2026-05-07_1419}{fig:ResultsFig6}{Identical to parameters of Figure~\ref{fig:ResultsFig5} except splitting attenuation is applied to both polarization modes rather than only $\perp\rightarrow\parallel\parallel$. High energy flux and (late-time) polarization are strongly modified, as only a narrow annulus at higher energy is capable of escape (e.g., 108~keV panel). The overall burst spectrum is also modified, with a steeper splitting-induced cutoff.}

\ResultsFigure{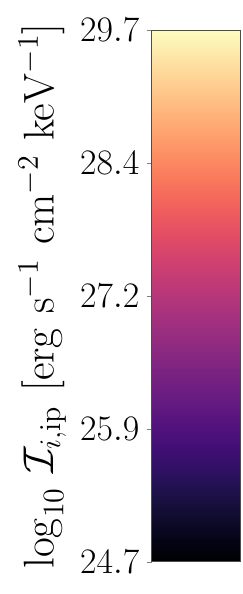}{2026-05-07_1541}{fig:ResultsFig7}{A quasi-polar fireball which attains higher altitudes. This viewing angle produces longer multi-branch IR and larger $Q$--$U$ loops than the case in Figure~\ref{fig:ResultsFig2}. The larger variation of emission heights also results in a more spectrally-extended burst with significant energy-dependent polarization variation.}

\ResultsFigure{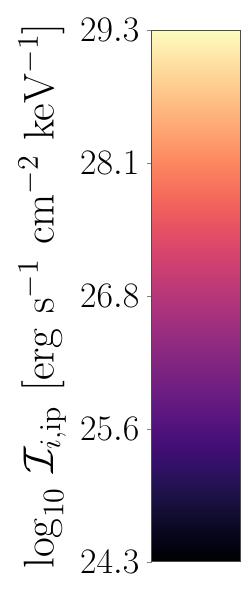}{2026-05-07_1845}{fig:ResultsFig8}{A pole-on view of flux tube at high compactness ($M=2.0 \,M_\odot$). The limited geometric extent of the flux tube results in a large net PD (in contrast to Figure~\ref{fig:ResultsFig4}) in both VB off and on cases. A lensed portion results in a disconnected branch in the IRs, the gap being Shapiro and R{\o}mer crossing timescale across roughly the NS radius.}

\ResultsFigure{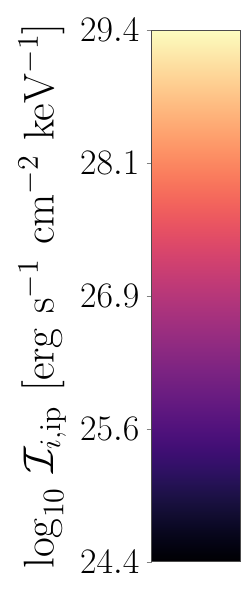}{2026-05-07_2033}{fig:ResultsFig9}{Identical parameters to Figure~\ref{fig:ResultsFig8} but with lower NS mass. The disconnected branch in the IRs is no longer present.}

\ResultsFigure{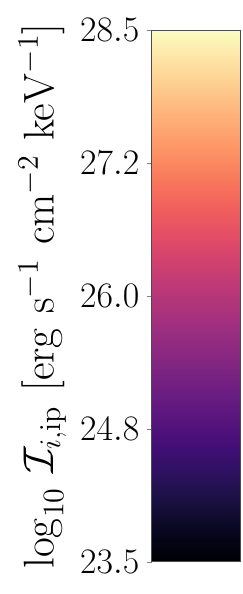}{2026-05-07_2225}{fig:ResultsFig10}{An edge-on flux tube exactly behind the star for the observer. There is little possibility to hide low energy portions of bursts due to enhanced visibility from lensing. In this case, the burst is also gravitationally focused with caustics in the IR at $\approx 75 \mu$s and $85 \mu$s from the inner and outer edges of the lensed flux tube. }

\bibliographystyle{aasjournalv7}
\bibliography{magnetarFRBrefs}

\end{document}